\documentclass[
12pt, 
oneside, 
english, 
onehalfspacing, 
 nolistspacing, 
parskip, 
headsepline, 
chapterinoneline, 
]{MastersDoctoralThesis} 

\usepackage[utf8]{inputenc} 
\usepackage[T1]{fontenc} 

\usepackage{mathpazo} 
\usepackage{amsmath}
\usepackage{amssymb}
\usepackage{amsfonts}
\usepackage{amsthm}
\usepackage{tikz}
\usepackage{caption}
\usepackage{subcaption}
\usepackage{physics}
\usepackage{algpseudocode}
\usepackage{algorithm}

\usepackage{graphicx}
\usepackage[export]{adjustbox}

\newcommand{\argmin}{\text{argmin}}
\newcommand{\argmax}{\text{argmax}}

\newtheorem{defn}{Definition}

\newtheorem{thrm}{Theorem}

\usepackage[backend=biber,style=nature,citestyle=nature,natbib=true]{biblatex} 
\usepackage[autostyle=true]{csquotes} 

\usepackage{fancyvrb}

\usepackage{tikz-cd} 

\thesistitle{Applications of Computational Topology to Fusion Science} 
\supervisor{A/Prof Vanessa \textsc{Robins}} 
\examiner{} 
\degree{Bachelor of Engineering (Research and Development)} 
\author{Nicholas \textsc{Bohlsen}} 
\addresses{} 

\subject{Applied Mathematics} 
\university{Australian National University} 
\group{Applied topology and geometry\\
Fusion plasma theory and modelling} 

\AtBeginDocument{
\hypersetup{pdftitle=\ttitle} 
\hypersetup{pdfauthor=\authorname} 
\hypersetup{pdfkeywords=\keywordnames} 
}

\AtBeginDocument{\RenewCommandCopy\qty\SI}

\begin{document}
\VerbatimFootnotes

\frontmatter 

\pagestyle{plain} 


\begin{titlepage}
\begin{center}

\vspace*{.06\textheight}
{\scshape\LARGE \univname\par}\vspace{1.5cm} 
\textsc{\Large Honours Thesis}\\[0.5cm] 

\HRule \\[0.4cm] 
{\huge \bfseries \ttitle\par}\vspace{0.4cm} 
\HRule \\[1.5cm] 
 
\begin{minipage}[t]{0.4\textwidth}
\begin{flushleft} \large
\emph{Author:}\\
{\authorname} 
\end{flushleft}
\end{minipage}
\begin{minipage}[t]{0.4\textwidth}
\begin{flushright} \large
\emph{Supervisor:} \\
{\supname} 
\end{flushright}
\end{minipage}\\[3cm]
 
\vfill

\large \textit{A document submitted in partial fulfillment of the requirements\\ for the degree \degreename}\\[0.3cm] 
\textit{Research Group:}\\[0.4cm]
\groupname
 
\vfill

{\large \today}\\[4cm] 

\vfill
\end{center}
\end{titlepage}


\vspace*{0.2\textheight}

\noindent\enquote{\itshape It is not uncommon that a new mathematical tool contributes to applications not by answering a pressing question-of-the-day but by revealing a different (and perhaps more significant) underlying principle.}\bigbreak

\hfill R. Ghrist


\begin{abstract}
\addchaptertocentry{\abstractname} 
Three novel applications of computational topology in the field of fusion science are developed. A procedure for the automatic classification of the orbits of magnetic field lines into topologically distinct classes using Vietoris-Rips persistent homology is presented and tested for a toy model of a perturbed tokamak. A method for estimating the distribution of the size of islands in the phase space of a Hamiltonian system or area-preserving map by sub-level set persistent homology is explored. This method is used to analyse the case of an accelerator mode island in the phase space of Chirikov's standard map and the possibility of detecting the self-similar island hierarchy responsible for anomalous transport in this model is investigated. Finally, it is suggested that TDA provides a toolset for the detection and characterisation of renormalisation group transformations which leave structures in Hamiltonian phase spaces invariant. The specific example of detecting the transform which leaves the neighborhood of a hyperbolic fixed point of the perturbed pendulum invariant is investigated using two different TDA approaches. Both of which are found to be partially sensitive to the symmetry in question but only weakly.
\end{abstract}


\begin{acknowledgements}
\addchaptertocentry{\acknowledgementname} 
This thesis is the largest and most complete piece of science research I have yet to complete and it would not have been possible without the support of numerous different people. 

I would like to thank both the Applied topology and geometry group and the Fusion plasma theory and modelling group generally. My colleagues in both of these groups have all been very extraordinarily helpful and friendly. They were willing to listen to my ideas and assist me when I asked, no matter how long my science and mathematics rants ran on for, and this patience is much appreciated. 

I would like to specifically thank my direct supervisor Associate Professor Vanessa Robins whose guidance was invaluable and her suggestions very helpful. Similarly, I must thank Professor Matthew Hole for advising me when asked and for his assistance in getting a paper on our previous work published while I was working on this thesis. I must also thank Vanessa, Matthew, and Bob Dewar, jointly for allowing me to slowly steal their libraries piece by piece over two years. 

Further I would like to thank the Mathematical Sciences Institute for tolerating me squatting in their building for two years. Also, I should thank the National Computational Infrastructure for providing the computing power which was needed to perform the calculations presented in this thesis. 
\end{acknowledgements}


\tableofcontents 
\listoffigures 

\mainmatter 

\pagestyle{thesis} 



\chapter{Introduction} 

\label{Chapter1} 


\newcommand{\keyword}[1]{\textbf{#1}}
\newcommand{\tabhead}[1]{\textbf{#1}}
\newcommand{\code}[1]{\texttt{#1}}
\newcommand{\file}[1]{\texttt{\bfseries#1}}
\newcommand{\option}[1]{\texttt{\itshape#1}}

\begin{figure}
    \centering
    \includegraphics[width = 0.5\textwidth]{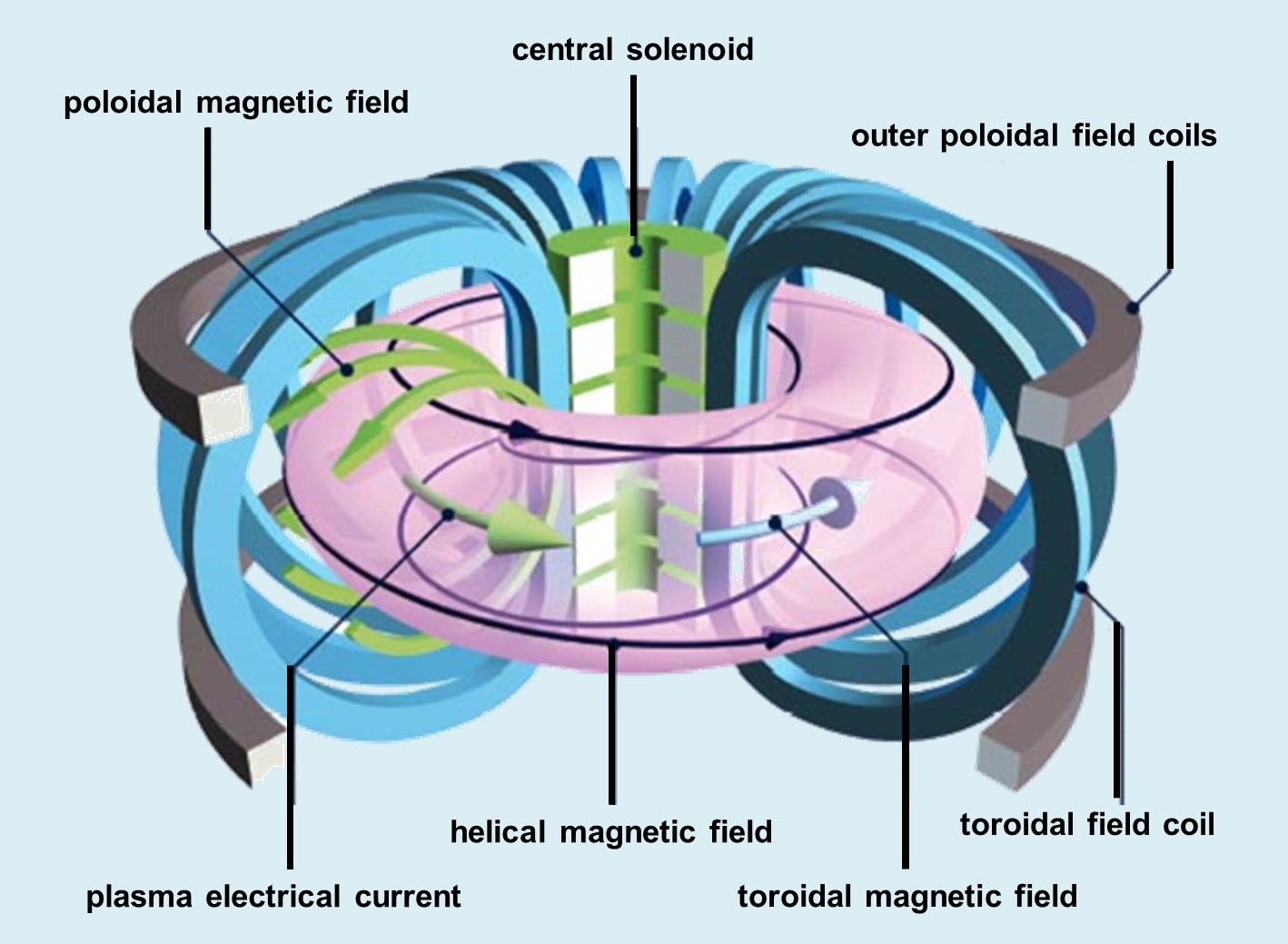}
    \caption{Diagram of a tokamak device. Courtesy of EUROfusion \cite{EUROfusion_2023}.}
    \label{fig:tokamak_field_lines}
\end{figure}

The development of a practical magnetic confinement device which can be used for the purposes of confining the burning plasma in a nuclear fusion device remains a task of great scientific and engineering interest. This problem has encouraged developments in many areas of physics and mathematics. For example: in Hamiltonian chaos and nonlinear dynamics \cite{mackay1983renormalization,greene1979method}; differential geometry \cite{PerellaExistence}; and computational topology and geometry \cite{deshmukh2021toward,robins2000computational,garland2016exploring}. This thesis is concerned with one such area, the physics of the orbits of charged particles and magnetic fields inside a tokamak machine, a diagram of which is presented as Figure \ref{fig:tokamak_field_lines}. This is chiefly a problem of classical Hamiltonian mechanics, and therefore the underlying dynamics is well understood. However, the magnetic fields inside fusion devices can be very complicated and this contributes to the creation of chaotic trajectories for the particles confined in the device. Such chaotic trajectories cannot be studied purely with the usual methods of classical mechanics, that is by solving differential equations. This is for two major reasons: the chaotic dynamics is, by definition, highly sensitive to the initial conditions and therefore it is only possible to make useful predictions on dynamics for short periods of time; and secondly because the phase space of general Hamiltonian systems is not usually totally chaotic and instead contains a hierarchy of ``islands'' whose existence substantially influences the dynamics of particles on chaotic trajectories. 

Beginning with Poincare, a program developed to study chaotic Hamiltonian mechanics with geometric and topological tools rather than the purely analytic methods which are preferred for simpler problems in classical mechanics \cite{mackay2020survey}. This thesis follows this program and attempts to extend it with modern methods from computational topology and its sub-field Topological Data Analysis (TDA). 

TDA is a mathematical field which consists of a set of tools for the extraction of shape, geometric and topological information, from data. We will demonstrate that geometric features of orbits in Hamiltonian systems and of hierarchies of phase space islands can be extracted from numerical simulations of Hamiltonian systems using TDA and used to characterise the motion of the chaotic system.  

\section{Physics of charged particles}

\subsection{Hamilton's equations}

Hamiltonian mechanics models the dynamics of classical objects whose state space we consider to be a $2n$ dimensional manifold $\mathcal{M}$ we call the phase space. This phase space is equipped with a choice of local coordinates $(\mathbf{p},\mathbf{q})$ called the generalised momenta $\textbf{p}$ and position $\textbf{q}$ respectively. All dynamical information is encoded in a scalar function $H:\mathcal{M}\cross \mathbb{R}\rightarrow\mathbb{R}$ which we call the Hamiltonian\footnote{The explicit separation of an$\mathbb{R}$ in the domain of $H$ is a physics convention indicating the possibility of explicit dependence upon time.}. Letting $(\textbf{p}(t),\textbf{q}(t))\in\mathcal{M}$ be the location of the particle in the phase space at time $t$, the particle's dynamics is defined by the Hamilton equations
\begin{align}
    \dv{\textbf{q}}{t} &= \pdv{H}{\textbf{p}}(\textbf{p},\textbf{q},t)\,, \nonumber\\ 
    \dv{\textbf{p}}{t} &= -\pdv{H}{\textbf{q}}(\textbf{p},\textbf{q},t)\,, \label{Hamilton}
\end{align}
where we adopt the notation $\pdv{H}{\textbf{q}}$ to refer to the vector of derivatives $(\pdv{H}{q_1},\ldots, \pdv{H}{q_n})$ \cite{arnol2013mathematical}. Any model of the type above is said to be of Hamiltonian form. This includes the classical mechanics of particles and rigid bodies but also includes the geometry of magnetic field lines which will be discussed below. In the context of classical mechanics $H$ often obtains the physical interpretation as the total energy function. It is by calculating said energy function that we determine the Hamiltonian which models a given system.

Hamiltonian systems are, by construction, dynamical systems, usually nonlinear ones, and so we are free to directly adapt concepts and notation from the mathematics of dynamical systems. We recall some relevant definitions now for later reference. 

Firstly, all dynamical systems define a map $\phi_t:\mathcal{M}\rightarrow\mathcal{M}$ called the flow on the phase space \cite{guckenheimer2013nonlinear}. We will state the definition for autonomous systems, that is systems for which the Hamiltonian does not depend explicitly upon time, but it can be generalised to non-autonomous systems.
\begin{defn}
    Given a time $t\in\mathbb{R}$ define $\phi_t:\mathcal{M}\rightarrow\mathcal{M}$ by $\phi_t(\textbf{p}_0,\textbf{q}_0) = (\textbf{p}(t),\textbf{q}(t))$ where $(\textbf{p}(t),\textbf{q}(t))$ is the solution to \eqref{Hamilton} with initial condition
    \begin{equation}
        (\textbf{p}(0),\textbf{q}(0)) = (\textbf{p}_0,\textbf{q}_0)\,.
    \end{equation}
\end{defn}

Note that in the definition above we have assumed that solutions to the autonomous Hamilton equations exist for all time, both forwards and backwards. This is acceptable because the Hamilton equations are automatically $H$ preserving. So the flow of a single phase space point is restricted to a level set of $H$ specified by the initial condition. These level sets can be assumed to be compact without loss of modelling flexibility. So the Hamilton equations define the flow of a vector field and so exist for all time, since continuous vector fields on compact manifolds are always complete \cite{guckenheimer2013nonlinear}.

We will often discuss the orbits of phase space points under the flow and invariant sets, usually invariant manifolds, of the flow. These are defined as follows. 

\begin{defn}
    The curve defined by $\{\phi_t(\textbf{p}_0,\textbf{q}_0)|t\in \mathbb{R}\}$ is the     \textit{trajectory}, or \textit{orbit} of the point $(\textbf{p}_0,\textbf{q}_0)$ under  Hamiltonian evolution. We can equivalently consider this as a parameterised curve defined as the map $\phi_\cdot(\textbf{p}_0,\textbf{q}_0):\mathbb{R}\rightarrow\mathcal{M}$.
\end{defn}

\begin{defn}
    A set $U\subset \mathcal{M}$ is invariant iff $\phi_t(U)\subset U$ for all $t\in \mathbb{R}^+$.
\end{defn}

For discrete dynamical systems, which are defined by iterating a map rather than integrating a differential equation, there are analogous definitions of the orbits of point and invariant sets \cite{natiello2007user}. 

\subsection{Maxwell's equations and the charged particle Hamiltonian}

In plasma physics we are chiefly concerned with modelling how charged particles interact with ElectroMagnetic (EM) fields. Adopting Euclidean coordinates $\textbf{x}$ for $\mathbb{R}^3$ we recall that the electric $\textbf{E}$ and magnetic $\textbf{B}$ fields are modelled by vector fields mapping $\mathbb{R}^3\cross\mathbb{R}\rightarrow\mathbb{R}^3$. These fields are defined as solutions to the Maxwell equations
\begin{align}
    &\nabla\cdot\textbf{E}(\textbf{x},t) = \frac{1}{\epsilon_0}\rho(\textbf{x},t),\            \,\nabla\cdot\textbf{B}(\textbf{x},t) = 0\,\nonumber\\
    &\nabla\cross\textbf{E}(\textbf{x},t) = -\dv{\textbf{B}}{t}\,,\            \,\nabla\cross\textbf{B}(\textbf{x},t) = \mu_0\textbf{J}(\textbf{x},t)+\mu_0\epsilon_0\dv{\textbf{B}}{t}\,,
\end{align}
where $\mu_0$ and $\epsilon_0$ refer to physical constants, and $\rho$ and $\textbf{J}$ define the electric charge and current densities respectively \cite{Jackson}. Since $\textbf{B}$ is divergence free it is possible to construct a scalar potential $\varphi$ and a vector potential, $\textbf{A}$,  such that $\textbf{B}=\nabla\cross\textbf{A}$ and $\textbf{E}=-\pdv{\textbf{A}}{t}-\nabla\varphi $. Adopting the Euclidean coordinates $\textbf{x}$ as our generalised position coordinates the Hamiltonian for a charged particle in an EM field can be modelled by
\begin{equation}\label{ChargedParticleHamiltonian}
    H(\textbf{x},\textbf{p},t) = \frac{|\textbf{p}(t)-q\textbf{A}(\textbf{x},t)|^2}{2m}+q\varphi(\textbf{x},t)\,,
\end{equation}
where $q$ and $m$ refer to the charge and mass of the particle respectively and $\textbf{p}$ is the generalised momentum \cite{hazeltine2003plasma}. This Hamiltonian represents the total energy of our charged particle, as was noted above. Solving the Hamilton equations generated by \eqref{ChargedParticleHamiltonian} provides a description of the dynamics of our particle. In the general case the dynamics will be chaotic and we are primarily concerned with the phase space geometry associated with this chaotic evolution. 


\section{Hamiltonian phase space}

 Hamiltonian systems in one position dimension (two phase space dimensions) are always integrable\footnote{By ``integrable'' we mean that the dynamics is exactly solvable. Formally a system with $2n$ phase space dimensions is Liouville integrable if there exists $n$ ``independent'' conserved quantities of the motion \cite{mackay2020survey}.} \cite{arnol2013mathematical}. However in higher dimensions the generic behaviour is chaos. Consider for example the three body problem, the system of three masses interacting under gravitation. It was shown by Poincare that this problem is not integrable and instead displays what is now known as Hamiltonian chaos. However, it was also shown that there exists an infinite number of periodic solutions to the three body problem. These periodic solutions form islands of order embedded within a chaotic sea and their geometry affects the dynamics of particles both inside the islands themselves and on chaotic trajectories \cite{mackay2020survey}. This is fundamentally the motivation for using geometric tools to study Hamiltonian dynamical systems. We introduce two such tools, the Poincare map and stroboscopic map, below and will then present the simple example of a perturbed nonlinear pendulum to demonstrate the island structure of a Hamiltonian phase space.

\subsection{Poincare return maps}

We are interested in analysing dynamical systems geometrically and furthermore are interested in periodic and quasiperiodic orbits. This suggests that we should study how the dynamics of our system recurs in time and space. This is the purpose of the Poincare section and Poincare return map which provide a geometric tool for analysing how our dynamical system returns in space. The fundamental idea of a Poincare map is that if we choose a surface $\Sigma$ within our phase space such that points on $\Sigma$ flow away from it under phase flow but also eventually return to $\Sigma$ then this induces a map from the surface to itself generated by the phase flow. This map encodes information about how different regions of phase space are transported by the phase flow. 

\begin{figure}
    \centering
    \includegraphics[width=0.6\textwidth]{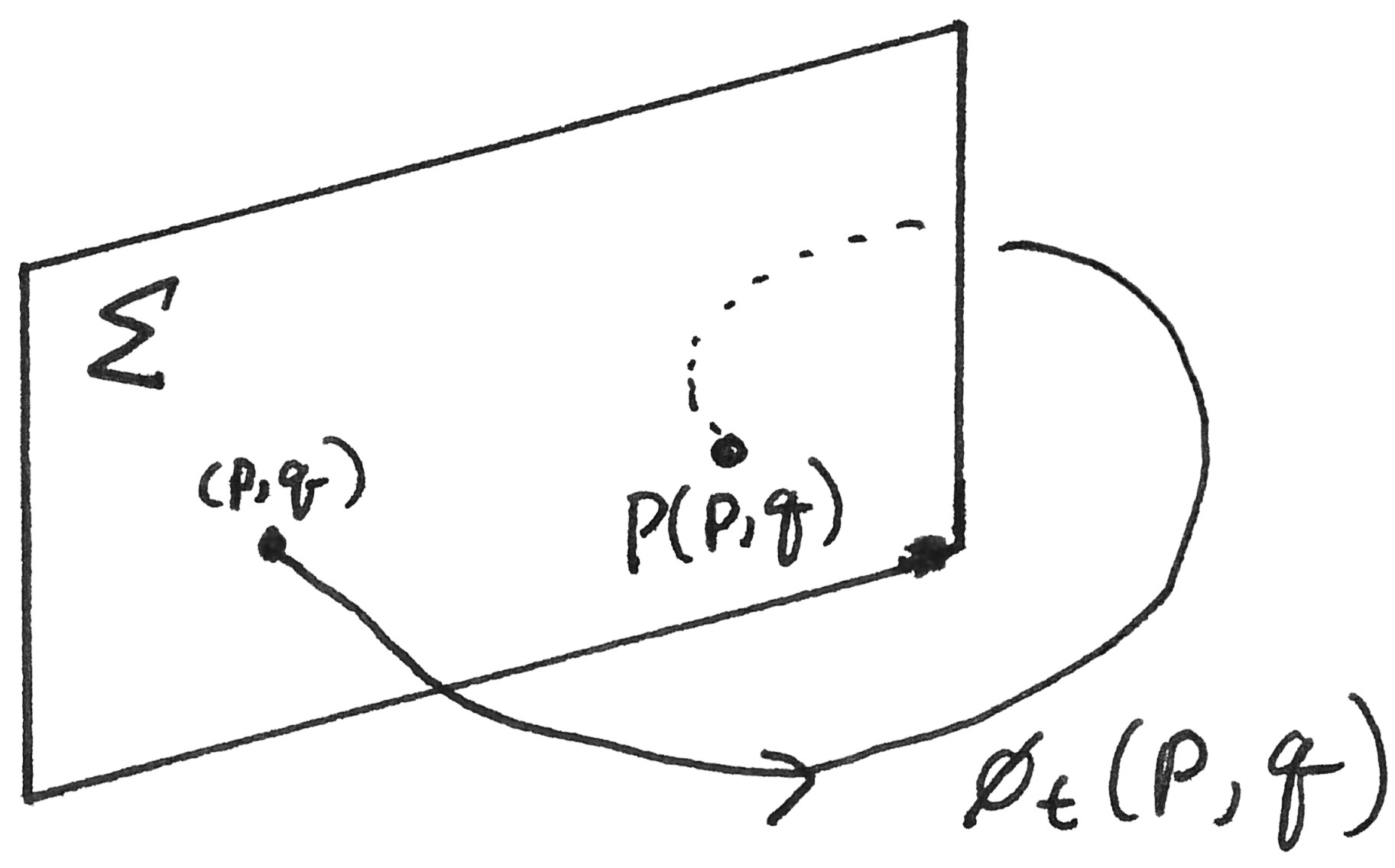}
    \caption{Cartoon depicting the construction of a Poincare return map of a dynamical system.}
    \label{fig:PoincareMap}
\end{figure}

The Poincare return map $P$ generated by our phase space flow $\phi_t$ is defined below and presented diagrammatically as Figure \ref{fig:PoincareMap}.
\begin{defn}
    Let $\Sigma \subset M$ be a codimension-$1$ surface transverse to the flow $\phi$. Then on an open subset $U\subset \Sigma$ we define the Poincare map $P:U\rightarrow\Sigma$ to map $P(\textbf{p},\textbf{q}) = \phi_\tau(\textbf{p},\textbf{q})$ where $\tau>0$ is the time taken for the orbit $\phi_t(\textbf{p},\textbf{q})$ based at $(\textbf{p},\textbf{q})\in\Sigma$ to first intersect $\Sigma$.
\end{defn}

We refer to the surface $\Sigma$ as the Poincare section and its selection is a choice we must make. Note that it is not immediately clear that $P$ as defined above is a well defined map. It is possible that the orbit $\phi_t(\textbf{p},\textbf{q})$ of a point $(\textbf{p},\textbf{q})\in \Sigma$ never returns to $\Sigma$ in finite time. However, Poincare's recurrence theorem states that as long as our Hamiltonian evolution has bounded orbits, which is equivalent to requiring compact level sets of $H$, then for almost any open set any trajectory which intersects it will intersect it infinitely often. The Poincare sections we construct will never belong to the very small set of open sets for which this theorem fails and so it ensures that our phase flow will return to $\Sigma$ and therefore $P$ is a well defined map from $\Sigma$ to itself \cite{arnol2013mathematical}. 

\subsection{Stroboscopic maps}

\begin{figure}
        \centering
        \includegraphics[width=0.8\textwidth]{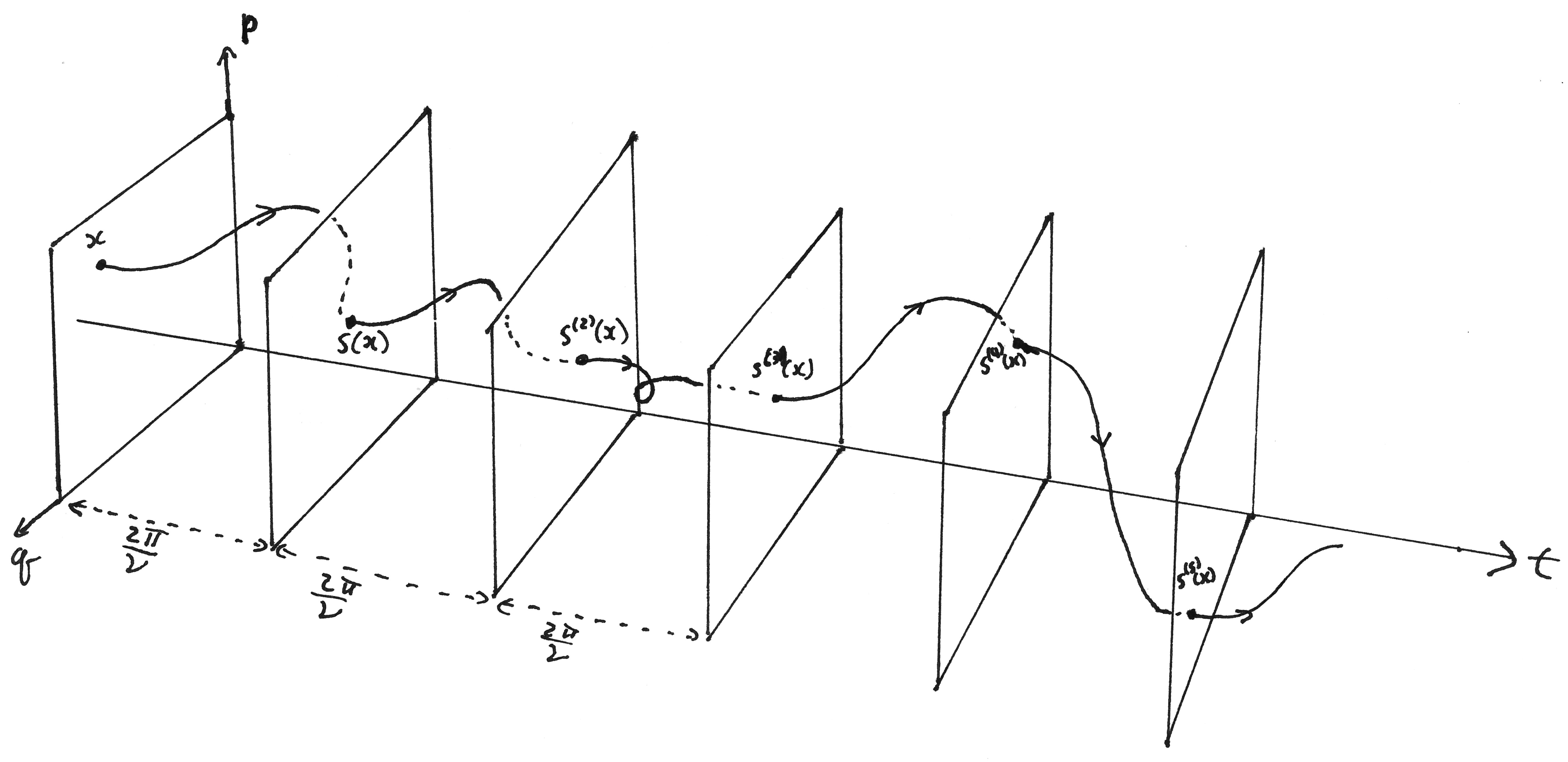}
        \caption{Cartoon depicting the construction of a stroboscopic map of a dynamical system. Note that the planes indicate the points in time at which the system is sampled.}
        \label{fig:StroboscopicMap}
\end{figure}

The Poincare map is a natural tool for analysing autonomous Hamiltonians. However, we will also be concerned with non-autonomous Hamiltonians when we study systems subjected to time dependent perturbations. An alternative construction which is more appropriate for non-autonomous Hamiltonians is the stroboscopic map. The stroboscopic map constructs a discrete dynamical system, a map of phase space to itself, out of our Hamiltonian evolution by imagining looking at our system as under a strobing light where we can only see the state of the system for a periodic sequence of brief instants when the light is flashed. The stroboscopic map $S$ generated by our phase flow and with frequency $\nu\in\mathbb{R}$ is defined below and a cartoon diagram of the construction is presented as Figure \ref{fig:StroboscopicMap}.

\begin{defn}
    If $\nu>0$ and $\phi_t(x)$ denotes the orbit of $x=(\textbf{p},\textbf{q})\in M$ for time $t$ then $S:M\rightarrow M$ is defined by $S(\phi_t(x))=\phi_{t+\frac{2\pi}{\nu}}(x)$.
\end{defn}

\subsection{Structures in the phase space}\label{StructuresInPhaseSpaces}

To understand the phase space structures with which this thesis is concerned it is best to consider a specific example. Here we will look briefly at the case of a nonlinear pendulum under a periodic perturbation. Consider the Hamiltonian 
\begin{equation}\label{NonlinearPendulum}
    H(p,q,t) = \frac{1}{2}p^2-\omega^2\cos(q)+\varepsilon \cos(kq - \nu t)\,,
\end{equation}
which models a unit mass pendulum in a uniform gravitational field with a natural frequency $\omega$. Note $k,\nu,$ and $\varepsilon$ are the wave number, frequency, and perturbation strength of the perturbing wave respectively. The phase space of this system is an infinite cylinder $(p,q)\in \mathcal{M} = S^1\cross\mathbb{R}$. Note that when $\varepsilon=0$ \eqref{NonlinearPendulum} has only one position degree of freedom, or two phase space degrees of freedom and so would be referred to as a $1D$ system. The addition of time dependence when $\varepsilon>0$ provides more freedom but not a full positional degree of freedom and so it is typical to refer to non-autonomous systems of this type as $1\frac{1}{2}D$ Hamiltonian systems.

\begin{figure}
    \centering
    \begin{subfigure}[b]{0.4\textwidth}
        \centering
        \includegraphics[width=\textwidth]{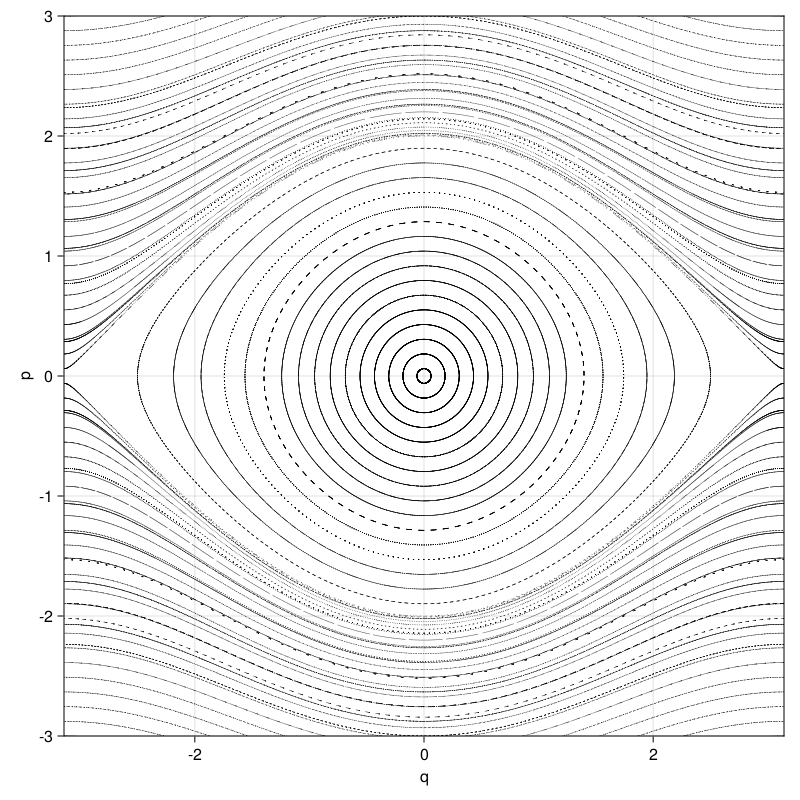}
        \subcaption{$\varepsilon=0$}
        \label{fig:integrable_pendulum}
    \end{subfigure}
    \begin{subfigure}[b]{0.4\textwidth}
        \centering
        \includegraphics[width=\textwidth]{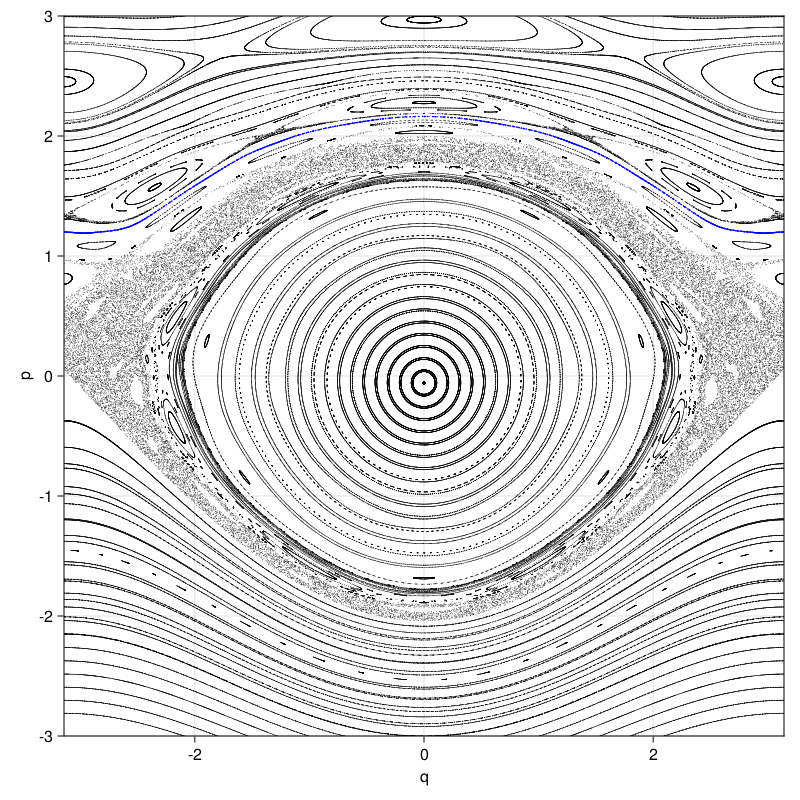}
        \subcaption{$\varepsilon=0.3$}
        \label{fig:weakly_perturbed_pendulum}
    \end{subfigure}
    \caption{Trajectories of the perturbed nonlinear pendulum stroboscopic map for $k=1,\omega=1,\nu=5.4$.}
    \label{fig:nonlinear_pendulum_phase_spaces}
\end{figure}

Figure \ref{fig:nonlinear_pendulum_phase_spaces} presents the orbits of many initial conditions under the Hamiltonian evolution of \eqref{NonlinearPendulum} as imaged by a stroboscopic map with the strobing frequency chosen to be identical to the perturbation frequency.

When $\varepsilon=0$, as in subfigure \ref{fig:integrable_pendulum} the phase space has the structure of an integrable system. All trajectories live on one of three classes of orbit of the stroboscopic map: fixed points, periodic orbits, and invariant circles. There exists an elliptic fixed point at $(p,q) = (0,0)$ and the orbits of initial conditions in a neighborhood of this point live on nested circles which surround the fixed point. There is a hyperbolic fixed point at $(p,q)= (0,-\pi)=(0,\pi)$ and an important trajectory which links this fixed point to itself forming what is called a homoclinic orbit \cite{guckenheimer2013nonlinear}. This orbit separates the other orbits into two classes, those which enclose the origin, called ``trapped'', and those which wrap around the cylinder, called ``passing'', and for this reason the homoclinic orbit is said to form a separatrix. Certain initial conditions lie on periodic orbits of the stroboscopic map, forming sets of disjoint points with a global circular structure. In subfigure \ref{fig:integrable_pendulum} one such periodic orbit is shown which surrounds the elliptic fixed point. These periodic orbits occur where the frequency of oscillation of the underlying Hamiltonian evolution is a rational multiple of the strobing frequency $\frac{2\pi}{\nu}$. Finally, the orbits of the majority of points densely fill circles, called invariant circles.

When $\varepsilon>0$ the structure of the phase space changes as shown in subfigure \ref{fig:weakly_perturbed_pendulum}. For small perturbations the elliptic fixed point is preserved as are the invariant circles which surround it. Rather than nested invariant circles extending all of the way out to the separatrix, we see chains of islands surrounding the primary elliptic island, both growing in number and decreasing in size further away from the elliptic fixed point. The separatrix breaks up into a positive measure fractal region in which the trajectories move around chaotically, called a stochastic layer. We observe further island chains in the region that previously contained passing trajectories. Each island chain forms around an elliptic periodic orbit, which lies at the centre of the islands, and is accompanied by a hyperbolic periodic orbit, in the ``X'' points between the islands, from which stochastic layers grow. Note that the largest island chains are created along trajectories of the unperturbed system with low order rational frequency \cite{Zaslavsky}. Highlighted in blue is an invariant circle which still exists despite the perturbation. The ``survival'' of invariant circles under small perturbations is guaranteed by the Kolmogorov-Arnold-Moser (KAM) theorem and as such these trajectories are called KAM toruses \cite{arnold2009small}. It is a famous conjecture of Greene's that the invariant toruses which are most robust to perturbation are those with highly irrational frequency\footnote{Highly irrational in this context means that the frequencies cannot be well approximated by rational numbers with small denominator and so only have large denominator Diophantine approximations.} \cite{greene1979method}. Generically, KAM toruses separate island layers and it is impossible for a trajectory to pass through a KAM torus and hence they separate the stochastic domain of Hamiltonian systems into disconnected regions \cite{mackay1984transport}. 

Observe that for perturbed $1\frac{1}{2}D$ Hamiltonian systems we have several topologically distinct classes of trajectory for our Hamiltonian system separated by the behaviour of their orbit under the Poincare map. We have invariant circles, KAM toruses, which form topological circles in phase space possibly enclosing an elliptic fixed point but this is not formally required. We have orbits which form island chains. These are disconnected sets of several circles in the plane. Finally we have stochastic regions which are connected fat fractals of topological dimension two. Many islands are inside these stochastic regions and can be considered as ``holes'' in the fractal. Since these classes are distinguished by the topology of their orbit we expect that it is possible to construct an algorithm which identifies the class of an orbit using only the information contained in the topology and geometry of the orbit on the Poincare section. We will demonstrate in Chapter \ref{Chapter2} of this thesis that this can be achieved using TDA. 

\subsection{Field line chaos}

In plasma physics we are concerned with both the dynamics of charged particles but also with the geometry of the fields. In the context of magnetic confinement devices we restrict our scope to the analysis of of static\footnote{Static in this context means constant in time.} magnetic fields. Under a static assumption the Maxwell equations reduce to the magnetostatic equations
\begin{align}
    \nabla\cdot \textbf{B} &= 0\,,\\
    \nabla\cross\textbf{B} &= \mu_0\textbf{J}\,.
\end{align}
This thesis is concerned with the magnetic field lines of $\textbf{B}$ which are defined as the solution curves $\textbf{x}(s)$ to the differential equation
\begin{equation}
    \dv{\textbf{x}}{s} = \frac{\textbf{B}(\textbf{x})}{||\textbf{B}(\textbf{x})||}\,,
\end{equation}
where $s$ is the arc length of the curve in the Euclidean metric.

\begin{figure}
    \centering
    \includegraphics[width = 0.6\textwidth]{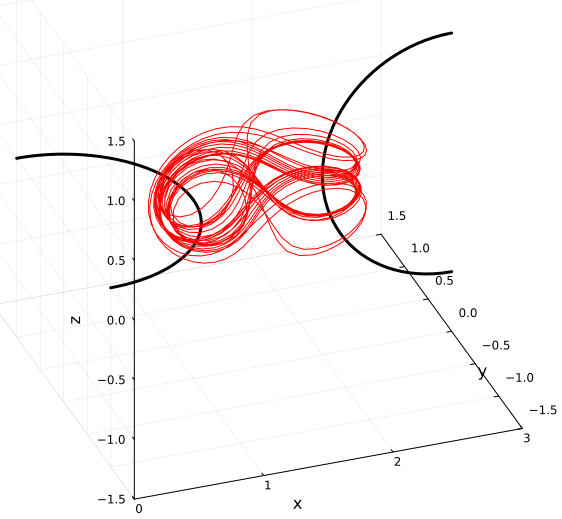}
    \caption{Chaotic field line near two circular current loops. The magnetic field line is shown in red and the currents in black. Note that the currents are complete circles and extend beyond the region of space shown.}
    \label{fig:chaotic_field_of_two_loops}
\end{figure}

A magnetostatic field will be produced by any distribution of currents. A classic problem of theoretical physics is the computation of the fields produced by a given current. Usually, the standard configurations of currents which are introduced in textbooks on electrodynamics are chosen because they are symmetric. This symmetry causes the field lines to close on themselves, that is, the field lines form complete loops in space. For example, this is the case for an infinite line current, an ideal dipole field, and a circular current loop \cite{Griffiths,Jackson}. If the current distribution does not exhibit a ``nice'' geometric symmetry, as is the generic case, then the field lines will not necessarily close on themselves and can even exhibit chaotic behaviour \cite{Morrison2000,ChaoticFields}. The study of field line chaos is of particular interest to us here because it is another example of Hamiltonian chaos. The field line flow as defined above can be interpreted as a $1\frac{1}{2}D$ Hamiltonian system. We will not prove this here but derivations of the ``field line Hamiltonian'' and its relationship to the vector potential can be found in standard texts on plasma confinement \cite{hazeltine2003plasma}.

As a demonstrative example of a chaotic field consider Figure \ref{fig:chaotic_field_of_two_loops} which presents a field line created by two circular current loops. The field line does not close on itself. Instead it circles around one of the currents for a short while before switching to the other loop. The switching occurs aperiodically.

\begin{figure}
    \centering
    \includegraphics[width = 0.6\textwidth]{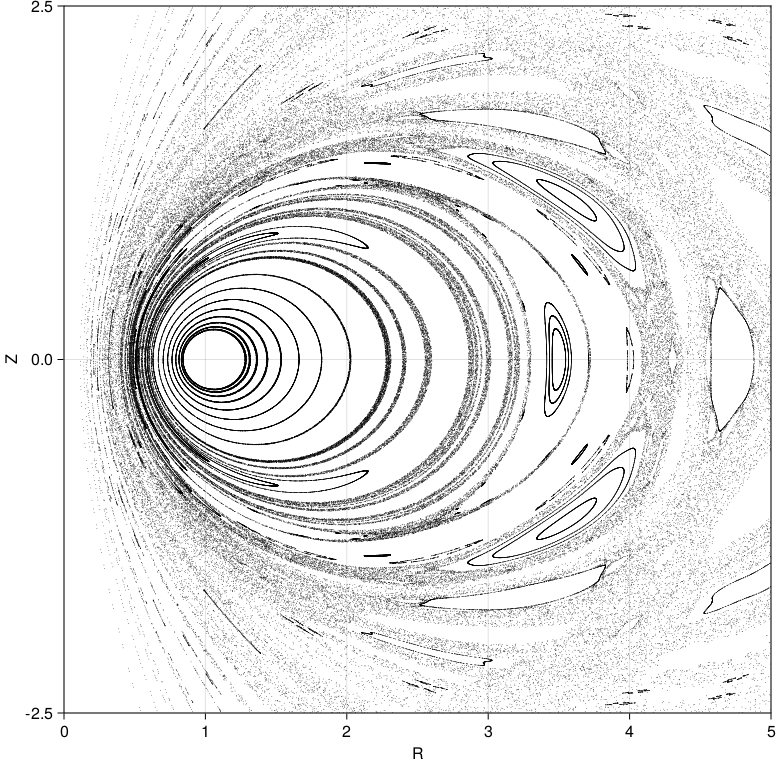}
    \caption{Poincare map orbits of several magnetic field lines in a toy model of a tokamak.}
    \label{fig:MagneticFieldPoincarePlot}
\end{figure}

To see empirically that field line chaos is an example of Hamiltonian chaos consider Figure \ref{fig:MagneticFieldPoincarePlot}. This presents the Poincare map orbits of many magnetic field lines for a toy model of a perturbed tokamak magnetic field. The specific toy model and Poincare section used here will be described in Chapter \ref{Chapter3}. We observe similar invariant circles, island chains, and stochastic layers to those we saw in the perturbed pendulum. This similarity is a result of the Hamiltonian structure of the field line flow and so we expect that any methods of analysis developed to study Hamiltonian evolution will automatically be applicable to the study of field line flow as well \cite{Morrison2000}.

\section{Topological methods in Hamiltonian chaos}

\subsection{Prior work}

Research in computational topology and geometry has regularly been inspired by its applications to the study of dynamical systems and chaos. For example, in one of the earliest work in the field Muldoon \textit{et al} investigated the extraction of the topological invariants of orbits from experimental time series \cite{muldoon1993topology}. This helped to introduce simplicial homology as a tool for the study of dynamics. In recent years there has been substantial development in this program. Tempelton and Khasawneh demonstrated that the shape of projections of chaotic trajectories to subspaces of the phase space appear geometrically distinct from those of ordered trajectories. They used this observation to construct a TDA based tool for the detection of chaotic trajectories by analysing the sub-level set persistent homology of kernel density estimates of their orbits \cite{tempelman2020look}. They demonstrate the application of their detection scheme for several dynamical systems but none of Hamiltonian type. Tymochko \textit{et al} demonstrated that Hopf bifurcations in one parameter families of dynamical systems can be detected with zig-zag persistent homology \cite{tymochko2020using}. Again, their analysis is concerned with dissipative systems not Hamiltonian ones. 

TDA, or more specifically persistent homology, has been applied in contexts more closely related to the charged particle and magnetic field trajectories which concern us here. Kramar \textit{et al} have demonstrated that the structures of convecting fluids can be computed, and the qualitative similarities of otherwise quantitatively distinct fluid flows can be observed in the topological data \cite{KRAMAR201682}. Similarly, it has been suggested that the lagrangian orbits of fluid packets can be classified by the topological structure of their intersections with a Poincare section and preliminary evidence has been created by Nunez \textit{et al} that this classification is computable with TDA \cite{nunez2022topological}. Lagrangian orbits of fluid flows are very similar to magnetic field lines, in the sense that they are also of Hamiltonian form when the fluid is steady state and incompressible. As such our work on field line classification, which will be presented in Chapter 3, is heavily inspired by the work of Nunez \textit{et al}. In a specifically plasma physics context, Banesh \textit{et al} demonstrated that TDA can be used to detect reconnection events in simulations of astrophysical magnetic fields \cite{banesh2020topological}. Their method does allow for the detection of topological change in the evolution of the field but it is restricted to the case of a planar magnetic field and hence is $2D$. It is also formally dissipative and therefore also not of Hamiltonian form. 

\subsection{Structure and contribution of this thesis}

This thesis presents a preliminary investigation into the characterisation of geometric and topological structures in chaotic Hamiltonian systems using TDA. Chapter \ref{Chapter2} presents some mathematical background explaining the TDA methods which we will apply. Some other relevant mathematical tools are also discussed. Chapters \ref{Chapter3}, \ref{Chapter4}, and \ref{Chapter5} each present a different novel application of TDA to chaotic Hamiltonian systems. Specifically, Chapter \ref{Chapter3} presents an automated procedure for the classification of the orbits of magnetic field lines in a model tokamak magnetic field based upon the Vietoris-Rips filtration. Chapter \ref{Chapter4} demonstrates that using the Signed Euclidean Distance Transform (SEDT) and sub-level set persistent homology the distribution of the number of islands of each size in a phase space can be computed and used to detect the existence of the island hierarchies which generate anomalous transport in Hamiltonian systems. Finally Chapter \ref{Chapter5} explores the possibility of using TDA to detect the existence of renormalisation group transforms which preserve the topology of the phase space of a Hamiltonian system and tests this possibility for the example of the periodically perturbed pendulum discussed above. Chapter 6 discusses avenues for possible future research and concludes the thesis. 

\chapter{Mathematical preliminaries} 

\label{Chapter2} 

Here we will discuss the major mathematical constructions and results upon which the remainder of this document relies. Firstly, we will discuss Topological Data Analysis (TDA) which is the primary mathematical tool we will apply to study the geometry of Hamiltonian phase spaces. Then, we will discuss the connection between Hamiltonian dynamics and symplectic maps and present some common examples of symplectic maps which we will use later. Finally we will introduce a method of chaos detection based on Weighted Birkhoff Averaging which we will use to construct high resolution images of phase space islands in Chapter \ref{Chapter4}.


\section{The topology of data}

\begin{figure}
    \centering
    \begin{subfigure}[b]{0.35\textwidth}
        \includegraphics[width = \textwidth]{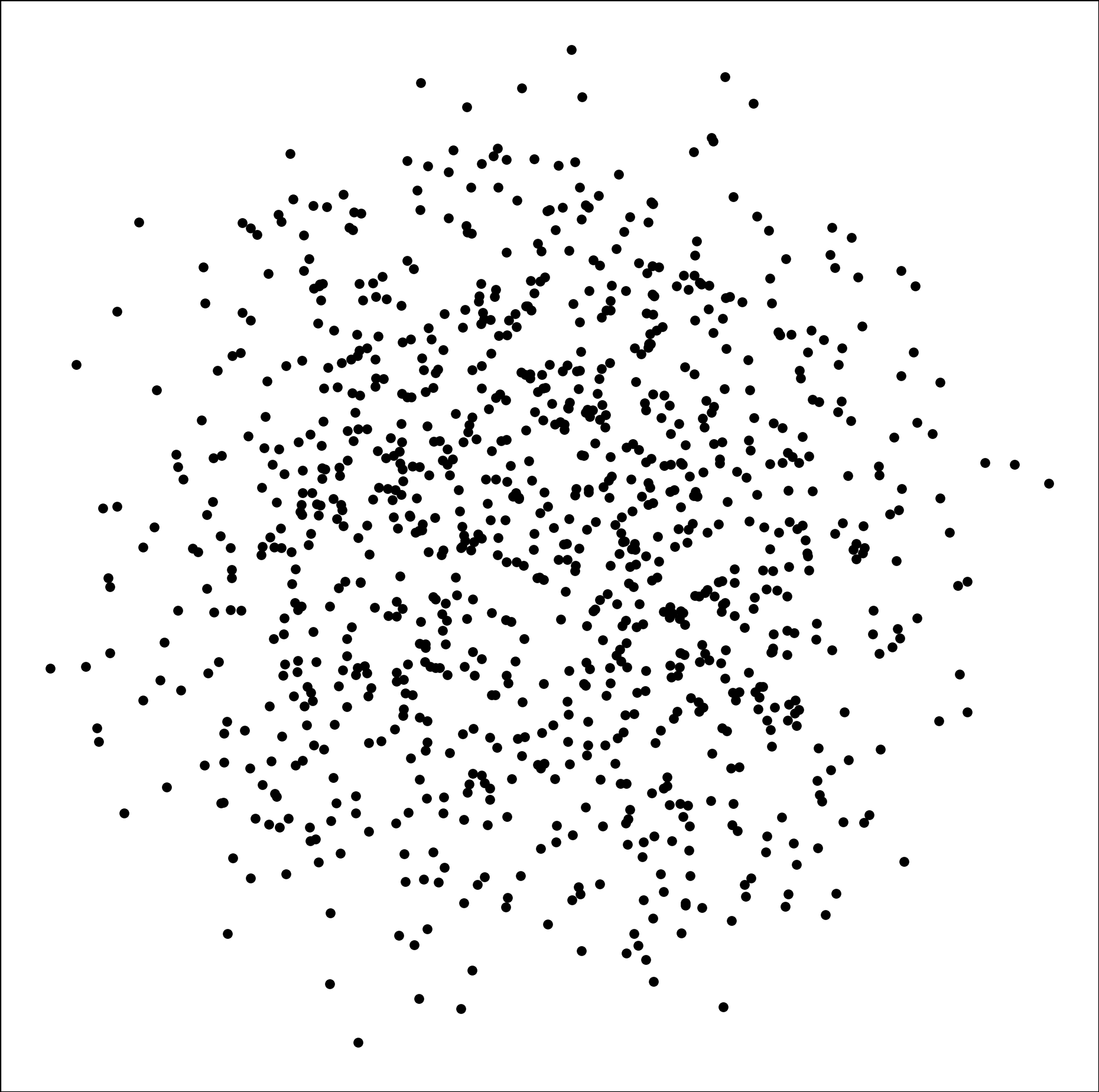}
    \end{subfigure}
    \begin{subfigure}[b]{0.35\textwidth}
        \includegraphics[width = \textwidth]{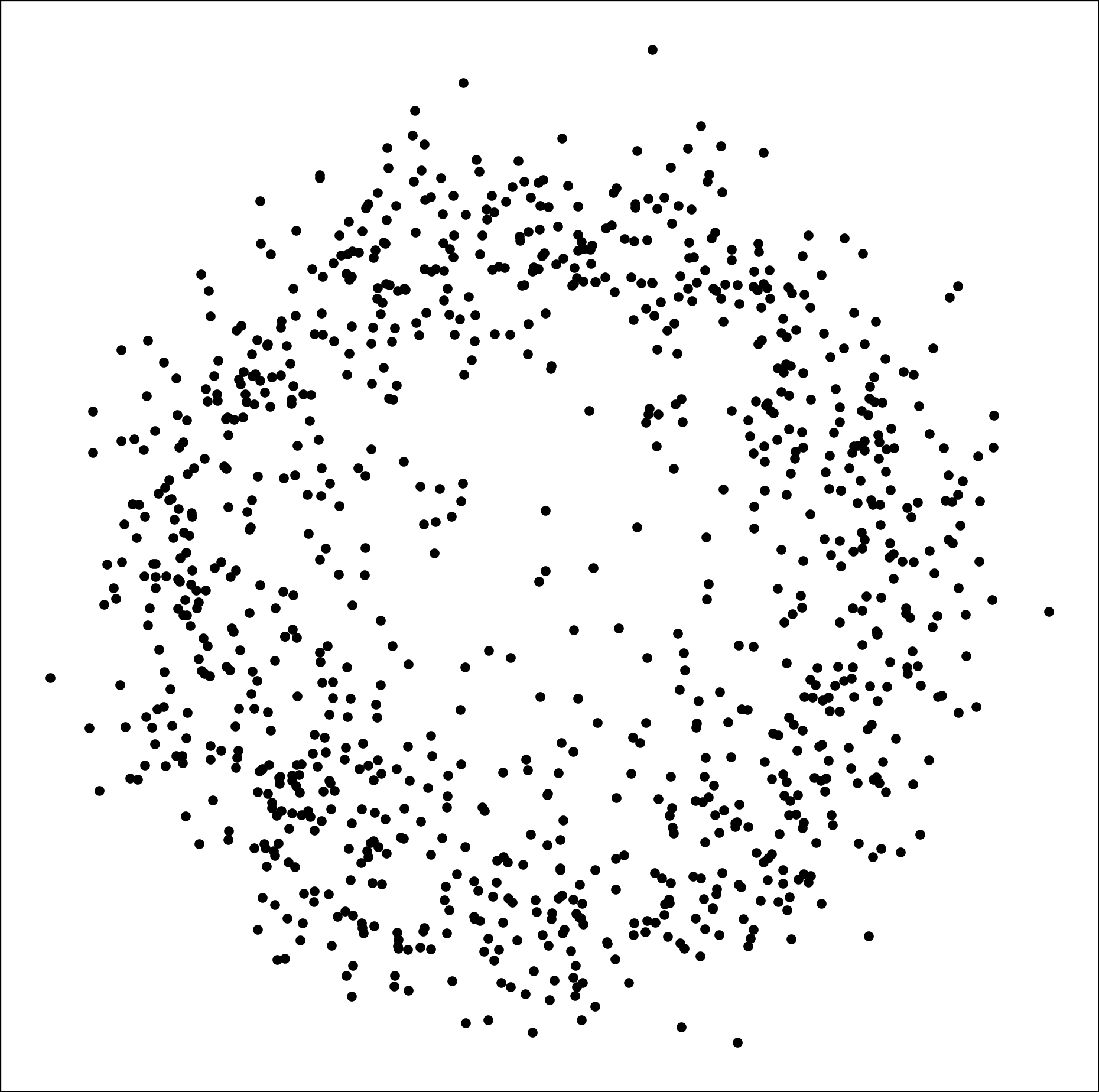}
    \end{subfigure}
    \caption{Two example point clouds}
    \label{fig:PointClouds}
\end{figure}

At its fundamental level, Topological Data Analysis (TDA) is concerned with quantifying the ``shape'' of datasets. To see the meaning of this consider Figure \ref{fig:PointClouds} which presents images of two hypothetical datasets in the plane. While both datasets are roughly circular it is obvious to a person that they are qualitatively different because there is a ``hole''in the cloud on the right which is not present in the left one. The aim of TDA is to make this notion mathematically rigorous so that a computer could detect this ``hole''. Holes in spaces are topological features not geometric ones and so we would initially expect that they can tell us very little about the ``shape'', that is geometry, of a dataset. However, by studying how the number of holes in our data changes with the length scale on which we examine the data we can determine geometric features.


\section{Simplicial complexes}

Many datasets are, at a fundamental level, just a finite set of points $X$ which belong to a metric space $M$. We refer to such a set as a \textit{point cloud}. $X$ inherits the topology of the metric space but is also topologically trivial as it is just a set of disconnected points. So, if we want to study the topology of data we need associate $X$ with a topological space which is non-trivial. The standard tool to accomplish this is the \textit{simplicial complex}. 

To define simplicial complexes we must first define the \textit{simplex} which generalises the notion of a triangle to higher dimensions. An $r$-simplex can be thought of as the analog of a triangle in $r$ dimensions. $0$-simplices are points, $1$-simplices are lines connecting two points, $2$-simplices are triangles defined by three vertices, $3$-simplices are tetrahedrons and so on. We are interested in the case of combinatorial, rather than geometric, simplices here which do not need to be embeddable in $\mathbb{R}^n$. Each $r$-simplex comes with subsets of itself which are lower dimension simplices called its faces. Precise definitions of the simplices can be found in any text on computational or algebraic topology, see \cite{Hatcher,EDELSBRUNNER,GHRIST,Nakahara}. We adopt the following slightly informal definition.
\begin{defn}
    An $r$-simplex $\sigma^r = (x_0x_1\ldots x_r)$ is a hyper-triangle with vertex set $V=\{x_0,x_1,\ldots,x_r\}$. All hyper-triangles with vertex sets $V'\subset V$ are subsets of $\sigma^r$ and define its set of faces. Note $\sigma'\leq \sigma$ means $\sigma'$ is a face of $\sigma$.
\end{defn}

A simplicial complex is a generalisation of a graph to higher dimensions. Formally it is a topological space $K$ which is made of a collection of $r$-simplices connected by attaching maps. A simplex can only be included in a complex if all of its faces are also included as simplices in the complex \cite{EDELSBRUNNER}. This statement is written formally as
\begin{defn}
If $K$ is a finite collection of simplices. Then we call it a simplicial complex if
        \begin{equation}
            \forall \sigma \in K, \sigma'\leq \sigma \implies \sigma' \in K\,.
        \end{equation}
\end{defn}


The topological dimension of $K$ is given by the largest $r$ of its simplices. We call such a complex a simplicial $r$-complex. A graph is an example of a simplicial $1$-complex and a triangulation of a $2$-manifold is a simplicial $2$-complex. 

%
If we can associate an abstract simplicial complex to a dataset then studying the topology of said complex provides us information about the geometry of the dataset. This is a fundamental idea in TDA and it is how we will describe the shape of data. For different types of data we use different methods to connect our data with a complex. We will discuss the methods applicable to point clouds and images respectively below.

\subsection{The Vietoris-Rips complex}

If we have a point cloud $X$ in a metric space $M$ then we can construct a simplicial complex by connecting points which are near each other with respect to the distance on $M$. We can then add triangles to our complex as soon as all of their faces are also in the complex. This procedure forms what is called the Vietoris-Rips (VR) complex of the point cloud at a given diameter $\epsilon>0$. It is formally defined as follows. 
\begin{defn}
    Let $S$ refer to the set of all simplices whose vertex set is a subset of $X$. Then, for $\epsilon>0$ define the VR complex of $X$ as 
    \begin{equation}
        VR_\epsilon(X) = \{\sigma \in S| \, \text{diam}(\sigma) \leq \epsilon\}\,.
    \end{equation}
\end{defn}

\begin{figure}
    \centering
    \includegraphics[width = \textwidth]{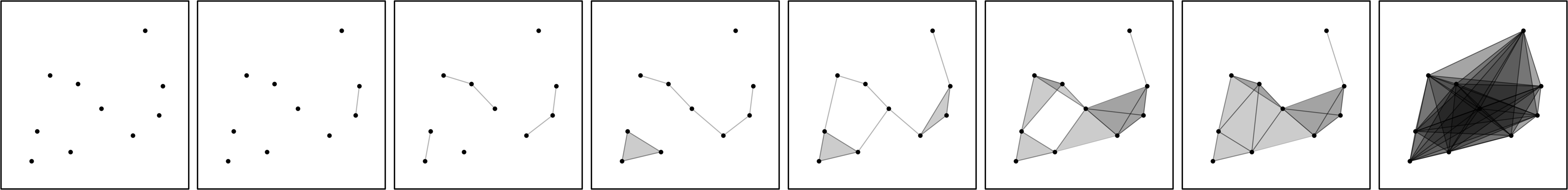}
    \caption{Image of Vietoris-Rips complexes of the same point cloud across a range of diameters showing only vertices, edges, and triangles. Note that the diameter increases from left to right.}
    \label{fig:RipsFiltration}
\end{figure}

Figure \ref{fig:RipsFiltration} presents images of the VR complex of a point cloud for a series of diameters $\epsilon$. For small diameters $VR_\epsilon(X) = X$ that is, only the actual points are included in the complex. However, for larger radii we see that edges and eventually triangles are added too. Also we see from this example that for the larger diameters the VR complex cannot be embedded in $\mathbb{R}^2$ without self-intersection. This is why we adopt an abstract combinatorial definition for the simplicial complexes rather than a geometric one in terms of geometrically realisable triangles. The complexes we associate to datasets will often not be embeddable. By varying the diameter $\epsilon$ we can investigate how the connectivity of our dataset changes with length scale.

\section{Simplicial homology}

Within the field of computational topology we are concerned with describing the topological properties of simplicial complexes. Mostly we will study a famous algebraic topological invariant of spaces called homology. The fundamental idea of this construction is to associate to a topological space $X$ a family of abelian groups, called the homology groups $H_n(X)$, which encode the number of $n$-dimensional "holes" in the space. We will briefly review the construction of simplicial homology. In the following, assume that $K$ always refers to a simplicial complex of interest and we always work over the finite field $\mathbb{Z}_2$. 

\begin{defn}
    Let the $n$-chain group $C_n(K)$ be the free abelian group generated by the $n$-simplices of $K$ with coefficients in the field $\mathbb{Z}_2$.
\end{defn}

That is, if we consider a complex containing $1$-simplices $\{\sigma^1_1,\sigma^1_2,\sigma^1_3\}$, a $1$-chain is a formal sum of $1$-simplices of the form $c = a_1\sigma_1^1+a_2\sigma_2^1+a_3\sigma_3^1$, where $a_1,a_2,$ \& $a_3 \in \mathbb{Z}_2$.

Homology theory tells us that we can construct a chain complex 
\begin{equation}
    \cdots \rightarrow C_n(K,\mathbb{Z}_2) \overset{\partial_n}{\longrightarrow}C_{n-1}(K,\mathbb{Z}_2)\overset{\partial_{n-1}}{\longrightarrow}\cdots\,,
\end{equation}
where the boundary maps are defined by summing the codimension-$1$ faces of a $n$-simplex $\sigma^n$ \cite{Nakahara,Hatcher}.

\begin{defn}
    We define a homomorphism from $n$-chains to $(n-1)$-chains called the boundary homomorphism $\partial_n:C_n(K)\rightarrow C_{n-1}(K)$. If $\sigma^n = (p^0p^1\cdots p^n)$ is an $n$-simplex in $K$ the the action of $\partial_n$ on $\sigma^n$ is given by the following sum over $(n-1)$-simplices
    \begin{equation}
        \partial_n \sigma^n = \sum_{i=0}^{n}(p^0\cdots \hat{p}^i\cdots p^n)\,,
    \end{equation}
    where the $\hat{p}^i$ indicates the removal of $p^i$ from $(p^0p^1\cdots p^n)$. The definition of $\partial_n$ is then extended linearly over all $n$-chains.
\end{defn}

\begin{figure}
        \centering
        \includegraphics[width = 0.6\textwidth]{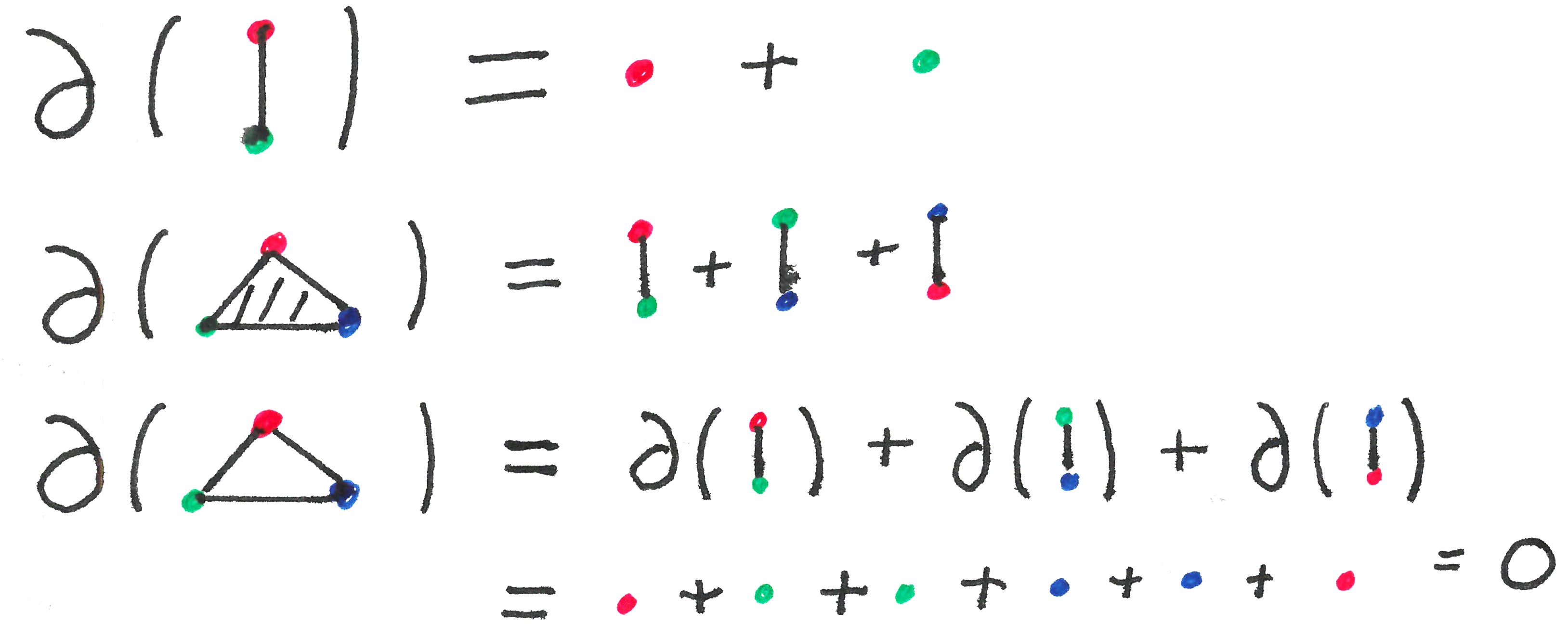}
        \caption{Diagrammatic representation of the action of the boundary operator on simplicial complexes. Note that a shaded triangle indicates that the triangle is included in the complex.}
        \label{fig:Boundary1}
\end{figure}

As examples of the action of the boundary map consider Figure \ref{fig:Boundary1} which presents the action on small simplicial complexes diagrammatically. Note that our diagrams describe the action of the boundary on the sum of simplices with dimension equal to the dimension of the complex.

In Figure \ref{fig:Boundary1} we specifically included the case of a loop of $1$-simplices which the boundary operator sent to zero in $C_0(K)$. This is a general observation, that structures which geometrically represent loops will vanish under the action of the boundary operator. This leads us naturally to the following definition

\begin{defn}
    If $\sigma^n\in C_n(K)$ satisfies $\partial_n \sigma^n=0$ then we call $\sigma^n$ an $n$-cycle.
\end{defn}

A key theorem of homology is that all boundaries are closed. This is formalised in the following.

\begin{thrm}
    If $\sigma^n\in C_n(K)$ then $\partial_{n-1}\partial_n \sigma^n =0$, that is $\partial_{n-1}\partial_n$ is the trivial homomorphism for all $n\in \mathbb{N}$.
\end{thrm}

The fact that all boundaries are closed means we cannot use the closure of chains alone to detect ``holes'' in spaces. The fundamental observation of homology is that a true ``hole'' in a space is represented by a chain which is closed but is not a boundary. So, we define the following.

\begin{defn}
    The $n$-homology group of $K$ is defined as 
    \begin{equation}
        H_n(K) = \frac{\ker{\partial_n}}{\Im\partial_{n+1}}\,.
    \end{equation}
\end{defn}

As a quotient group, $H_n(K)$ is made of equivalence classes of cycles in $C_n(K)$ where the equivalence relation is that $c \sim c'$ if $c=c'+\partial_{n+1}d$ for some $d\in C_{n+1}(K)$. That is, classes are sets of cycles whose elements differ by the boundaries of $(n+1)$-chains. Each class in $H_n$ is associated geometrically to a single $n$-dimensional ``hole'' in the simplicial complex $K$.

\subsection{Induced maps}
 It is important in TDA that maps between simplicial complexes generate induced maps on the homology groups of those complexes. Suppose we have simplicial complexes $K$ and $K'$ and a continuous map $\phi:K\rightarrow K'$. Then for any chain $c\in C_n(K)$ we define an induced map on chains $\phi_\#:C_n(K)\rightarrow C_n(K')$ by linearly extending the action of $\phi$, which is defined on simplices, over sums of simplices. It can be shown that $\phi_\#$ commutes with the boundary operator and as a result always maps cycles to cycles and boundaries to boundaries \cite{Hatcher}. For this reason $\phi_\#$ can be used to construct a map on the quotient group, that is the homology groups. If $[c]$ is an equivalence class in $H_n(K)$ described by the representative $c\in C_n(K)$, we can define a map $\phi_*:H_n(K)\rightarrow H_n(K')$ by the following
 \begin{equation}
     \phi_*([c]) = [\phi_\#(c)]\,.
 \end{equation}
 It follows that this map $\phi_*$ is an homomorphism between the homology groups and so we call it the induced homomorphism on homology \cite{EDELSBRUNNER}. If we have a sequence of continuous maps $\phi,\psi,\ldots$ between several simplicial complexes $K,K',K'',\ldots$ that is
\begin{center}
\begin{tikzcd}
\cdots \arrow[r] & K \arrow[r, "\phi"] 
& K' \arrow[r, "\psi"] & K'' \arrow[r] & \cdots \,,
\end{tikzcd}
\end{center}
then there will exist a sequence on induced maps between the Homology groups
\begin{center}
\begin{tikzcd}
\cdots \arrow[r] & H_n(K) \arrow[r, "\phi_*" ] & H_n(K') \arrow[r, "\psi_*" ] & H_n(K'') \arrow[r] & \cdots\,.
\end{tikzcd}
\end{center}
It is by studying such a sequence of homology groups that we will be able to determine how the topology of a dataset changes with scale.


\section{Persistent homology}

\subsection{The evolution of topological features}

We now develop a core tool of TDA called Persistent Homology. Suppose we have a finite sequence of simplicial complexes $K_0,K_1,\ldots K_m$ such that $K_i\subset K_j$ for all $i<j$. We can construct inclusion maps from one complex to another to form a sequence of inclusion maps
\begin{equation}
\begin{tikzcd}
0 \arrow[hookrightarrow]{r} & K_0 \arrow[hookrightarrow]{r} 
& K_1 \arrow[hookrightarrow]{r} & \cdots \arrow[hookrightarrow]{r} & K_m\,.
\end{tikzcd}
\end{equation}
Such a sequence is referred to as a filtration over the set of complexes. 

\begin{figure}[H]
        \centering
        \includegraphics[width = 0.8\textwidth]{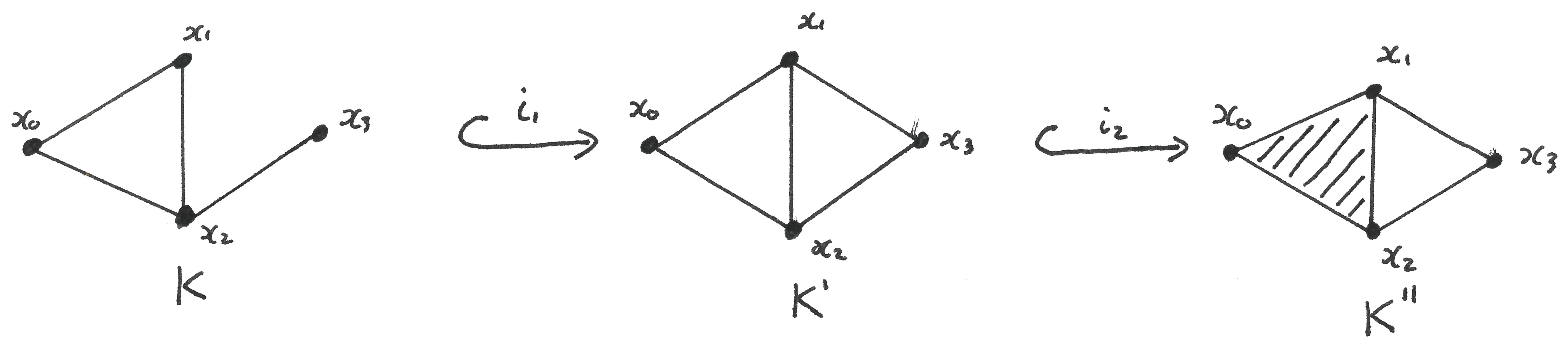}
        \caption{Example inclusion maps between simplicial complexes.}
        \label{fig:inclusion}
\end{figure}

Consider Figure \ref{fig:inclusion} which presents a small sub-sequence of complexes from  a filtration. Choosing representatives for the homology classes we can explicitly write the first homology groups of these complexes. 
\begin{align*}
    &H_1(K) = \{ a[(x_0x_1)+(x_1x_2)+(x_0x_2)] \,|\, a \in \mathbb{Z}_2\}\,,\\
    &H_1(K') = \{ a[(x_0x_1)+(x_1x_2)+(x_0x_2)]+b[(x_1x_3)+(x_2x_3)+(x_1x_2)] \,|\, a,b \in \mathbb{Z}_2\}\,,\\
    &H_1(K'') = \{ a[(x_1x_3)+(x_2x_3)+(x_1x_2)] \,|\, a \in \mathbb{Z}_2\}\,.
\end{align*}
From the definition of the induced map we can then write out the image of each of the homology groups under the induced maps under inclusion. For example
\begin{align*}
    {i_1}_*(H_1(K)) &= \{ a {i_1}_*([(x_0x_1)+(x_1x_2)+(x_0x_2)]) \,|\, a \in \mathbb{Z}_2\}\,,\\
    &= \{ a [i_1((x_0x_1)+(x_1x_2)+(x_0x_2))] \,|\, a \in \mathbb{Z}_2\}\,,\\
    &= \{ a [(x_0x_1)+(x_1x_2)+(x_0x_2)] \,|\, a \in \mathbb{Z}_2\}\,,
\end{align*}
where in the final line $(x_0x_1)+(x_1x_2)+(x_0x_2) \in C_1(K')$ instead of the $C_1(K)$ group in which is started. We can now make an important observation ${i_1}_*(H_1(K)) \subset H_1(K')$ but $H_1(K')\neq {i_1}_*(H_1(K))$. That is, there are classes in $H_1(K')$ which are not in the image of $H_1(K)$ induced by inclusion. This suggests that there is a topological distinction between $K$ and $K'$. This corresponds to the fact that $K$ contains one hole in it where $K'$ has two. In the language of persistent homology we would say that a $1$-cycle is \textit{born} when the $(x_1x_2)$ edge is added to the complex.

In contrast consider the induced map on $H_1(K')$ under inclusion which we compute as 
\begin{align*}
    {i_2}_*(H_1(K')) &= \{ a {i_2}_*([(x_0x_1)+(x_1x_2)+(x_0x_2)])+b{i_2}_*([(x_1x_3)+(x_2x_3)+(x_1x_2)]) \,|\, a \in \mathbb{Z}_2\}\,,\\
    &= \{ a [i_2((x_0x_1)+(x_1x_2)+(x_0x_2))]+b [i_2((x_1x_3)+(x_2x_3)+(x_1x_2))] \,|\, a \in \mathbb{Z}_2\}\,,\\
    &= \{a [\partial ((x_0x_1x_2))]+b [(x_1x_3)+(x_2x_3)+(x_1x_2)]  \,|\, a \in \mathbb{Z}_2\}\,,\\
    &= \{b [(x_1x_3)+(x_2x_3)+(x_1x_2)]  \,|\, a \in \mathbb{Z}_2\}\,,
\end{align*}
where the third line holds because the $(x_0x_1x_2)$ simplex is in the collection $K''$ where it was not for $K$ and $K'$, and the fourth line holds because all boundaries of $2$-chains are equivalent to the identity in $H_1$. So, we now observe that $ {i_2}_*(H_1(K'))=H_1(K'')$ so no classes are born when the $(x_0x_1x_2)$ triangle was added but we note that
\begin{equation}
    \ker{{i_2}_*} = \{ a[(x_0x_1)+(x_1x_2)+(x_0x_2)] \,|\, a \in \mathbb{Z}_2\}\cong \mathbb{Z}_2\,.\\
\end{equation}
That is, a class in $H_1(K')$ is sent to zero by ${i_2}_*$, so we say that a $1$-cycle \textit{dies} when the triangle $(x_0x_1x_2)$ is added to the complex. Note that a class can die without the selected representative being sent to zero when it merges with another class. The kernel of the induced map under inclusion will still be non-trivial in this case.

To make the observation above into a general, formal theory we define the \textit{persistent homology groups}. 
\begin{defn}
    If $f^{i,j}_n:H_n(K_i)\rightarrow H_n(K_j)$ is the homomorphism induced by the inclusion $K_i\hookrightarrow K_j$ for $0\leq i\leq j\leq m$, then the $n$-th persistent homology groups are $H_n^{i,j} = \text{im}{f^{i,j}_n}$. 
\end{defn}
We observe that $H^{i,j}_n$ describes the $n$-cycles from stage $K_i$ in the filtration which are still alive at $K_j$. We now define the birth and death of classes as follows
\begin{defn}
    If $\alpha \in  H_n(K_i)$ but $\alpha \notin H_n^{i-1,i}$ we say that the class is \textit{born} at $K_i$. Similarly, we say the class \textit{dies} entering $K_j$ if $f_n^{i,j-1}(\alpha) \notin H^{i-1,j-1}_n$ but $f_n^{i,j}(\alpha) \in H^{i-1,j}_n$. The persistence of the class is defined to be $p(\alpha) = j-i$, the difference of the death and birth times.
\end{defn}
The definition of birth provided above follows our discussion from earlier, that a class is born when it is in the homology of a complex but not in the image of the homology of the previous complex in the filtration. The definition of death is somewhat subtle and includes both the possibility that classes are sent to zero as well as merged with older classes. For a detailed discussion we refer to Chapter 7 of \cite{EDELSBRUNNER}.

For a given filtration, we can compute the set of all birth and death times of $n$-th homology classes. We refer to this set as the $n$-th Persistent Homology (PH) of the filtration. 

The \textit{modus operandi} of TDA is then as follows: we associate to our dataset a sequence of nested simplicial complexes; we construct a filtration from said sequence of complexes; we compute the PH for that filtration; and finally we study the PH itself to infer the shape of the dataset.

\subsection{Filtrations for VR complexes}

Consider Figure \ref{fig:RipsFiltration} again. Observe that each complex is a sub-complex of those to the right of it in the diagram. Generically we have that for a point cloud $X$, if $\epsilon<\epsilon'$ then
\begin{equation}
    VR_\epsilon(X) \subset VR_{\epsilon'}(X)\,.
\end{equation}

\begin{figure}
    \centering
    \includegraphics[width = \textwidth]{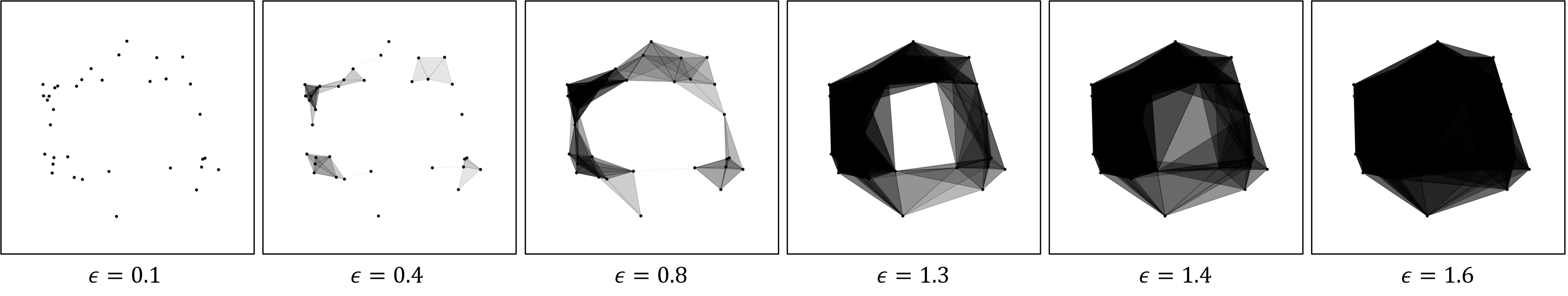}
    \caption{Image of Vietoris-Rips complexes of the point cloud sampled from the unit circle with normally distributed radial noise. The diameter ($\epsilon$) of each complex is indicated under each image.}
    \label{fig:RipsFiltration2}
\end{figure}

This implies that we can construct a filtration from the $VR$ complexes by the inclusion maps between complexes them. Hypothetically, this is a real valued filtration in the sense that we have a different VR complex for each $\epsilon\in \mathbb{R}^+$. However, if $X$ is a finite point cloud then the VR complex for each $\epsilon$ will not be unique and instead only changes at diameters which are lengths of edges of the complex. Therefore, we will have a series of intervals of the form $[\epsilon_i,\epsilon_j)$ over which the VR complex does not change and so we only need to consider a finite sequence of diameters $\epsilon_0 < \epsilon_1 <\ldots < \epsilon_m$ which we take as the lower bound of each of the intervals. Then our filtration is 
\begin{equation}
\begin{tikzcd}
0 \arrow[hookrightarrow]{r} & VR_{\epsilon_0}(X) \arrow[hookrightarrow]{r} 
& VR_{\epsilon_1}(X) \arrow[hookrightarrow]{r} & \cdots \arrow[hookrightarrow]{r} & VR_{\epsilon_m}(X)\,.
\end{tikzcd}
\end{equation}
We refer to this as the Vietoris-Rips (VR) filtration and its persistent homology as the Vietoris-Rips persistent homology of the point cloud $X$, and notate said persistent homology as $PH_n(X)$ where $n$ indicates which dimension of the homology we are considering. 

In this document we will visualise $PH_n(X)$ for the VR complex using Persistence Diagrams (PD). The PD is a scatter plot of the birth-time and death-time pairs $(b,d)$ on the plane. Since all classes must die after they are born we have $d>b$ and so all persistence pairs appear above the line $y=x$. Note that the vertical distance between any point and this line is the persistence of the class. So classes which are very persistent appear well above the line. 

\begin{figure}
    \centering
    \includegraphics[width = 0.5\textwidth]{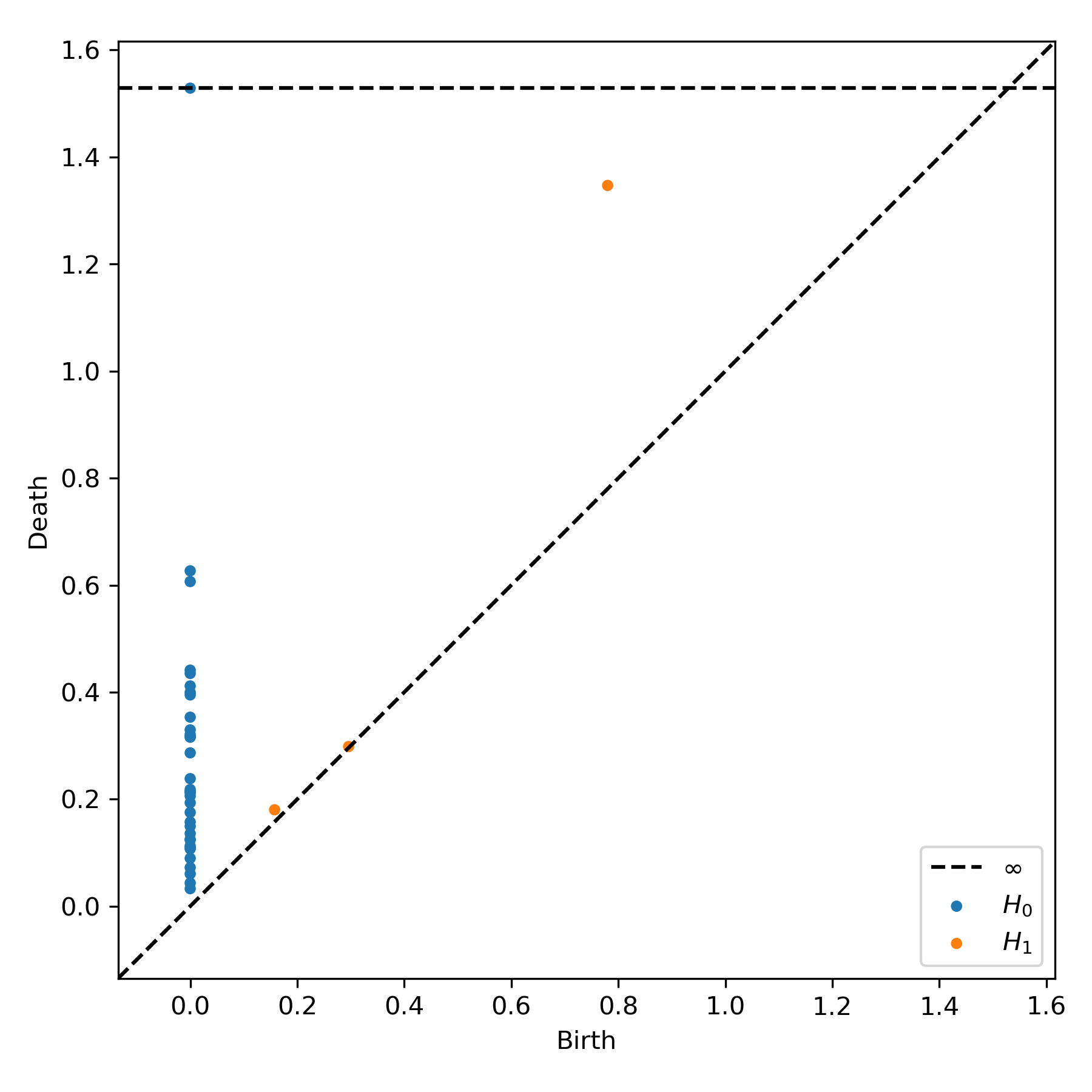}
    \caption{Persistence diagram of the VR filtration shown as Figure \ref{fig:RipsFiltration2}.}
    \label{fig:RipsFiltration2PD}
\end{figure}

Consider Figure \ref{fig:RipsFiltration2} which presents a sample of the complexes from the VR filtration of a point cloud which is approximately circular we will call $X$. For very small $\epsilon$ the complex is largely a collection of disconnected clusters. As $\epsilon$ is increased the clusters grow until there is only one connected component left. At $\epsilon\approx 0.8$ an edge is added which creates $1$-cycle in the VR complex corresponding to the hole in the center of the point cloud. At this point a homology class is born. This class dies again at $\epsilon \approx 1.4$ when an edge and two triangles are added simultaneously which closes the hole. For very large $\epsilon$ each point is connected to every other point and our render of the complex degenerates into a black mass.

To study the VR persistent homology of the $X$ cloud directly, we computed $PH(X)$ using the software program \verb!Ripser!. This produces a persistence diagram which is presented as Figure \ref{fig:RipsFiltration2PD}. The classes from $PH_0(X)$ are shown as blue dots and those of $PH_1(X)$ as orange dots. We observe that there is one $H_1$ class well above the $y=x$ line, that is a highly persistent class. This corresponds to a loop in our dataset which persists across many scales and indicates the existence of a geometric hole in the dataset. It corresponds specifically to the fact that our dataset is approximately circular in shape. Note that the death time of the class $\epsilon_{\text{death}}\approx 1.3$ provides an estimate for the geometric size of the hole. The other classes in $PH_1$ are very close to the $y=x$ line and the convention in TDA is to attribute them to ``noise'' or more accurately ``sampling error''.

\subsection{Filtrations for images}

We saw in the above that for point clouds the shape of the dataset is encoded in the change in the topology of our dataset over many scales. However, not all datasets are point clouds and so not all datasets admit the construction of a VR filtration. We expect that for other types of data, by developing an appropriate notion of \textit{change of scale} for which we can study the change in topology, we can similarly describe the structure of the dataset. Here we look at the case of grayscale images, which are a common form of dataset to encounter in practice and one we will analyse in Chapters \ref{Chapter4} and \ref{Chapter5}.

Grayscale images are approximations to functions from $\mathbb{R}^p\rightarrow \mathbb{R}$. We will restrict to the case in which these images are restricted to the vertices of a \textit{mesh} $K$, which is a triangulation of $\mathbb{R}^p$. So our image can be interpreted as a map $f:V(K)\rightarrow \mathbb{R}$, where $V(K)$ refers to the vertices of the complex $K$. On functions, the notion of, change with scale, which is appropriate is change with increasing function value. We aim to construct a nested sequence of complexes indexed by the function value. We cannot use the level sets of the function as $f^{-1}(a)$ is not necessarily a subset of $f^{-1}(b)$ even if $a<b$ and hence these sets would not be nested. However, the sub-level sets of the function do have the correct nesting property. That is if $a<b$ then $f^{-1}(-\infty,a] \subset f^{-1}(-\infty,b]$. Unfortunately, as constructed, each sub-level set is not a full simplicial complex since the function is only defined on the vertices of $K$.  However, we can construct full simplicial complexes from the sub-level sets of $f$ as follows
\begin{equation}
    K_h(f) = \{ \sigma \in K \,|\, \forall \, v\in V(\sigma) \,, f(v) \leq h\}\,,
\end{equation}
where $V(\sigma)$ refers to the vertex set of the simplex $\sigma$. That is, we can build up a complex $K_h(f)\subset K$ which represents the mesh at level $h$ by adding all vertices of $K$ whose image value is less that $h$ and then adding all edges, triangles, and so on from $K$ which connect included vertices. Note that this addition of simplices ``as soon as possible'' is similar to the VR complex but the distinction is that for images we only add simplices which exist in the underlying mesh $K$.

It is immediate that if $h <h'$ then $K_h \subset K_{h'}$ so we can construct a filtration over the sub-level complexes as follows
\begin{equation}
\begin{tikzcd}
0 \arrow[hookrightarrow]{r} & K_{h_0}(f) \arrow[hookrightarrow]{r} 
& K_{h_1}(f) \arrow[hookrightarrow]{r} & \cdots \arrow[hookrightarrow]{r} & K_{h_m}(f) = K\,,
\end{tikzcd}
\end{equation}
where as before we note that if the mesh only contains a finite number of simplices then the sub-level complexes will only change at a finite number of function values. Specifically, we recognise that $\{h_0,h_1,\ldots h_m\} = f(V(K))$.

Formally, in the above we have described the sub-level set filtration for images assuming a simplicial approximation to $\mathbb{R}^p$. In practice a cubical approximation is used instead. The definitions for this case are largely identical if we swap in-place ``triangle'' for ``square'' and and ``tetrahedron'' for ``cube'' and so on. For details on cubical complexes we refer to \cite{kaczynski2004computational}.

\begin{figure}
    \centering
    \includegraphics[width = 0.7\columnwidth]{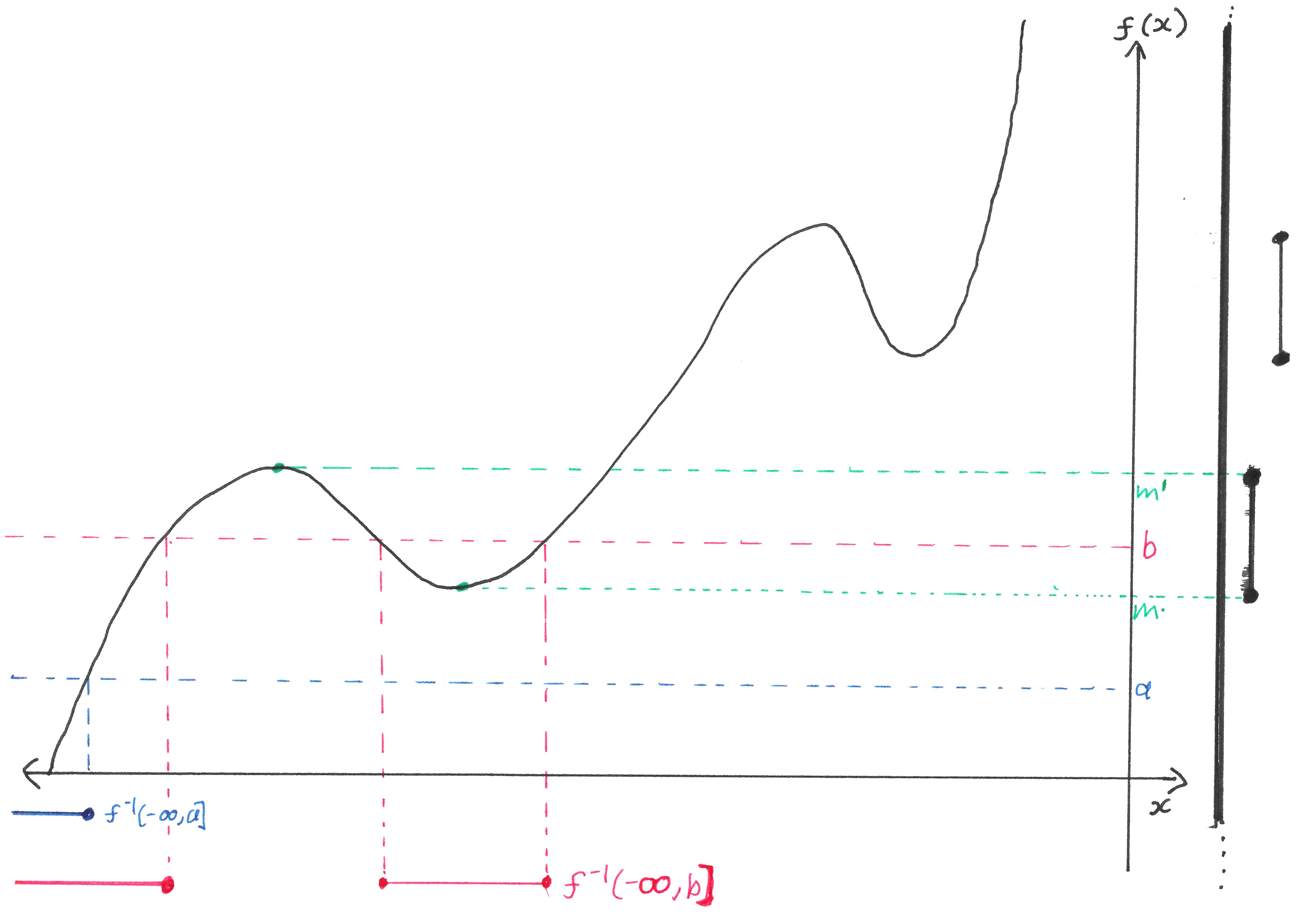}
    \caption{Diagram of the sub-level set persistent homology of a one dimensional function.  The sub-level sets at levels $a$ and $b$ are presented as the blue and red horizontal subsets of the line respectively. The birth and death locations of a connected component are presented by the dotted green lines connecting the associated minimum and maximum of $f$ to the bar of the connected component in the barcode.}
    \label{fig:diagram}
\end{figure}

We now know that we can construct a filtration using our image data and can then compute the persistent homology of the filtration. Before we can actually use this to analyse any images we need to understand what information about the image is actually encoded in the persistent homology. To determine this consider the case of a function $f:\mathbb{R}\rightarrow\mathbb{R}$ presented as Figure \ref{fig:diagram}. Note that it is assumed that the function decreases and increases monotonically to the left and right respectively outside of the region shown. 

The sub-level set of $f$ at level $a$, $f^{-1}(\infty,a]$ is presented as the blue horizontal line. This is an interval subset of $\mathbb{R}$ and we note that it contains only one connected component. In contrast $f^{-1}(\infty,b]$ contains two connected components as shown. This second connected component appears in $f^{-1}(\infty,h]$ when $h=m$, that is at the minimum of the function shown with a green horizontal line. This second connected component persists as we increase the function value above $b$ until at $m'$ it merges with the older connected component. This corresponds to a death in the sub-level set $PH_0$. However we can observe that the $m'$ is not an arbitrary value of $f$ and is instead a local maximum. This argument will work in general and so we now see that the births and deaths of classes in the sub-level set persistent homology correspond to critical points of the function. That is, the critical points of the function are the topological features which are described by the sub-level set filtration on functions and images. For images in two dimensions the persistent homology will describe the minimums, maximums, and saddle points of the function.

Note that Figure \ref{fig:diagram} includes, on the right edge, a set of vertical black lines. Each of these lines corresponds to a class in the $PH_0$ persistent homology. Note that the infinite vertical line shown corresponds to the connected component which will exist at all function values. The lines encode all of the information about the persistent homology and are referred to as the \textit{barcode}. For functions or images under the sub-level set filtration we will refer to the set of such bars as $PH_n(f)$ \cite{EDELSBRUNNER}. We are being technically informal here but this is sufficient for our purposes. 

In Chapter \ref{Chapter4} of this thesis we will compute the $PH_n(f)$ for several two-dimensional images. These calculation were performed with a software program called \verb|Diamorse|, which calculates the sub-level set persistent homology for images using Discrete Morse Theory applied to the cubical complex filtration mentioned above. A detailed discussion this calculation is unnecessary for our purposes but can be found in \cite{DIAMORSE,DIAMORSE1}.

\section{Absolute and relative persistence}

\begin{figure}
    \centering
    \begin{subfigure}[b]{0.49\textwidth}
        \includegraphics[width = \textwidth]{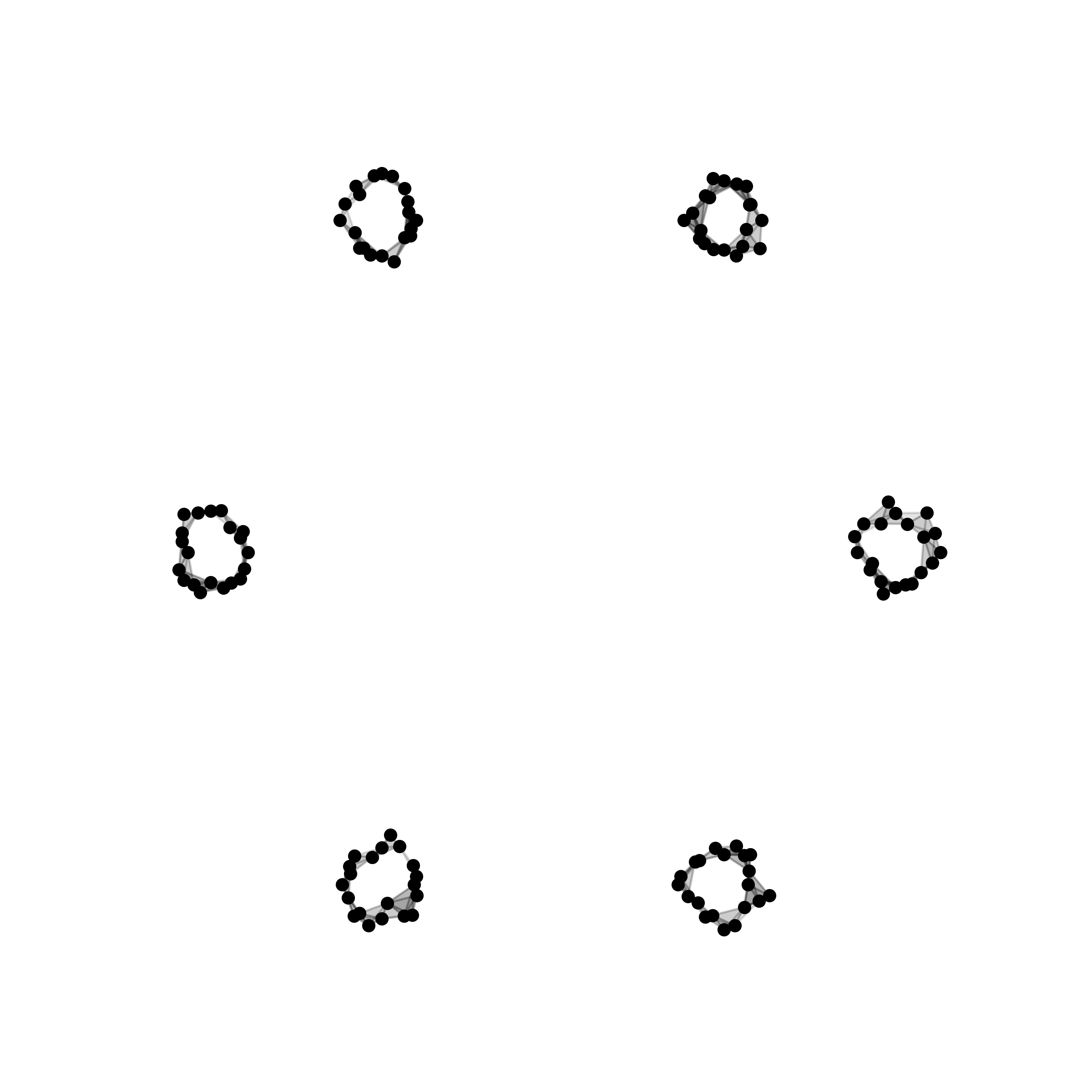}
        \subcaption{}
    \end{subfigure}
    \begin{subfigure}[b]{0.49\textwidth}
        \includegraphics[width = \textwidth]{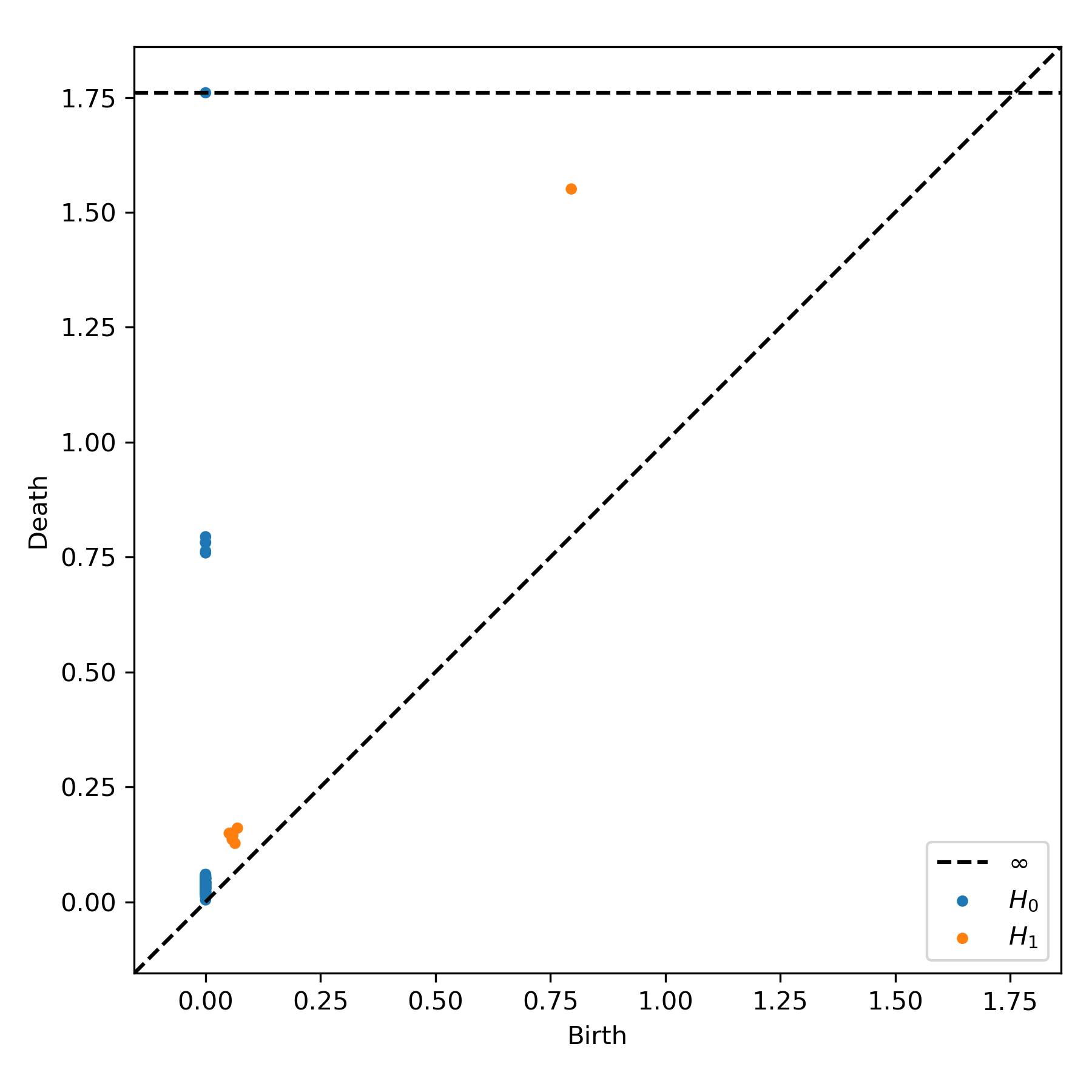}
        \subcaption{}
    \end{subfigure}
    \caption{Diameter $\epsilon = 0.1$ VR complex for a point cloud of small loop clusters and $PH(X)$ for said point cloud}
    \label{fig:loops}
\end{figure}

In TDA it is common to use the difference between the death and birth times of a class, defined above as the \textit{persistence}, to measure the statistical significance of said class. However, this can have disadvantages when we deal with point clouds and complexes with geometric structures on several length scales \cite{bobrowski2023universal}. Consider Figure \ref{fig:loops} which presents the $VR_{\epsilon= 0.1}(X)$ for a point cloud $X$, which is approximately an arrangement of six circles on the vertices of a regular hexagon, and its $PH(X)$. From the persistence diagram we note that there is one very persistent $H_1$ class and six $H_1$ classes with low persistence. The persistence of these classes is so small that without further investigation one might choose to consider them as less statistically significant and even ignore them in further analysis. Ignoring the low persistence $H_1$ classes we could conclude that our dataset is topologically close to a circle. The $H_0$ information tells us that the global circle is built from six tight clusters but nothing about those cluster. However, looking at the dataset geometrically we would argue that this is not representative. By completely ignoring the low persistence classes any information about the small loops is lost. 

\begin{figure}
    \centering
    \includegraphics[width = 0.5\textwidth]{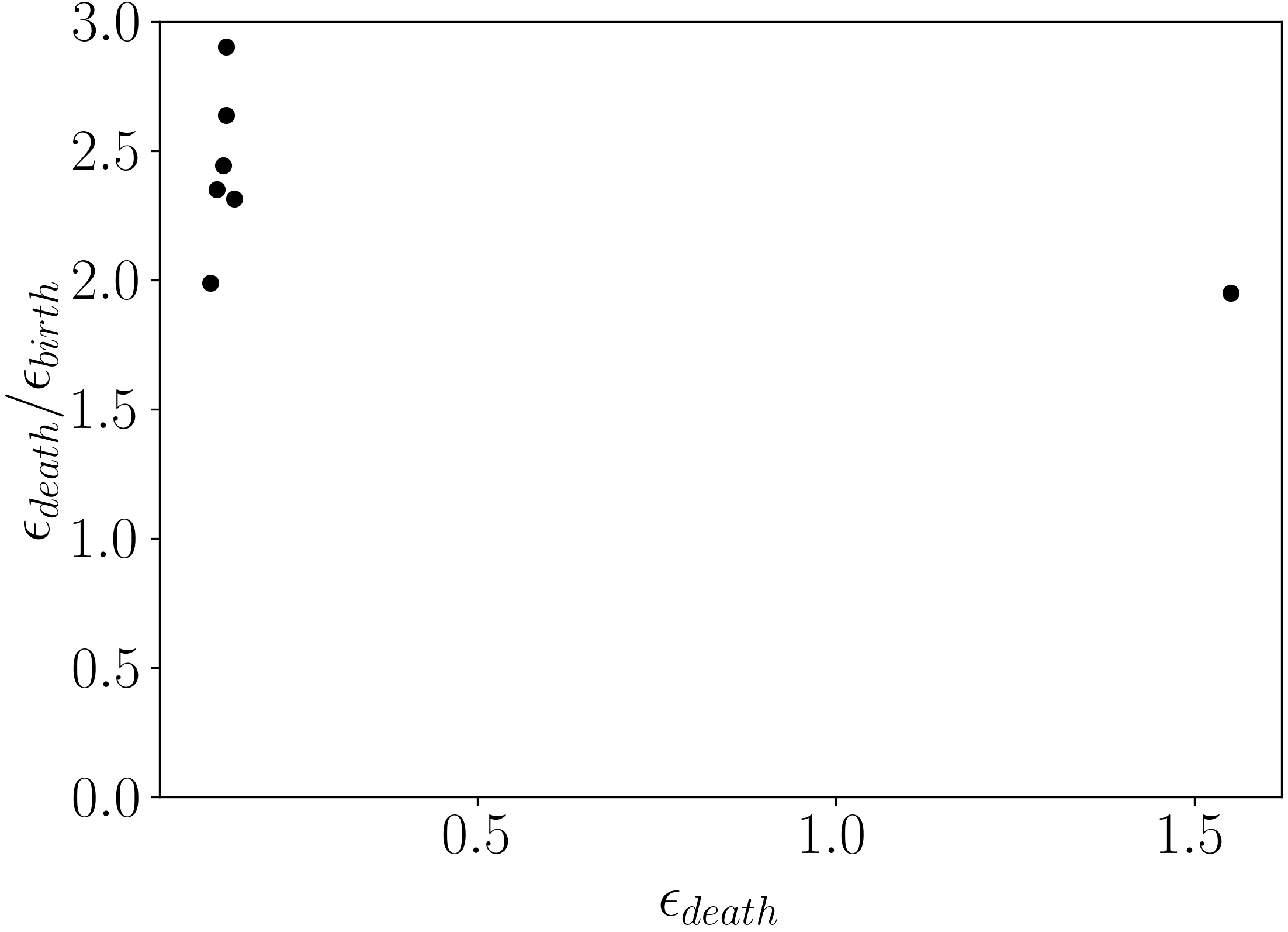}
    \caption{Scatter plot of the relative persistence against the death time for each each class in Figure \ref{fig:loops}.}
    \label{fig:LoopsRelativerPersistence}
\end{figure}

The problem in the above is that our dataset has topological structures on two geometrically distinct size scales. We will see in Chapter \ref{Chapter3} that the same problem emerges when we study the orbits of Hamiltonian flows under Poincare maps. Therefore, we would prefer to construct a tool which can measure the statistical significance of classes but without causing geometrically small features to be ignored. This can be achieved by looking at the ratio of the death and birth times instead of their difference. Figure \ref{fig:LoopsRelativerPersistence} presents a scatter plot of the ratio of death time to birth time, that is $d/b$, against the death time itself for the $H_1$ classes in $PH_1(X)$. Note that the birth and death times are indicated by the diameter of birth and death in the VR complex. We see that the geometrically large loop, which has the greatest death time, has a very similar death time to birth time ratio as the six classes which die earlier. Hence, adopting $d/b$ as our measure of statistical significance we would conclude that the seven classes in $PH_1(X)$ are similarly statistically significant and hence can recognise that the point cloud is not a single circle formed from six small clusters but instead is a single circle, or hexagaon, formed from six smaller circles.

In Chapter \ref{Chapter3} we will make repeated use of this alternative measure of class significance and so define it as follows.
\begin{defn}
    If $(b,d)$ are the birth and death times of a PH class respectively. Then $d-b$ is defined to be its absolute persistence and $d/b$ its relative persistence.
\end{defn}
Note that other authors have referred to $d/b$ as the \textit{multiplicative persistence} instead of relative persistence \cite{bobrowski2023universal}. We choose to adopt the terminology above as it follows the experimental science conventions for measurement errors.


\section{Some model symplectic maps}

The key feature which separates chaotic Hamiltonian systems from those of regular nonlinear dynamics is that they are area preserving \cite{arnol2013mathematical}. This fact is traditionally presented as Liouville's theorem
\begin{thrm}[Lioville's Theorem]
    If $U\subset \mathcal{M}$ is a region of phase space and $\phi_t:\mathcal{U}\rightarrow\mathcal{M}$ denotes the flow induced by Hamilton's equations then for all $t\in \mathbb{R}$
    \begin{equation}
        \text{Area}(U) = \text{Area}(\phi_t(U))\,,
    \end{equation}
\end{thrm}
Note that the Poincare recurrence theorem mentioned earlier is a consequence of Liouville's theorem. The preservation of phase space areas under Hamiltonian evolution is the key property from which Hamiltonian mechanics and chaos derives its behaviour.


Since both Poincare and Stroboscopic update maps are generated by the flow itself it follows that they are symplectic, that is area preserving, as well. For the stroboscopic maps, which is generated by the flow $\phi_{2\pi/\nu}$ of a strobing period, this fact follows directly.\footnote{The justification for why the Poincare map $P$ is area-preserving is more complicated due to the fact that points on $\Sigma$ return to $\Sigma$ at different times. A detailed explanation can be found in \cite{siegel2012lectures}.} The physics we are concerned with is dominated by this area preservation property and so it is common in the literature to study discrete symplectic maps on manifolds as toy models of the Poincare and stroboscopic maps of Hamiltonian flows. We will follow this convention and so introduce some simple symplectic maps here which we will use later. 

\subsection{Chirikov's standard twist map}

The canonical example in Hamiltonian chaos is called Chirikov's standard twist map \cite{CHIRIKOV1979263}. It can be constructed as the stroboscopic map of a particular Hamiltonian system which models a kicked rotator \cite{ott2002chaos}. Suppose we have a generalised position $q$ which we interpret as an portion of an angle. That is we take $2\pi q$ modulo $2\pi$. We also have a generalised momentum $p$. This forms a phase space $S^1\cross \mathbb{R}$ which we recognise as a cylinder.

The Hamiltonian for our system is 
\begin{equation}\label{KickedRotatorHamiltonian}
    H(p,q;t) = \frac{1}{2}p^2-\frac{k}{(2\pi)^2}\cos(2\pi q)\sum_{n=-\infty}^{\infty} \delta(nT-t)\,,
\end{equation}
where $\delta$ represents the Dirac delta function, $T=2\pi/\nu$ is a perturbation period, and $k$ is a perturbation strength. This is the energy of a free particle $\frac{1}{2}p^2$ which is periodically perturbed by an instantaneous force which varies sinusoidally around the cylinder. The Hamilton equations for $H$ are then
\begin{align}
    &\dv{p}{t} = -\pdv{H}{q} = -\frac{k}{2\pi}\sin(2\pi q)\sum_{n=-\infty}^{\infty} \delta(nT-t)\,,\\
    &\dv{q}{t} = \pdv{H}{p} = p\,.
\end{align}
Now, we construct a stroboscopic map for this system by taking samples of $(p,q)$ at each time $t_n = nT-\epsilon$, for $\epsilon>0$. We define $(p_n,q_n)=(p(t_n),q(t_n))$ and will solve the Hamilton equations to determine $(p_{n+1},q_{n+1})$ in terms of $(p_n,q_n)$. We first integrate over the interval $[t_n,nT+\epsilon]$. In the limit that $\epsilon\rightarrow0$ this is just an integral across the delta function $-\frac{k}{2\pi}\delta(nT-t)$ which contributes an instantaneous jump in $p$ of $\Delta p = -(k/2\pi)\sin(2\pi q_n)$. Over the time interval $t \in [nT+\epsilon,t_{n+1}]$ the Hamilton equations are $\dot{p}=0$ and $\dot{q}=p$. Integrating to $t_{n+1}$ initial condition $(q_n,p_n+\Delta p)$ gives 
\begin{equation*}
    p(t_{n+1}) = p_n-(k/2\pi)\sin(2\pi q_n)\,,\,    \  q(t_{n+1}) = T(p_n-(k/2\pi)\sin(2\pi q_n))+q_n\,.
\end{equation*}
Finally we choose to take $T=1$ to obtain Chirikov's standard map
\begin{align}
    &p_{n+1} = p_n-\frac{k}{2\pi}\sin(2\pi q_n)\,,\nonumber\\
    &q_{n+1} = q_n+p_{n+1}\, \hspace{13mm} (\text{mod } 1)\label{stdmap}\,.
\end{align}

\begin{figure}
    \centering
    \begin{subfigure}[b]{0.4\textwidth}
        \includegraphics[width = \textwidth]{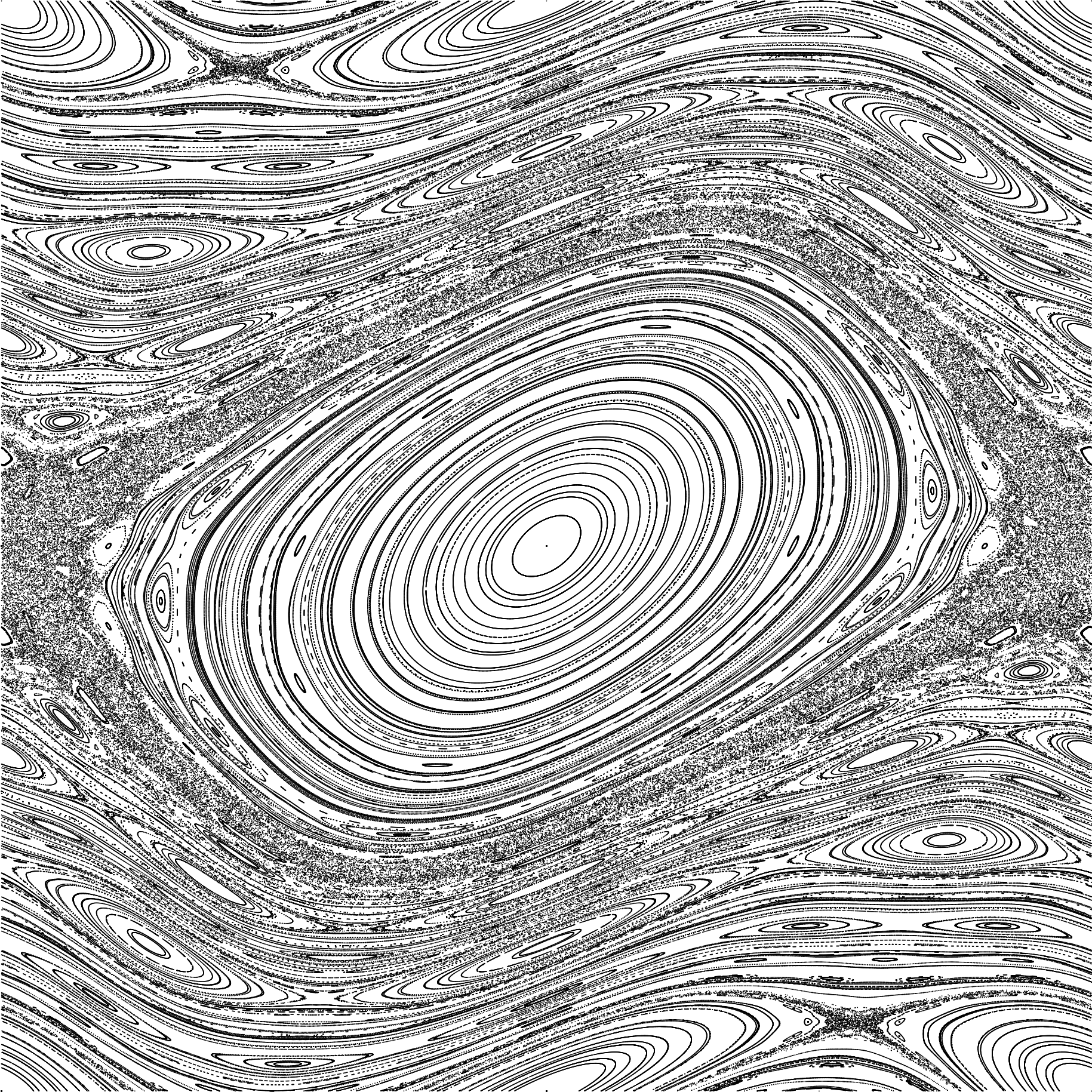}
        \subcaption{Standard map orbits with $k=0.8$.\\ \phantom{Plaeholder.}}
        \label{subfig:SymplecticOrbitsTwist}
    \end{subfigure}
    \begin{subfigure}[b]{0.4\textwidth}
        \includegraphics[width = \textwidth]{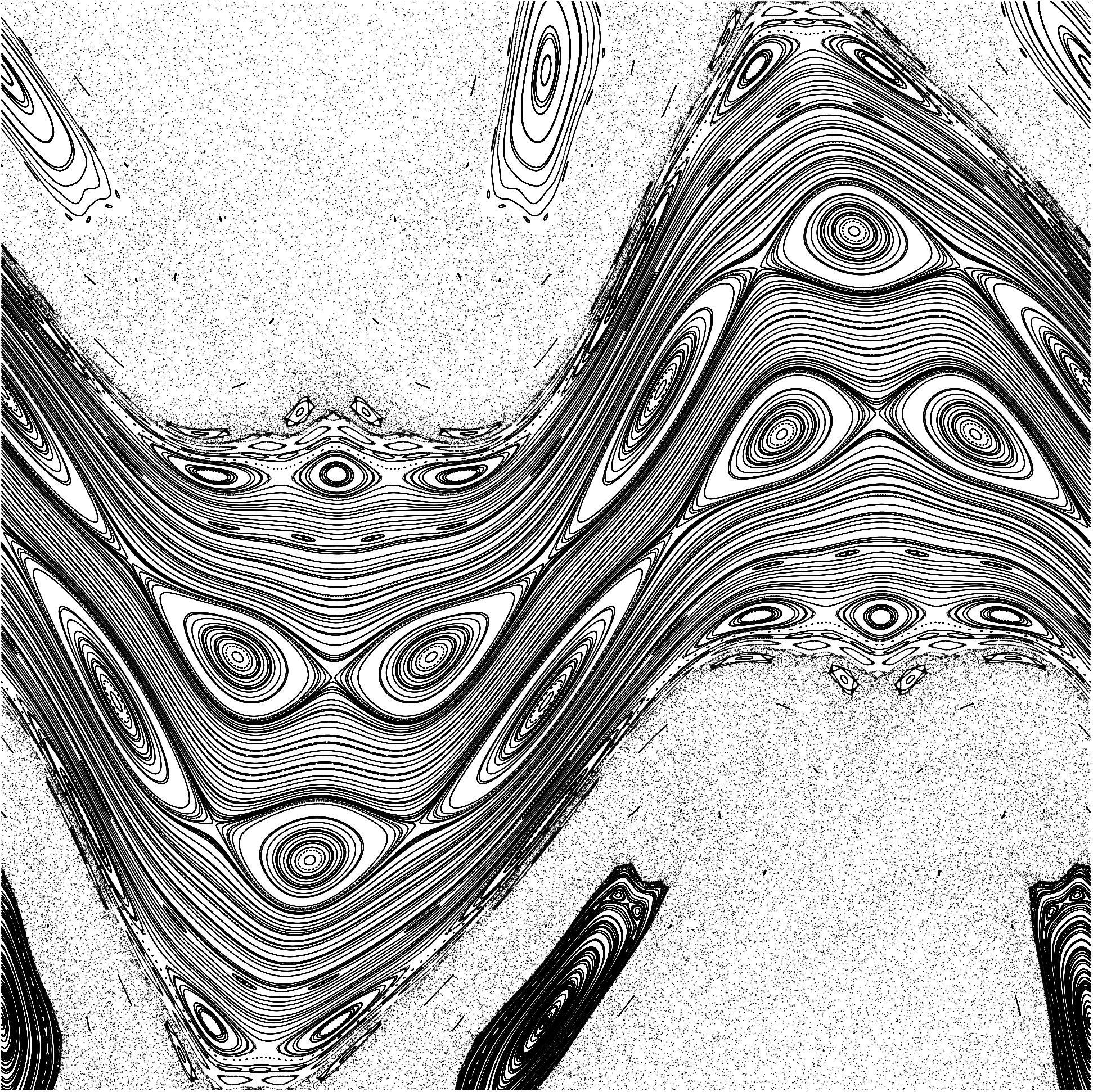}
        \subcaption{Standard non-twist map orbits with $a=0.618$ and $b=0.4$.}
        \label{subfig:SymplecticOrbitsNonTwist}
    \end{subfigure}
    \caption{Results of computing many orbits for our two model symplectic maps.}
    \label{fig:SymplecticOrbits}
\end{figure}

This is a symplectic map as can be seen by computing the determinant of the Jacobian of the map. The standard map is commonly used as a local model for the evolution of $1\frac{1}{2}$D Hamiltonian system in the neighborhood of an elliptic fixed point. When $k=0$ the model is integrable and each point in the phase space is either on a periodic orbit or an invariant circle. As $k$ is increased the invariant circles breakup and we observe a similar island structure in the phase space as was discussed for Poincare maps of other Hamiltonian flows in Section \ref{StructuresInPhaseSpaces}. A plot of many orbits of the standard map is presented a subfigure \ref{subfig:SymplecticOrbitsTwist}.

\subsection{The standard non-twist map}

Chirikov's map is representative of area preserving maps in the wide class of \textit{twist} maps. This class refers to maps $(p_n,q_n)\rightarrow(p_{n+1},q_{n+1})$ such that for all points in the phase space $\partial q_{n+1}/\partial p_n \neq 0$. It is common for this to be true in physical systems but it not universal, so it is useful to study examples of maps which are area-preserving but do not satisfy the twist condition. Our example of choice is Morrison's standard nontwist map \cite{del1996area,Morrison2000}. Which is defined as
\begin{align}
    &p_{n+1} = p_n-b\sin(2\pi q_n)\,,\nonumber\\
    &q_{n+1} = q_n+a(1-p_{n+1}^2)\,,
\end{align}
where $q$ is taken as $q\in [-0.5,0.5]$ and treated $\text{mod } 1$. This is an area-preserving map but it fails the twist condition. The map is interesting for several reasons, for example the invariant circles of the map breakup differently than in twist maps. Specifically, they converge to continuous but not differentiable curves in the plane at the point of breakup \cite{del1996area}. A plot of the orbits many points in the plane under this map is presented as subfigure \ref{subfig:SymplecticOrbitsNonTwist}.


\section{Fast chaos detection with the Weighted Birkhoff Average}

In Chapter \ref{Chapter4} we will require a fast method to compute binary images of the phase space of an area preserving map with white and black pixels indicating whether the points belong on stochastic orbits or not respectively. This will allow us to image the islands in the phase space. The method we adopt is based on a recent approach to chaos detection using the Weighted Birkhoff Average (WBA) first presented by Sanders and Meiss \cite{SandersAndMeiss}. Here we will present an informal introduction to the WBA method.

Let $f$ be the update map for a discrete dynamical system on a phase space $\mathcal{M}$. Let $T \in \mathbb{N}$ correspond to some number of iterations and $h:\mathcal{M} \rightarrow \mathbb{R}$ be some observable of the dynamics. For our purposes an ``observable'' is just a continuous function from the phase space to $\mathbb{R}$. As an example, for the standard map $h(p,q) = \cos(q)$ is a possible observable function. For a point $x\in \mathcal{M}$ in the phase space define the finite time WBA of the observable $h$ to be 
\begin{equation}
    WBA_T(x;h) = \frac{1}{T}\sum_{t=0}^{T-1}h(f^t(x))g(t/T)\,,
\end{equation}
where $g\propto e^{(t(1-t))^{-1}}$ is a normalised bump function and $f^t$ refers to the composition of $t$ copies of the $f$ function. As an average of an observable, if we consider $WBA_T(x;h)$ for larger and larger $T$ we expect that $WBA_T(x;h)$ will converge to a stable long-run value. This will be true for all points in the phase space but points will not necessarily converge at equal rates. It was shown by Das and Yorke that for periodic and quasi-perioidic orbits of Hamiltonian systems\footnote{Their proof actually covers the more general case of arbitrary ergodic dynamical systems of which Hamiltonian flows are a subset.} the rate of convergence is super-polynomial in $T$ \cite{Das_2018}. In contrast, the convergence rate for stochastic points are not guaranteed to be super-polynomial and is found in practice to be much slower than that of the quasiperiodic points \cite{SandersAndMeiss}. To check if a point $x$ is stochastic we attempt to determine the convergence rate of the WBA for a point $x$. We do this by computing both $WBA_T(x;h)$ and $WBA_T(f^T(x;h))$ and measuring how similar these approximations to the WBA of the are. 

So, we select a particular $T\in\mathbb{N}$ and define the function
\begin{equation}
    D(x;h) = -\log_{10} |WBA_T(x;h)-WBA_T(f^T(x);h)|\,,
\end{equation}
which is the number of equivalent digits in $WBA_T(x;h)$ and $WBA_T(f^T(x);h)$. For points on periodic or quasiperiodic orbits, $D$ will be large and for points of chaotic orbits it will be small regardless of the choice of observable $h$. Hence checking the value of $D$ for a point with a given observable provides a method of determining whether the function lies on an ordered or a chaotic orbit.  

For convenience we will define a normalised $D$ function by
\begin{equation}
    \tilde{D}(x;h) = \frac{D(x;h)}{\max_{x'\in \mathcal{M}}D(x';h)}\,.
\end{equation}
If we compute $\tilde{D}(x;h)$ for many $x$ taken from a subset of $\mathcal{M}$ we typically observe that the values $\tilde{D}$ from two clusters. One cluster at small $\tilde{D}$, and one at large $\tilde{D}$, that is $\tilde{D}\approx1$, corresponding to the stochastic points and islands respectively. 

\begin{figure}
    \centering
    \begin{subfigure}[b]{0.4\textwidth}
        \includegraphics[width = \textwidth]{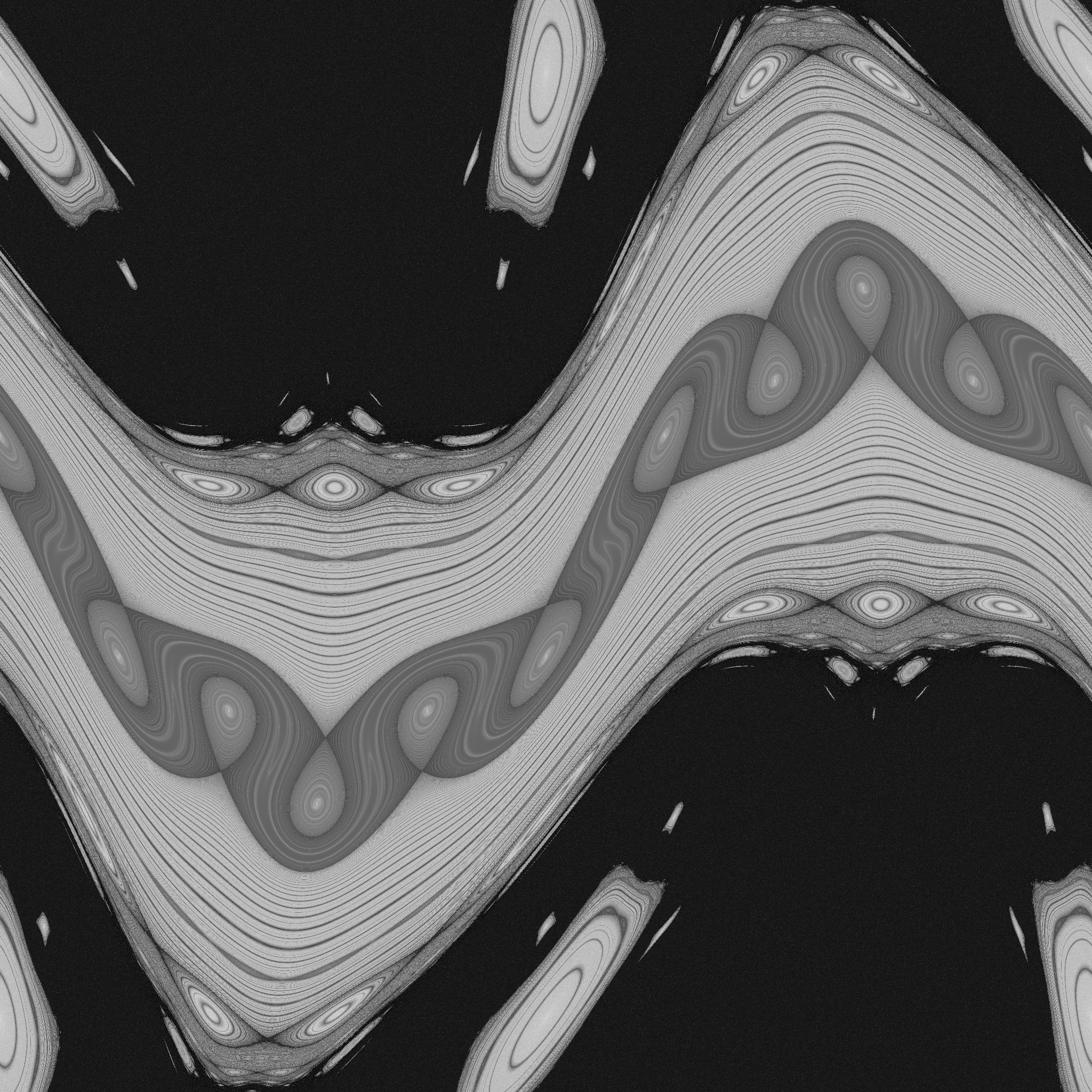}
        \subcaption{}
        \label{subfig:DimageNontwist}
    \end{subfigure}
    \begin{subfigure}[b]{0.4\textwidth}
        \includegraphics[width = \textwidth]{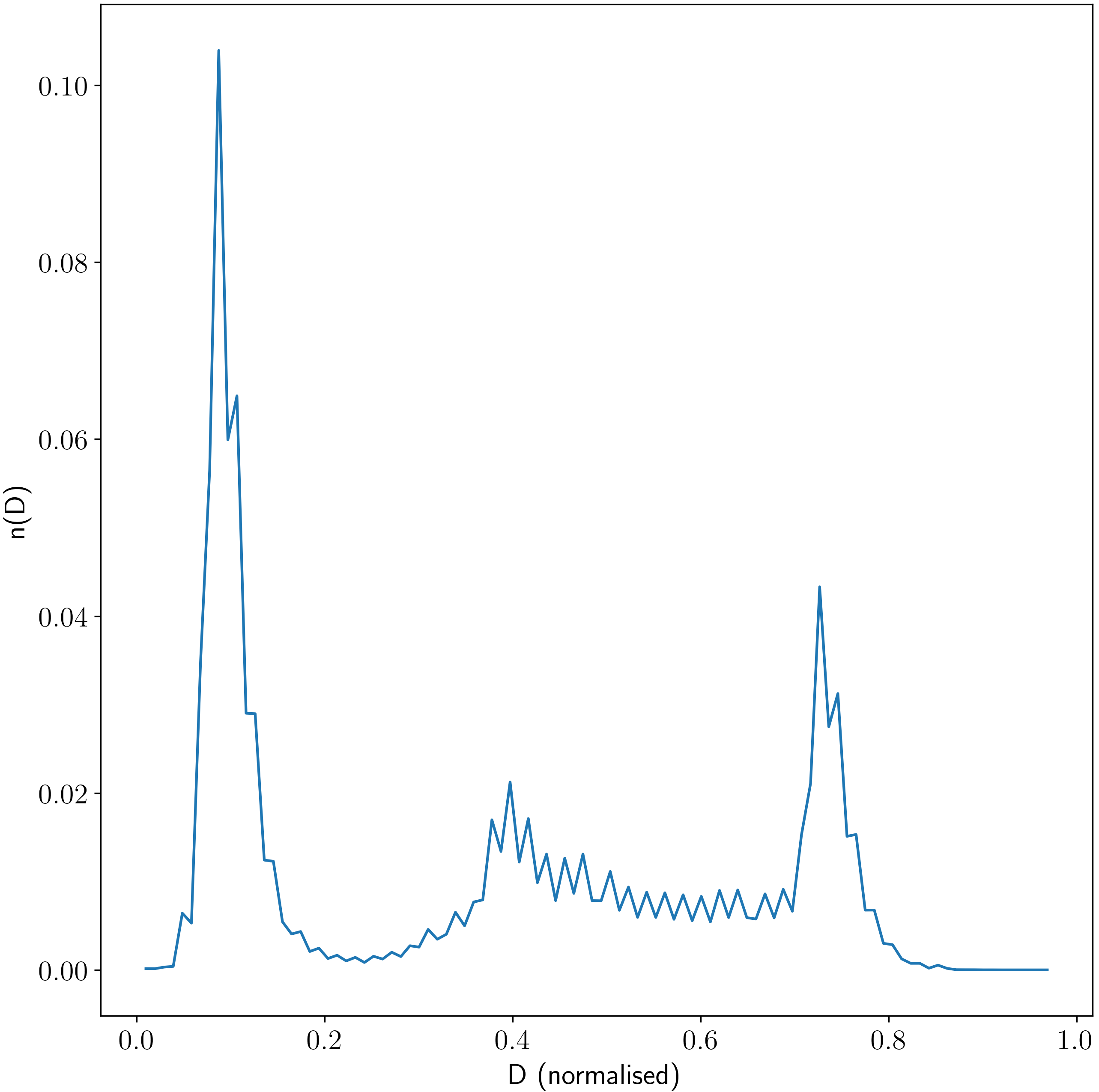}
        \subcaption{}
        \label{subfig:DimageNontwistHistogram}
    \end{subfigure}
    \caption{Grayscale WBA image for the standard non-twist map and the associated histogram of pixel brightness.}
    \label{fig:standard_nontwist_wba}
\end{figure}

We are interested in the case of imaging the islands in the phase space and so will compute $\tilde{D}$ across a uniform grid and interpret the value as the $\tilde{D}$ function as the brightness of an image. To demonstrate that the brightness usually forms two clusters we will consider the example of the standard non-twist map. Computing $\tilde{D}$ at every point in a $2048^2$ lattice over the domain $(x,y)\in[-0.5,0.5]^2$ for a time of $T=5000$ yields the grayscale image presented as subfigure \ref{subfig:DimageNontwist}. Displayed as subfigure \ref{subfig:DimageNontwistHistogram} is the associated normalised\footnote{That is with an $L^1$ integral of unity.} histogram of pixel brightness values $n(\tilde{D})$. There are two clusters of $n(\tilde{D})$ values in the histogram, one peaked near $\tilde{D}\approx0.08$ corresponding to the points in the stochastic region, and one peaked at $\tilde{D}\approx0.7$ associated to points which are periodic or quasi-periodic and hence not stochastic. There is also a third smaller peak at $\tilde{D}\approx0.4$, but the pixels in this cluster are associated with those in the cluster at $\approx0.7$ as they both correspond to non-stochastic orbits. Specifically, the $0.4$ peak is due to the band of darker grey pixels visible in the middle of subfigure \ref{subfig:DimageNontwist} which are on non-chaotic orbits but whose WBA converges more slowly because they have highly irrational frequency.

The location of the minimum value between stochastic and non-stochastic peaks can be used as a cutoff to distinguish stochastic and ordered points. In the case of subfigure \ref{subfig:DimageNontwistHistogram} the minimum is located at $\tilde{D} \approx 0.232$ and so we say that all points whose pixel value is less than this are stochastic and the others not. This gives us an approximation to an indicator function which identifies stochastic points. We will return to this discussion in more detail in Chapter \ref{Chapter4}.

\section{The Signed Euclidean Distance Transform}

In Chapter \ref{Chapter4} of this thesis we will use the indicator function discussed above to render binary images of the islands in the phase space of area-preserving maps. We will then want to quantify the ``size'' of each island and we can do this with the Signed Euclidean Distance Transform (SEDT) \cite{ye1988signed}. The SEDT is a function we can apply to a binary image which produces a new grayscale image of the same dimensions whose values indicate the size of connected black and white regions in the input image. The operation is defined as follows. 

\begin{defn}
    For a binary image with pixel coordinates $(x,y)$ the SEDT of a pixel $(x_p,y_p)$ is the euclidean distance $\pm\sqrt{(x_p-x_b)^2+(y_p-y_b)^2}$ to the nearest border $(x_b,y_b)$ where the positive sign is taken if $(x_p,y_p)$ has pixel value $1$ and the negative sign if the opposite.
\end{defn}

Note that ``border'' in the above refers to a transition line between a black and white pixel.


\chapter{Orbit classification} 

 \label{Chapter3} 

The automated characterisation of the orbits of points under iterated maps, or flows of vector fields, is a problem of practical scientific value. For example, magnetic field lines confined to toroidal surfaces pose a barrier to transport for particles orbiting in a tokamak \cite{joffrin2003internal}. Therefore by detecting these field lines it is possible to partition the domain of the tokamak into regions between which particle transport, and therefore energy transport, is suppressed and hence characterise this transport. Here we will present a TDA approach to the characterisation of magnetic field line orbits in a tokamak from their Poincare section through the use of Vietoris-Rips persistent homology. We will test our method numerically using a toy model of a tokamak magnetic field. 

We will demonstrate empirically that the different classes of field line orbit in a tokamak are topologically distinct. Our classification procedure leverages this fact to determine the class of each field line by approximating the topology of an orbit from the topology of the point cloud it forms on a Poincare section.

Computing the topology of a trajectory of a dynamical system using TDA requires that we have a finite representation for that trajectory. To obtain such a representation for a Poincare update map, we pick an initial point and then compute the trajectory for only a finite number of iterations in the forward time direction. For field lines this means we should pick an initial point on the Poincare section $\Sigma$ and then integrate the field line flow, but stop integrating when the number of intersections between the field line and $\Sigma$ reaches a desired threshold. This gives a finite set of points in $\Sigma$ which admit the computation of the VR persistent homology. It will be useful later to define some notation 
 
\begin{defn}
    Suppose we have a magnetic field, or general dynamical system, and have selected a Poincare section $\Sigma$. Take that $P:\Sigma\rightarrow\Sigma$ is the Poincare update map induced on $\Sigma$ by field line flow. Given an integer $T$ and a point $x\in \Sigma$ we define the \textit{trajectory} of the point $x$ for finite time $T$ to be the set 
    \begin{equation}
        X_T(x) = \left\{ P^{(t)}(x),\text{ for } t \in 0,1,2,\ldots T\right\}\subset \Sigma\,.
    \end{equation}
\end{defn}

\section{Toy model for a perturbed tokamak field}

As our test magnetic field we will use a toy model for a tokamak field formed from a circular current loop and an infinite straight line current. This model is convenient theoretically because it has the expected qualitative structure of a tokamak field, that is of helical field lines confined to toroidal surfaces, but can be analytically calculated and therefore the computation of the field lines for this field is fast \cite{Morrison2000}. The exact analytic expressions for the required fields are presented in Appendix \ref{AppendixA}. So, we write our toy tokamak field as
\begin{equation}
    \textbf{B}(\textbf{x}) = \textbf{B}_t(\textbf{x})+\textbf{B}_p(\textbf{x})\,,
\end{equation}
where $\textbf{B}_t$ is the field generated by a straight line current $I_z$ and $\textbf{B}_p$ is the field generated by a circular current $I_\phi$. The labels $t$ and $p$ refer to the \textit{toroidal} and \textit{poloidal} components of the field and are conventional in the discussion of the toroidal geometry of a tokamak. Also, we will refer to the location of the current loop as the \textit{magnetic axis}. Again, this follows another convention of tokamak geometry.

\begin{figure}
    \centering
        \includegraphics[width = 0.7\textwidth]{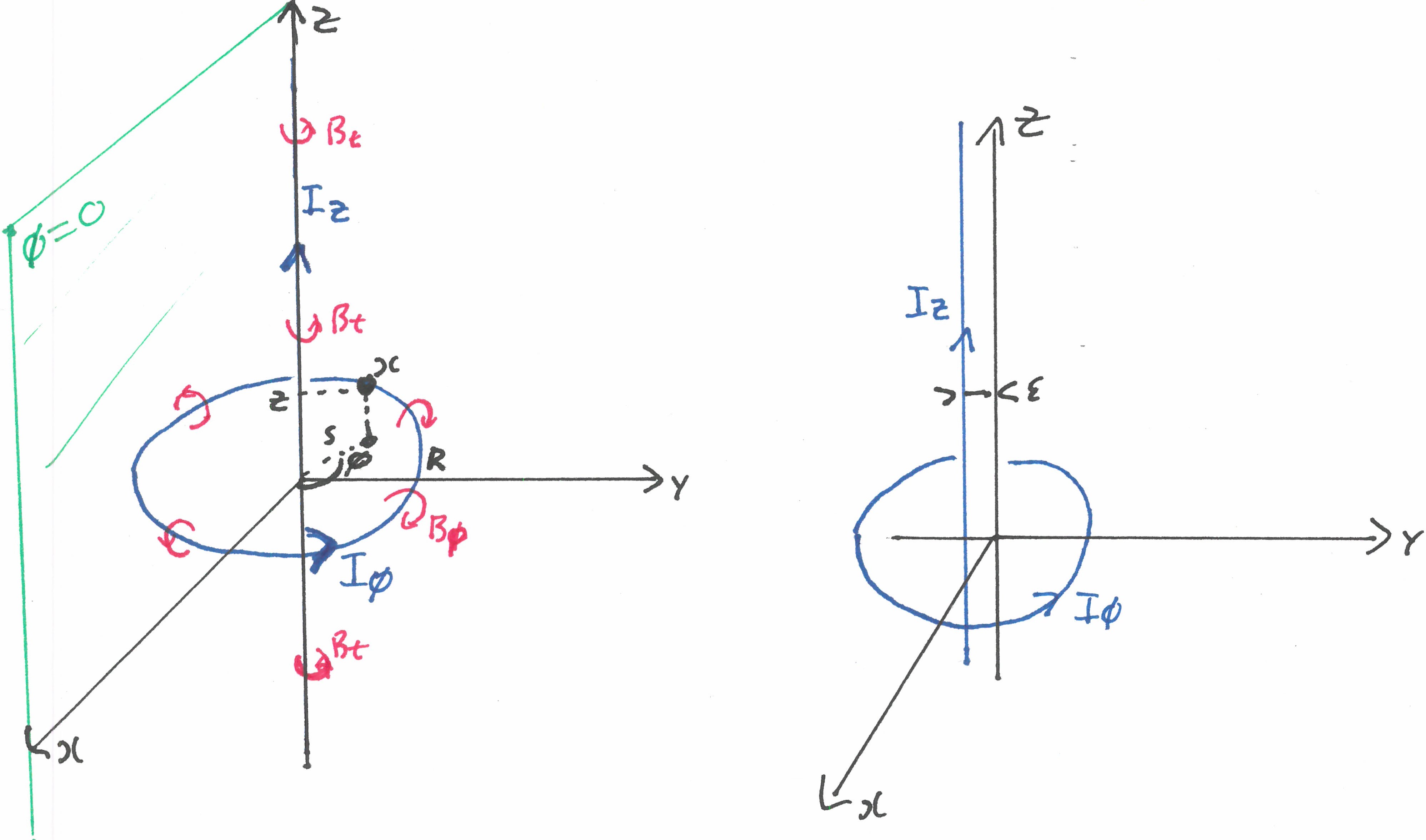}
    \caption{Cartoon of the current configuration for our toy tokamak model. Note that, for our simulations $R=1$}
    \label{fig:ToyTokamakModel}
\end{figure}

The specific field generated by these currents depends upon the geometry of the current configuration. We adopt the geometry presented in Figure \ref{fig:ToyTokamakModel}. Note that if we adopted only the symmetric configuration presented on the left our magnetic field lines would be an integrable Hamiltonian system. That is, all field lines would lie on invariant circles in the Poincare section and therefore there would not be any different classes of field lines to classify \cite{Morrison2000}. Therefore, to accurately represent the chaotic field line case, in which we are interested, we must explicitly break the symmetry of the setup. This is demonstrated in Figure \ref{fig:ToyTokamakModel} by the $\varepsilon$ displacement of $I_z$ away from the $z$-axis. This perturbation breaks the symmetry of our setup and causes the field lines to become chaotic, generating the same islands and island chains we saw in Chapter \ref{Chapter1}. 

Note that all Poincare sections presented below were computed for the case where 
\begin{equation}
    \frac{I_z}{I_\phi} = 1\,, \,   \ \frac{\varepsilon}{R} = 0.005\,.
\end{equation}
It is only these ratios which meaningfully effect the field lines. This is because $\textbf{B}$ is linear in the current density and so increasing both currents by a uniform factor increases $\textbf{B}$ globally by the same factor but does not change the geometry of its field lines. Similarly, increasing both $\varepsilon$ and $R$ by a constant factor corresponds only to rescaling the dimensions of length, a dilation transformation, and so similarly does not affect the geometry. We also adopt the half-plane defined by $\phi=0$ in cylindrical coordinates as our Poincare section $\Sigma$. This plane is shown explicitly in Figure \ref{fig:ToyTokamakModel}.


\section{VR persistent homology of magnetic field line orbits}

Here we consider the case of the VR persistent homology of the intersections of single magnetic field lines with $\Sigma$. We will compute the persistent homology and analyse the relationship between the geometric features of the orbit and both the persistence diagram and the relative persistence of the classes. 

\subsection{A KAM torus}

\begin{figure}
    \centering
        \includegraphics[width = 0.4\textwidth]{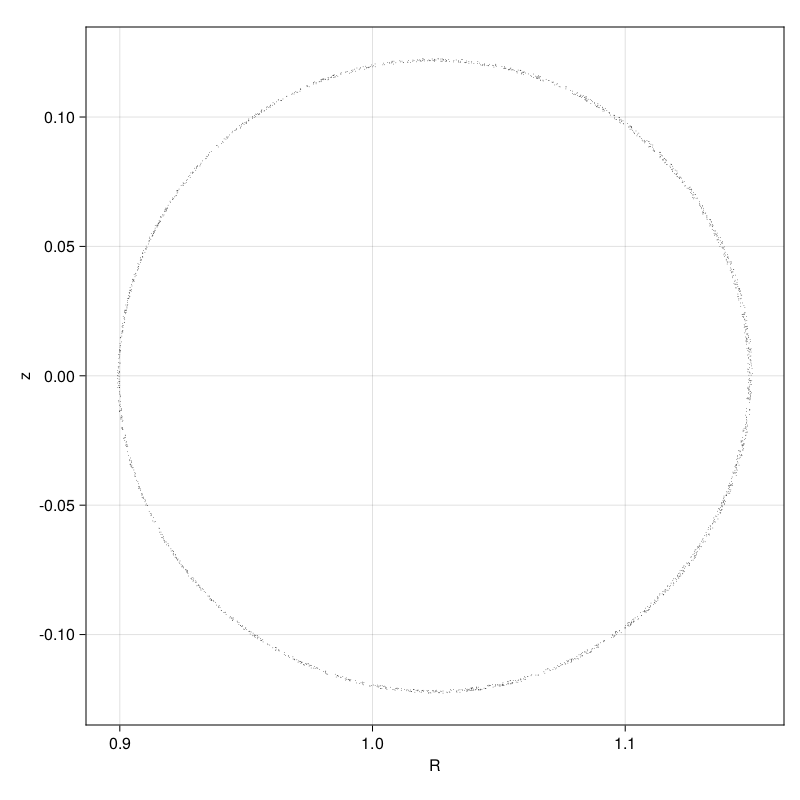}
    \caption{Point cloud of a field line on a KAM torus.}
    \label{fig:kamtorus_section}
\end{figure}

\begin{figure}
    \centering
    \begin{subfigure}[b]{0.39\textwidth}
        \centering
        \includegraphics[width = \textwidth]{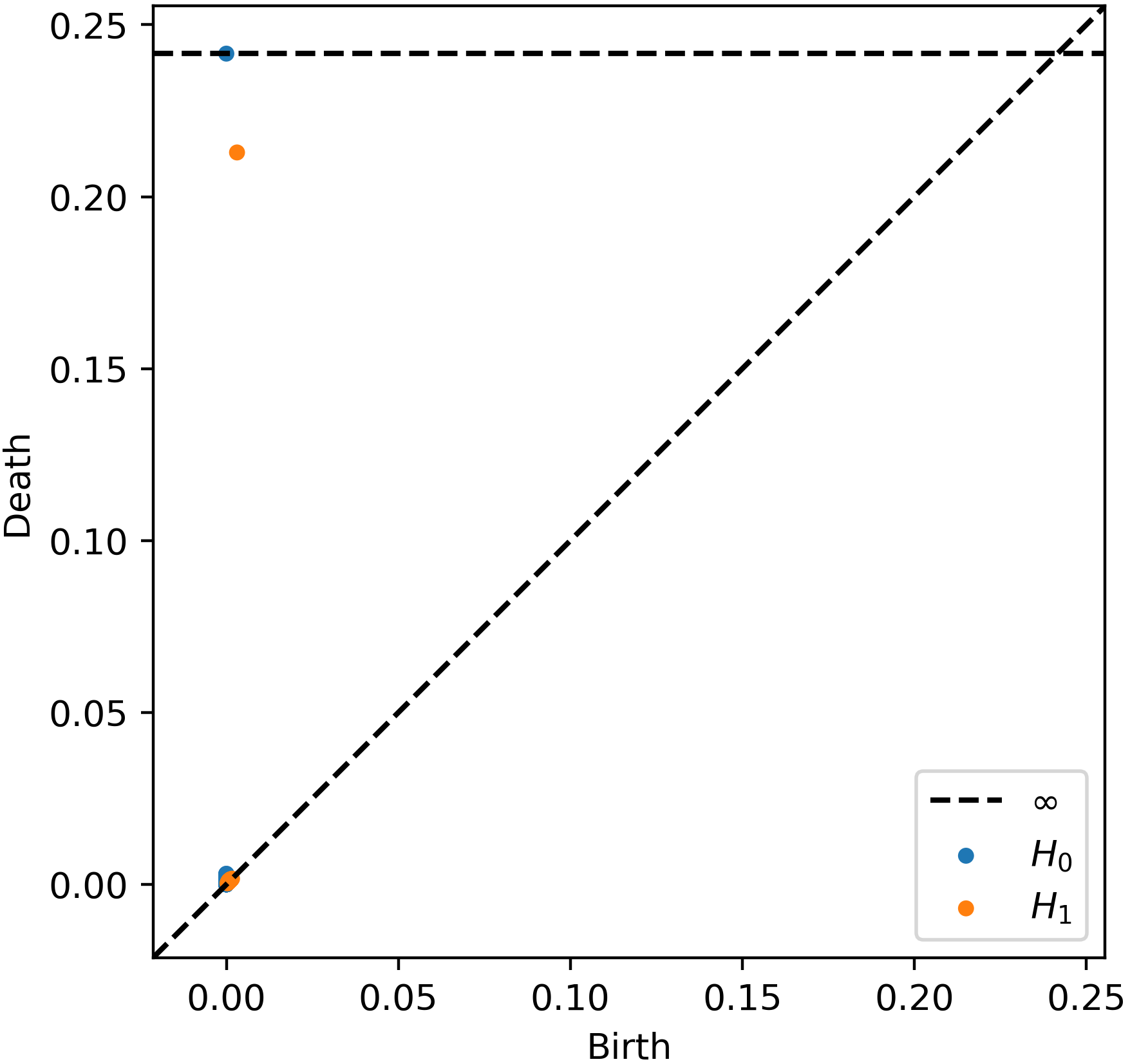}
        \subcaption{VR persistence diagrams.}
        \label{subfig:torusRips}
    \end{subfigure}
    \begin{subfigure}[b]{0.50\textwidth}
        \centering
        \includegraphics[width = \textwidth]{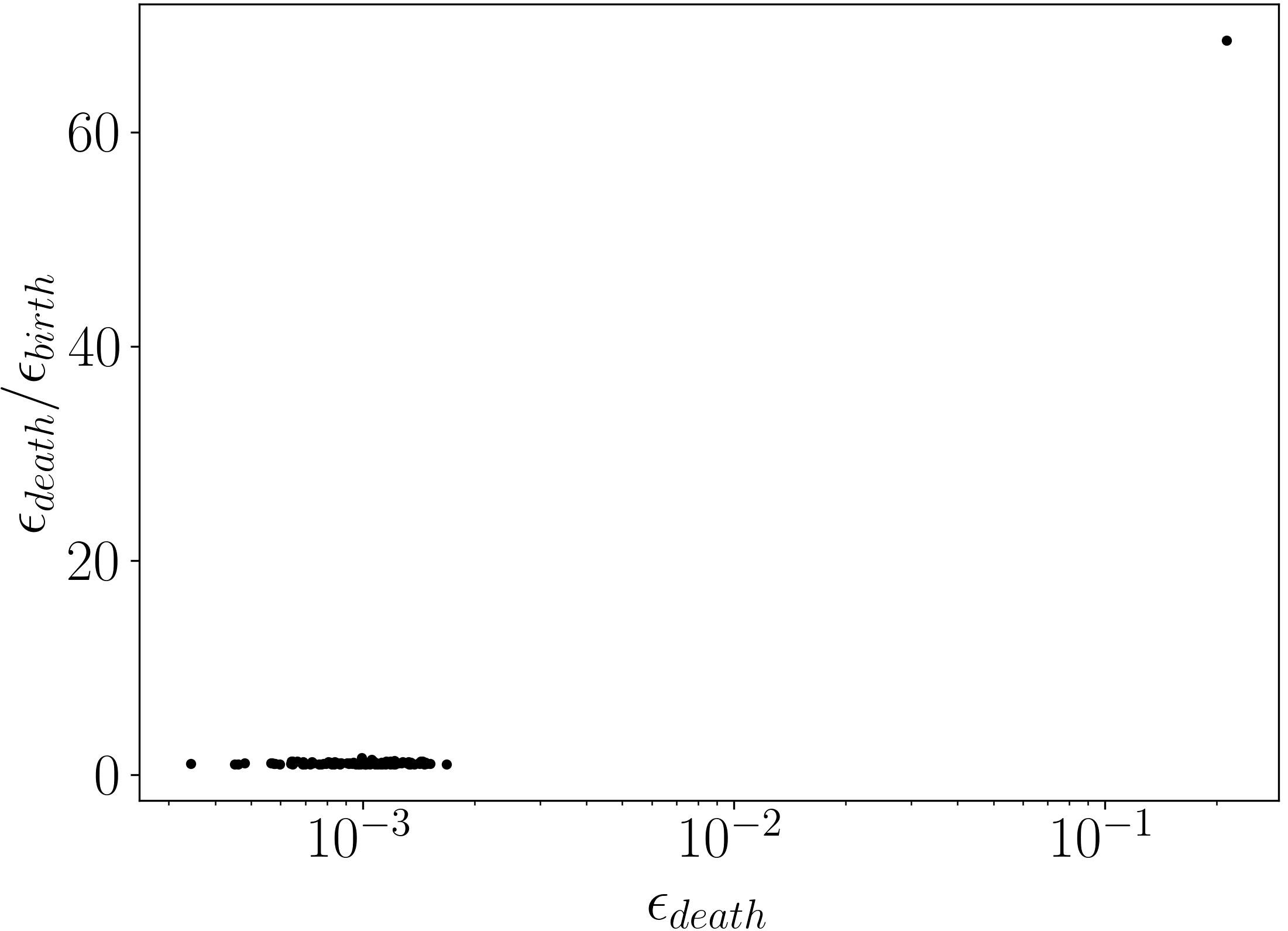}
        \subcaption{$PH_1$ class death to birth ratio.}
        \label{subfig:torusDB}
    \end{subfigure}
    \caption{VR homology results of the KAM torus shown in Figure \ref{fig:kamtorus_section}.}
    \label{fig:kamtorus_section_rips}
\end{figure}

The simplest field line orbit we can consider is that of a KAM torus. In the 3-d space in which the field lines live such an orbit densely covers a 2-torus surrounding the magnetic axis. We therefore expect that after projecting to $\Sigma$ the orbit will form a topological circle containing the magnetic axis. Figure \ref{fig:kamtorus_section}, which presents the orbit of a magnetic field line on a KAM torus, confirms this expectation. Computing the VR persistent homology of this point cloud with the Python software \verb!Ripser! yields the persistence diagram presented as subfigure \ref{subfig:torusRips}. We observe a single $H_0$ class which exists for all diameters\footnote{The horizontal dotted line indicates that the class does not die by the end of the scan.} corresponding to the fact that for $\epsilon$ larger than the diameter of the point cloud the VR complex is a connected simplicial complex. 

We also observe a very long lived $H_1$ class, which encodes the circular shape of the point cloud. A similar feature will appear in many of the cases which follow and its persistence is associated to whether our field line encloses the magnetic axis. Note that by ``encloses'' here we are referring to whether or not the surface, or volume, filled by the field line separates the 3-d space into two disconnected regions; one containing the magnetic axis and the other not. The 2-torus surfaces of KAM toruses do separate space into two such regions and so we say they enclose the axis. In this case the point is very persistent, with a large death time to birth time ratio of $\epsilon_{\text{death}}/\epsilon_{\text{birth}}\approx 70$, and this suggests that it is highly likely that our field line encloses the magnetic axis, which it actually does in this case. 

\subsection{A magnetic island}

The next simplest type of field line orbit to consider after KAM toruses are those associated to islands and island chains. The field lines of magnetic islands islands do not separate the magnetic axis from the rest of the space\footnote{Really from the point at infinity.}. This means that their Poincare section can be made of several connected components and does not enclose the axis. For low order resonant perturbations, such as ours, the islands have a characteristic ``banana'' shape \cite{Zaslavsky}.

\begin{figure}
    \centering
    \begin{subfigure}[b]{0.4\textwidth}
        \centering
        \includegraphics[width = 1.0\textwidth]{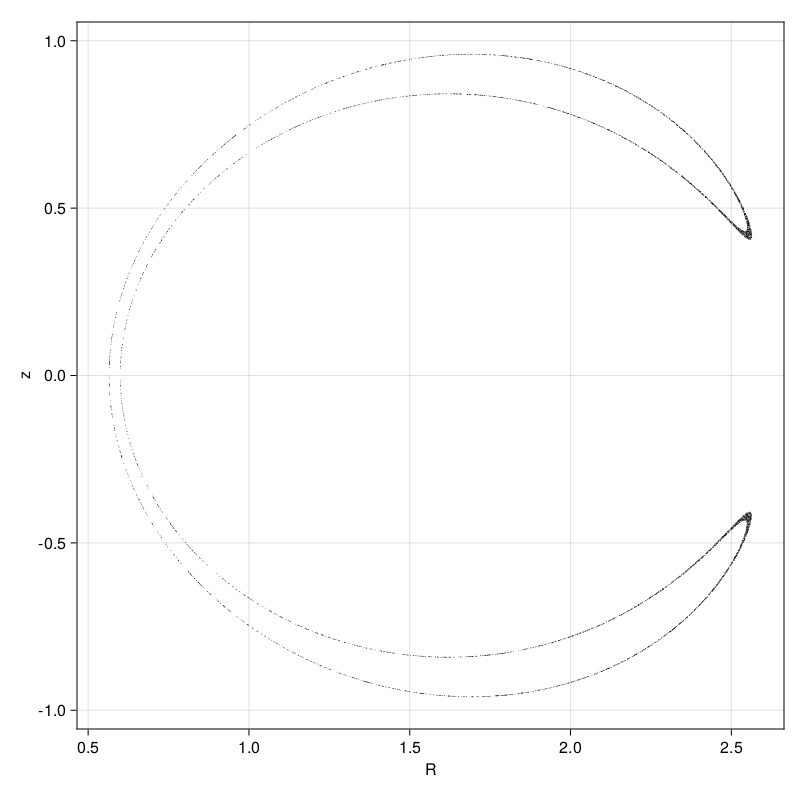}
        \subcaption{Poincare section.}
        \label{subfig:islandSection}
    \end{subfigure}
    \begin{subfigure}[b]{0.39\textwidth}
        \centering
                \includegraphics[width = \textwidth]{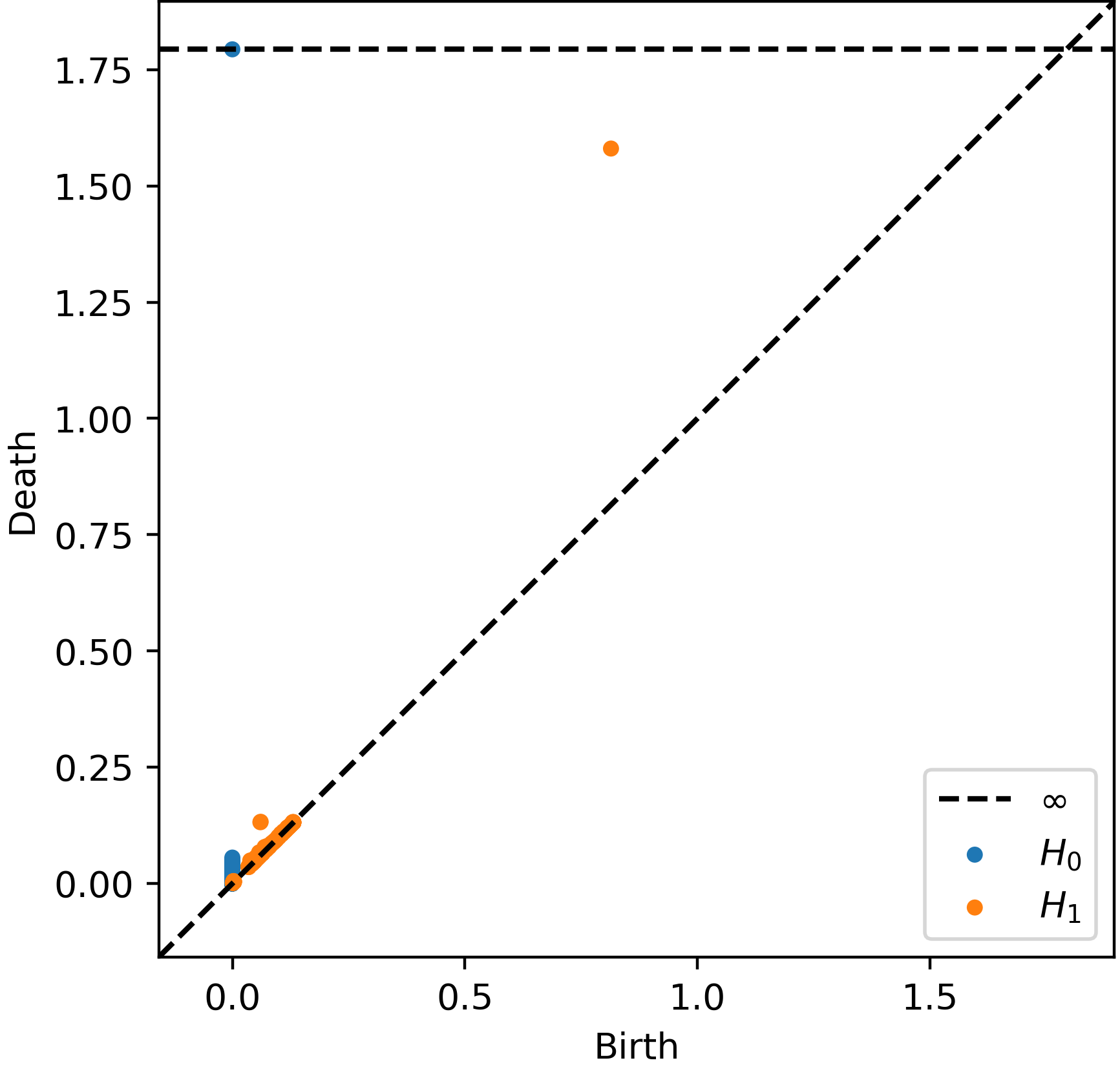}
        \subcaption{VR persistence diagrams.}
        \label{subfig:islandRipsPH}
    \end{subfigure}
    \caption{Point cloud and $PD$ for an order one island.}
    \label{fig:island_section_and_rips}
\end{figure}

An example of such an island in the field of our perturbed toy tokamak is presented as subfigure \ref{subfig:islandSection}. The point cloud appears as an approximately closed loop which does not enclose the magnetic axis. The persistence diagram, shown as subfigure \ref{subfig:islandRipsPH}, again includes the expected infinite lifetime $H_0$ class corresponding to connectivity. We can observe two interesting features in the $H_1$ persistence diagram: there is an $H_1$ class born at $\epsilon_{\text{birth}}\approx 0.7$ which is associated to the fact that our island nearly closes around the axis; and there appears to be another $H_1$ class born at $\epsilon_{\text{birth}}\approx0.1$ which we recognise is associated to the fact that our field line is a single closed island, however this class dies early $\epsilon_{\text{death}}\approx 0.15$ because the banana is a very narrow. These two features have very different absolute persistence $\epsilon_{\text{death}}-\epsilon_{\text{birth}}$ but similar relative persistence $\epsilon_{\text{death}}/\epsilon_{\text{birth}}$. This is a good example of why the relative persistence is the preferred measure of the statistical significance of a topological feature when the shape in question contains structures on several length scales since it does not weight geometrically larger features more heavily \cite{bobrowski2023universal}.

\begin{figure}
    \centering
        \includegraphics[width = 0.6\textwidth]{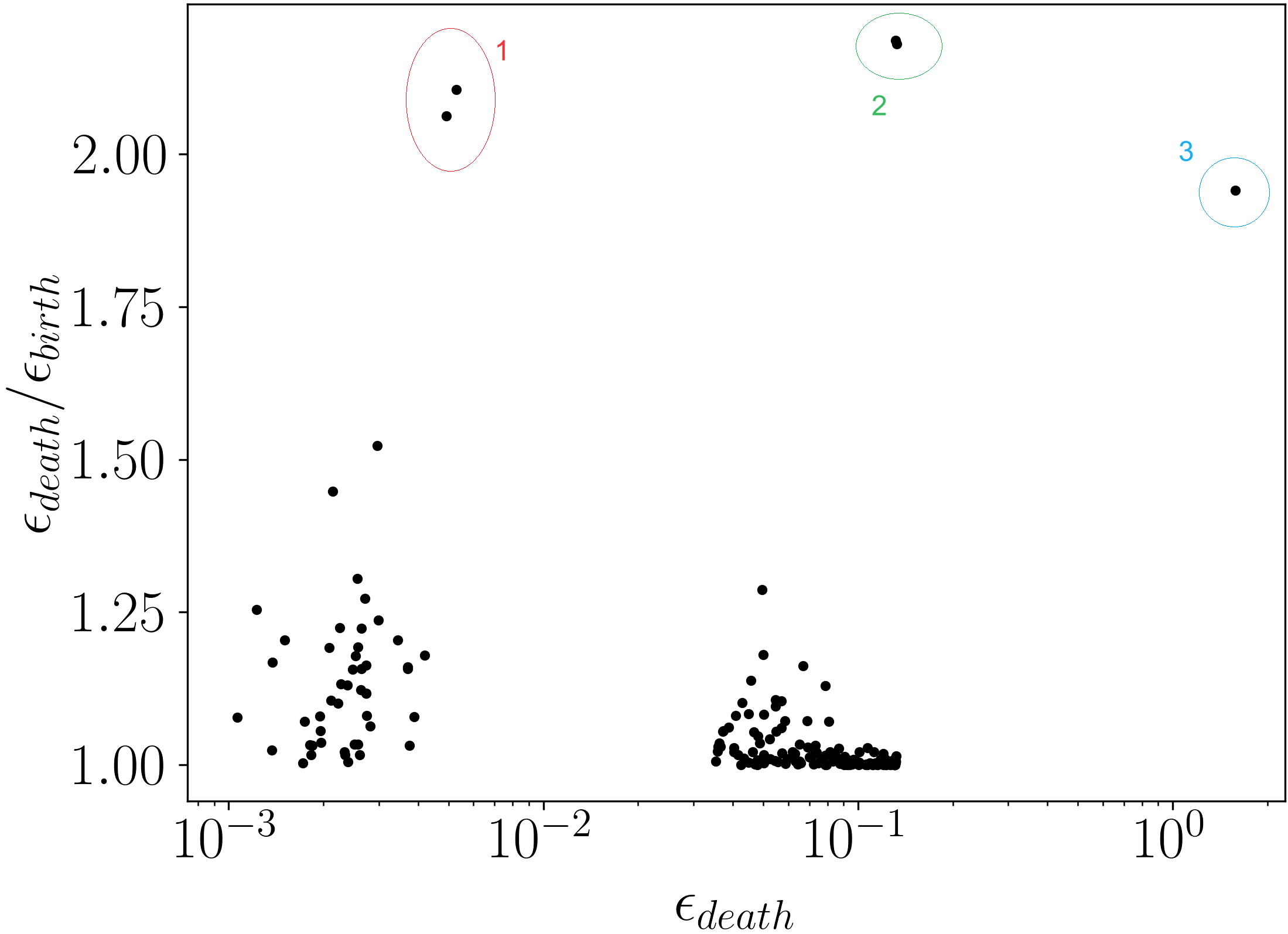}
    \caption{$PH_1$ relative persistence of the island shown as subfigure \ref{subfig:islandRipsPH}. The clusters $1,2,$ and $3$ are indicated in red, green, and blue respectively.} 
    \label{fig:island_section_bdrat}
\end{figure}

To see that the relative persistence is similar we consider Figure \ref{fig:island_section_bdrat} which presents the relative persistence for the classes in subfigure \ref{subfig:islandRipsPH}. We observe three clusters of features with high relative persistence at $\epsilon_{\text{death}} \approx 5(10^{-3})$, $10^{-2}$, and $2$. Which we will refer to as clusters $1$, $2$, and $3$ respectively. 

There are two classes in cluster $1$ and they are each associated to one of the two darker regions on the outboard side of the toy tokamak. The point cloud is ``thick'', fills a two-dimensional area, in these regions. The exact orbit of the field line would not be ``thick'' here but our computation of the field line orbit is not exact and so the integration error compounds at these points, where the field line changes direction quickly, and the field lines therefore start to fill out a $2$D region. There exists small gaps between the points in the Poincare section at the filled in regions, allowing for the formation of closed loops in the Vietoris-Rips complex, and this contributes a cluster of small diameter $H_1$ classes to $PH_1$ which we recognise as cluster $1$. 

Cluster $2$ is associated with the simple closed loop shape of the banana orbit but the reasonably low death time to birth time ratio of $\approx 2.2$ indicates that from only the specific point cloud examined we cannot be completely confident that the field line forms a simple closed curve. This is because near to the left edge, the Poincare section is somewhat sparse with an average distance between individual points on the order of the minimum width of the banana. This contrasts with the case of Figure \ref{fig:kamtorus_section} where the average spacing between points was several orders of magnitude less than the minimum width of the loop and therefore were very confident that the trajectory is a simple closed curve. We would expect that if we followed the field line for longer and included more samples from the orbit the maximum relative persistence of a class in cluster $2$ would increase
and we would be more confident that the field line orbit is a closed loop in the Poincare section. 

Finally, cluster $3$ is associated with enclosure around the magnetic axis and the fact that its relative persistence is lower than that of clusters $1$ and $2$ indicates that from the point cloud data we should not be confident that the field line surrounds around the axis. It is good that we are not confident in this since the field line does not surround the axis.

\begin{figure}
    \centering
    \begin{subfigure}[b]{0.4\textwidth}
        \centering
        \includegraphics[width = 1.0\textwidth]{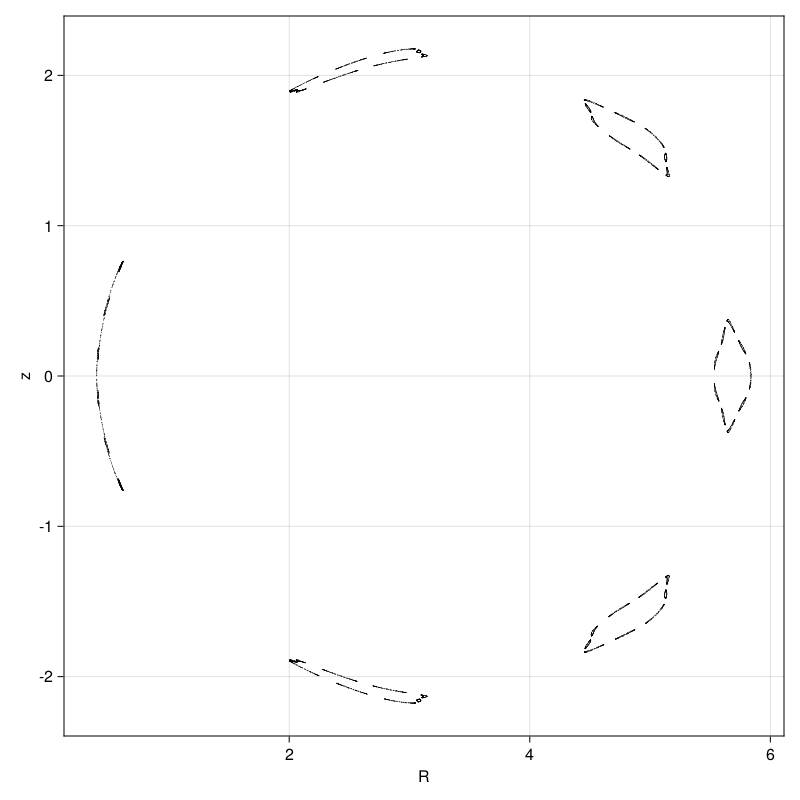}
        \subcaption{Poincare section.}
        \label{subfig:island_chainSection}
    \end{subfigure}
    \begin{subfigure}[b]{0.39\textwidth}
        \centering
                \includegraphics[width = \textwidth]{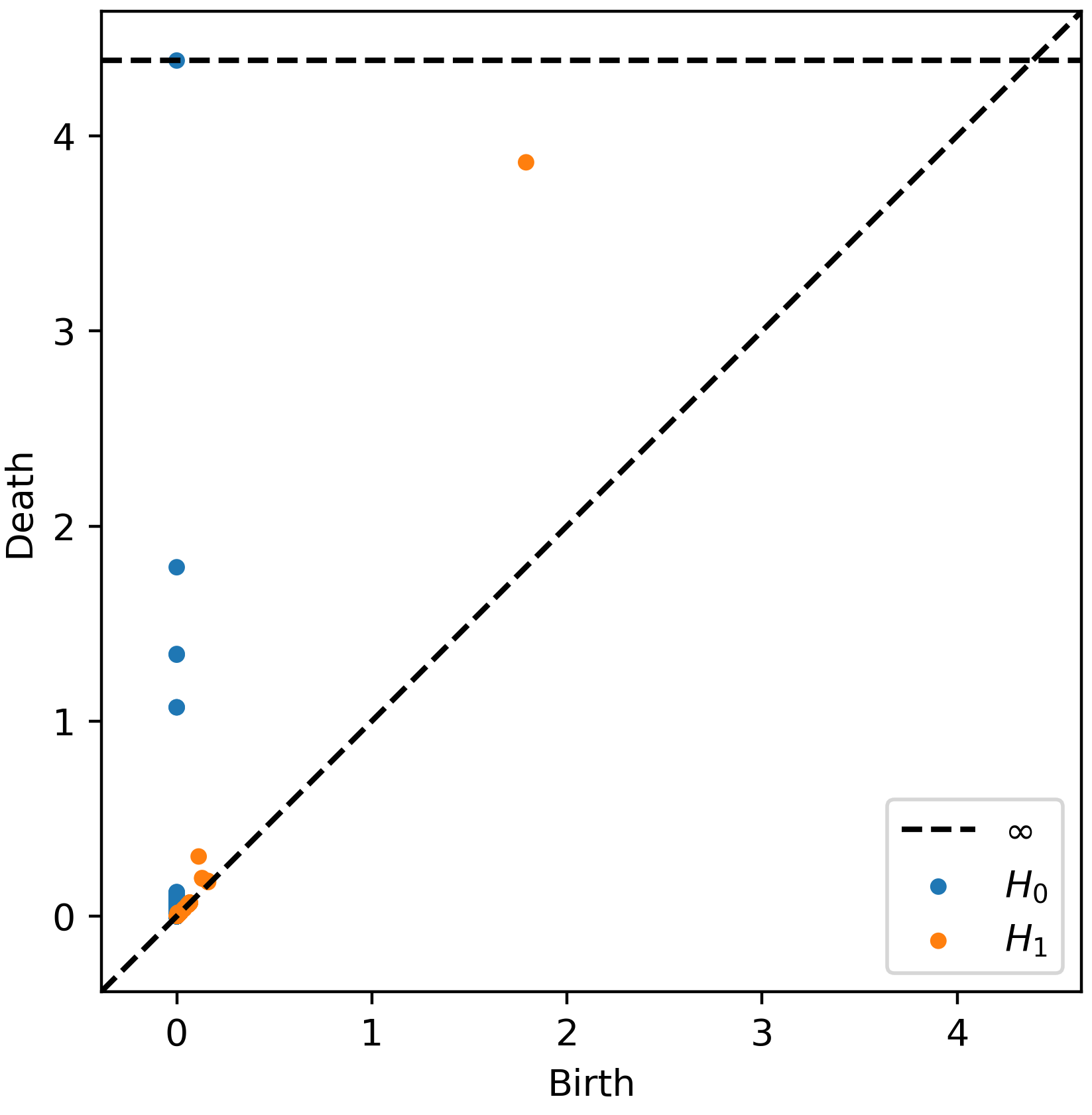}
        \subcaption{VR persistence diagrams.}
        \label{subfig:island_chainRips}
    \end{subfigure}
    \caption{Point cloud and persistent homology for an island chain.}
    \label{fig:island_chain_section_and_rips}
\end{figure}

\subsection{An island chain}

Next we will look at the case of the island chain presented as subfigure
\ref{subfig:island_chainSection}. We see that this is a chain of six major islands each of which is formed from many smaller islands. From the persistence diagram in \ref{subfig:island_chainRips} we observe what appears as four persistent $H_0$ classes. However, we would expect six persistent classes since there are six major islands\footnote{We specify ``major'' here because each of the disconnected lines that form the major islands are actually not linear features and instead are higher order islands. That is, the ``major'' islands are formed from sets of disconnected ``minor'' islands.} in the orbit. In fact there are six persistence classes in \ref{subfig:island_chainRips} but two of them have been plotted under two of the others. This is because our orbit has a vertical parity symmetry and therefore the uppermost two islands are almost identical to the lowermost two and hence are associated to classes with nearly identical birth and death times. This is a downside of using persistence diagrams as a visualisation technique, symmetries in the geometry of the point cloud will form coincident points on the diagram which then cannot be distinguished on inspection. 

\begin{figure}
    \centering
        \includegraphics[width = 0.6\textwidth]{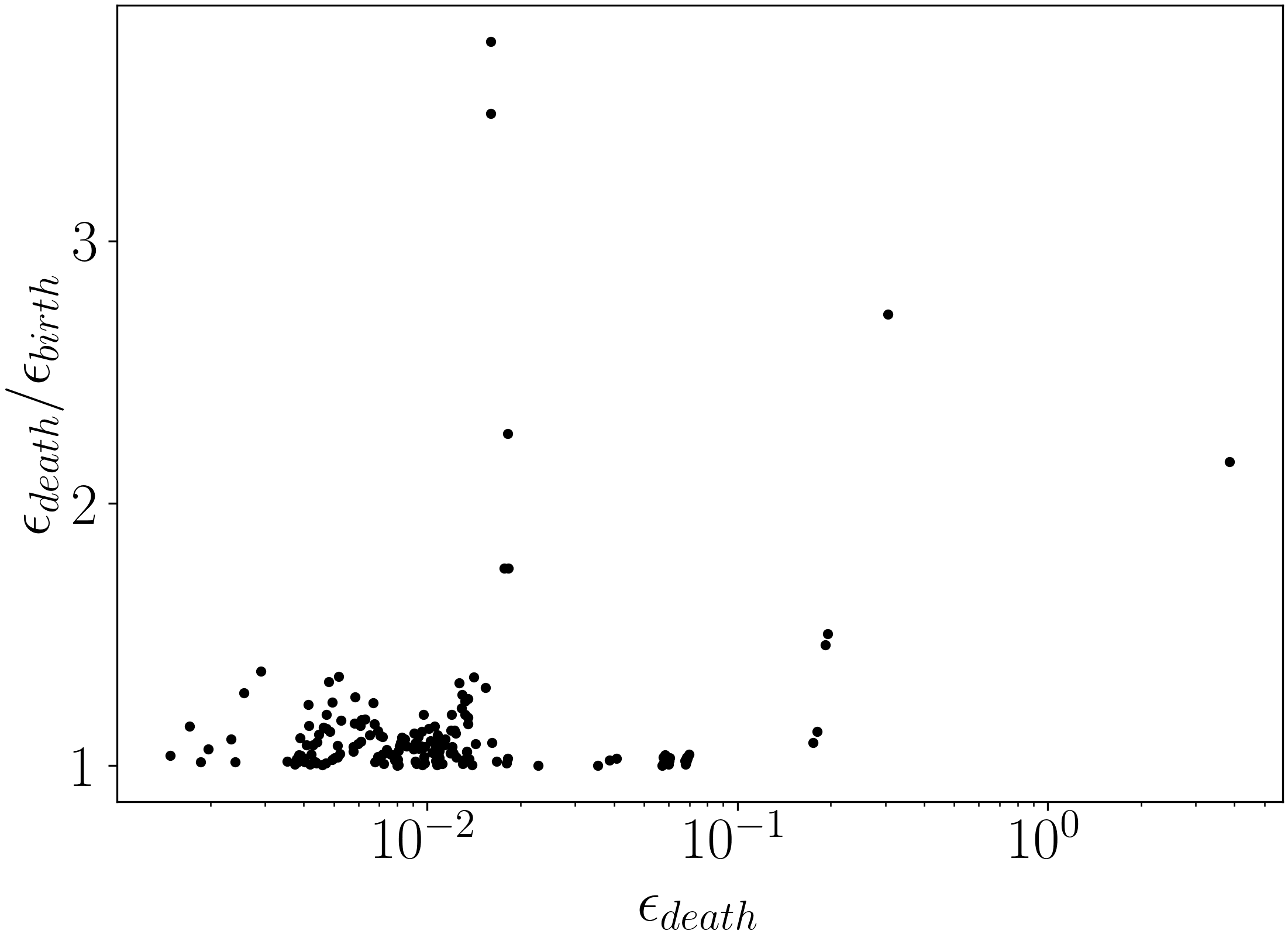}
    \caption{$PH_1$ relative persistence of the island chain in subfigure \ref{subfig:island_chainRips}}
    \label{fig:islandchain_section_bdrat}
\end{figure}

If we consider the relative persistence $\epsilon_{\text{death}}/\epsilon_{\text{birth}}$ of each of the $H_1$ classes we obtain Figure \ref{fig:islandchain_section_bdrat}. The point with largest $\epsilon_{\text{death}}$ corresponds, as before, to the enclosure of magnetic axis. We recognise that the weak relative persistence $\approx 2.25$ of this point corresponds to the fact that the island chain is composed of disconnected islands and so does not enclose the magnetic axis. 

In the same figure we observe a cluster of  five $H_1$ classes with $\epsilon_{\text{death}} \in (0.1,0.5)$. Each of these classes corresponds to a closed loop in the point cloud with a diameter in $(0.1,0.5)$. Observe from the Poincare section that the only structures in this range are the major islands. This indicates that each of the classes in the cluster corresponds to one of the major islands. Note that the death diameter $\epsilon_{\text{death}}$ describes the minimum width of the island. The fact that there are five classes in the cluster indicates that the VR persistence has detected five major islands. However, we can see from inspection of the Poincare section that there are actually six major islands in the chain. So, we conclude that an island was not detected by the $H_1$ information. This should be expected because the island at smallest $R$ is so narrow as to appear on inspection as a straight line. The width of this island is less than the spacing between the sub-islands which comprise it and consequently a closed loop in the VR complex is never formed during the filtration, hence no class appears in the $H_1$ associated to this island. A class associated to the narrow island does appear in the $H_0$ persistence diagram though, as was noted above.

\subsection{Thin stochastic layers}

\begin{figure}
    \centering
    \begin{subfigure}[b]{0.4\textwidth}
        \centering
        \includegraphics[width = \textwidth]{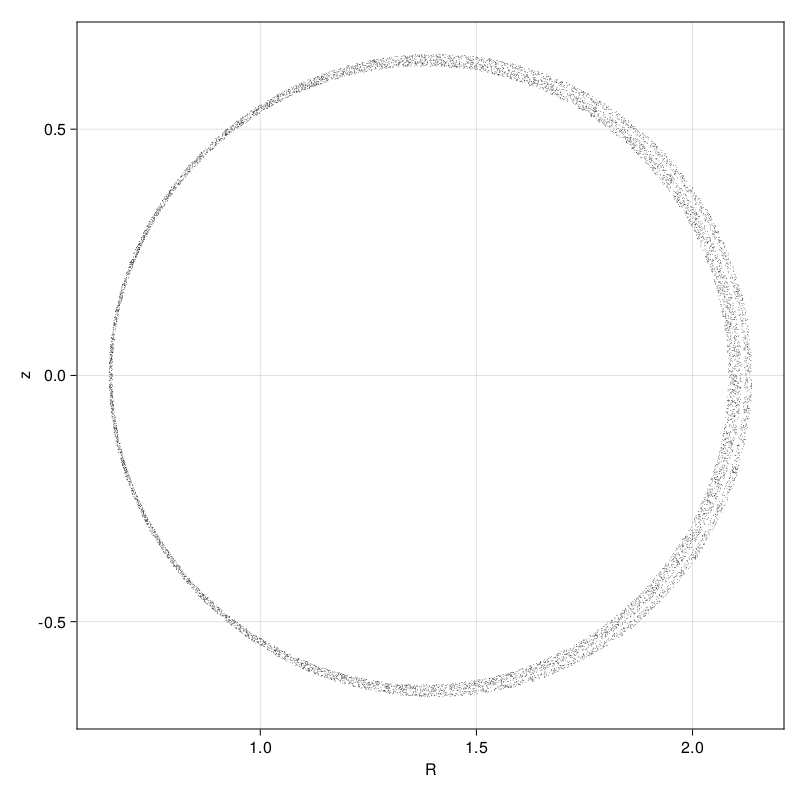}
        \subcaption{Poincare section.}
        \label{subfig:stochastic_thinSection}
    \end{subfigure}
    \begin{subfigure}[b]{0.39\textwidth}
        \centering
                \includegraphics[width = \textwidth]{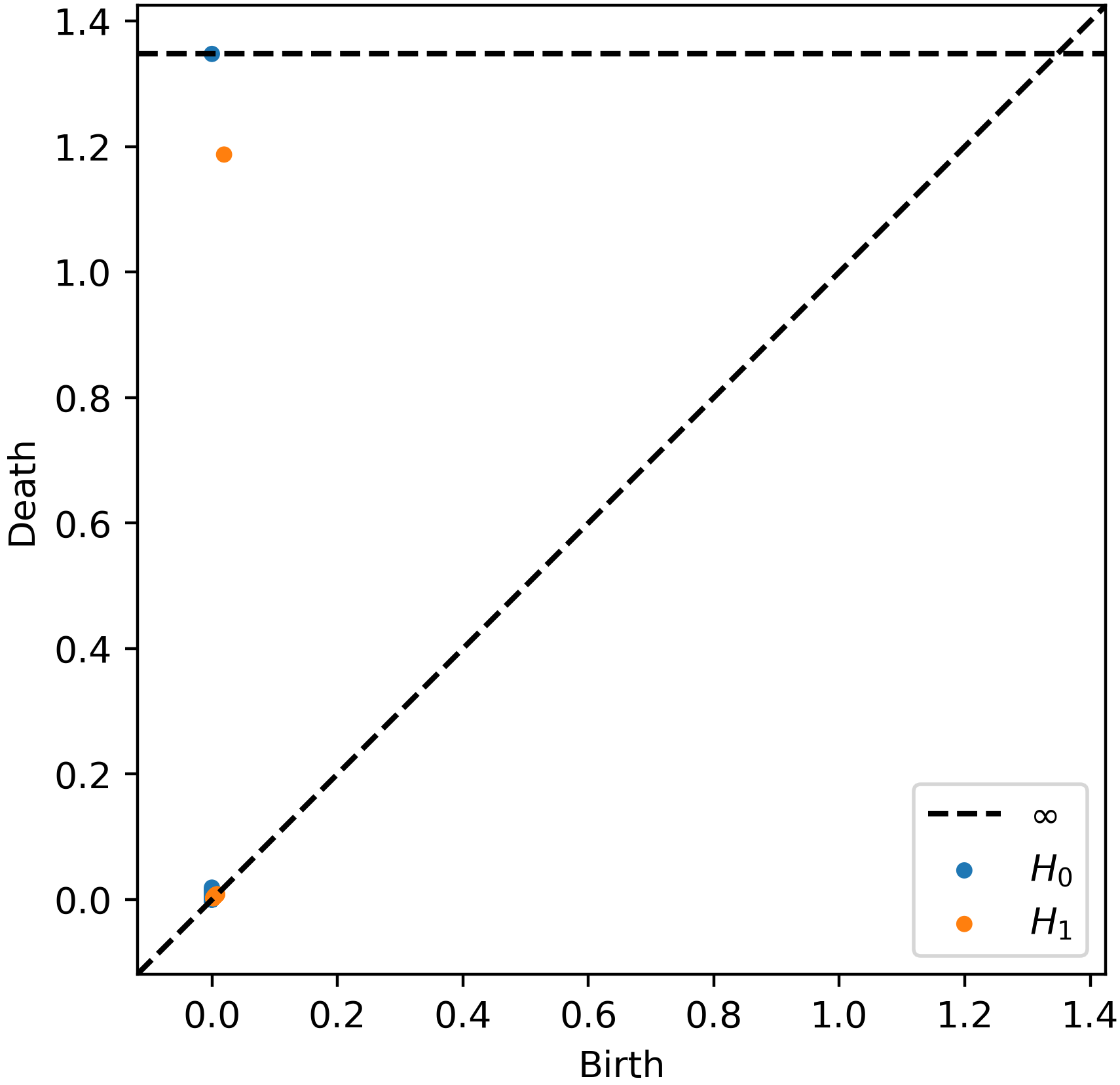}
        \subcaption{VR persistence diagrams.}
        \label{subfig:stochastic_thinRips}
    \end{subfigure}
    \caption{Point cloud and $PD$ for a thin stochastic layer}
    \label{fig:stochastic_thin_section_and_rips}
\end{figure}

\begin{figure}
    \centering
        \includegraphics[width = 0.6\textwidth]{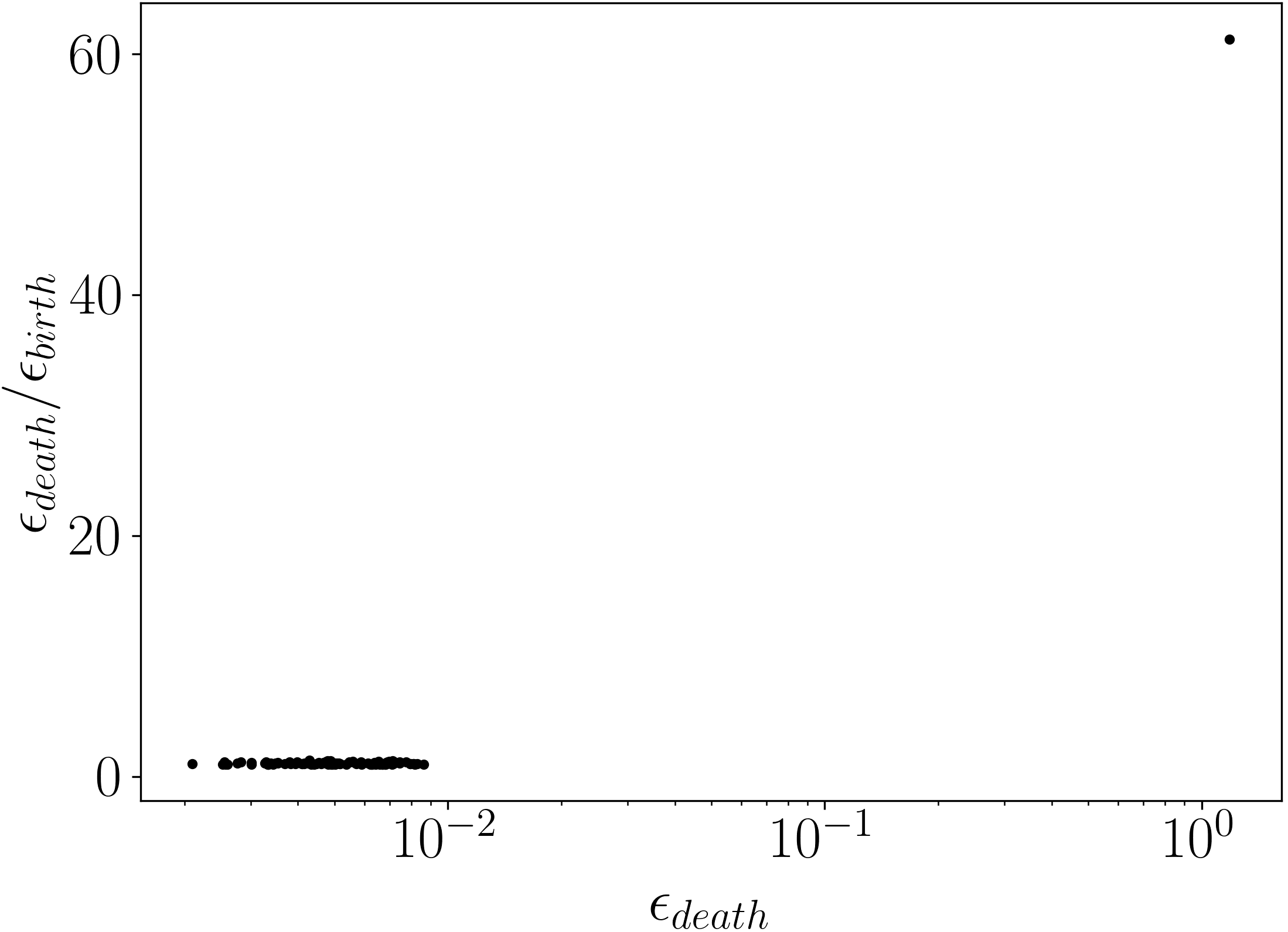}
    \caption{$PH_1$ relative persistence of the thin stochastic layer in subfigure \ref{subfig:stochastic_thinRips}}
    \label{fig:stochastic_thin_section_bdrat}
\end{figure}

We now look at a case in which the VR persistent homology struggles to delineate between two classes of structure. We consider a thin stochastic layer. These layers form between KAM toruses and are very hard to distinguish from KAM toruses both on inspection and numerically. Consider subfigure \ref{subfig:stochastic_thinSection} which presents one of these stochastic layers. This particular layer is bounded by KAM toruses and actually contains a very thin banana island inside it. Subfigure \ref{subfig:stochastic_thinRips} presents the persistence diagram for this layer and it is on inspection very similar to that of \ref{subfig:torusRips}. Similarly, the $H_1$ class death to birth ratio, as shown in Figure \ref{fig:stochastic_thin_section_bdrat} contains only a cluster of classes very close to zero below $\epsilon_{\text{death}} = 10^{-2}$ and a single class with relative persistence $\approx 60$ corresponding to the trajectory enclosing the axis. This is again very similar to the KAM torus case. It is evident that the VR persistent homology is not capable of clearly distinguishing a thin stochastic layer from a KAM torus and other methods are needed for this problem such as measurements of the Lyapunov exponent of the orbit, or the WBA chaos detection procedure discussed in Chapter \ref{Chapter2}.

\subsection{Large stochastic region}\label{LargeStochasticExample}

\begin{figure}
    \centering
        \includegraphics[width = 0.4\textwidth]{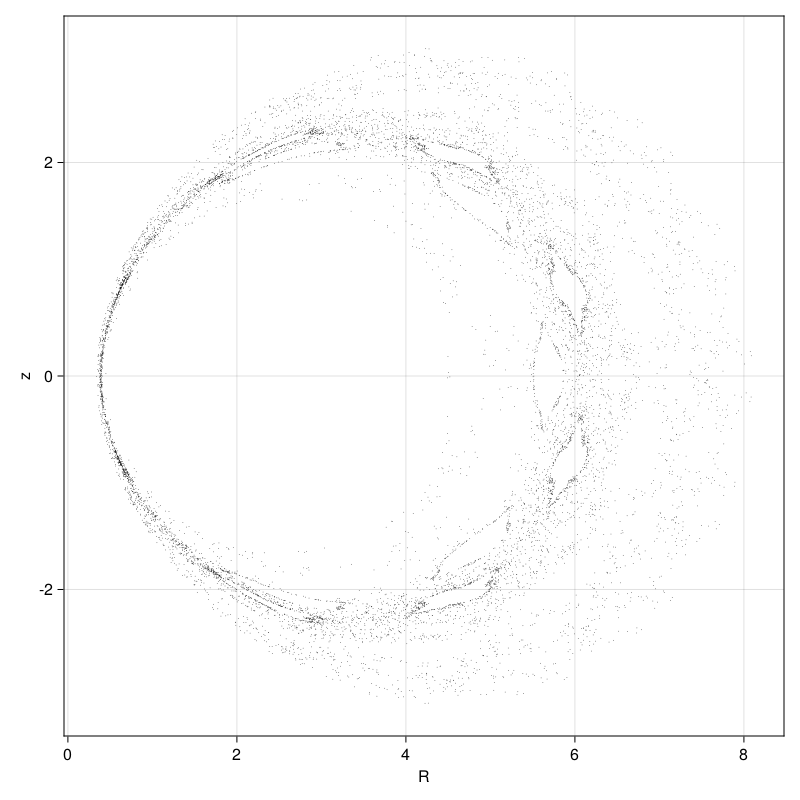}
    \caption{Point cloud of an field line orbit in a stochastic region}
    \label{fig:stochastic_section}
\end{figure}

For our final example we consider the case of a large stochastic region with many internal islands. Such an orbit is presented as Figure \ref{fig:stochastic_section}. We observe that the orbit encloses the magnetic axis and hence we should expect, as above, that the last $H_1$ class to die will have a high relative persistence. Since there are many islands enclosed by the stochastic region we should also expect that there will be many $H_1$ classes detected but few persistent $H_0$ classes. This is because the stochastic region itself is connected, and hence we should not expect persistent $H_0$ classes describing the presence of several connected components, but not simply connected. That is, there are holes in the stochastic region, the islands, which ensure the first homology is nontrivial.

\begin{figure}[t]
    \centering
    \begin{subfigure}[b]{0.39\textwidth}
        \centering
        \includegraphics[width = \textwidth]{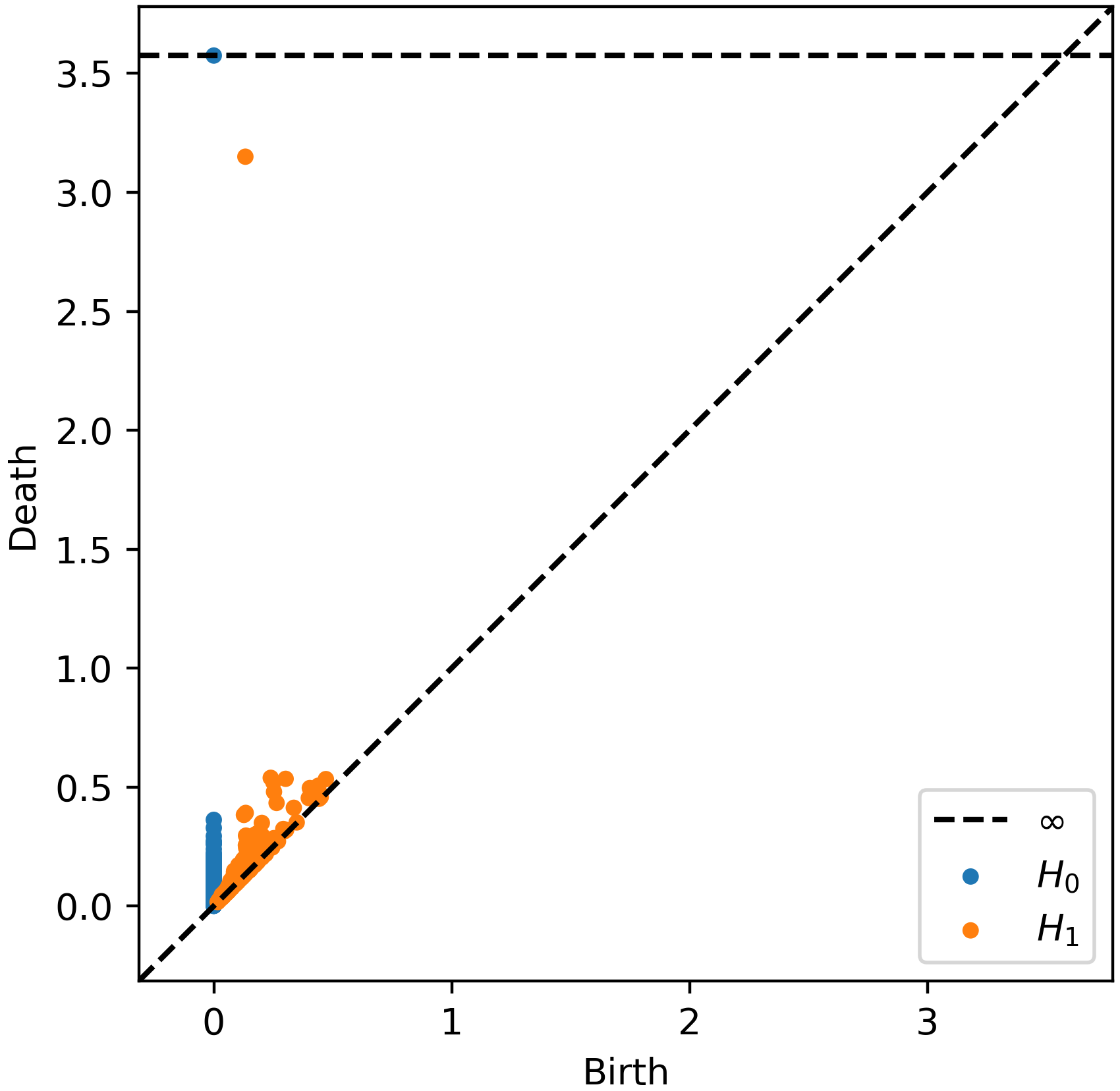}
        \subcaption{VR persistence diagrams.}
        \label{subfig:stochasticRipsPH}
    \end{subfigure}
    \begin{subfigure}[b]{0.49\textwidth}
        \centering
        \includegraphics[width = 0.95\textwidth]{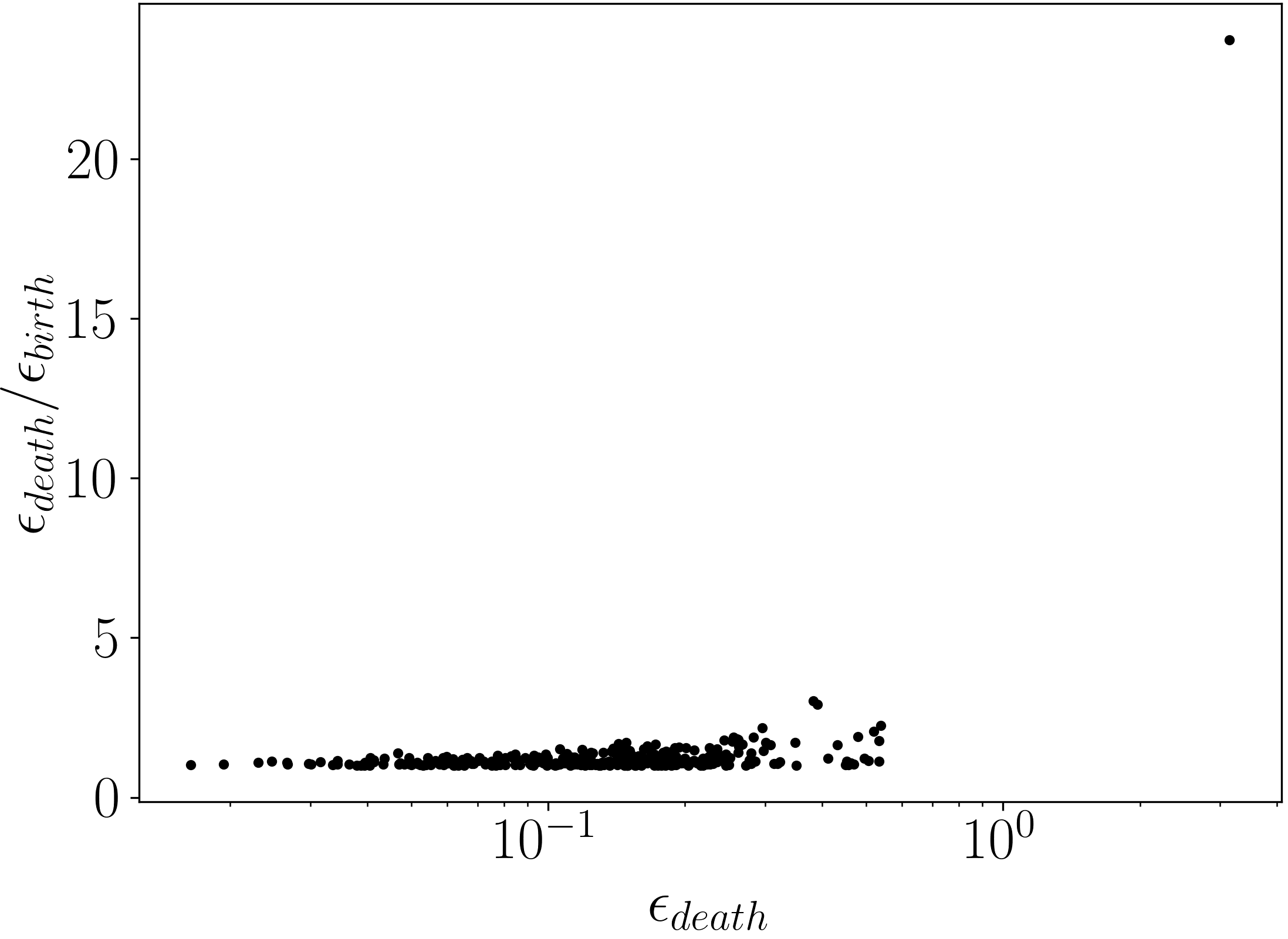}
        \subcaption{$PH_1$ class birth-death ratio.}
    \end{subfigure}
    \caption{VR homology results for the stochastic region in Figure \ref{fig:stochastic_section} . }
    \label{fig:stochastic_section_rips}
\end{figure}

Looking at the VR persistence diagram and death to birth ratio, which are presented as Figure \ref{fig:stochastic_section_rips}, we see that the topological data follows the expected behaviour noted above. We see the $H_1$ class we expect with high relative persistence confirming our first expectation. Also, there is only one persistent $H_0$ class, corresponding to global connectivity, however we see many $H_1$ classes with death time above $\epsilon_{\text{death}} = 10^{-1}$ - recalling that for our KAM torus and thin stochastic layer there were no $H_1$ classes above this range - and relative persistence of $\approx 2$ like those found in the island chains earlier. These $H_1$ classes are evidently associated to the many internal islands thus confirming our second expectation.  

\section{An automated classification procedure}

Having analysed the Vietoris-Rips peristent homology of different classes of magnetic field line orbit and connected geometric and topological features of the orbits to features in the persistent homology itself, we can proceed with developing a list of criteria which will allow us to automatically identify the class of an orbit only from its persistent homology. We will present methods based upon the observations made above for both distinguishing islands and islands chains from KAM toruses and stochastic regions, and then for distinguishing different order island chains. We will also present the results of applying our classification criteria to the case of our toy tokamak model. 

\subsection{Distinguishing islands}

Recall that in the above section we observed that for all orbits except island chains the last to die $H_1$ class has a high relative persistence. We related this to the observation that the islands are the only orbit to not form complete circles containing the magnetic axis in the Poincare section. This indicates that we can distinguish the Poincare section of an island chain from others by computing the relative persistence of the last $H_1$ class to die. If $\epsilon_{\text{death}}/\epsilon_{\text{birth}} < \text{threshold}$ for said class, and a selected threshold, then we claim that the orbit corresponds to an island chain. 

We define a function $c_l$ mapping point clouds $X\subset \Sigma$ to the class in $PH_1(X)$ with the largest death time. That is we define it by
\begin{equation}
    c_l(X) = \underset{c\in PH_1(X) }{\argmax}\epsilon_{\text{death}}(c)\,.
\end{equation}
The relative persistence of this class defines a number we will refer to as the \textit{enclosure}, $e$, of the  point cloud. That is we define
\begin{equation}
    e(X) = \frac{\epsilon_\text{death}(c_l(X))}{\epsilon_\text{birth}(c_l(X))}\,.
\end{equation}
To determine whether the orbit of a point belongs to the class of \textit{islands} or not we will check if the enclosure of the trajectory $X_T(x)$ is small, that is less than a chosen threshold. When it is small we declare the trajectory to lie in an island. 

Selecting a real number $e_\text{thresh}>0$, we define a set of points in $\Sigma$ called $IS(e_\text{thresh})$ by
\begin{equation}
    IS(e_\text{thresh}) = \left\{ x\in \Sigma |\, \ e(X_T(x))<e_\text{thresh} \right\}\,.
\end{equation}
We claim that if $e_\text{thresh}$ is selected correctly, $IS(e_\text{thresh})$ will be an acceptable approximation to the points whose orbits are islands. 

\begin{figure}
    \centering
        \includegraphics[width = 0.6\textwidth]{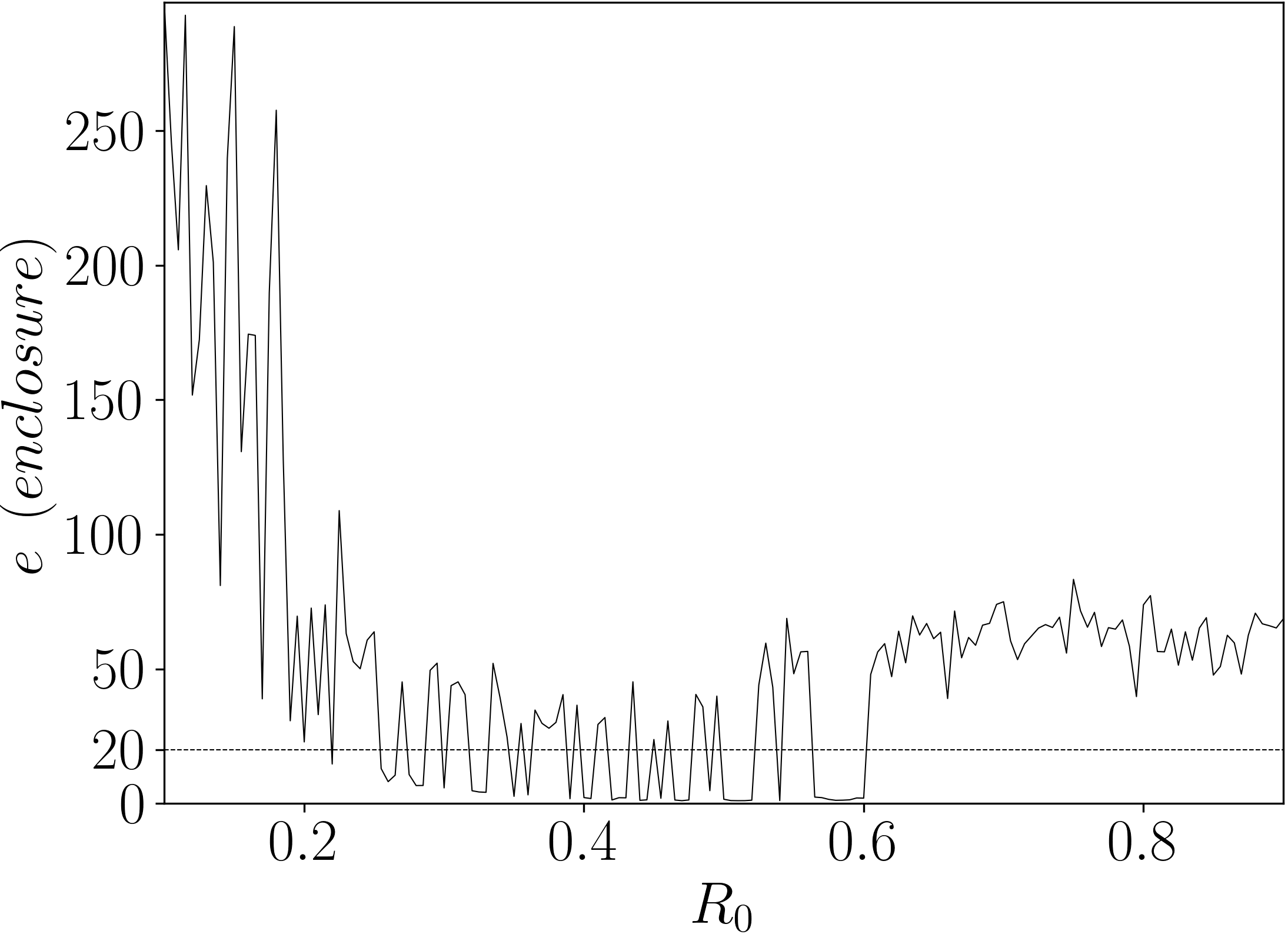}
    \caption{Enclosure as a function of the initial point $(R,Z)=(R_0,0)$ of the field line.}
    \label{fig:db_rat_radius}
\end{figure}

To demonstrate that this is a feasible concept we computed the enclosure $e(X_T(R,Z))$ for $T=2000$ point trajectories generated from $160$ initial points $(R,Z) = (R_0,0)$ with $R_0 \in [0.1,0.9]$. This set of initial conditions spans the left side of our toy tokamak. The plot of the calculated enclosure is presented as Figure \ref{fig:db_rat_radius}. We observe that the enclosure fluctuates as we vary $R_0$ but the values are all either above $e = 20$ or near zero. This indicates that if we choose our threshold at $\approx 20$ we can separate the trajectories into two classes which we recognise will be \textit{islands} and \textit{not islands}. Note that for $R_0>0.6$ the enclosure of all field lines appears to be large and so we would conclude that there are no magnetic islands near to the axis. This is expected as the magnetic axis is an elliptic fixed point of the Poincare map\footnote{Technically, the magnetic axis is a singular point for this solution but it is a removable singularity and whose removal leaves a Poincare map fixed point.} and so the field lines form an elliptic island around it comprised of KAM toruses separated by thin stochastic layers, both of which have, as observed above, high enclosure. 

\begin{figure}
    \centering
        \includegraphics[width = 0.6\textwidth]{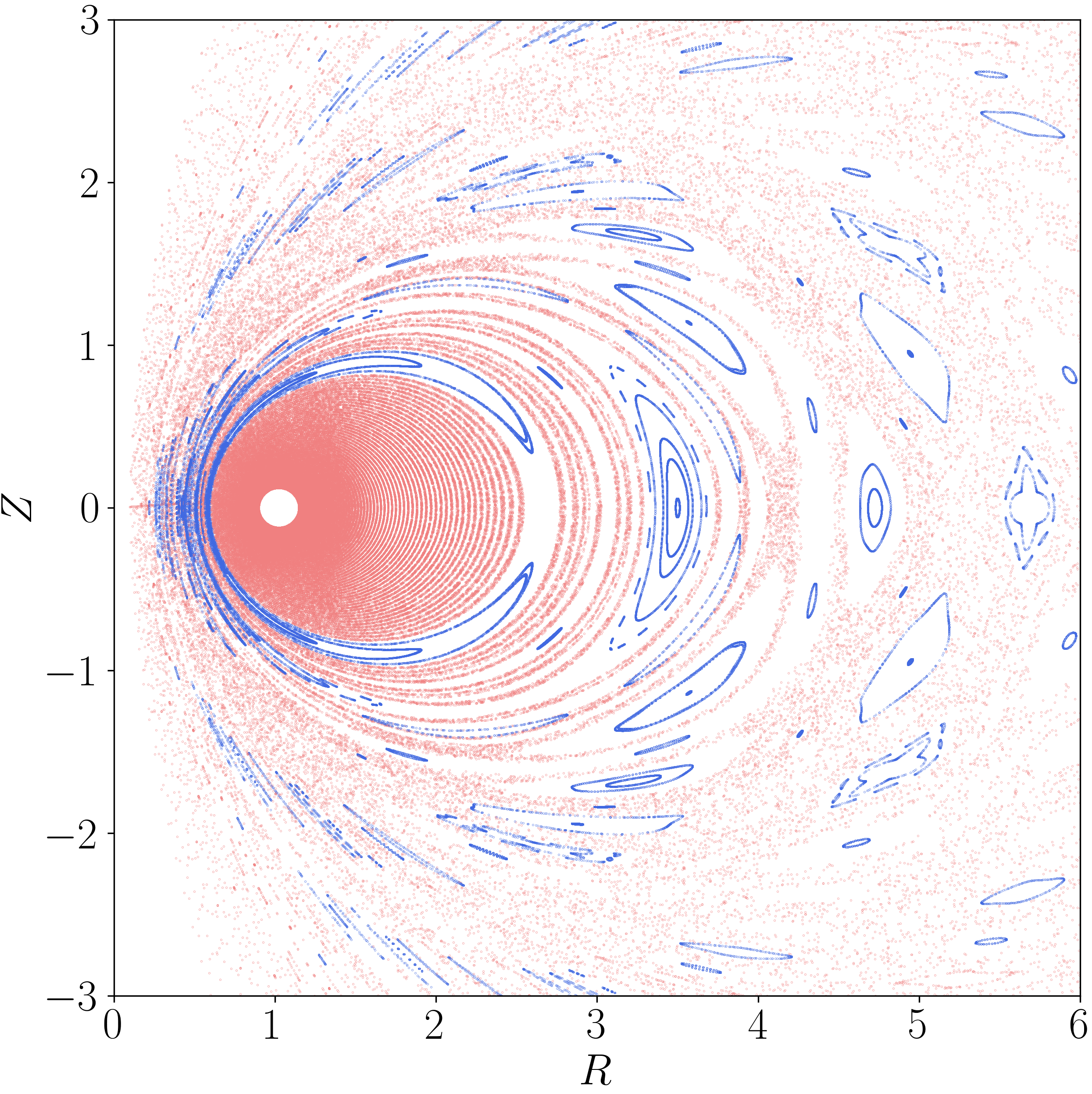}
    \caption{Orbit class map of islands detected by the condition $e_{thresh} = 20$. Islands are shown in blue and other classes in red.}
    \label{fig:toy_tokamak_islands_coloured1}
\end{figure}

To check that our classification scheme can correctly identify islands, at least up to the level of inspection, we now set our $e_\text{thresh} = 20$ and separate the initial conditions into those that belong to $IS(20)$ and those that do not. We now construct an image presenting the trajectories of all of our initial conditions. We colour the initial conditions in $IS(20)$ blue and the others in red. This produces the image presented as Figure \ref{fig:toy_tokamak_islands_coloured1}. Looking at this picture it appears, at least on inspection, that our classification scheme has acceptably separated the islands from the stochastic layers and KAM toruses. 

\subsection{An approach to island count and poloidal mode number}

Having demonstrated that we can identify the islands and island chains by their enclosure we now look at counting the number of major islands in each identified chain. We could achieve this by counting the number of persistent $H_1$ classes. We get better results however by counting the number of persistent $H_0$ classes. This is because we know that for an island chain, each connected component, and hence $H_0$ class, should correspond to a single island in the chain. 

\begin{figure}
    \centering
        \includegraphics[width = 0.65\textwidth]{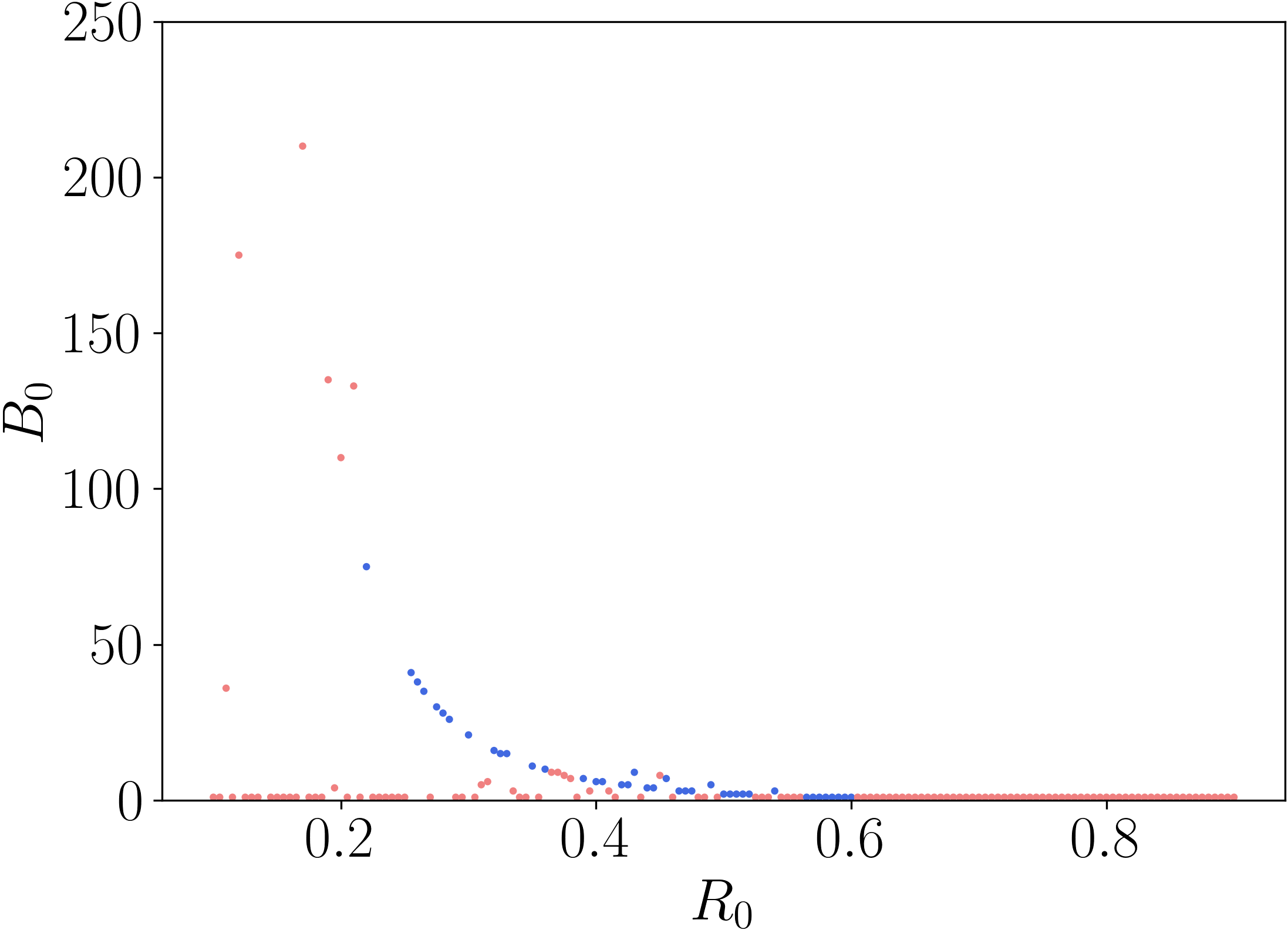}
    \caption{Scatter plot of the detected number of connected components in each orbit as a function of the initial radius for the case of $\epsilon_\text{thresh}  = 0.25$}
    \label{fig:betti0_radius}
\end{figure}

To count islands we cannot just compute the total number of $H_0$ classes in $PH_0(X_T(x))$. This is because $X_T(x)$ is, ignoring the case of periodic trajectories, a set of $T+1$ disconnected points and each of which provides a class to $PH_0(X_T(x))$ so the the total number of classes is always $T+1$. Instead we will compute the number of classes with absolute persistence greater than a chosen threshold. Specifically for a chosen $\epsilon_\text{thresh}>0$ define the $0$-th Betti number\footnote{We call this the Betti number because it describes an approximation to the true Betti number of the infinite time orbit of the field line.} of a point cloud $X$ as 
\begin{equation}
    B_0(X)  = \#\{ c \in PH_0(X) | \, \epsilon_\text{death}(c) > \epsilon_\text{thresh} \}\,.
\end{equation}

We can then compute $B_0(X_T(x))$ for each of our initial conditions along the $Z=0$ axis defined in the previous section. The result of this calculation for $\epsilon_{\text{thresh}}=0.25$ is presented as Figure \ref{fig:betti0_radius}. We have coloured points on this scatter plot blue if the underlying trajectory is classified as an island chain by its enclosure value, and red if not. We observe that the $B_0$ values for the islands appear to mostly lie on  a curve which increases as $R_0$ approaches zero. While the $B_0$ of non-islands is either close to zero, for thin stochastic layers, or effectively random for large stochastic layers. A few red dots at small $R_0$ seem to lie on the blue curve. This is a result of a misclassification error in the enclosure based island detection scheme. Specifically, island chains with a very high number of major islands will be identified as KAM toruses. 

Misclassification occurs when the ratio between the effective diameter of the island chain and the distance between adjacent major islands is greater than $e_{\text{thresh}}$. This allows us to estimate at what island count we should begin to see misclassification. As a rough approximation take that the point clouds are circles of radius $R$. Then let $n$ be the major island number and assume that the gaps between the islands are equal in size to the island themselves. This assumption enforces a ``worst case scenario'' where the islands are already close to enclosing the axis. Then we have an average distance between islands we can estimate as of order $2\pi R/2n$. Then, misclassification will occur when the diameter $2R$ divided by this distance is greater than $e_{\text{thresh}}$
\begin{equation*}
    \frac{2R}{2\pi R/2n} > e_\text{thresh}\,,
\end{equation*}
\begin{equation}
    n >\frac{\pi}{2} e_{\text{thresh}} \approx \frac{3}{2} e_{\text{thresh}}\,.
\end{equation}
For the $e_\text{thresh} = 20$ value adopted earlier we have $n>30$. In practice this will be an underestimate of the island number at which misclassification occurs because field line orbits have circumferences larger than $2\pi R$ since they are not circular. This agrees with the observation that, in Figure \ref{fig:betti0_radius}, all red points which lie on the blue island curve have $B_0\approx n$ well above $30$. 

\begin{figure}
    \centering
    \begin{subfigure}[b]{0.49\textwidth}
        \centering
        \includegraphics[width = \textwidth]{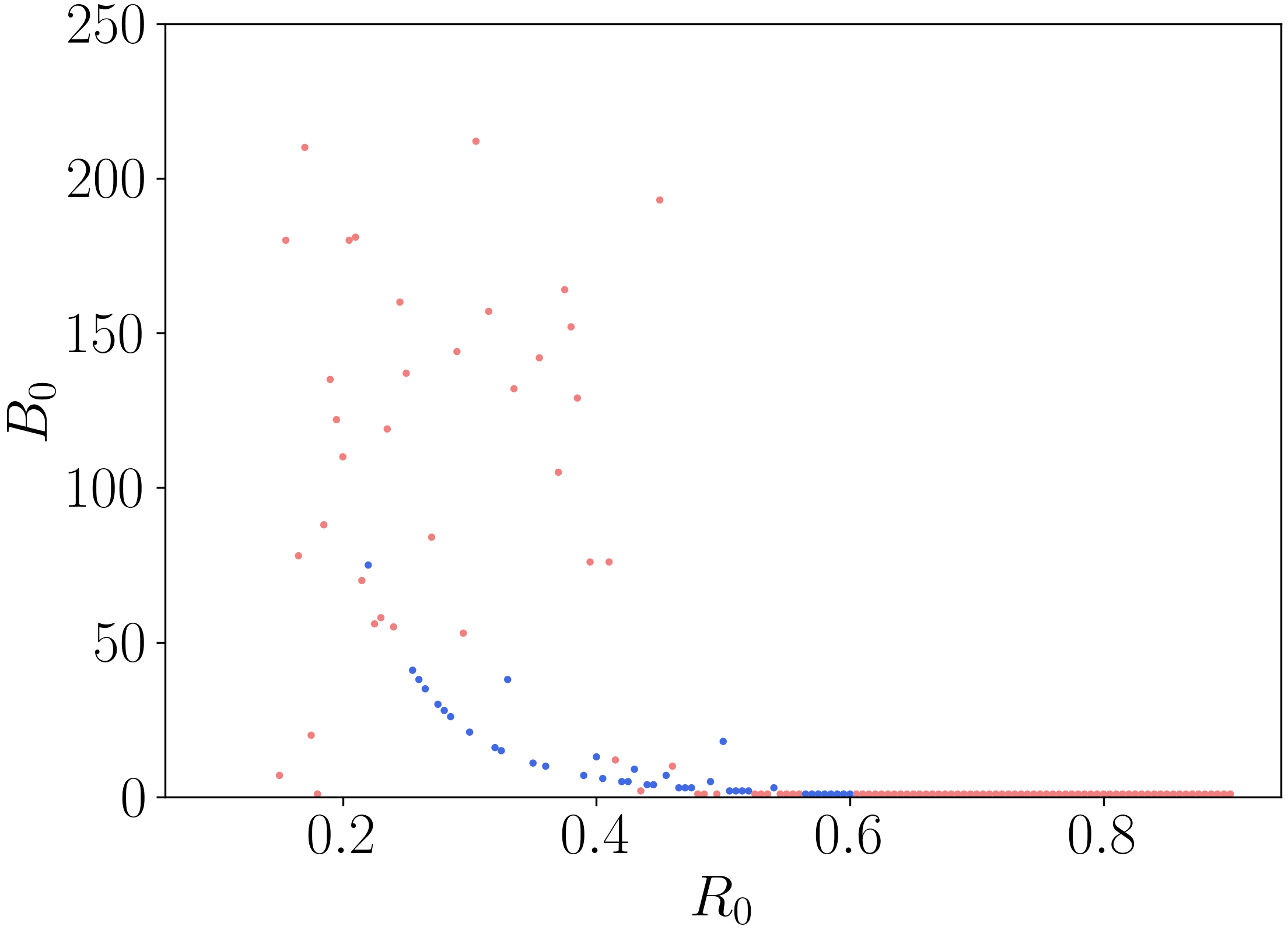}
        \subcaption{Case with $\epsilon_\text{thresh}  = 0.1$}
    \end{subfigure}
    \begin{subfigure}[b]{0.49\textwidth}
        \centering
        \includegraphics[width = \textwidth]{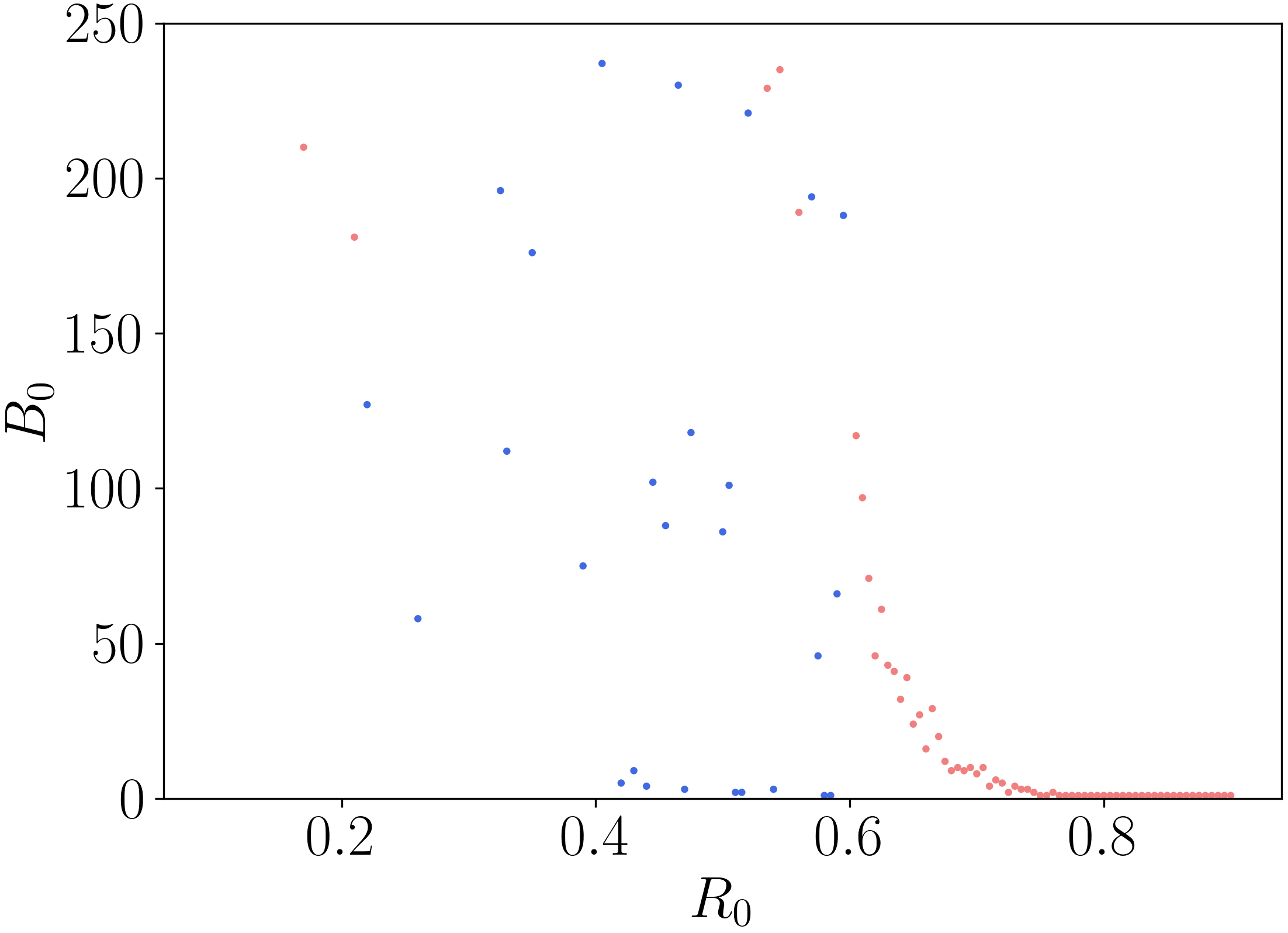}
        \subcaption{Case with $\epsilon_\text{thresh}  = 0.01$}
    \end{subfigure}
    \caption{Comparison of the Betti number plots with varying $\epsilon_{\text{thresh}}$.}
    \label{fig:betti0_radius2}
\end{figure}

We now ask the question of what happens when we lessen the threshold $\epsilon_{\text{thresh}}$. Consider Figure \ref{fig:betti0_radius2} which presents the same plot of $B_0$ for two smaller values of the threshold. We observe that for $\epsilon_{\text{thresh}}=0.1$ the spread of Betti numbers of stochastic layers, the red points, increases dramatically but the detected number of components in the island orbits, the blue dots, is consistent with Figure \ref{fig:betti0_radius}. This suggests that when restricted to the case of orbits which are known to form island chains the detected Betti number is relatively insensitive to the specific choice of threshold. For an even lower threshold of $\epsilon_{\text{thresh}}=0.01$ the island curve disappears and the KAM toruses near the magnetic axis, where $R_0\approx 0.7$, are detected as containing multiple components. This is inaccurate and suggest that the threshold has been selected too small. Clearly then as long as $\epsilon_{\text{thresh}}$ is chosen sufficiently large its precise value is not important.

\begin{figure}
    \centering
        \includegraphics[width = 0.6\textwidth]{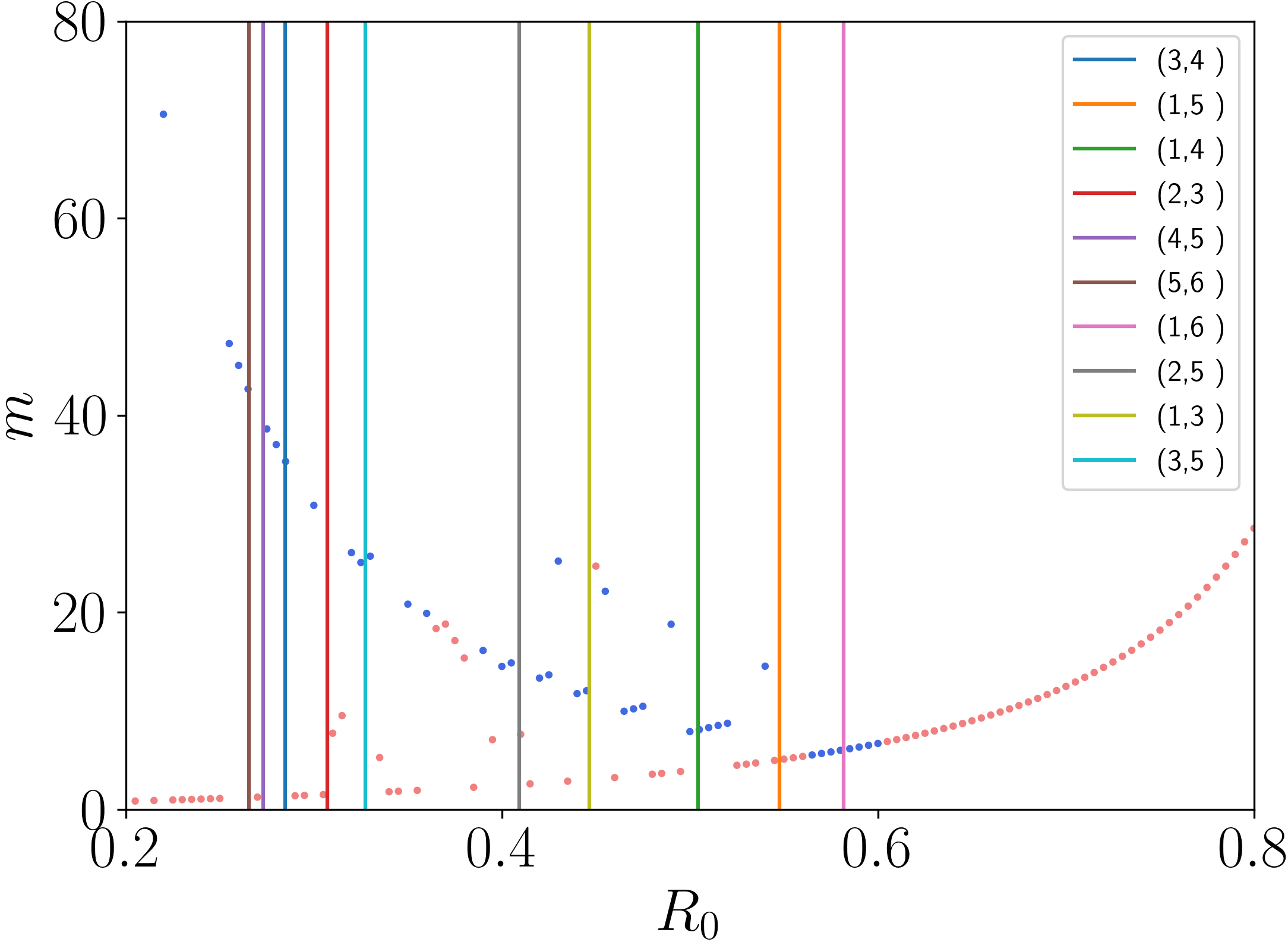}
    \caption{Case with $\epsilon_\text{thresh}  = 0.25$. $m$ derived from the unperturbed $q$ profile. Vertical $(n,m)$ indicates points where $q=n/m$.}
    \label{fig:betti0_radiusresonances}
\end{figure}

We are now in a position to perform one last test of the accuracy of our island detector and island counter by comparing the radial positions of the island chains we detected to the positions we should expect from fundamental theory. Our field lines perform both toroidal rotations, that is laps around the $z$ axis, and poloidal rotations, laps around the magnetic axis. We can define a quantity called by tokamak physicists the safety factor or $q$-profile of the orbit as the average number of toroidal rotations of the field line per poloidal rotation. For the periodic field line in the centre of each island chain this is a rational number $q=n/m$ where $n$ determines the number of toroidal islands \cite{hazeltine2003plasma}. 


It follows from fundamental theory that $q$ is a measure of how twisted the field lines are and therefore can be computed in terms of the toroidal and poloidal components of the magnetic field \cite{miyamoto1997fundamentals}. At position $\textbf{x} = (R_0,0,0)$ we have 
\begin{equation}
    q(R_0) = (1-R_0)\frac{B_t(R_0,0,0)}{B_p(R_0,0,0)}\,,
\end{equation}
where $B_t$ is just the $\hat{y}$ component of $\textbf{B}$ and $B_p$ is the sum in quadrature of the $\hat{x}$ and $\hat{z}$ components.

The largest island chains form where the field line is more resonant with the periodic perturbation and this occurs where $q$ is a rational with small denominator in the unperturbed field. Since we have an exact expression for the unperturbed field we can exactly calculate $q$, find the location of low order rational values, constructed as a Farey tree, and then check that our detected islands occur at the same locations. 


Also, note that for our island chains $n$ is approximated by our measured $B_0$ and so we are able to solve for the poloidal mode number $m$ using the information from the $q$-profile. We have
\begin{equation}
    m(R_0) = \frac{B_0(X_T(R_0))}{q(R_0)}\,.
\end{equation}

Figure \ref{fig:betti0_radiusresonances} presents the result of computing the $m$ value for each of the orbits in our toy tokamak. The vertical lines indicates surfaces of rational $q$. Note that the rational numbers of these lines were taken from the Farey tree layer $F_5$. Some lines for values in $F_5$ are omitted to reduce clutter on the diagram. We observe that our blue clusters, that is islands, usually occur in clusters around the vertical lines. This indicate that the islands we have detected are occurring in the expected location around surfaces of rational $q$. We can observe from the $m$ values that the islands mostly form linear clusters in the $m$ plot. However, at the edge of several of the clusters there is a single raised island chain with a higher $m$. This occurs because our scan over radius sometimes lands on a higher order island chain where we see twice as many island for the same $q$. That is, since $2n/2m = n/m = q$, at the same rational surfaces we can see island chains with any integer multiple of the number of islands in the fundamental chain.

\subsection{Finding stochastic layers filled with islands}\label{StochLayersWithIslands}

We saw above that we could discern the topological difference between island chains and stochastic layers using TDA but we also know that there are several different types of stochastic field lines, and we would prefer to also be able to distinguish them. This can be done and we demonstrate how in this subsection.

\begin{figure}
    \centering
        \includegraphics[width = 0.5\textwidth]{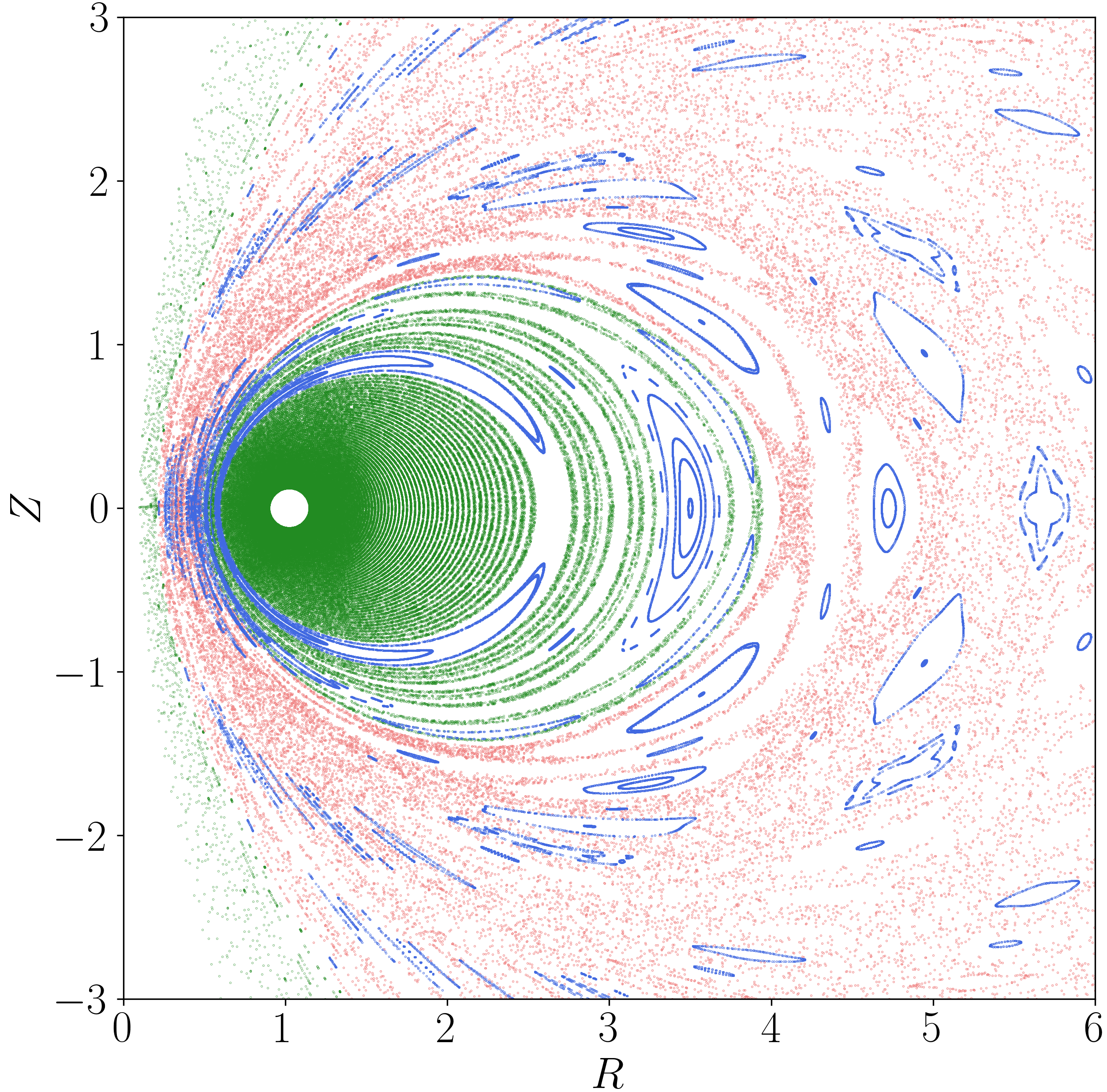}
    \caption{Orbit class map  for the case with threshold of $\epsilon_{\text{thresh}}=10^{-1}$. Blue, red, and green indicate islands, large stochastic layers, and thin stochastic layers respectively.}
    \label{fig:toy_tokamak_islands_coloured2}
\end{figure}


Recall from Section \ref{LargeStochasticExample} that $PH_1$ for a large stochastic trajectory, as opposed to a thin stochastic layer, contained many $H_1$ classes with death time $\epsilon_{\text{death}} \approx 10^{-1}$ well above the approximate spacing between points. This can be seen by analysing the persistence diagram presented as subfigure \ref{subfig:stochasticRipsPH} and noting that there exists many $H_1$ classes which die above the blue cluster which describes the deaths of $H_0$ classes. For all stochastic layers and KAM toruses, at least one class must exist above this threshold, with death time of order $1$, the diameter of the surfaces of magnetic flux. So we can conclude that if the $PH_1$ contains more than one $H_1$ class which dies after $10^{-1}$ then the point cloud in question is most likely not a thin stochastic layer and hence is a large stochastic layer. Adopting this line of reasoning we construct a classification scheme as follows.

\begin{figure}
    \centering
        \includegraphics[width = 0.5\textwidth]{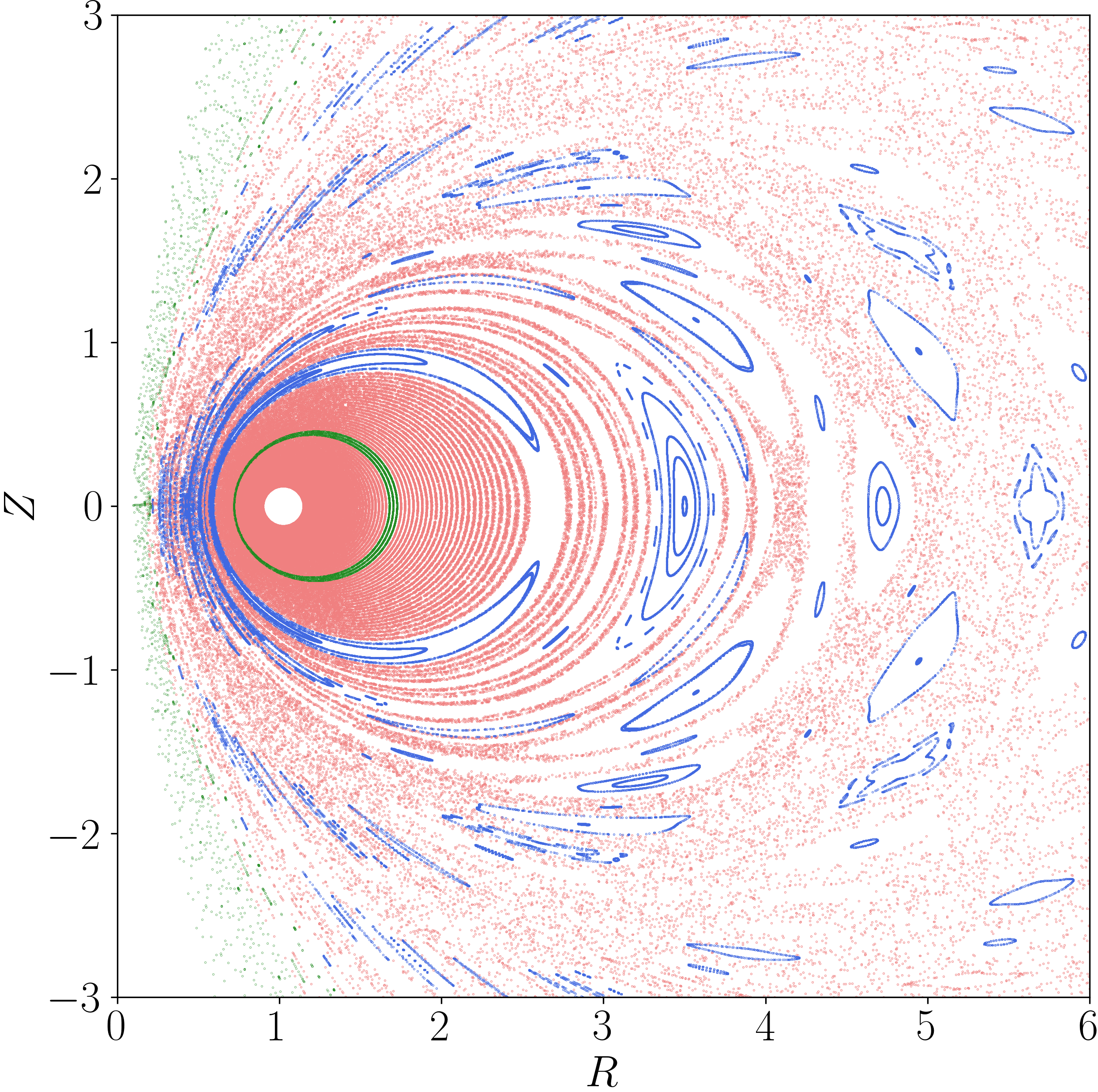}
    \caption{Orbit class map for the case $\epsilon_\text{thresh}$ is the mean death time of $H_0$ classes. Only the orbits in the neighborhood of a KAM torus are green.}
    \label{fig:toy_tokamak_islands_coloured3}
\end{figure}

We define a function $ihc:\Sigma \cross \mathbb{R}^+\rightarrow \mathbb{N}$, the \textit{internal hole count}, as the number of $H_1$ classes in the $PH_1$ of the trajectory generated by the input point. That is we define
\begin{equation}
    ihc(x,\epsilon_{\text{thresh}}) = \# \{ c \in PH_1(X_T(x))| \, \ \epsilon_{\text{death}}(c) >\epsilon_{\text{thresh}} \}\,.
\end{equation}
Then the set of points lying on large stochastic layers is defined to be 
\begin{equation}
    LS(e_{\text{thresh}},\epsilon_{\text{thresh}}) = \{x \in \Sigma |\, \ x \notin IS(e_{\text{thresh}})\, \ \text{and}\, \ ihc(x, \epsilon_{\text{thresh}}) > 1\}\,.
\end{equation}

We now produce a new version of the orbit class map from Figure \ref{fig:toy_tokamak_islands_coloured1} but with the orbits whose initial point is in $LS$ coloured differently. This is presented as Figure \ref{fig:toy_tokamak_islands_coloured2} where the points in $LS$ are presented in red and other stochastic layers are presented in green. We observe the expected behaviour that the large stochastic layers form a bounded layer between thin stochastic layers which lie near to the magnetic axis, $R_0 =1$ and KAM toruses near to what would serve as the effective boundary of a machine $R_0 < 0.1$. We see that the trajectories which are coloured green form disconnected thin layers each separated either by a KAM torus or an island chain. This confirms that the $ihc$ metric is capable of distinguishing the large stochastic layers from thin layers and KAM toruses. 

The method described above works but does not come equipped with a method for determining the best $\epsilon_\text{thresh}$. The threshold can be seen as a free parameter which can be adjusted by a user to select for thinner and thinner stochastic layers. However, we would prefer an automated procedure to obtain it. One possible approach is to determine $\epsilon_\text{thresh}$ using other information contained in the VR persistent homology of the orbit. We noted above that to distinguish stochastic layers we should search for $H_1$ classes representing holes which are larger than the spacing between points. This spacing can be estimated as the mean of the death times in $PH_0$, excluding the class which does not die. We refer to this mean as $\langle d(PH_0)\rangle$. Note that the death times of $H_0$ classes are edge lengths of a minimal spanning tree and their mean overestimates the average nearest neighbor distance. Thresholding with the $\langle d(PH_0)\rangle$ gives an automated procedure to generate $\epsilon_\text{thresh}$ but comes at the cost that the threshold will be different for each orbit, although this may be advantageous since the varying spacing of the points in different orbits is being compensated for. 

The results of constructing an orbit class map using $\langle d(PH_0)\rangle$ as $\epsilon_\text{thresh}$ is shown in Figure \ref{fig:toy_tokamak_islands_coloured3}. Many thin stochastic layers have been reclassified as large stochastic layers and the only remaining green point clouds form a very thin layer around what is actually the only true KAM torus in the set of orbits included in the diagram. Hence we observe that by $\langle d(PH_0)\rangle$ as a threshold we can partially distinguish KAM toruses from thin stochastic layers by their persistent homology.



\section{Orbit classification in parameter space}

In the above discussion of orbit classification we were concerned with automatically mapping the classes of all of the orbits in a given tokamak from the Poincare sections of its field line orbits. We now briefly consider a dual problem, where we ask how the orbit class of a specific point in space changes with the parameters of the tokamak. This question is motivated by the problem of designing a tokamak, where we are concerned with determining what parameters of the machine will produce desired behaviour. We can imagine a design constraint where we are interested in ensuring that a particular place in the machine lies in a magnetic island. Hence, we would want to determine how the orbit class of said place varies with the machine parameters. 

\begin{figure}
    \centering
        \includegraphics[width = 0.7\textwidth]{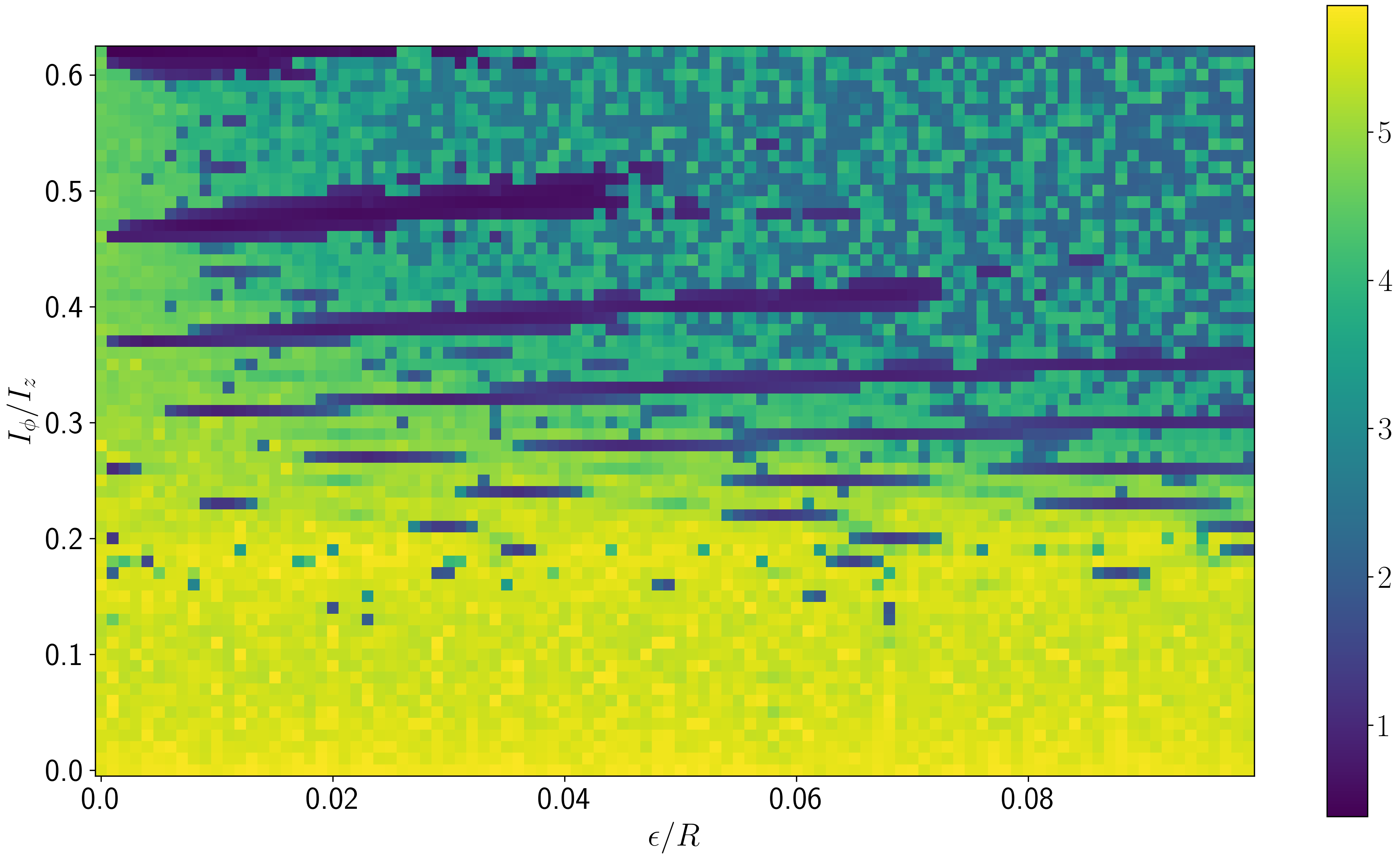}
    \caption{The enclosure $e$ of the orbit with initial condition $(R_0,Z_0) = (2,0)$ with as function of the parameters of the toy tokamak model. Note that the colour of each pixel is defined by $\log(e)$ not $e$ itself.}
    \label{fig:field_line_phase_diagram}
\end{figure}

In the case of our toy tokamak our machine parameters are $I_\phi/I_z$ and $\varepsilon/R$. We choose, for no particular reason, to focus on the orbit of the point $(R_0,Z_0) = (2,0)$. For each set of tokamak parameters we can calculate $X_T(R_0,Z_0)$ and then the enclosure of this point cloud. This gives the enclosure as a two dimensional function of the parameters which we can display as an image. 

Such an image is presented as Figure \ref{fig:field_line_phase_diagram}. We observe dark blue regions which occur in stripes. These are the parameters for which the orbit of $(2,0)$ is an island chain. The striped behaviour occurs because as the parameters of the tokamak is varied the point occurs in different generations of magnetic islands in sequence. The bright yellow region at the bottom, when the enclosure is high, is the set of parameters for which $(2,0)$ lies on a KAM torus or in a very thin stochastic layer. Then the fuzzy green region in the top right define the parameters in which the orbit lies in a large stochastic layer. The islands occur as a boundary between these two phases of dynamics. 

The above example, which demonstrates the generic behaviour, indicates that computing the enclosure of the orbit of a point as the parameters are varied provides a practical method for constructing a visualisation of the orbit class of a specific point as a function of external parameters. Note that this visualisation is akin to a phase diagram and could be generalised to as many parameter dimensions as required, although at nontrivial computational cost.


\chapter{Distribution of island Size} 

\label{Chapter4}

The phase space islands in a Hamiltonian system are ``sticky''. Stochastic trajectories which enter the boundary region of an island will remain in the boundary region for longer than would be expected if the motion were purely diffusive, by which we mean if each orbit had the statistics of a simple random walk. This is a consequence of the fact that islands come in hierarchies where each island has satellite sub-islands which each are accompanied by their own sub-islands and so on \textit{ad infinitum}. This means that the boundary region near a island can be thought of as a fractal labyrinth which the stochastic orbit must travel through. This process takes considerable time due to the restriction on the motion \cite{zaslavsky1994renormalization}.

 The precise statistics of transport properties of particles in the stochastic region depends strongly on the structure of the island hierarchy. That, is on how many islands of each diameter exist, how closely spaced they are, and how large their relative size is. Consequently, computing the distribution, that is the histogram, of the size of the islands within the phase space may provide insight into the particle transport. 
 
 In this chapter we propose and demonstrate a procedure using TDA for approximating the island size distribution inspired by a similar approach taken to compute the distribution of pore sizes in the porous rocks \cite{robins2016percolating}. We will restrict to the 2D case for simplicity but the method is sufficiently general to facilitate application in arbitrary dimensions. We will then apply our method to the case of an accelerator mode island in the standard map. This particular island is notable for generating anomalous transport due to its specific island hierarchy. We investigate whether our method can capture the existence of this hierarchy.


\section{Procedure for computing the radius distribution}

Suppose we have a phase space $M$ containing points with both ordered and chaotic orbits. To compute the distribution of radii of phase space islands, or magnetic islands, we follow a four step process. Firstly, we construct an indicator function $c:M\rightarrow \mathbb{Z}_2$ which outputs 1 on points whose orbit is chaotic and 0 otherwise and evaluate it over a square grid to construct a binary image of the stochastic region. We will use the WBA approach discussed in Chapter \ref{Chapter2} for this purpose. Secondly, we apply the SEDT to our binary image producing a new grayscale image whose local minimums describe the size and location of connected regions in our binary image. Thirdly we use \verb!DIAMORSE! to compute the birth times of each of the zero classes, that is, the values of the minimums in the image. Each of these minimums describes the size of a connected black region of the image, which we will recognise as islands. The negative of the birth time of a connected component is the radius of the largest inscribed circle which can be fit inside a connected black region. We adopt this as an approximation to the radius of the associated connected component, and so we extract these radii from the persistent homology. Finally we compute a histogram of the negative birth times, and normalise. An algorithmic description of the procedure discussed above is presented as algorithm \textit{BirthsDist}.

\begin{algorithm}
\caption{BirthsDist(N,T,h)}
\begin{algorithmic}[1]
\Require $N,T\in\mathbb{N}^+$ and $h$ an observable function.
\State Choose an $N\times N$ square grid $G$ of points $x_{ij}$ in the phase space $\mathcal{M}$.
\For{$x_{ij} \in G$}
    \State Calculate the $D$ function defined above for time $T$ orbits against observable $h$.
    \State $D_{ij} \leftarrow D(x_{ij},h)$ 
\EndFor
\State $\tilde{D} \leftarrow D/\underset{i,j}{\max}D_{ij}$ \Comment{Normalising image to max value $1$}
\State Compute the histogram of $\tilde{D}$ image pixel brightness $p(\tilde{D})$.
\State Find the center of the two chaotic and ordered clusters $\tilde{D}_c$ and $\tilde{D}_o$.
\State $\tilde{D}_{\text{thresh}} \leftarrow \underset{\tilde{D}\in [\tilde{D}_c,\tilde{D}_o]}{\argmin}p(\tilde{D})$ \Comment{Pull the minimum between clusters}
\For{$\tilde{D}_{ij}\in \tilde{D}$}
    \If{$\tilde{D}_{ij}<\tilde{D}_{\text{thresh}}$}
        \State $PSB_{ij} = 1$
    \Else
        \State $PSB_{ij} = 0$
    \EndIf
\EndFor
\State $PSB$ is now an $N\times N$ binary image of the phase space which is $1$ where the orbit is detected as stochastic and $0$ where it is not.
\State Compute $SEDT(PSB)$.
\State Using \verb!Diamorse! calculate the sub-level set persistent $0$-th homology of $SEDT(PSB)$ and call it $PH_0$.
\State Pull all birth times of classes $b = \{b_0,b_1,\ldots b_m\}$ from $PH_0$.
\State Calculate the normalised histogram of the set $r:=\{-b_0,-b_1,\ldots -b_m\}$ and call it $n(r)$. 
\State \Return $n(r)$
\end{algorithmic}
\end{algorithm}

Before we discuss the results of this calculation on relevant models it is helpful to elaborate upon the intermediate stages and discuss some technical problems which emerge. We will do this by considering the example of the standard non-twist map. 

\begin{figure}
    \centering
    \begin{subfigure}[b]{0.35\textwidth}
        \includegraphics[width = \textwidth,frame]{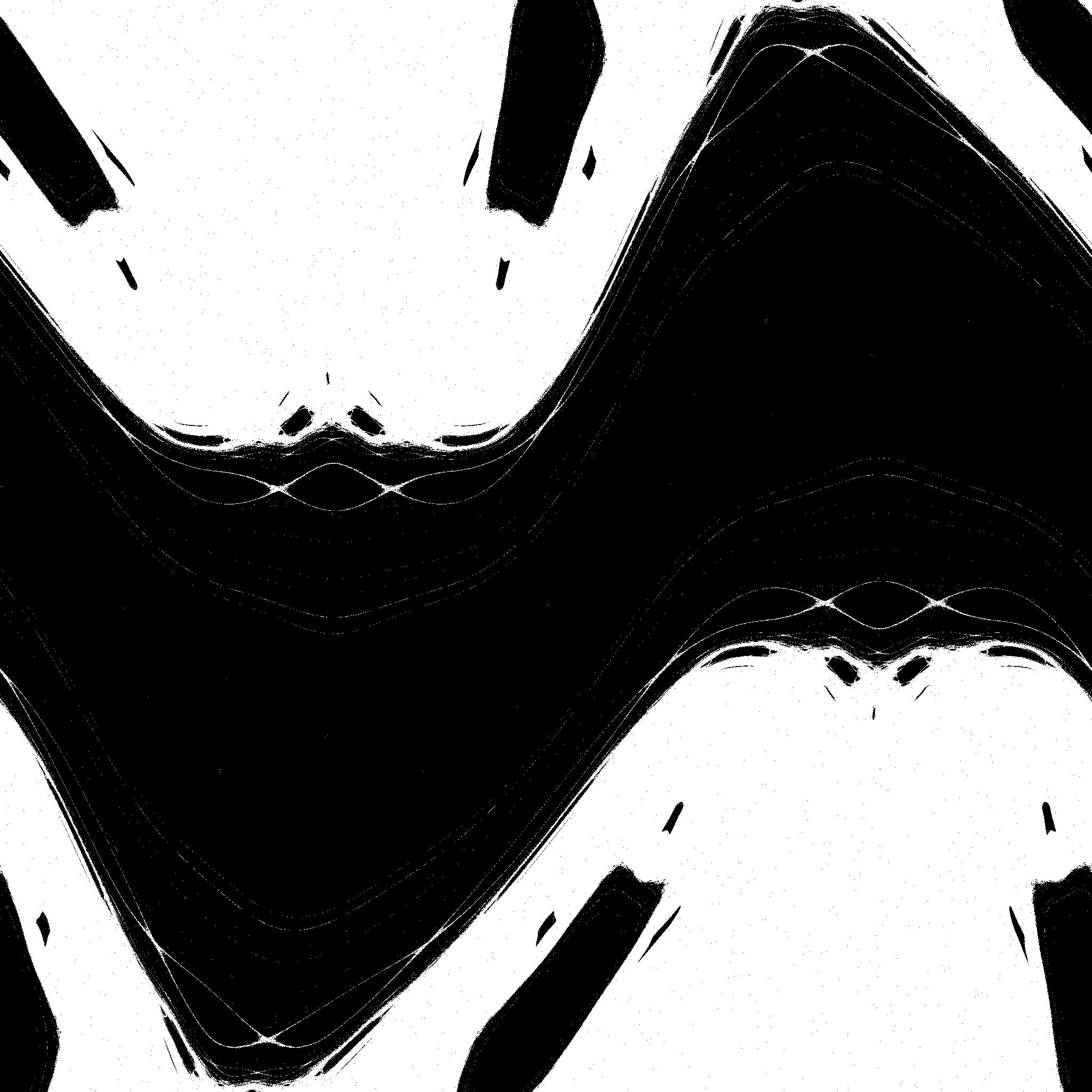}
        \subcaption{Phase space before closure.}
        \label{NontwistBinaryOpen}
    \end{subfigure}
    \begin{subfigure}[b]{0.35\textwidth}
        \includegraphics[width = \textwidth]{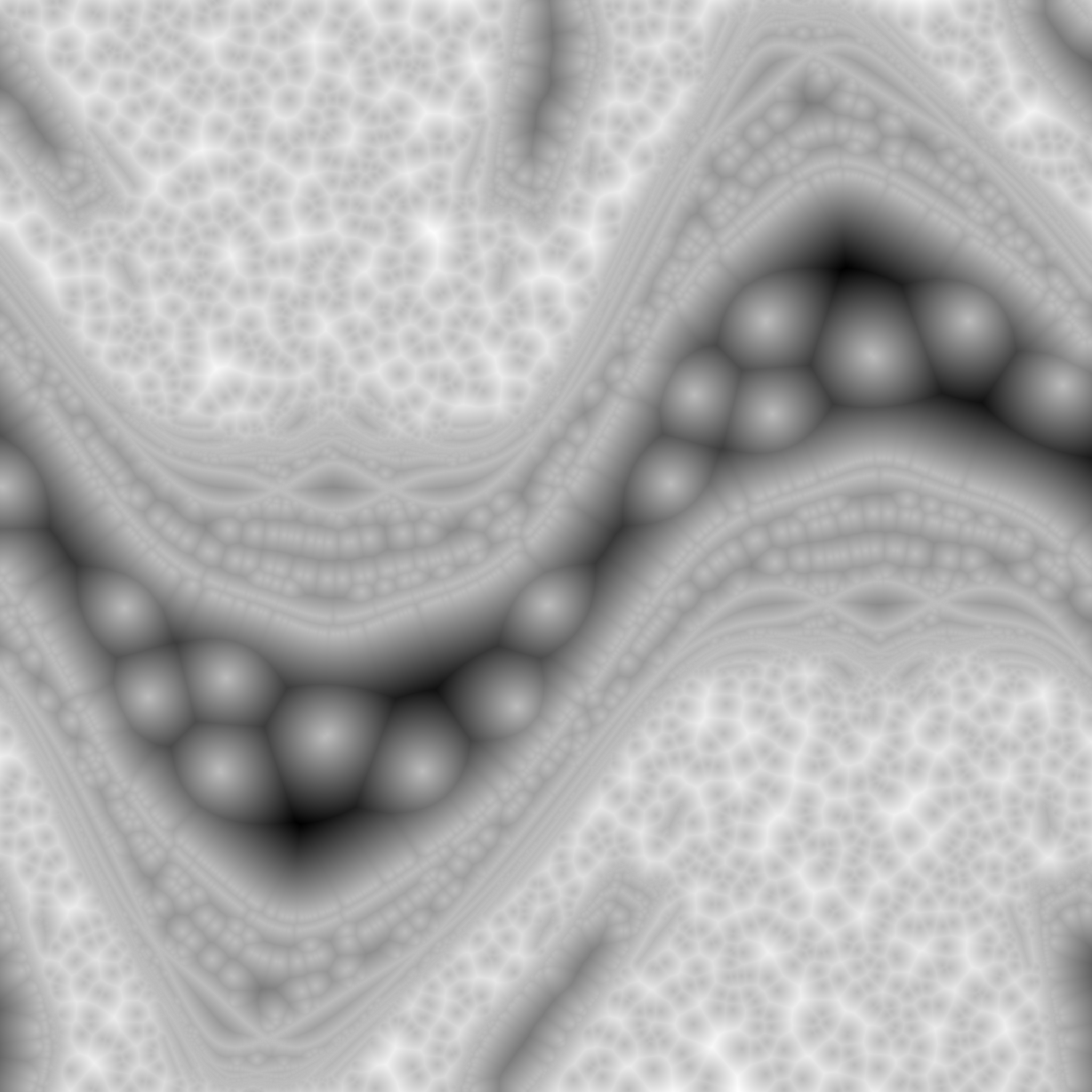}
        \subcaption{SEDT before closure.}
        \label{NontwistSEDTOpen}
    \end{subfigure}

    \begin{subfigure}[b]{0.35\textwidth}
        \centering
        \includegraphics[width = \textwidth,frame]{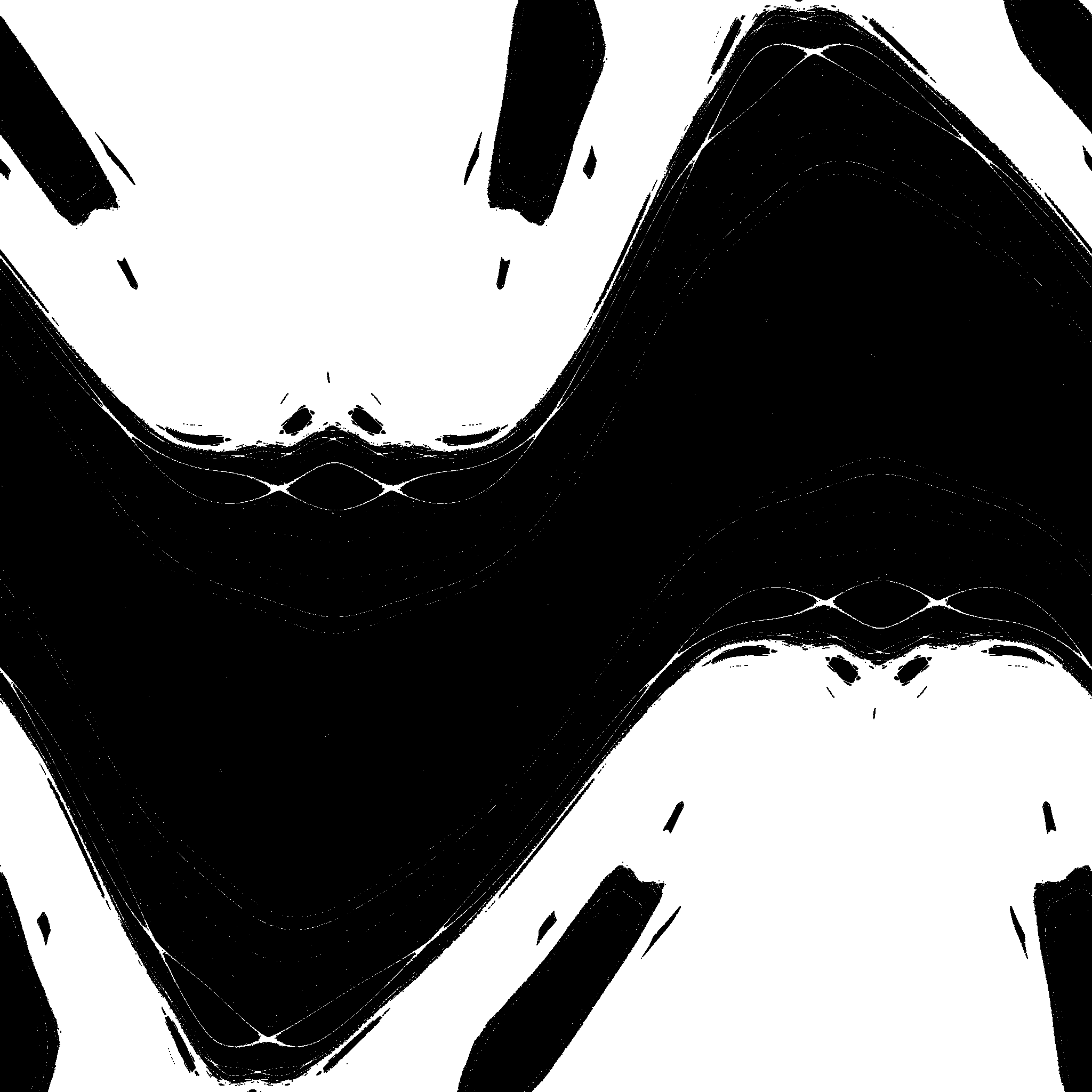}
        \subcaption{Phase space after closure.}
        \label{NontwistBinaryClosed}
    \end{subfigure}
    \begin{subfigure}[b]{0.35\textwidth}
        \centering
        \includegraphics[width = \textwidth]{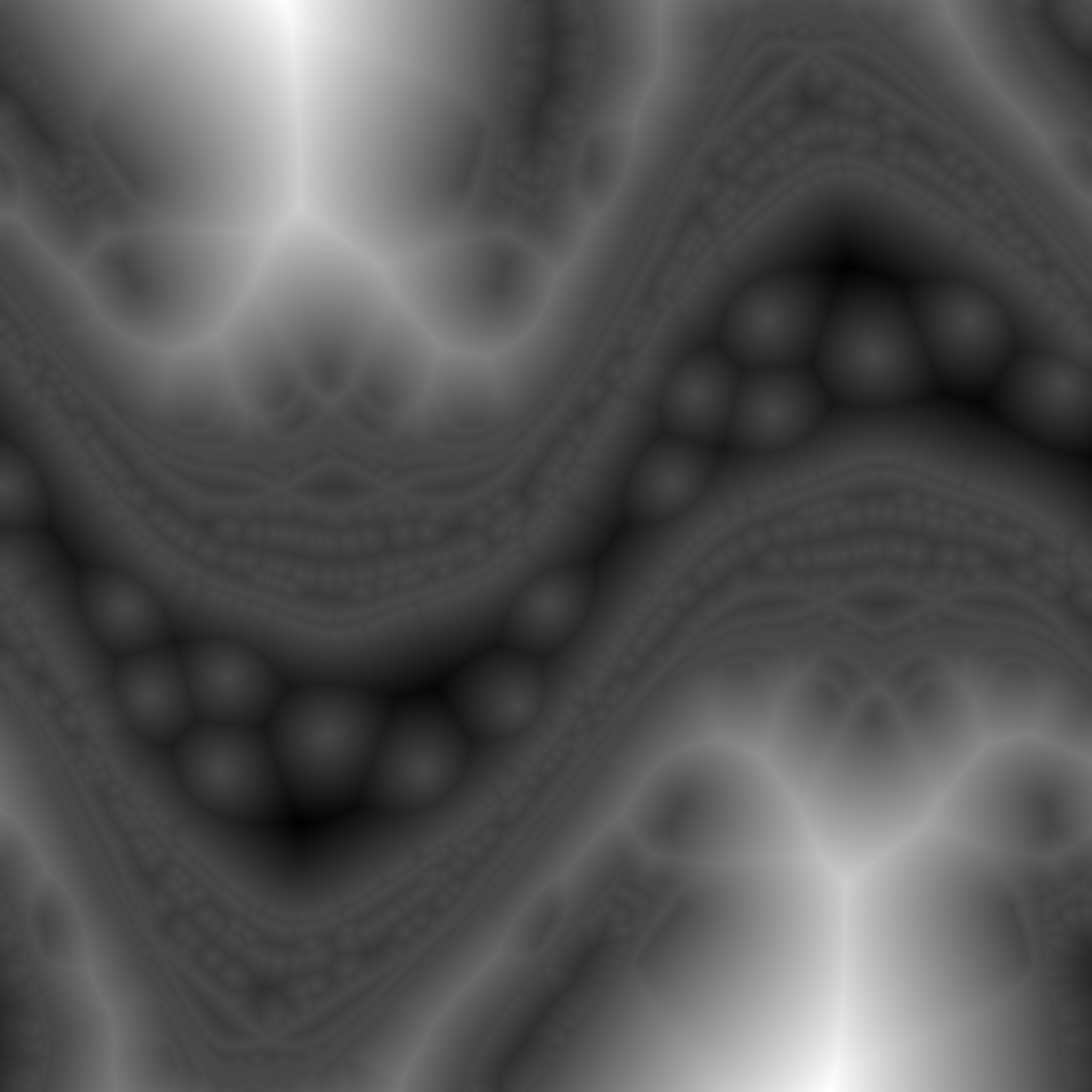}
        \subcaption{SEDT after closure.}
        \label{NontwistSEDTClosed}
    \end{subfigure}
    \caption{Binary images of the stochastic region, and its SEDT, of the standard non-twist map before and after closure}
    \label{fig:standard_nontwist_wba_binarySEDT}
\end{figure}

Recall Figure \ref{fig:standard_nontwist_wba} which presented and image of the $\tilde{D}$ values evaluated for the standard non-twistmap. Binarising this image with the threshold taken to be the minimum of the brightness histogram produces the binary image presented as subfigure \ref{NontwistBinaryOpen} the SEDT of which is shown in subfigure \ref{NontwistSEDTOpen}. We see that the islands, connected black regions, in the binary image correspond to locally dark regions in the SEDT. Importantly, each locally dark region is a local minimum and so by identifying each local minimum we can quantify the number and size of the islands in the binary image. Note though, that we see many small local minimums in what we expect to be the stochastic region. These are a numerical artifact caused by single black pixels which appear inside the stochastic region. These pixels are not black because the point on which they are centered has a quasi-periodic orbit but instead are stochastic points whose $\tilde{D}$ happened to be slightly above the binarisation threshold. This error is an unavoidable side-effect of the use of a finite time calculation for the WBA, and we would expect it to be less relevant for larger $T$. For now, we will just ignore these points but will see later below that they can be removed and it is best to do so.

\begin{figure}
    \centering
    \begin{subfigure}[b]{0.49\textwidth}
        \includegraphics[width = \textwidth]{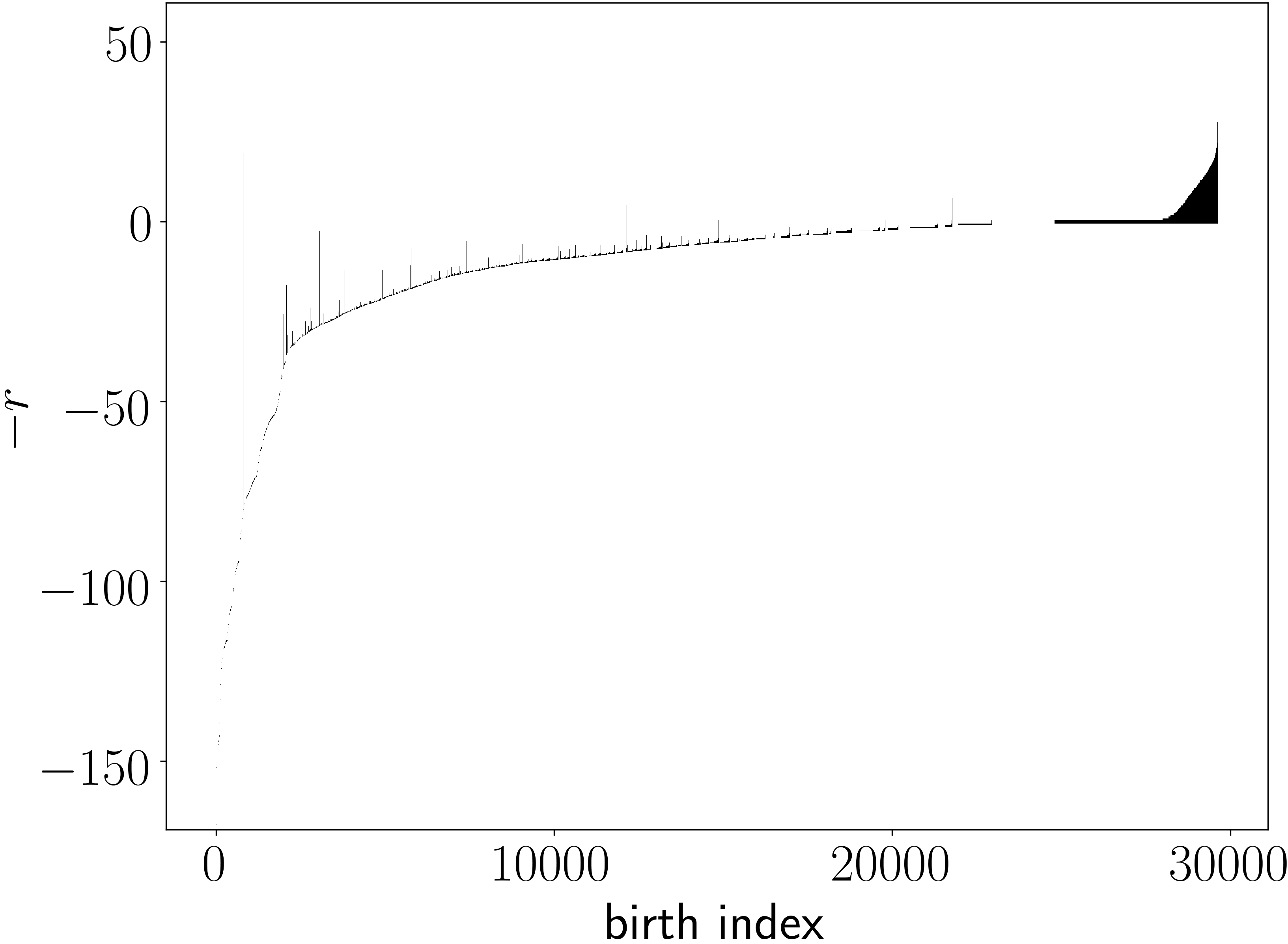}
        \subcaption{Barcode before closure.}
        \label{Nontwist13bw_no_closure_barcode_0}
    \end{subfigure}
    \begin{subfigure}[b]{0.49\textwidth}
        \includegraphics[width = \textwidth]{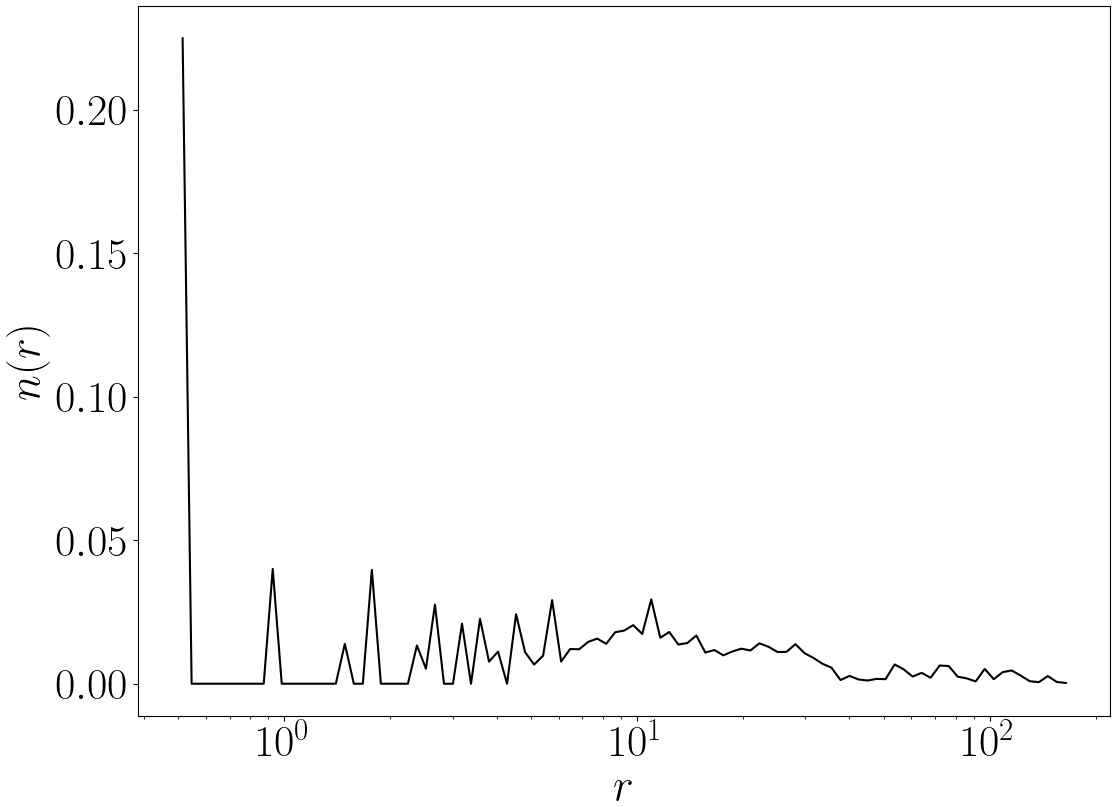}
        \subcaption{Island size distribution before closure.}
        \label{birth_dist_Nontwist13NoClosure}
    \end{subfigure}
    \begin{subfigure}[b]{0.49\textwidth}
        \includegraphics[width = \textwidth]{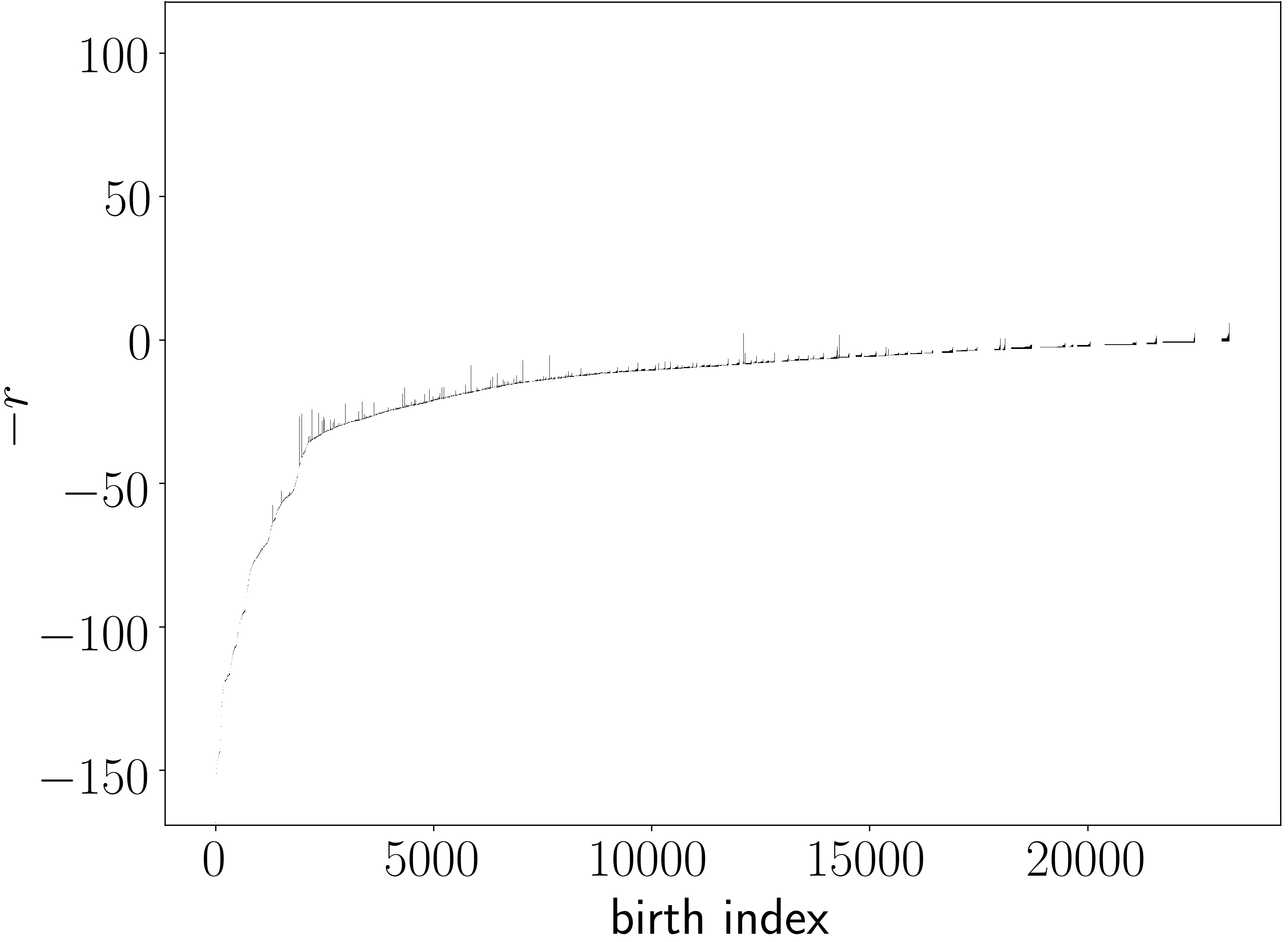}
        \subcaption{Barcode after closure.}
        \label{Nontwist13bw_2_closure_barcode_0}
    \end{subfigure}
    \begin{subfigure}[b]{0.49\textwidth}
        \includegraphics[width = \textwidth]{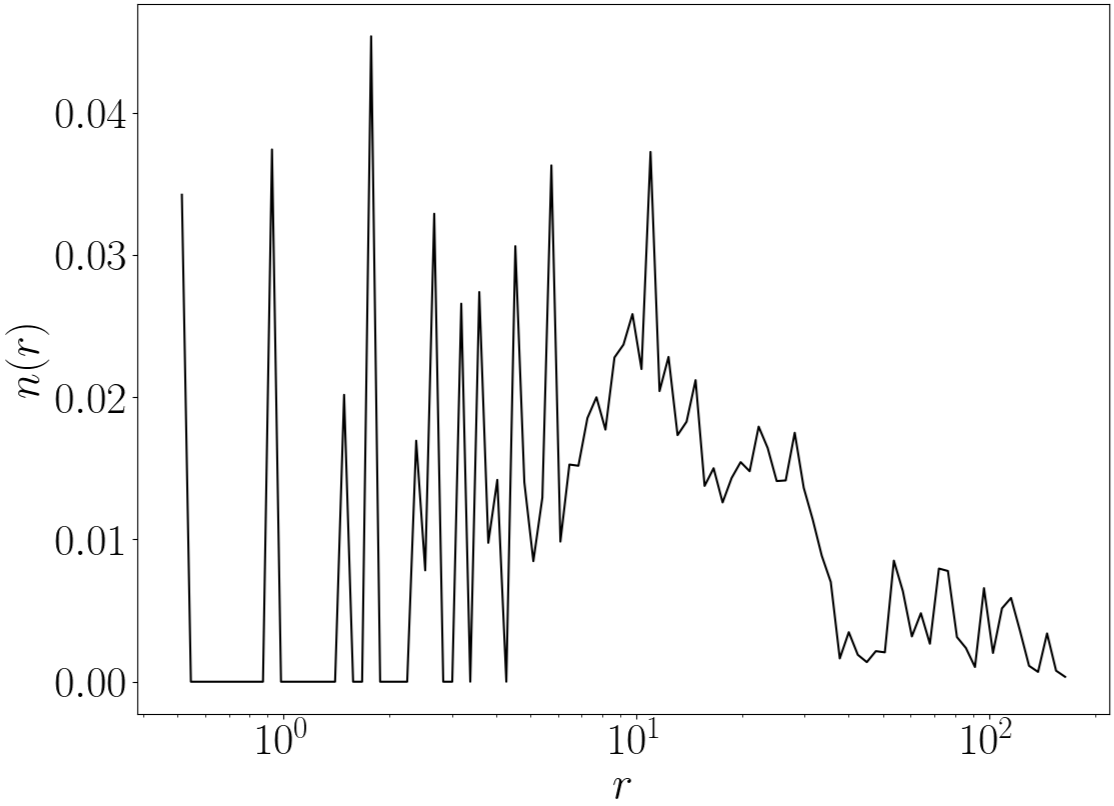}
        \subcaption{Island size distribution before closure.}
        \label{birth_dist_Nontwist13wClosure}
    \end{subfigure}
    \caption{Barcode and island size distribution of the standard non-twist map phase space shown as sub-figure \ref{NontwistSEDTOpen} before and after closure.}
    \label{fig:standard_nontwist_no_closure_island_distribution}
\end{figure}

The next step in the computation of the island distribution is to compute the sub-level set persistent homology of subfigure \ref{NontwistSEDTOpen} using \verb!DIAMORSE!. This will output a list of birth and death times for topological features of the image which we can display as a barcode. The barcode is presented as subfigure \ref{Nontwist13bw_no_closure_barcode_0}. Note that in this figure the bottom of each bar is located at the value in the SEDT image at which the associated connected component is born. Our approximation to the radius of the connected component is the negative of the birth time, as noted above, and so we label the vertical axis with $-r$. We observe that there is a very large number of very short bars, that is of connected components which are not persistent. This is because no \textit{pair cancellation} has been performed and so many of the bars do not correspond to complete islands. We will discuss this further below. 

Subfigure \ref{birth_dist_Nontwist13NoClosure} presents the histogram of connected component radii, that is the histogram of values $r$ taken from the barcode to its left, using a logarithmic scale of radii. This is a complete island radius distribution $n(r)$ and it is the output of the \textit{BirthsDist} algorithm. The notable feature of subfigure \ref{birth_dist_Nontwist13NoClosure} is that $\approx 20\%$ of all of the detected islands belong to one bin at the smallest possible $r$ size. These islands correspond to the single isolated black dots present within the white stochastic region of sub-figure \ref{NontwistBinaryOpen}. As such they are actually an image artifact and not a topological feature of interest. Therefore, it is preferable if we can remove these isolated dots from the input binary image, so that they do not contribute to the calculated $n(r)$ function. We can achieve this with tool from the study of mathematical morphology called \textit{closure}.

\subsection{Mathematical morphology and Closure}

Mathematical morphology is a theory concerned with analysing and processing geometrical objects using fundamental tools from set theory and topology and it is most commonly utilised to process binary images. Many different tools can be classed as morphological operations but we are concerned only with closure here. This operation is a method of removing small holes in binary images and it is constructed out of two simpler operations, \textit{dilation} and \textit{erosion} which are computed by scanning over the image with a structuring element \cite{szeliski2022computer}.  

\begin{defn}
    Suppose we have a binary image $I$ which we represent as an $n\times m$ binary matrix. Let $S$ be a $s\times s$ matrix of all $1$'s, called the structuring element. The matrix $c = I * S$, where $*$ indicates matrix convolution, is a matrix which counts the number of $1$'s in each $s\times s$ sized submatrix of $I$.
    
    Then we define $dilate(I)$ and $erode(I)$ as the matrices
    \begin{equation}
        dilate(I)_{ij} = \begin{cases}
            1\,, &\text{  if   } c_{ij} \geq 1\\
            0 \,, &\text{ otherwise}
        \end{cases}\,,\,      \ erode(I)_{ij} = \begin{cases}
            1\,, &\text{  if   } c_{ij} = s^2\\
            0 \,, &\text{ otherwise}
        \end{cases}\,.
    \end{equation}
    The closure of the image is then defined as 
    \begin{equation}
        close(I) = erode(dilate(I))\,.
    \end{equation}
\end{defn}

The definition above is formally incomplete as we have not detailed how to handle the edges of the image, where the sum in our convolutions will extend outside the image. There are many standard approaches to solve this problem and so we will not discuss this in detail here. Note that all closures on images presented in this thesis were computed with $2\times 2$ structuring elements using \textit{OpenCV} implementation of closure and therefore the edges were handled by reflection, which is the \textit{OpenCV} default.

As an example, we apply closure to subfigure \ref{NontwistBinaryOpen} and compute the SEDT of the \textit{closed} image. The output of this computation is presented as subfigures \ref{NontwistBinaryClosed} and \ref{NontwistSEDTClosed}. Comparing the binary image of the stochastic region before and after closure we can see that the large scale structure, and hence the global topology, of the image was not be substantially affected by the act of closure. However, the SEDT has changed dramatically. Many ``false'' local minimums corresponding to isolated black points have been removed and there is now a clear global maximum in each of the disconnected stochastic regions rather than many equivalent local maximums. Then, computing the sub-level set persistence of the SEDT image and the associated negative birth time histogram, that is $n(r)$, we obtain subfigure \ref{birth_dist_Nontwist13wClosure}. The removal of isolated black points has reduced the length of the barcode, from above $30000$ bars to less than $25000$ bars and importantly it has removed the large spike from the smallest radius bin while leaving the shape of the island size distribution unchanged at larger radius. This demonstrates that morphological closure is sufficient to remove the effect of the unwanted isolated points, which, to reiterate, are an image artifact, without affecting the rest of the distribution we are attempting to compute. 

\subsection{The effect of pair cancellation}

There is a major consideration when computing the sub-level set persistence barcode for an image, and the birth time distribution derived from it, which we have yet to discuss and that is the problem of finite size effects. We observe that in the barcode presented as subfigure \ref{Nontwist13bw_2_closure_barcode_0} there are many very short bars. Recall that short bars are associated to ``non-persistent'' topological features and are usually associated to error. In our case the main sources of error are the amorphous shape of the islands\footnote{A neck joining two connected black regions will not necessarily kill one of the minimums associated to a connected region.}, and the incomplete coverage of the stochastic region. These artifacts do not correspond to real topological features, they are not real islands in our case, and so we would prefer to remove them so that they do not contribute to $n(r)$.

\begin{figure}[H]
    \centering
    \begin{subfigure}[b]{0.45\textwidth}
        \includegraphics[width = \textwidth]{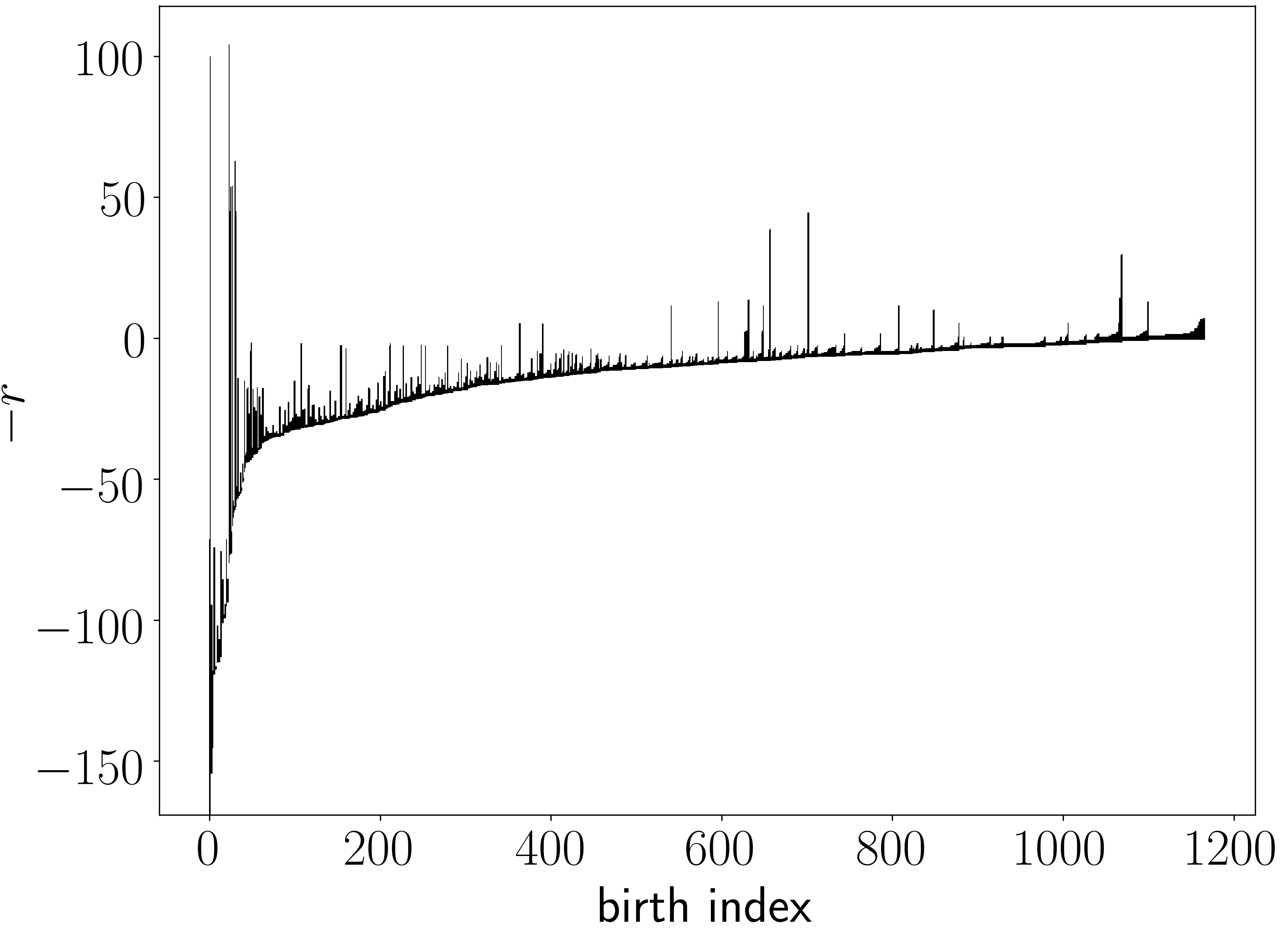}
        \subcaption{$t=1.0$}
    \end{subfigure}
    \begin{subfigure}[b]{0.45\textwidth}
        \includegraphics[width = \textwidth]{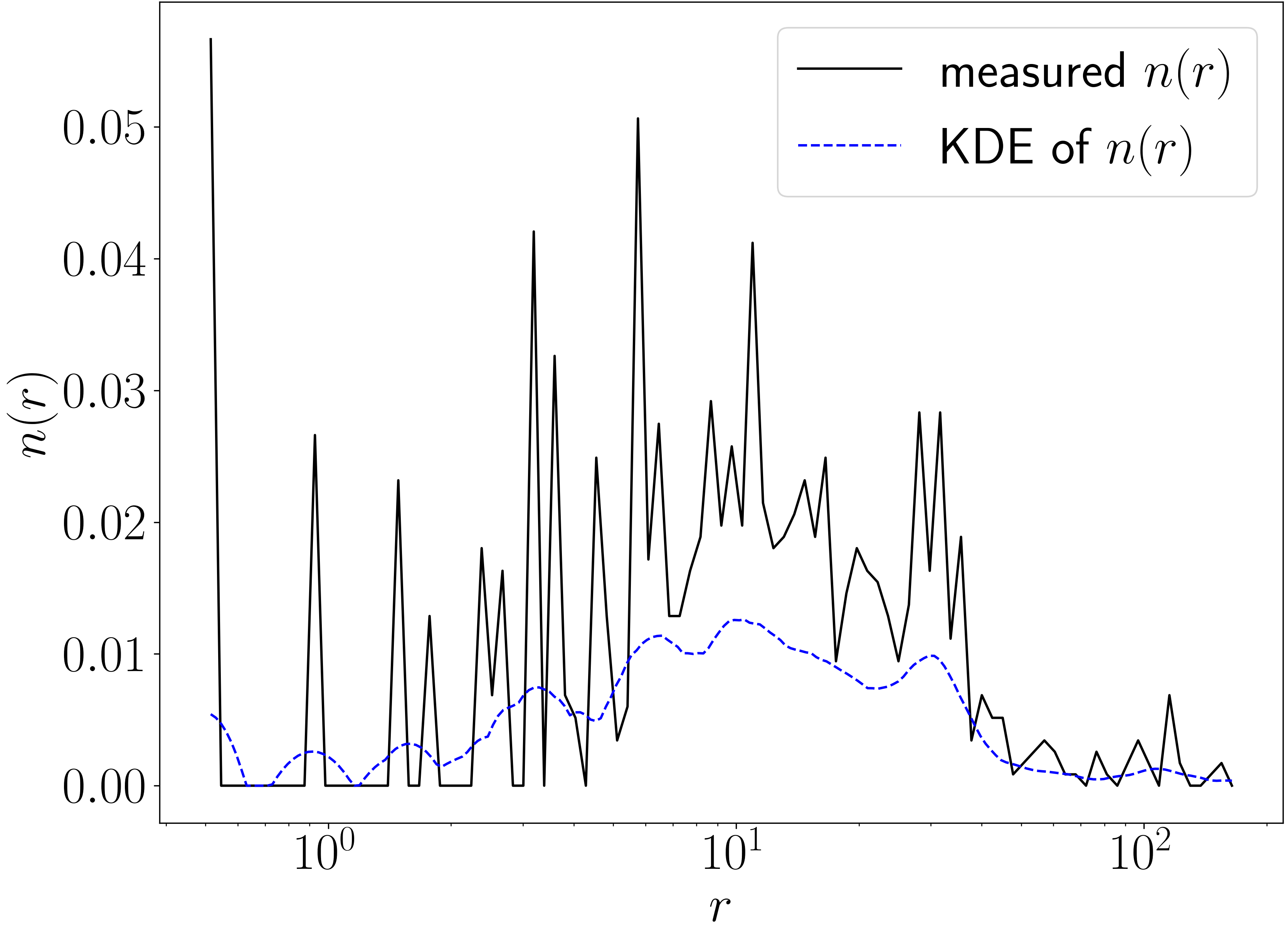}
        \subcaption{}
    \end{subfigure}
    \begin{subfigure}[b]{0.45\textwidth}
        \includegraphics[width = \textwidth]{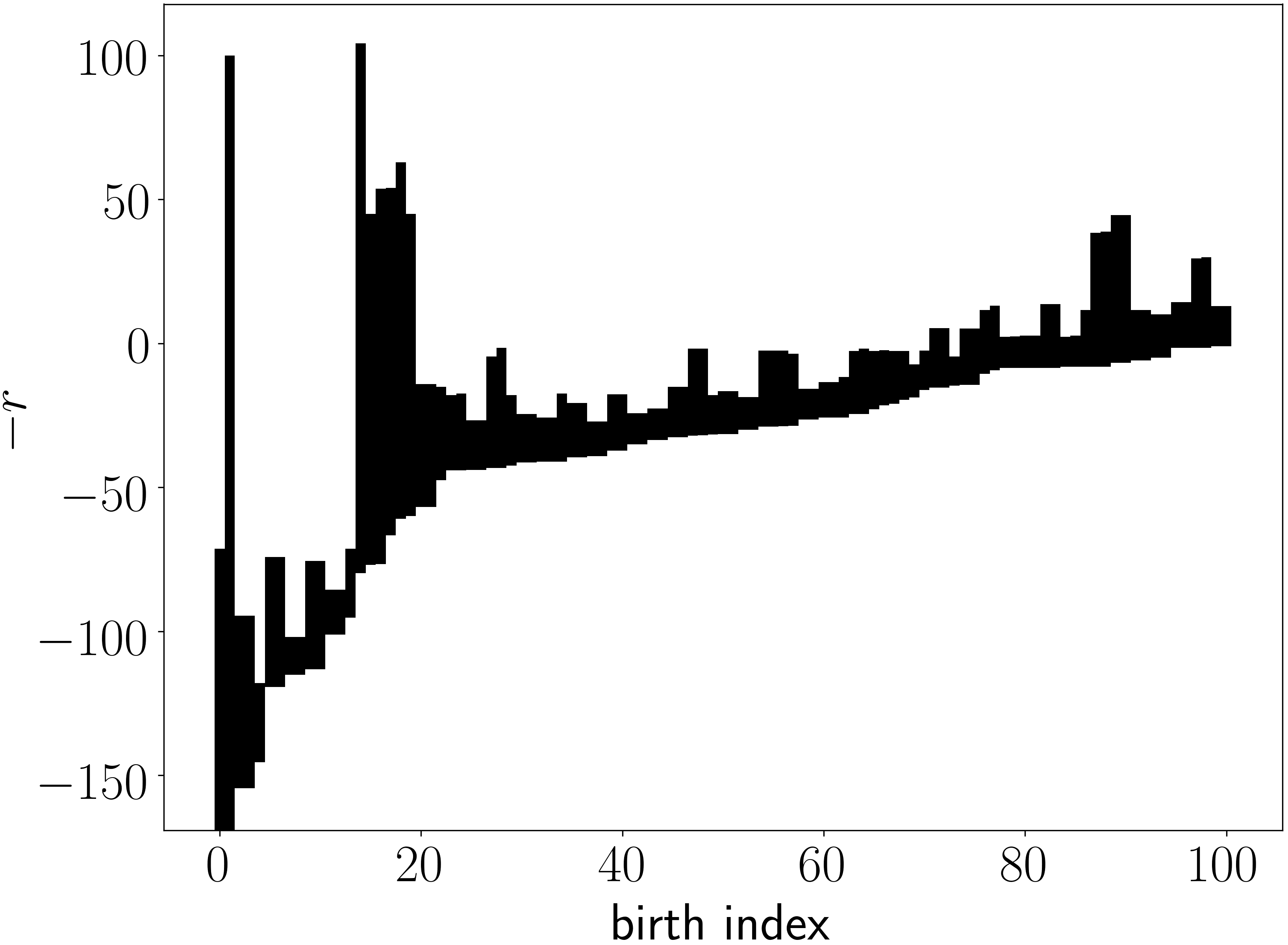}
        \subcaption{$t=10.0$}
    \end{subfigure}
    \begin{subfigure}[b]{0.45\textwidth}
        \includegraphics[width = \textwidth]{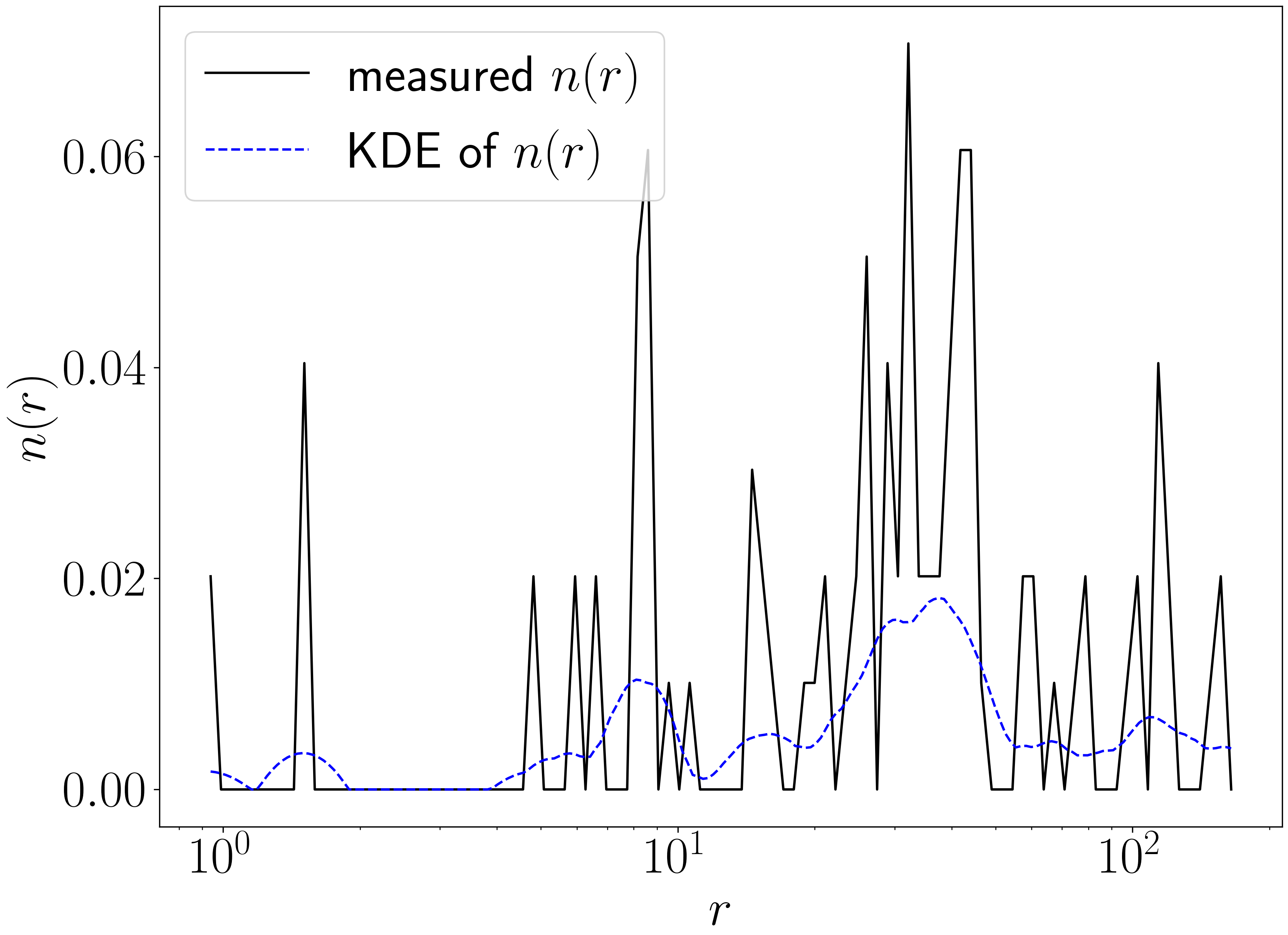}
        \subcaption{}
    \end{subfigure}
    \caption{Comparison of two barcodes and island size distributions for different levels of pair cancellation.}
    \label{fig:standard_nontwist_pair_cancellation_effect}
\end{figure}

We have a procedure for doing this, called close pair cancellation which is described in detail in \cite{DIAMORSE1}. It allows us to remove artifacts due to finite size effects but this comes at the cost of information loss. The more cancellation we perform, the more confident we can be that the topological features we detect are real but as we do this we will also lose information corresponding to real features which we misidentify as errors. 

Consider Figure \ref{fig:standard_nontwist_pair_cancellation_effect} which presents the barcodes and birth time distributions for the standard non-twist map computed after applying close pair cancellations at thresholds $t=1$ and $t=10$ respectively. We can see immediately from the barcodes that the close pair cancellation has removed the shortest bars from the barcode as expected. The longer bars have been unaffected and the ``contour'' of the barcode is similar. This confirms that close pair cancellation does allow us to perform error reduction on our topological data. Importantly this is controllable, by the value of $t$, and we observe that when $t$ is larger many fewer bars remain.  Notably the birth time distributions are quite distinct and this is due to the loss of relevant topological information associated with over-cancelling of close pairs. This suggests that $t=10$ is too high of a simplification threshold for this particular example. The optimal level of simplification cannot be determined in generality and instead needs to be identified for any given problem by experimentation or via heuristics. We mostly adopt $t=1$ as it is a commonly accepted standard value but will use $t=2$ where is it more appropriate. 


\section{Artificial examples}\label{ArtificialExamples}

\begin{figure}
    \centering
    \begin{subfigure}[b]{0.32\textwidth}
        \centering
        \includegraphics[width = 0.8\textwidth,frame]{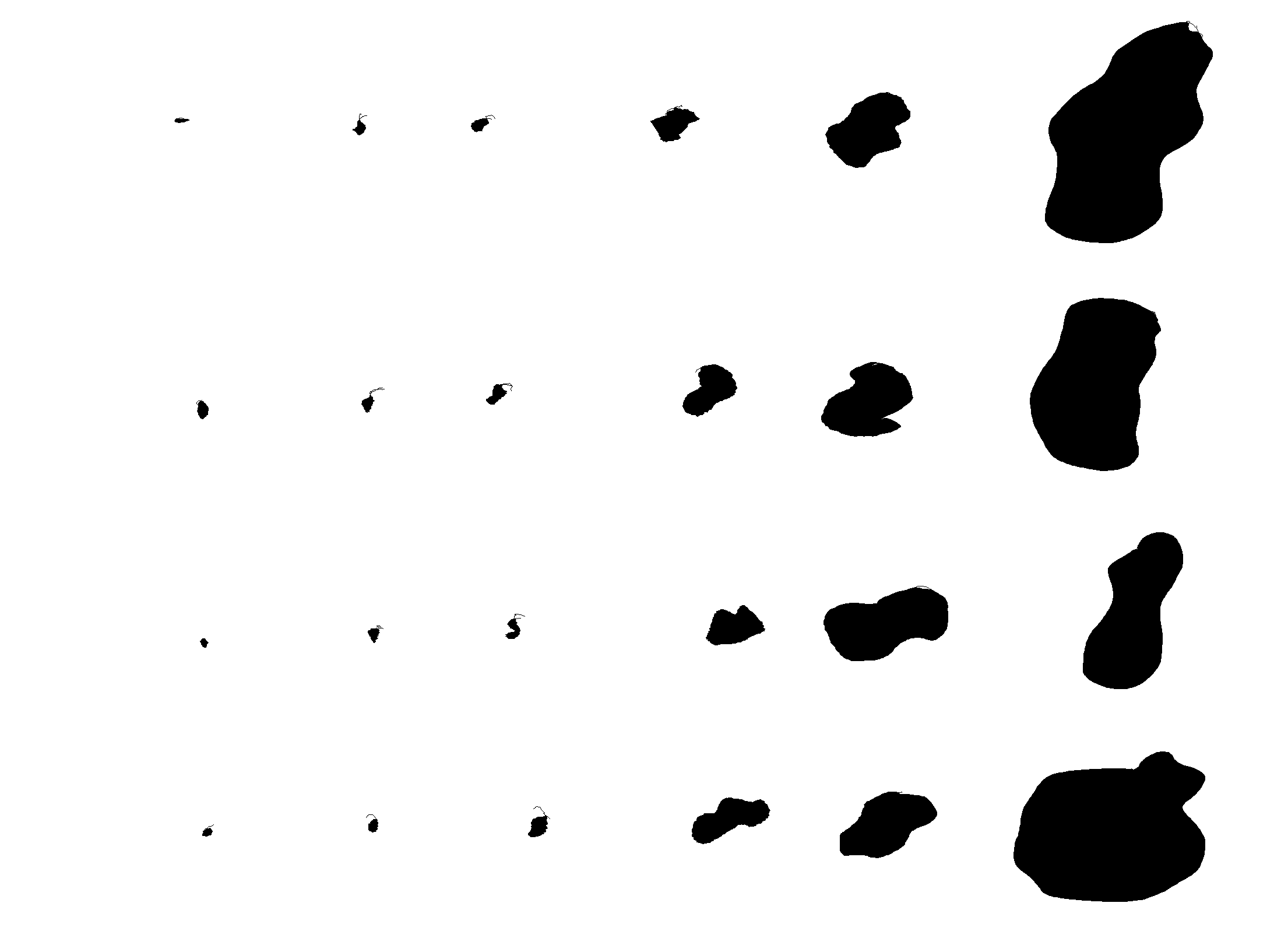}
        \subcaption{}
        \label{subfig:artificial_island_images_a}
    \end{subfigure}
    \begin{subfigure}[b]{0.32\textwidth}
        \centering
        \includegraphics[width = 0.58\textwidth,frame]{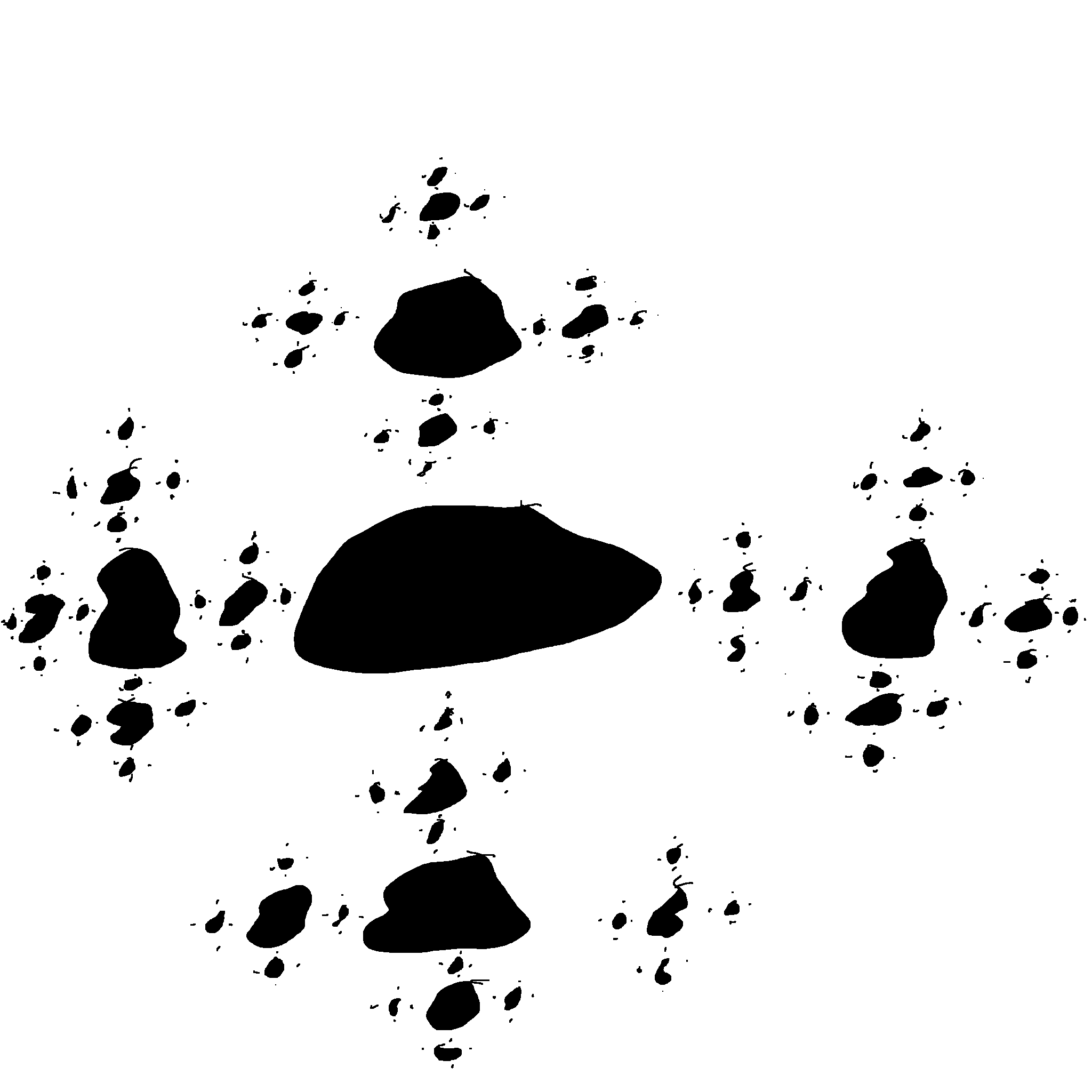}
        \subcaption{}
        \label{subfig:artificial_island_images_b}
    \end{subfigure}
    \begin{subfigure}[b]{0.32\textwidth}
        \centering
        \includegraphics[width = 0.83\textwidth,frame]{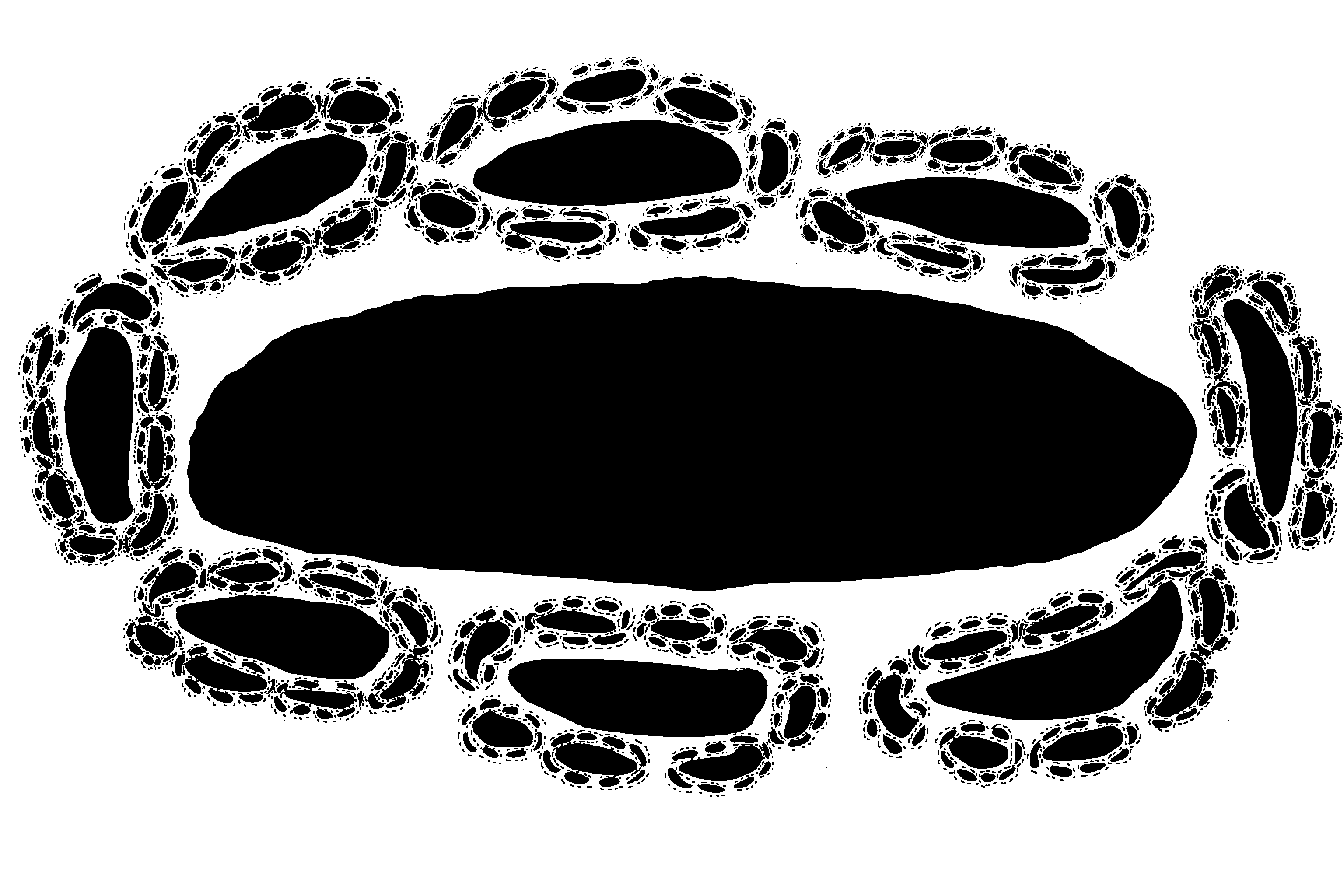}
        \subcaption{}
        \label{subfig:artificial_island_images_c}
    \end{subfigure}
    \caption{Binary images of artificial island hierarchies}
    \label{fig:artificial_island_images}
\end{figure}

It is interesting to look at a few artificial examples to demonstrate how the island size distribution changes with the image, and to also illustrate that the birth time distribution we are computing is indicative of the true spectrum of island radius. Consider then the three binary images presented in Figure \ref{fig:artificial_island_images}. Each image illustrates a different hierarchy of islands. In subfigure \ref{subfig:artificial_island_images_a} we have an equal number of islands in each generation, in subfigure \ref{subfig:artificial_island_images_b} each island generation contains a factor of four more islands, and for subfigure \ref{subfig:artificial_island_images_c} the equivalent factor is eight. Therefore, for these images we expect that the number of islands at smaller radii will get progressively larger and therefore $n(r)$ should rise at smaller $r$, while for subfigure \ref{subfig:artificial_island_images_a} $n(r)$ should be approximately uniform in $r$. 

\begin{figure}
    \centering
    \begin{subfigure}[b]{0.45\textwidth}
        \includegraphics[width = \textwidth]{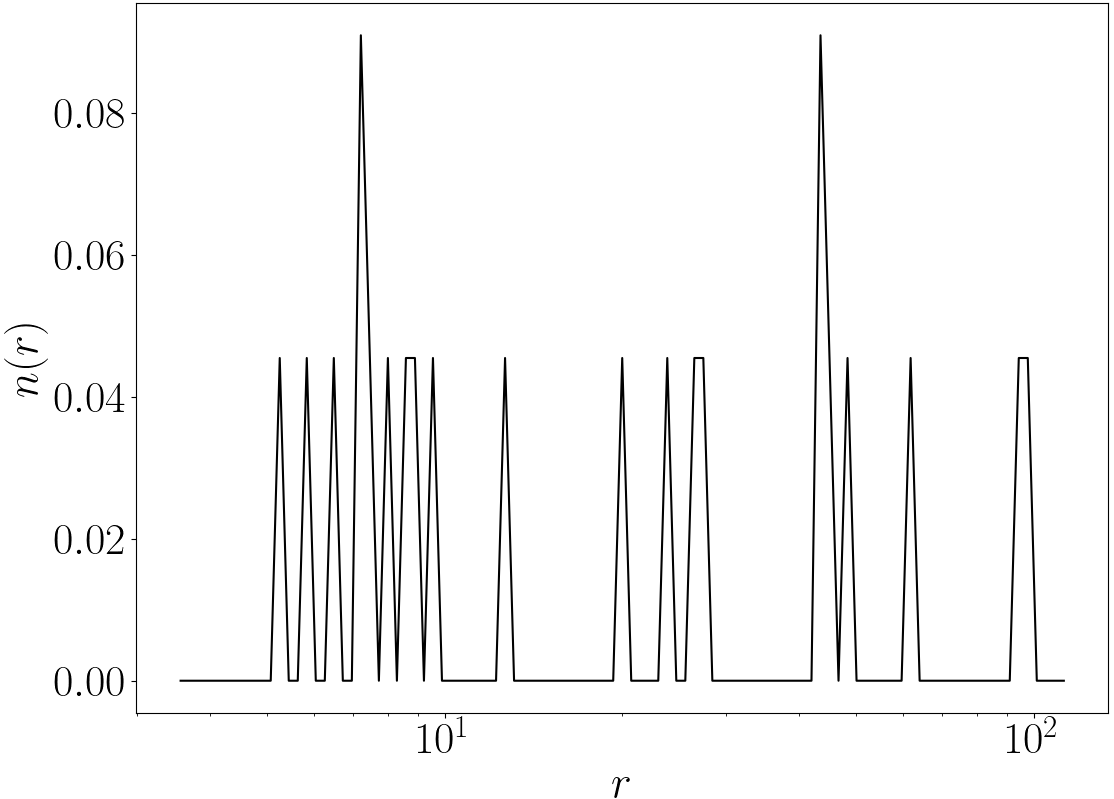}
        \subcaption{}
        \label{subfig:artificial_island_radii_distributions_a}
    \end{subfigure}
    \begin{subfigure}[b]{0.45\textwidth}
        \includegraphics[width = \textwidth]{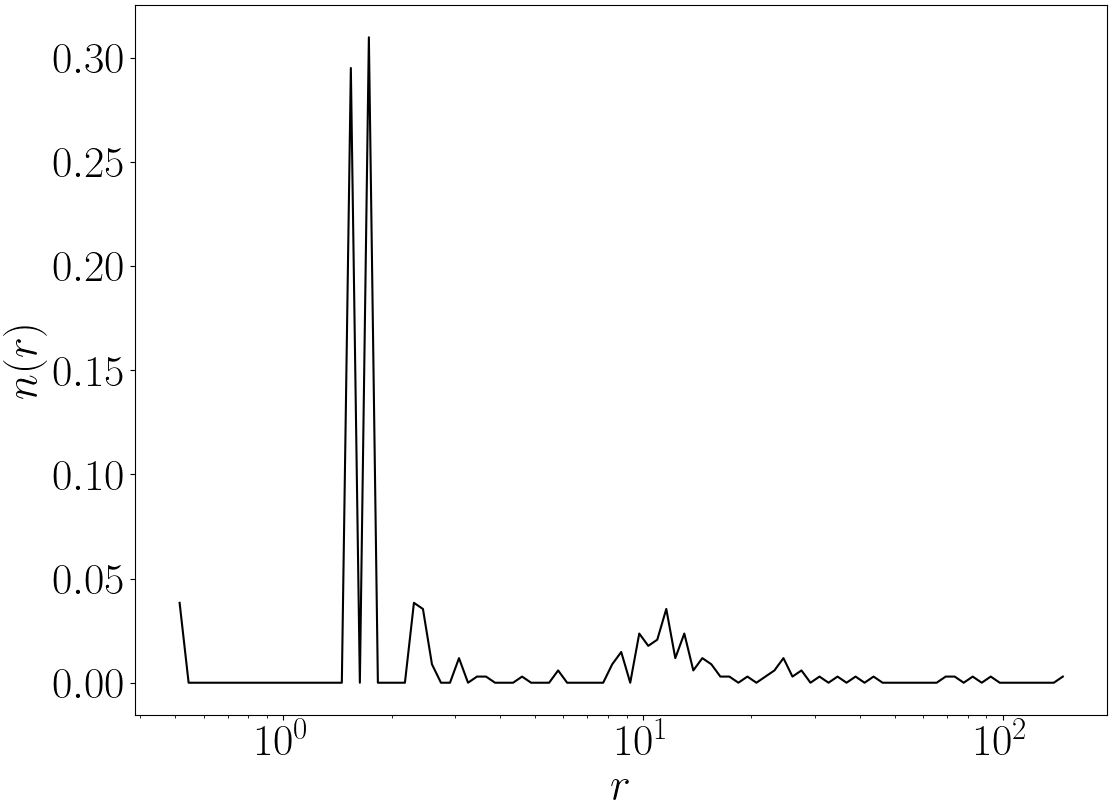}
        \subcaption{}
        \label{subfig:artificial_island_radii_distributions_b}
    \end{subfigure}
    \begin{subfigure}[b]{0.45\textwidth}
        \includegraphics[width = \textwidth]{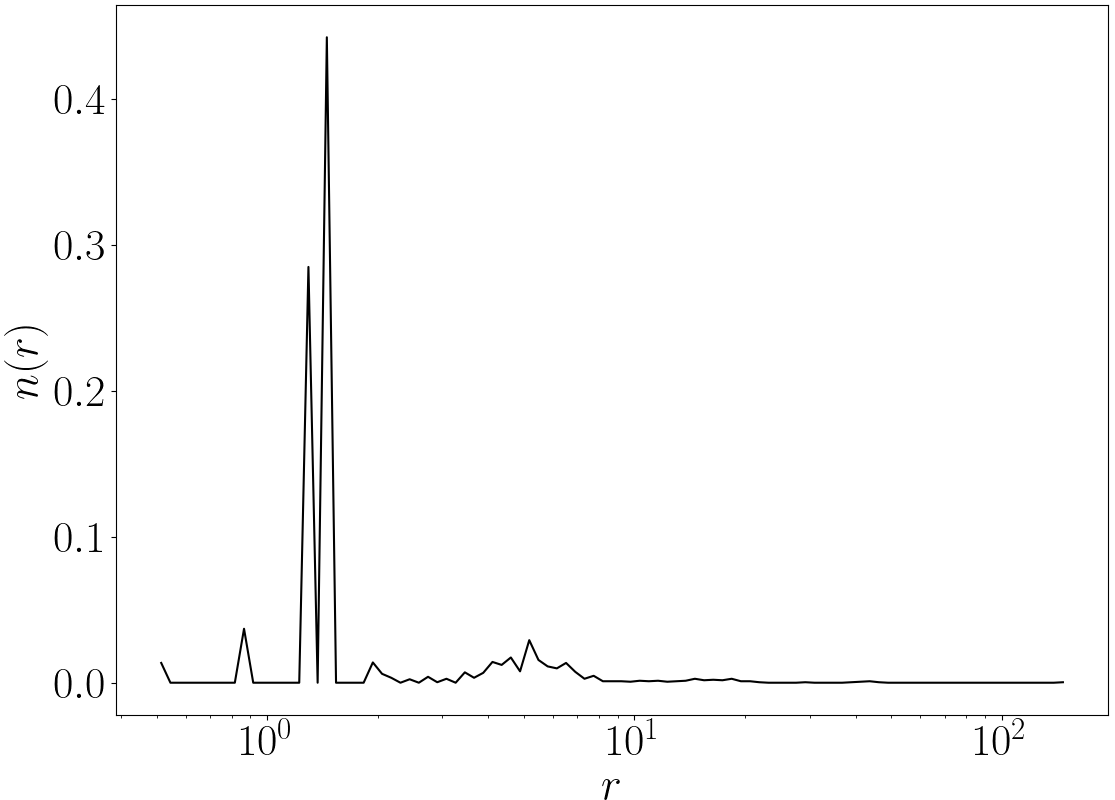}
        \subcaption{}
        \label{subfig:artificial_island_radii_distributions_c}
    \end{subfigure}
    \caption{Estimated island radius distributions for subfigures \ref{subfig:artificial_island_images_a} , \ref{subfig:artificial_island_images_b}, and \ref{subfig:artificial_island_images_c} respectively.}
    \label{fig:artificial_island_radii_distributions}
\end{figure}

To confirm the above hypotheses we compute $n(r)$ for each image. Note that we will use the simplification threshold of $t=2$ in this case. Performing this computation we obtain the $n(r)$ graphs presented as Figure \ref{fig:artificial_island_radii_distributions}. We observe that the subfigure \ref{subfig:artificial_island_radii_distributions_a} distribution is approximately flat as we expected. The subfigure \ref{subfig:artificial_island_radii_distributions_b} and \ref{subfig:artificial_island_radii_distributions_c} distributions each contain a very large spike at a radius of order $O(10^0)$. This indicates that both subfigures \ref{subfig:artificial_island_images_b} and \ref{subfig:artificial_island_images_c} contain a plurality of islands which are only a few pixels in radius. We recognise that this spike is associated with the ``final'' generation of the islands, which are necessarily the most numerous and also the smallest. 

Subfigures \ref{subfig:artificial_island_radii_distributions_b} and \ref{subfig:artificial_island_radii_distributions_c} highlight a general problem that when we image the island distribution of ``fractal'' structures we are dominated by the smallest scales and therefore, at least visually, it is hard to resolve the features in the distribution by simply displaying them as we have. Instead, it is more natural to adopt a logarithmic vertical axis. 

\begin{figure}
    \centering
    \begin{subfigure}[b]{0.49\textwidth}
        \includegraphics[width = \textwidth]{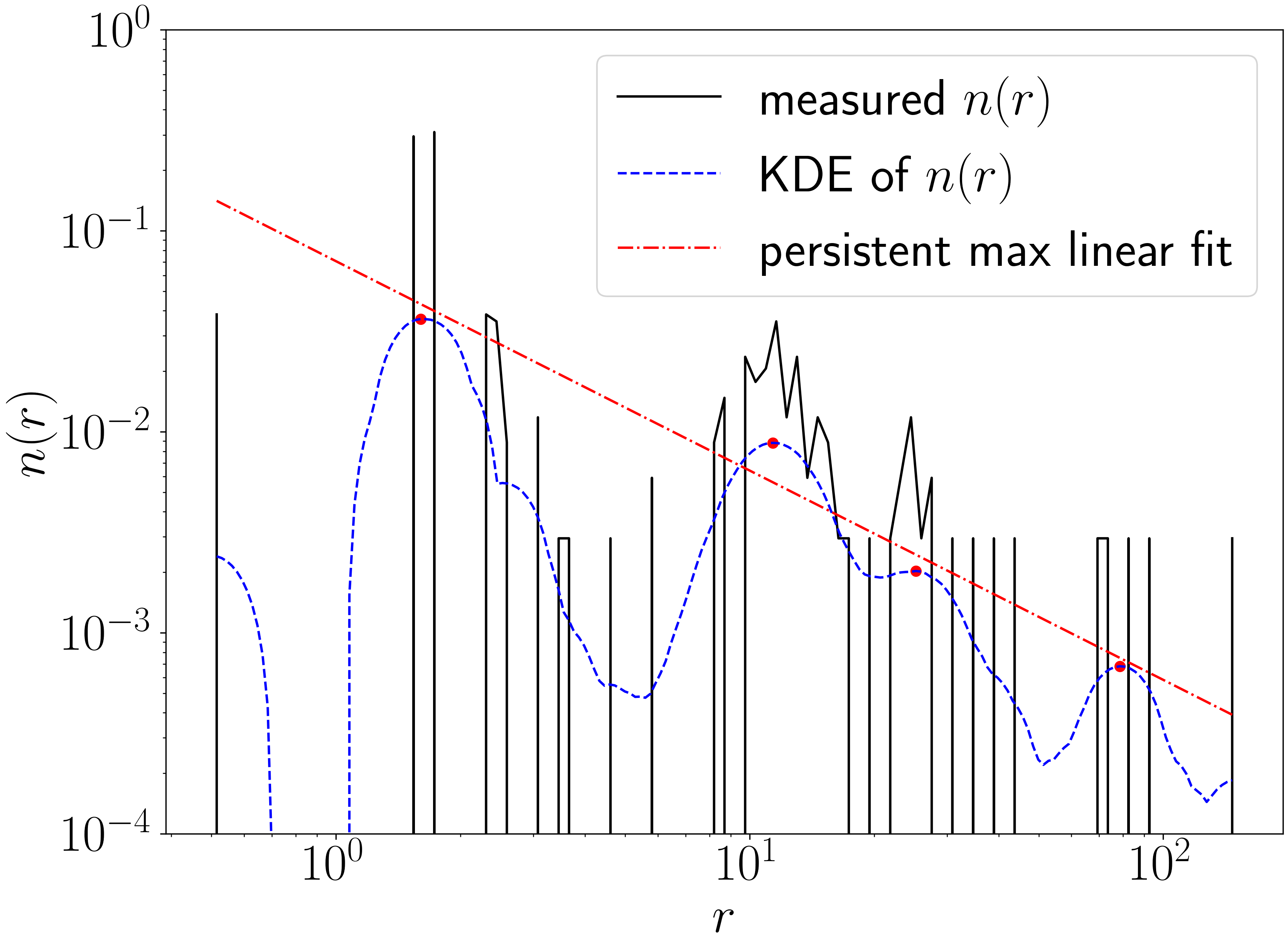}
        \subcaption{}
        \label{fig:artificial_island_radii_distributions_log_a}
    \end{subfigure}
    \begin{subfigure}[b]{0.49\textwidth}
        \includegraphics[width = \textwidth]{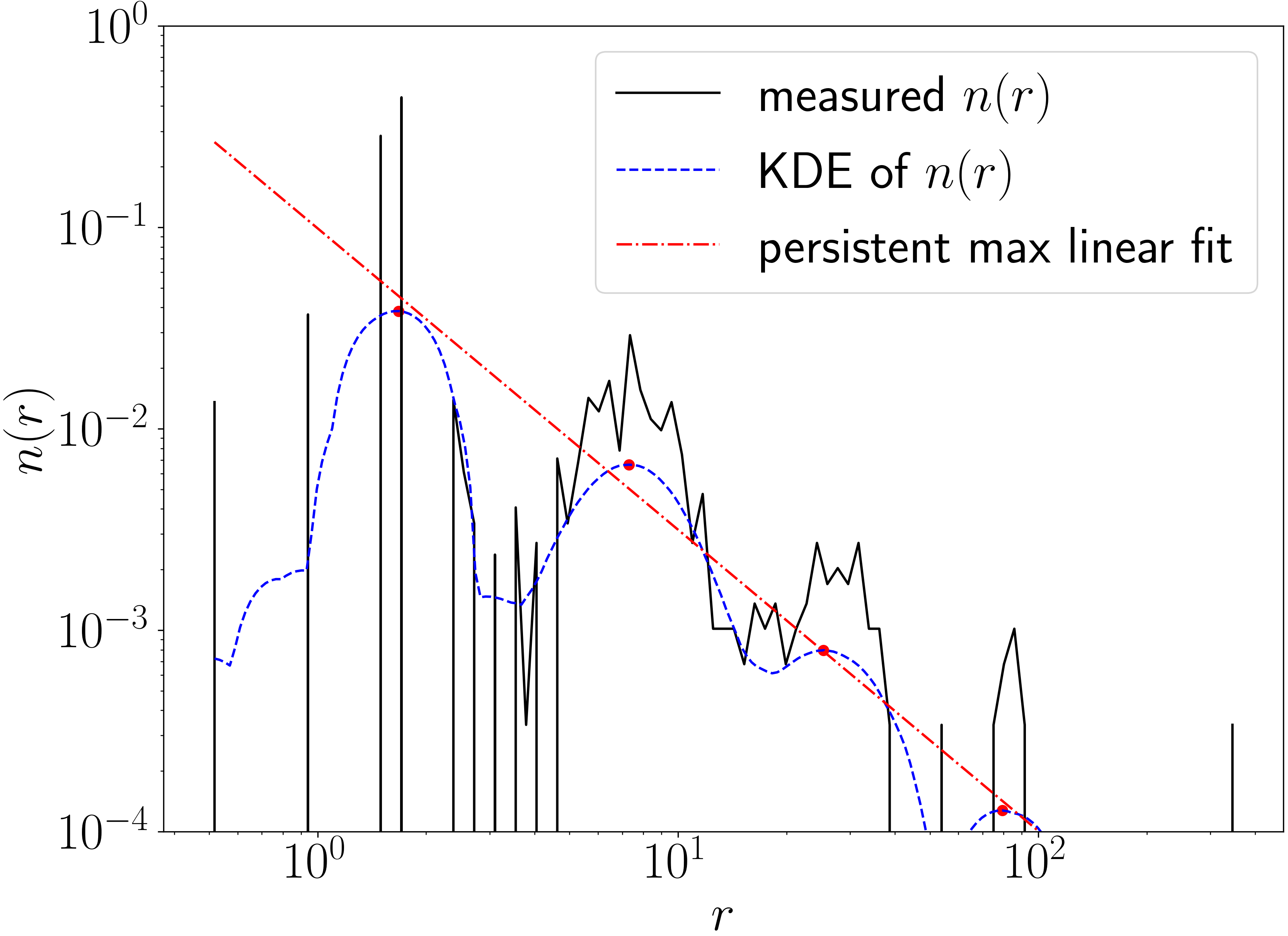}
        \subcaption{}
        \label{fig:artificial_island_radii_distributions_log_b}
    \end{subfigure}
    \caption{Estimated island radius distributions for subfigures \ref{subfig:artificial_island_images_b} and \ref{subfig:artificial_island_images_c} respectively with a logarithmic vertical axis. The red points indicate the location of the four most persistent maxima in the kernel density estimate of $n$, and the red line in the linear fit to those points.}
    \label{fig:artificial_island_radii_distributions_log}
\end{figure}

Figure \ref{fig:artificial_island_radii_distributions_log} presents the same data as subfigures \ref{subfig:artificial_island_radii_distributions_b} and \ref{subfig:artificial_island_radii_distributions_c} but using a logarithmic vertical axis. Included on this plot is a Kernel Density Estimate (KDE) of the underlying $n(r)$ histogram, which was fit with a bandwidth of $0.15$ using the epanechnikov kernel. This kernel is chosen only because it is a standard choice in the \verb!sklearn! kernel density estimation routine. Upon a visual inspection the maximas of this KDE appear to align with the clusters in the $n(r)$ histogram and their peak values seem to fall on a line. We would expect this linear arrangement of clusters on a log-log plot because the number of islands $n$ and radius of each island $r$ in generation $k$ both scale as power laws in subfigures \ref{subfig:artificial_island_images_b} and \ref{subfig:artificial_island_images_c}. That is if we have that $n=n_0^k$ and $r= r_0^k$ then it follows that $\log{n} = \frac{\log n_0}{\log r_0}\log{r}$ and so we expect a linear relationship between $n$ and $r$ on a log-log plot. There are five generations of islands in both images so we would expect five major peaks in both of their island spectra. However, the first generation contains only a single island and does not contribute meaningfully. So we instead search for four peaks. 

Since the power law scaling of the island generations in the images is not exact and the sampling is incomplete we expect that there will be false peaks in both the measured $n(r)$ and its KDE, and we do observe this in both subfigures \ref{fig:artificial_island_radii_distributions_log_a} and \ref{fig:artificial_island_radii_distributions_log_b}. Therefore, we need a criteria to select the four most major peaks, which we can associate with the island generations. We choose to do this with persistent homology, specifically we can consider 1D sub-level set persistent homology and choose our clusters by identifying the four maxima which kill the four longest lived classes\footnote{This computation of persistent homology was performed with the open source \verb!Persistence1D! Python program written by Toni Weinkauf.}. Performing this analysis on the KDEs we obtain the red dots presented on the same plots which when fit with a linear regression produce the red lines shown. The slope of the linear fit in subfigure \ref{fig:artificial_island_radii_distributions_log_a} is $\approx -1.04$ and for subfigure \ref{fig:artificial_island_radii_distributions_log_b} is $\approx -1.495$ with this steeper slope being associated to the observation that the number of islands grows faster relative to the rate at which the radius decreases in subfigure \ref{fig:artificial_island_radii_distributions_log_b} when compared to subfigure \ref{fig:artificial_island_radii_distributions_log_a}. This confirms that the birth time distribution we computed successfully manages to capture the statistics of the distribution of island size across different scales. 

\section{Growth of chaos}

Now that we have seen that the island size distribution can encode the structure of the island hierarchy on some artificial examples we can begin to analyse what the size distribution can say about chaos onset. We will look specifically at the phase space of the standard map \eqref{stdmap} here. 

\begin{figure}[t]
    \centering
    \begin{subfigure}[b]{0.24\textwidth}
        \includegraphics[width = \textwidth,frame]{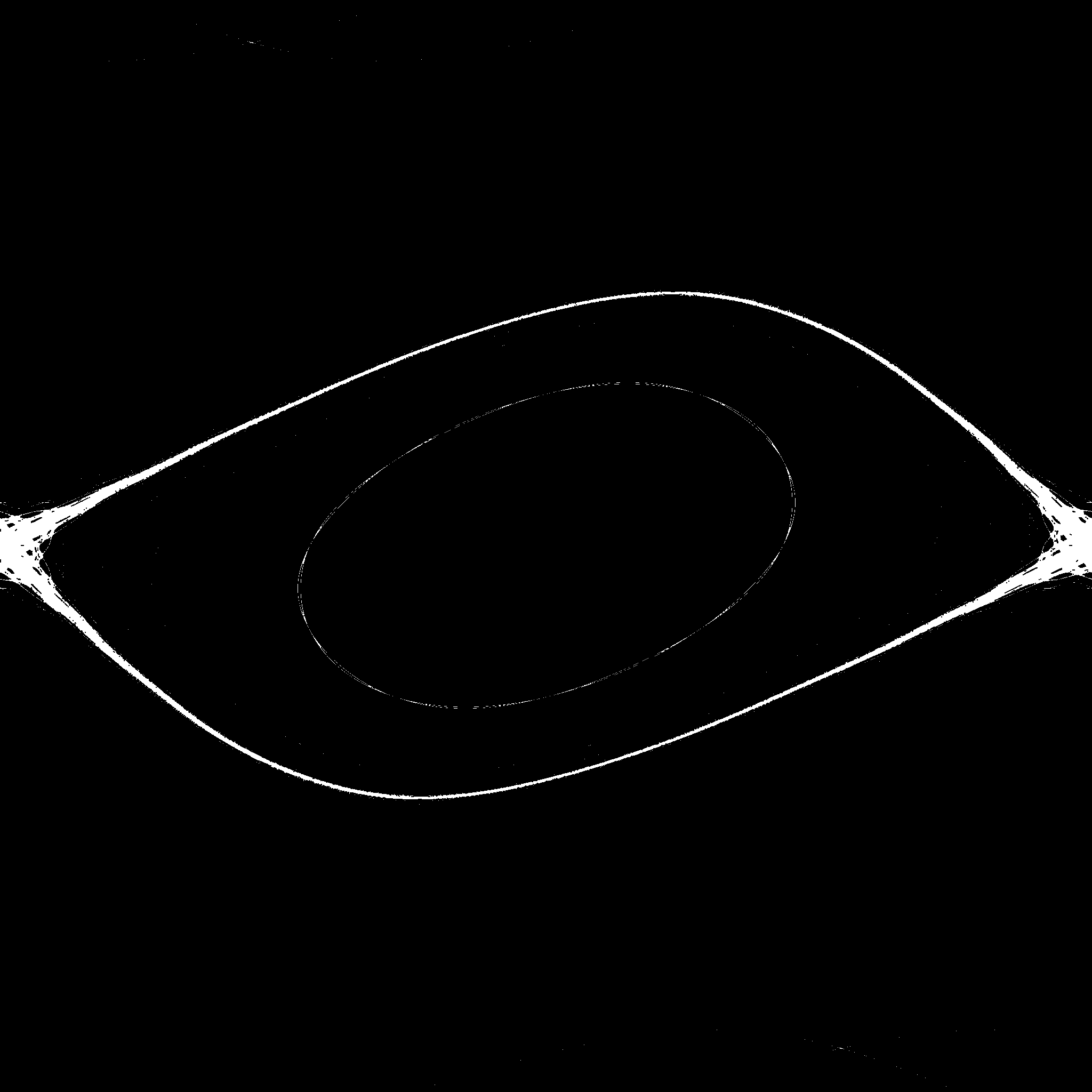}
        \subcaption{$k=0.5$}
        \label{subfig:standard_map_phase_space_a}
    \end{subfigure}
    \begin{subfigure}[b]{0.24\textwidth}
        \includegraphics[width = \textwidth,frame]{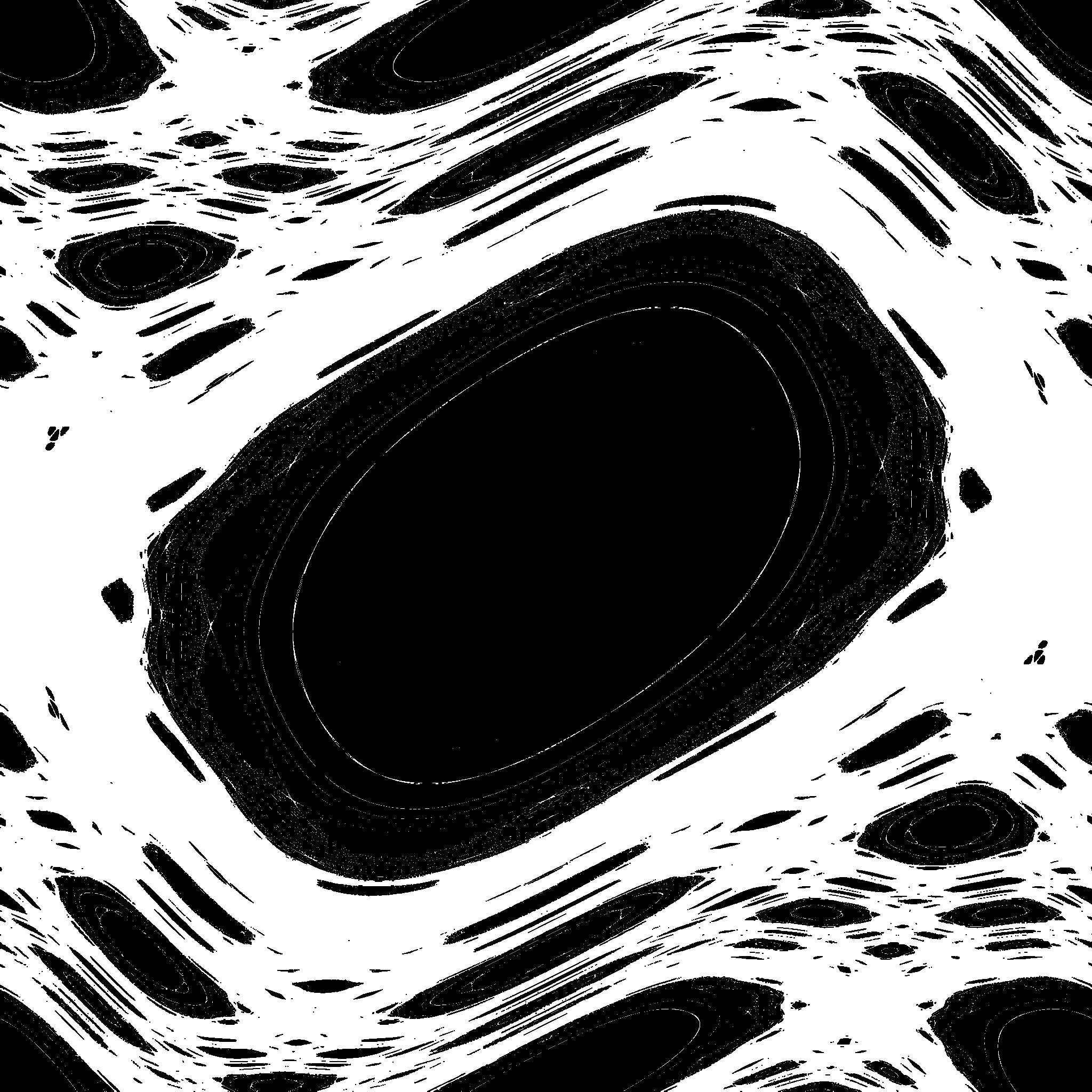}
        \subcaption{$k=1.0$}
        \label{subfig:standard_map_phase_space_b}
    \end{subfigure}
    \begin{subfigure}[b]{0.24\textwidth}
        \includegraphics[width = \textwidth,frame]{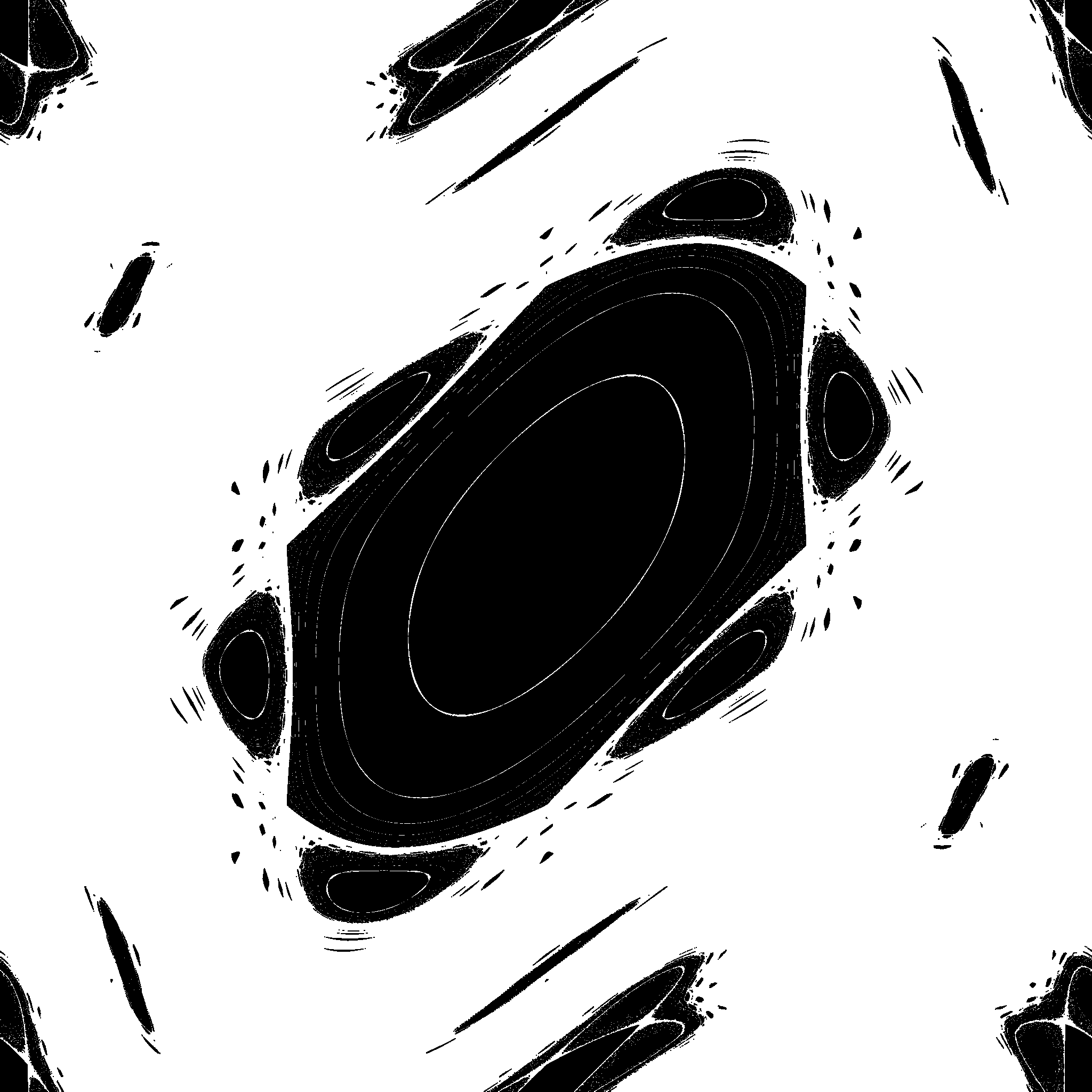}
        \subcaption{$k=1.5$}
        \label{subfig:standard_map_phase_space_c}
    \end{subfigure}
    \begin{subfigure}[b]{0.24\textwidth}
        \includegraphics[width = \textwidth,frame]{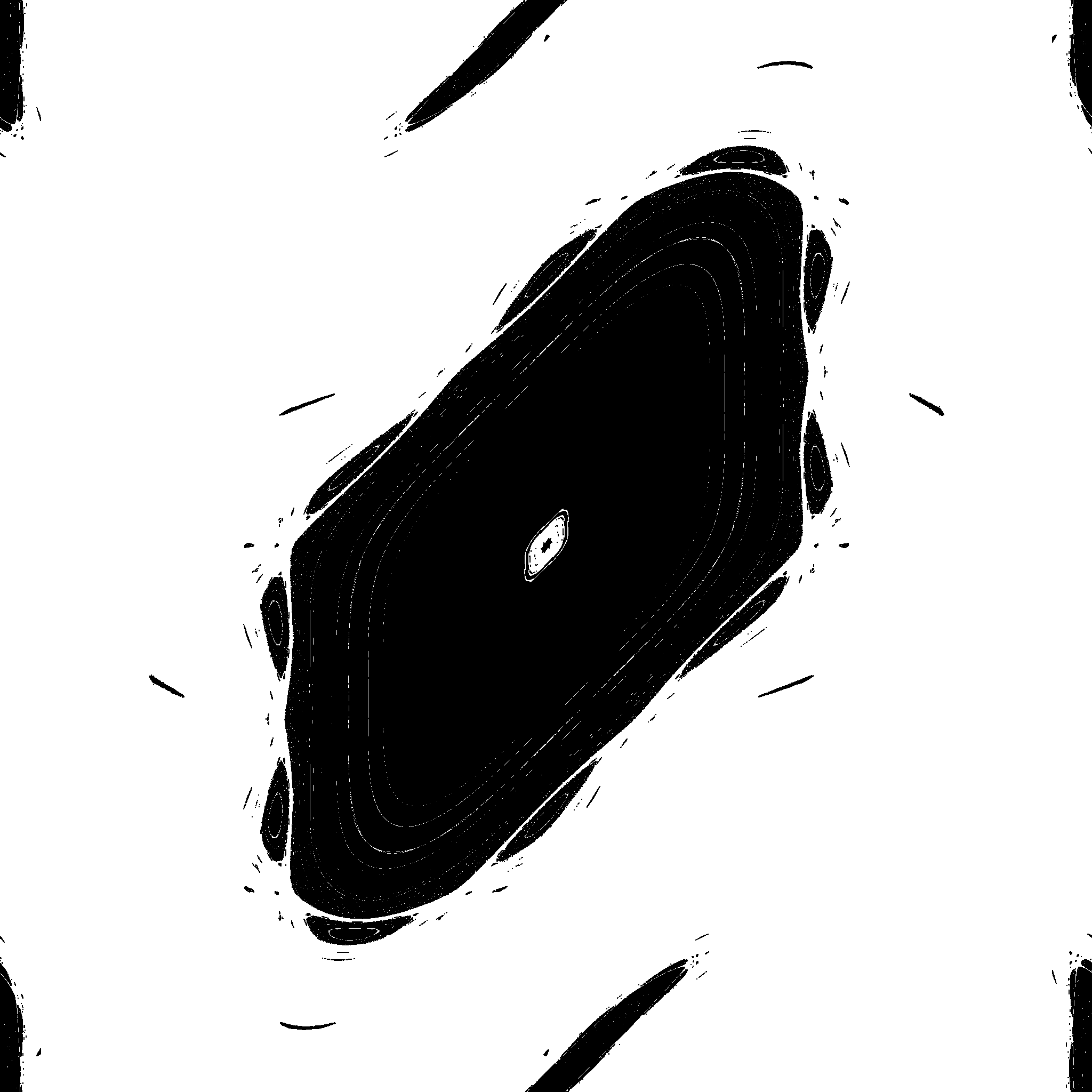}
        \subcaption{$k=2.0$}
        \label{subfig:standard_map_phase_space_d}
    \end{subfigure}
    \caption{Sample images of the global phase space of the standard map at varying $k$.}
    \label{fig:standard_map_phase_space}
\end{figure}

We image the stochastic region of the standard map for a range of 16 values of the nonlinearity parameter $k\in[0.5,2.0]$. Note that we take the number of points per row in the grid $N$, and the length of the orbits $T$ to be $N=2048,T=2500$. A representative sample of the binary images of the phase space we obtain is presented as Figure \ref{fig:standard_map_phase_space}. 

\begin{figure}[t]
    \centering
    \includegraphics[width = 0.5\textwidth]{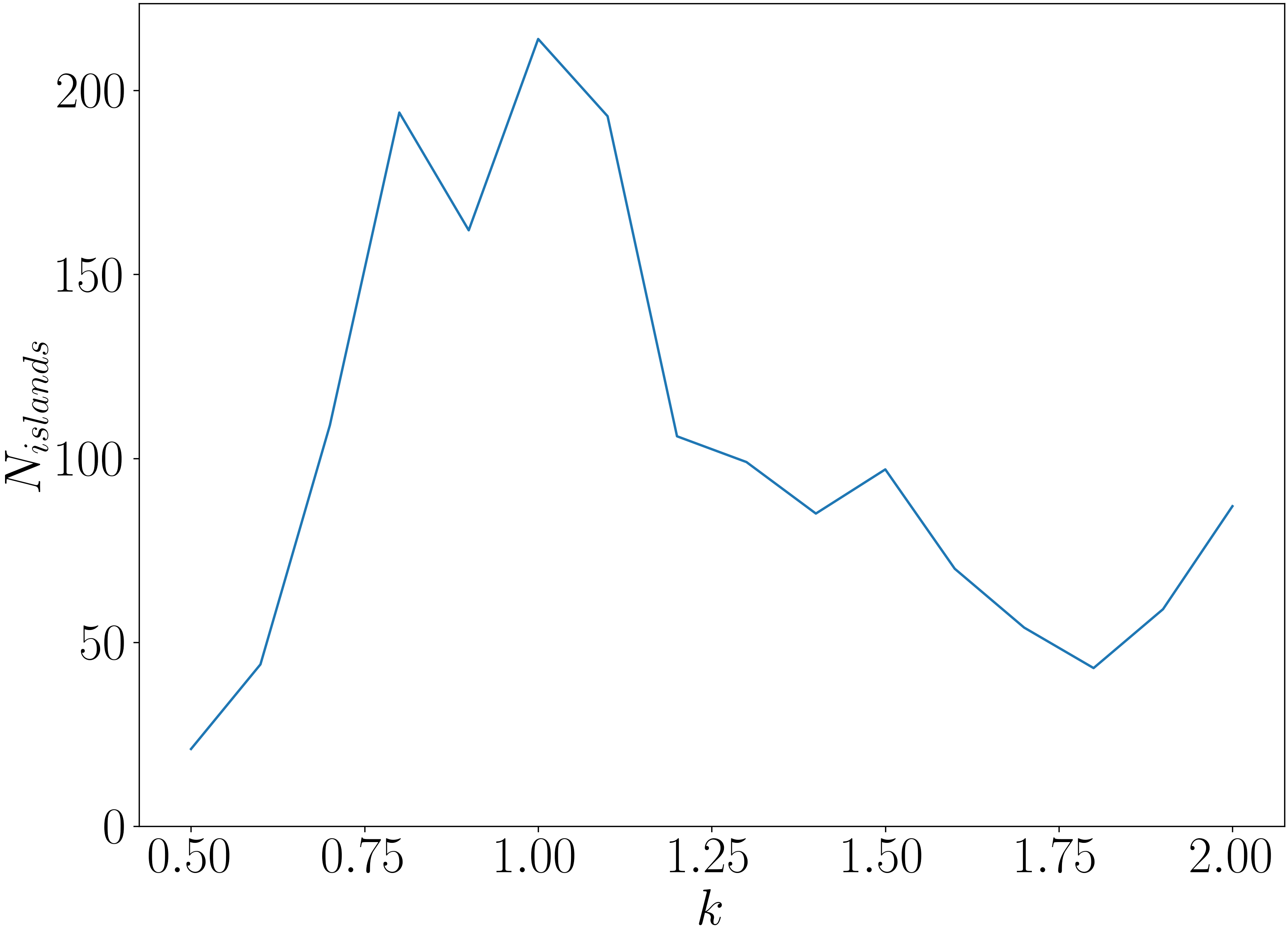}
    \caption{Variation of the island radius distribution normalisation of the standard map with $k$.}
    \label{fig:island_count_function}
\end{figure}

\begin{figure}[t]
    \centering
    \includegraphics[width=0.8\textwidth]{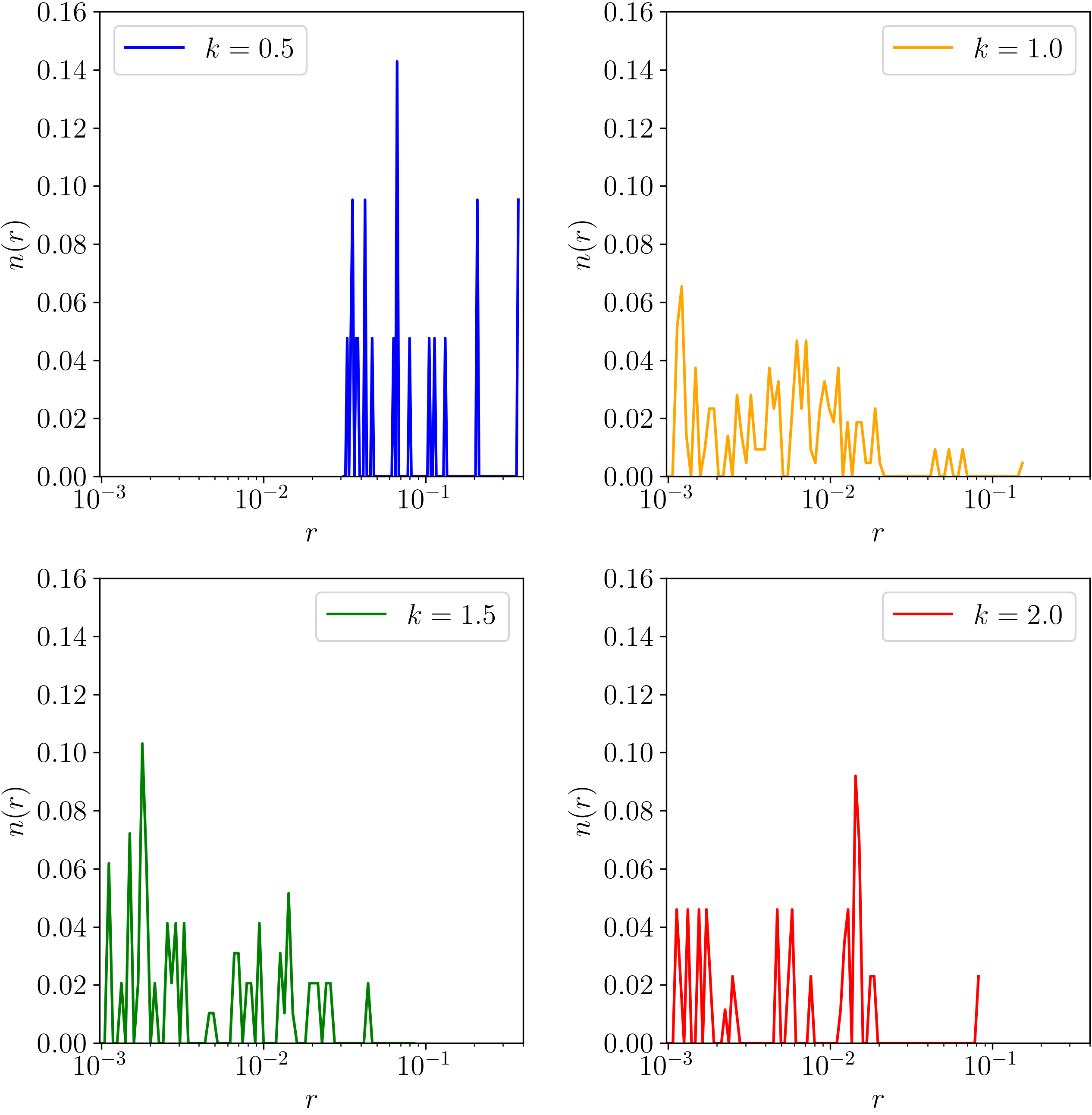}
    \caption{Island distribution for the images of the standard map stochastic region included as Figure \ref{fig:standard_map_phase_space}.}
    \label{fig:standard_map_island_distribution}
\end{figure}

We then compute the island distribution at each $k$ value using the \textit{BirthsDist} algorithm. The full results of this calculation are somewhat impractical to display here in full, although a representative sample is presented as Figure \ref{fig:standard_map_island_distribution} for later analysis. We can however discuss some relevant properties. The island radii distributions included above were all normalised, but there is information encoded in the normalisation factor itself. Specifically, the normalisation corresponds to the total number of islands in the phase space, at least which are large enough for us to detect. Figure \ref{fig:island_count_function} presents the normalisation of the standard map island radii distribution as a function of the nonlinearity $k$. We observe that the normalisation achieves a maximum close to the critical nonlinearity $k\approx 0.97\ldots$. That is, as the nonlinearity increases, islands form as the stochastic region around chains of resonant islands ``thickens'' and KAM toruses break-up. After the last KAM torus is broken, at the critical nonlinearity, the stochastic region can only expand by making the islands smaller, effectively ``absorbing'' them and so the number of islands begins to decrease. This creates the peak near $k\approx 1.0$ we observe in the normalisation factor of the island radii distribution. Our conclusion agrees with those of White, Rax, and Wu \cite{WhiteRaxWu} who computed the total area of islands for the standard map and observed that the area is maximised around the critical threshold, when there are the most possible islands. 

Regarding Figure \ref{fig:standard_map_island_distribution} we observe that at $k=0.5$ the islands are concentrated at very large radius. This agrees with a visual analysis of subfigure \ref{subfig:standard_map_phase_space_a} which contains a single major island. As the nonlinearity is increased the island distribution shifts to lower $r$ and as such we obtain the distributions for $k=1.0,1.5,$ and $2.0$ which are both less concentrated than $k=0.5$ and also demonstrate that the islands are in general of smaller radius.


\section{The case of an accelerator mode}

In the above we analysed the distribution of islands in the standard map phase space for small $k$, that is near to the point of breakup of the last KAM torus. The distribution of island size continues to evolve as $k$ is increased, and interesting phase space structures begin to emerge. Here we will look at an accelerator mode island, the definition of which will follow. We choose such an island because they are known to exhibit a self-similar structure and so we expect that we will obtain a similar power-law behaviour in the island size distribution to that observed for the artificial examples of Section \ref{ArtificialExamples}.

One physical system which can be described, at lowest order, by the standard map is a charged particle in a circular accelerator. To see this, suppose we have a particle moving in uniform circular motion, such as a charged particle in a uniform magnetic field whose velocity is perpendicular to that field. In coordinates $(\theta,v)$ such a particle, assuming unit mass, has Hamiltonian
\begin{equation}
     H_0 = \frac{1}{2}v^2\,.
\end{equation}
In a circular accelerator an electric field is pulsed periodically to accelerate the particle. If, at the moment of the pulse, the particle is moving parallel to the applied electric field it will be accelerated, and if it is moving anti-parallel it will be decelerated by the field. We can model this as providing a instantaneous change in energy with period $T$ of $\Delta E = -(k/(2\pi)^2)\cos(\theta)$. Therefore, we can model our particle motion with the time dependent Hamiltonian
\begin{equation}
H(\theta,v,t) =  \frac{1}{2}v^2-\frac{k}{(2\pi)^2}\cos(\theta)\sum_{n=-\infty}^{\infty}\delta(t-nT)\,.
\end{equation}
We recognise that this is the Hamiltonian for a kicked rotator, which as demonstrated in Chapter 2 has stroboscopic dynamics modelled by the standard map. Thus we have confirmed that the standard map provides a simple model for the dynamics of a particle in a circular accelerator after we make the identification that $(x,p) = (\theta/2\pi,v)$ 

Suppose that our objective is to design a circular accelerator. To do this we will need to choose the magnitude of the electric field with which we pulse our particle, that is, $k$. We cannot use an arbitrary value because for our particle to accelerate it must always be synchronised with the electric field pulses, even as it accelerates. That is, we want to ensure that our particle will always be parallel to the field at the moment of each pulse. If it is not parallel then we cannot guarantee that the particle will accelerate. To achieve this design goal we seek to find an initial condition $(x_0,p_0)$ and a nonlinearity $k$ such that on each iteration the particle gains a constant impulse. That is, we want
\begin{equation*}
    p_{n+1}-p_n = \Delta p \in \mathrm{R}\,.
\end{equation*}

\begin{figure}
    \centering
    \begin{subfigure}[b]{0.49\textwidth}
        \includegraphics[width = \textwidth]{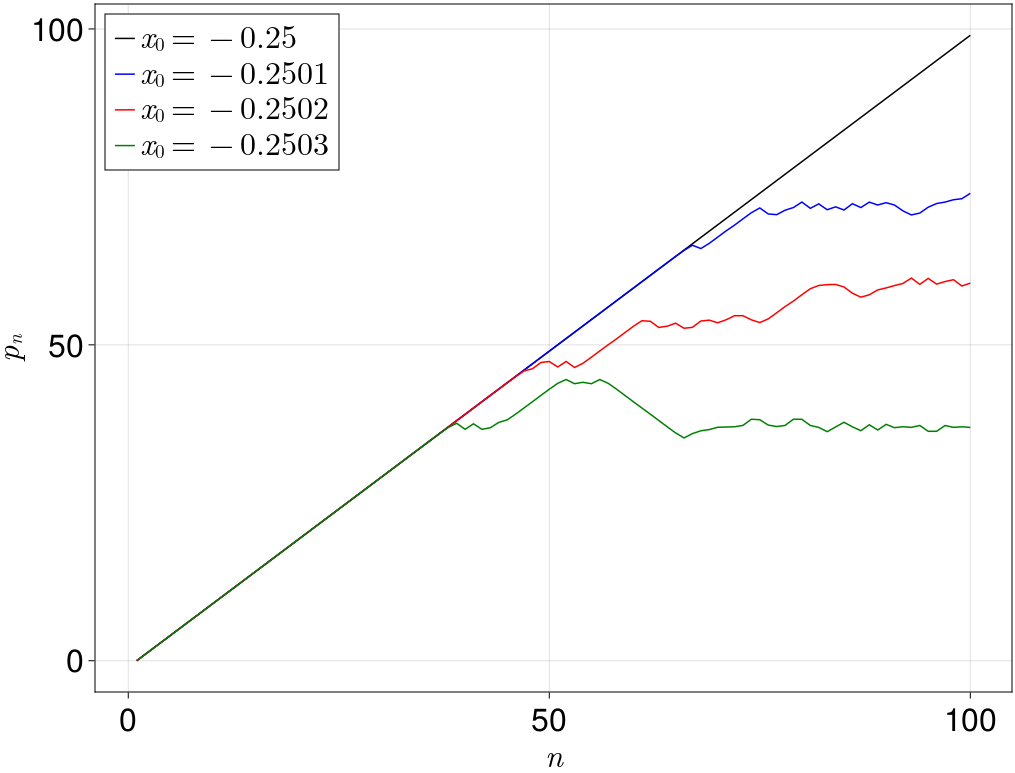}
        \subcaption{$k=6.283$}
        \label{subfig:accelerator_mode_trajectories_a}
    \end{subfigure}
    \begin{subfigure}[b]{0.49\textwidth}
        \includegraphics[width = \textwidth]{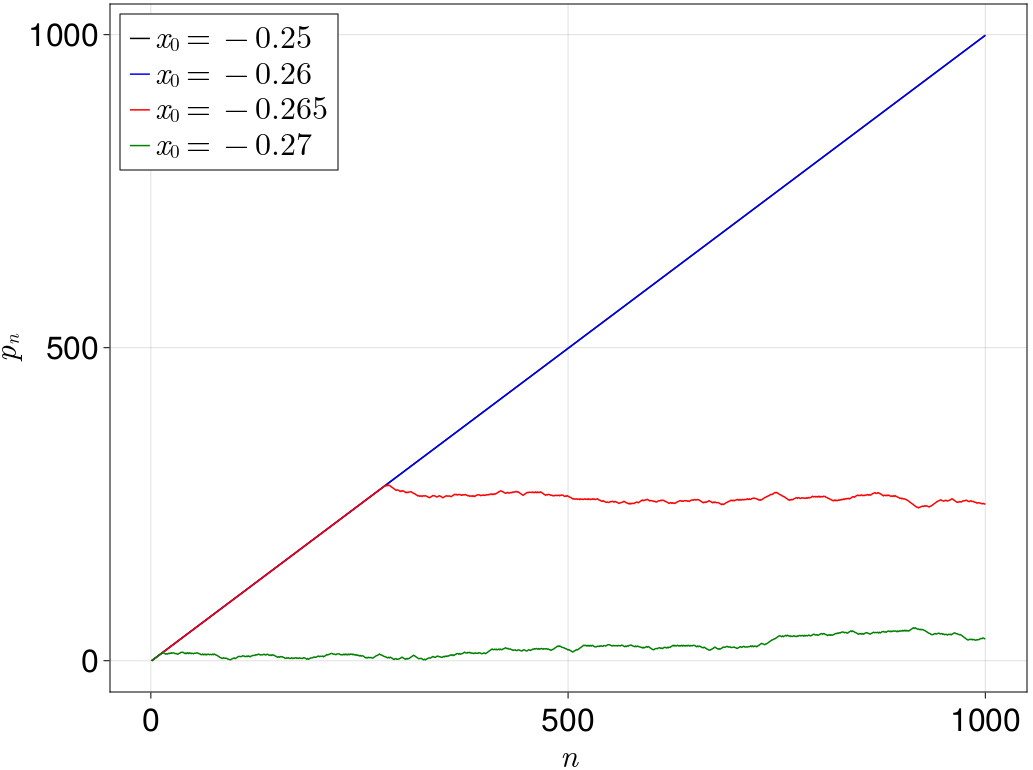}
        \subcaption{$k=6.383$}
        \label{subfig:accelerator_mode_trajectories_b}
    \end{subfigure}
    \caption{Orbits of phase space points of the form $(x_0,0)$ near to an accelerator mode solution}
    \label{fig:accelerator_mode_trajectories}
\end{figure}

The change in momentum is given by $-\frac{k}{2\pi}\sin (2\pi x_n)$ and so to ensure that this is constant we require that $x_n$ be constant, since the $\sin$ function is a bijection on the circle. Therefore, we need $x_{n+1} = x_n$ and so $p_{n+1}$ must equal an integer since in that case $x_{n+1} = x_n + p_{n+1} \mod{1} =x_n$ as required. Now, if $p_n$ is an integer for all $n$ then the impulse across an iteration $\Delta p$ must also be an integer. Hence we have that the condition required to create an accelerator mode is
\begin{equation}
    \Delta p = -\frac{k}{2\pi}\sin(2\pi x_n) = l \in \mathbb{Z}\,.
\end{equation}
Here we will focus on the simplest solution to this equation. We choose $x_n=-1/4$ so that we have $\sin(2\pi x_n)=-1$ in which case we can solve for $k$ to find
\begin{equation}
    k_l = 2\pi l\,.
\end{equation}
At this point we have constructed an infinite family of accelerator modes the dynamics of which are constant acceleration at a fixed $x_n=-1/4$. Again we will focus on the first solution, that is when $l=1$ and $k=k_1= 2\pi$. 

\begin{figure}
    \centering
    \begin{subfigure}[b]{0.49\textwidth}
        \includegraphics[width = \textwidth]{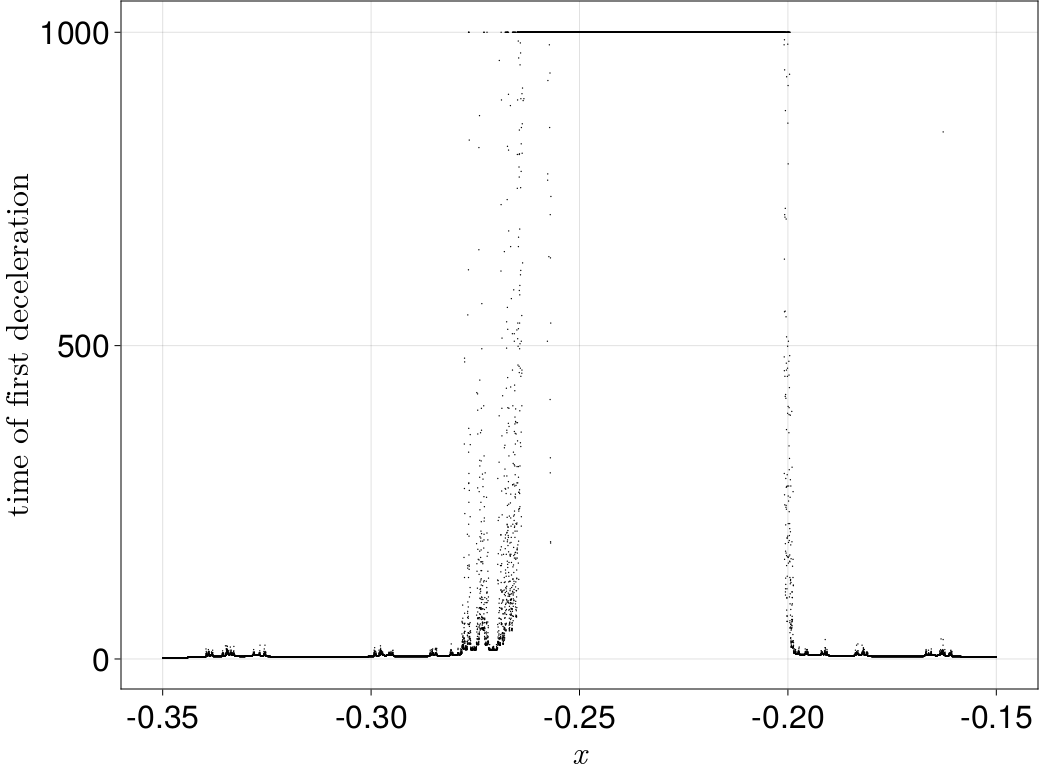}
        \subcaption{Time of first deceleration for $k=6.383$ as a function of $x_0$.}
        \label{subfig:accelerator_mode_growth_a}
    \end{subfigure}
    \begin{subfigure}[b]{0.49\textwidth}
        \includegraphics[width = \textwidth]{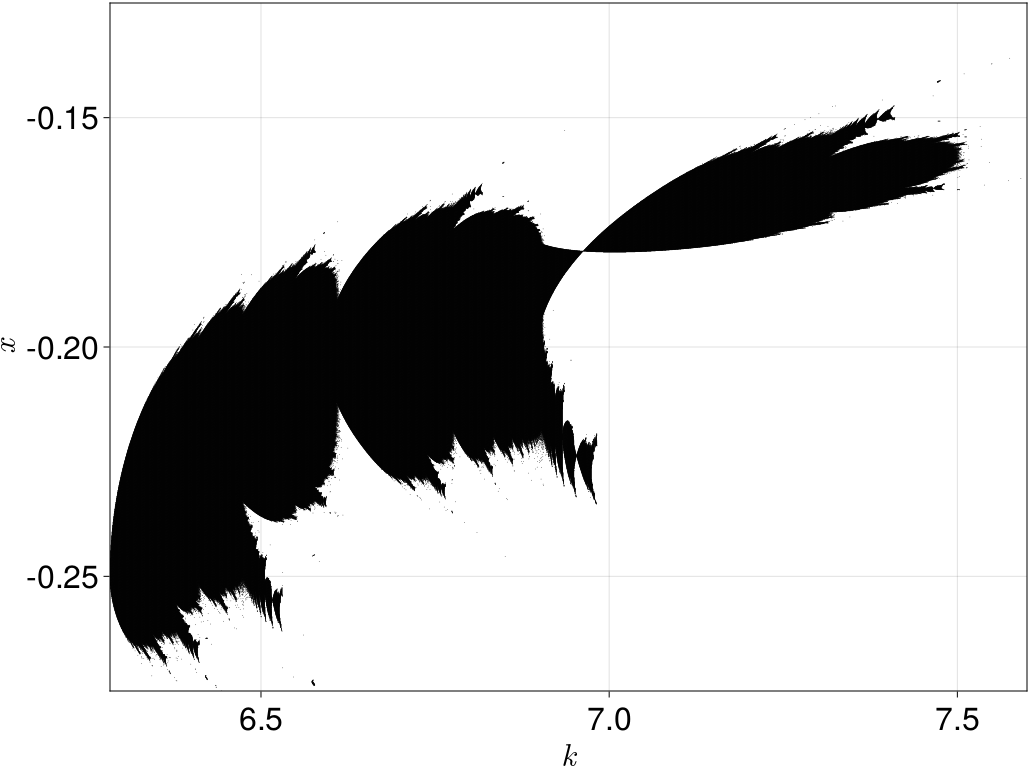}
        \subcaption{$p_0=0$ slice of the accelerator mode as a function of $k$.}
        \label{subfig:accelerator_mode_growth_b}
    \end{subfigure}
    \caption{Images depicting the growth of an accelerator mode island.}
    \label{fig:accelerator_mode_growth}
\end{figure}

To check that at $k_1$ an accelerator mode actually exists we will compute the orbit of the point $(x_0,p_0)=(-1/4,0)$ as well as some nearby points. The $p_n$ component of these trajectories for $100$ iterations is presented as subfigure \ref{subfig:accelerator_mode_trajectories_a}. From these plots we observe that the orbit of $x_0=-1/4$ accelerates uniformly as expected. The points which started nearby follow the accelerating trajectory for a short time but eventually deviate from the linear trajectory and begin to randomly walk. This indicates that when $k=k_1$ the accelerator mode exists but only at a single point. Note that even after the red, blue, and green trajectories have deviated from the primary the linear one they still contain irregular linearly increasing and decreasing sub-trajectories visible as medium length straight line segments. This is a non-brownian effect, attributable to the ``stickiness'' of the accelerator mode itself. Each straight line flight\footnote{``Flight'' is technical term for such straight line subtrajectory.} corresponds to a time when the trajectory approaches an accelerator mode island and gets trapped in the boundary layer around it approximately orbiting the island for a finite period. These long flights are the cause of anomalous transport, which is discussed below, near an accelerator mode \cite{Zaslavsky}. 

\begin{figure}
    \centering
    \begin{subfigure}[b]{0.33\textwidth}
        \includegraphics[width = \textwidth,frame]{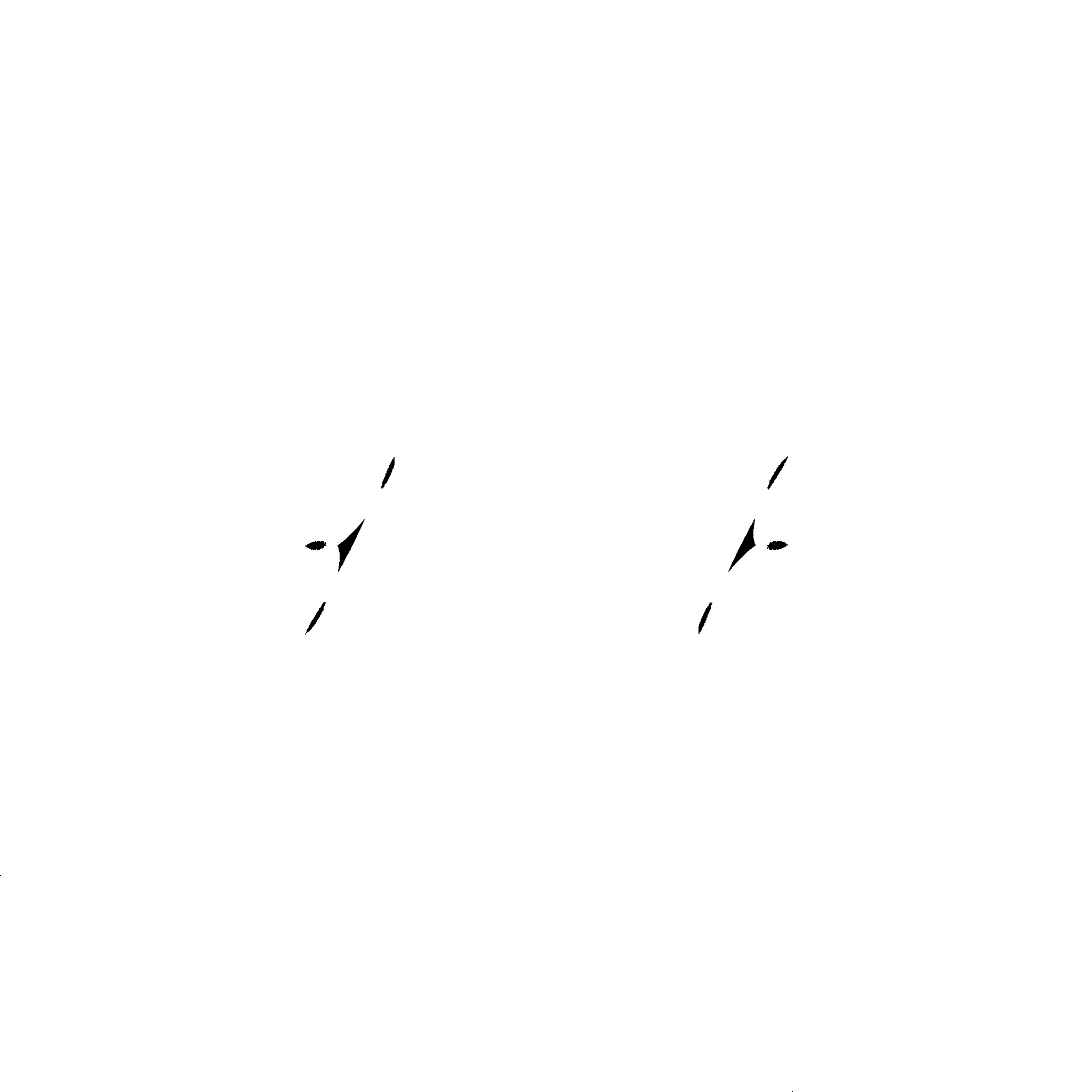}
        \subcaption{}
        \label{subfig:accelerator_mode_phase_space_a}
    \end{subfigure}
    \begin{subfigure}[b]{0.33\textwidth}
        \includegraphics[width = \textwidth,frame]{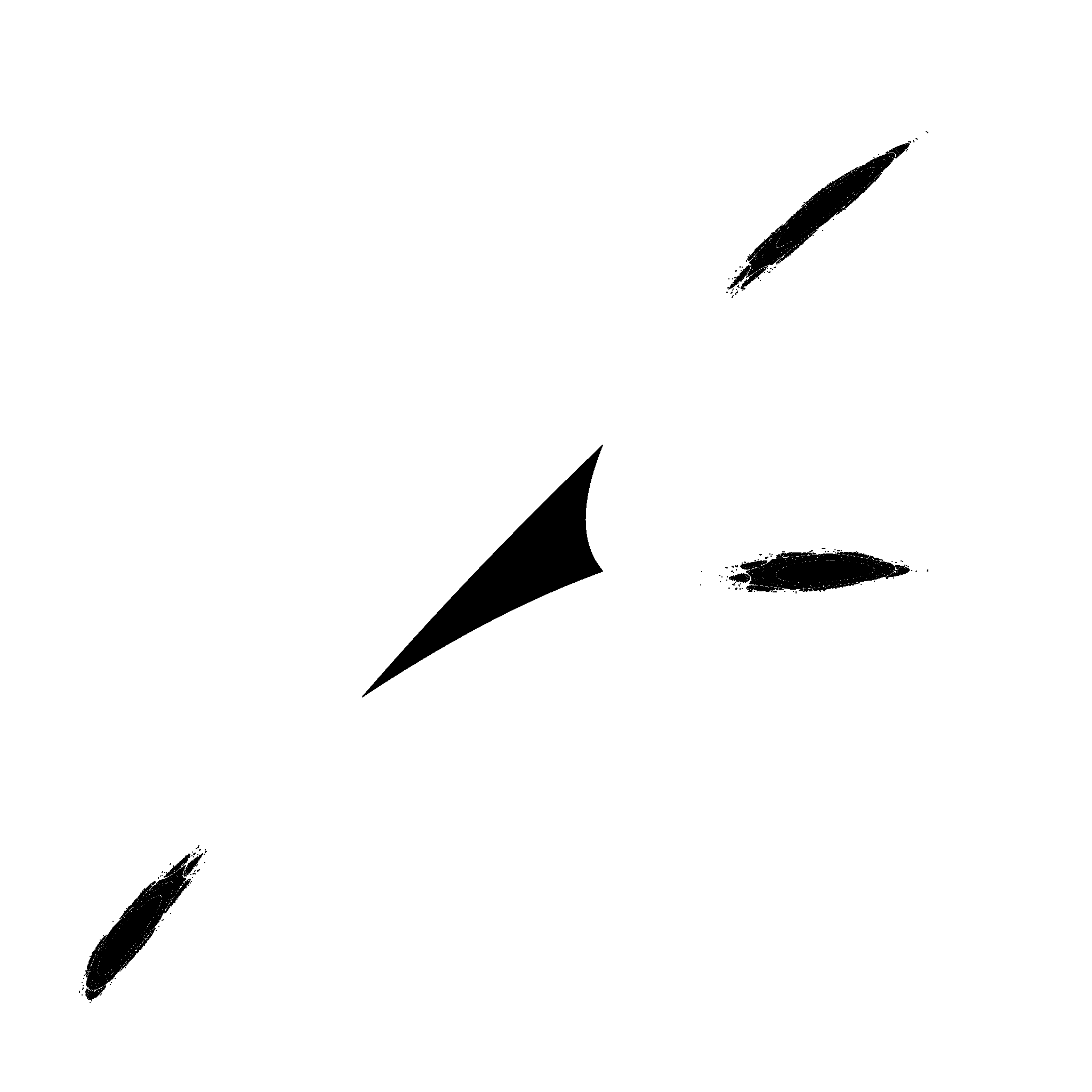}
        \subcaption{}
        \label{subfig:accelerator_mode_phase_space_b}
    \end{subfigure}
    \begin{subfigure}[b]{0.33\textwidth}
        \includegraphics[width = \textwidth,frame]{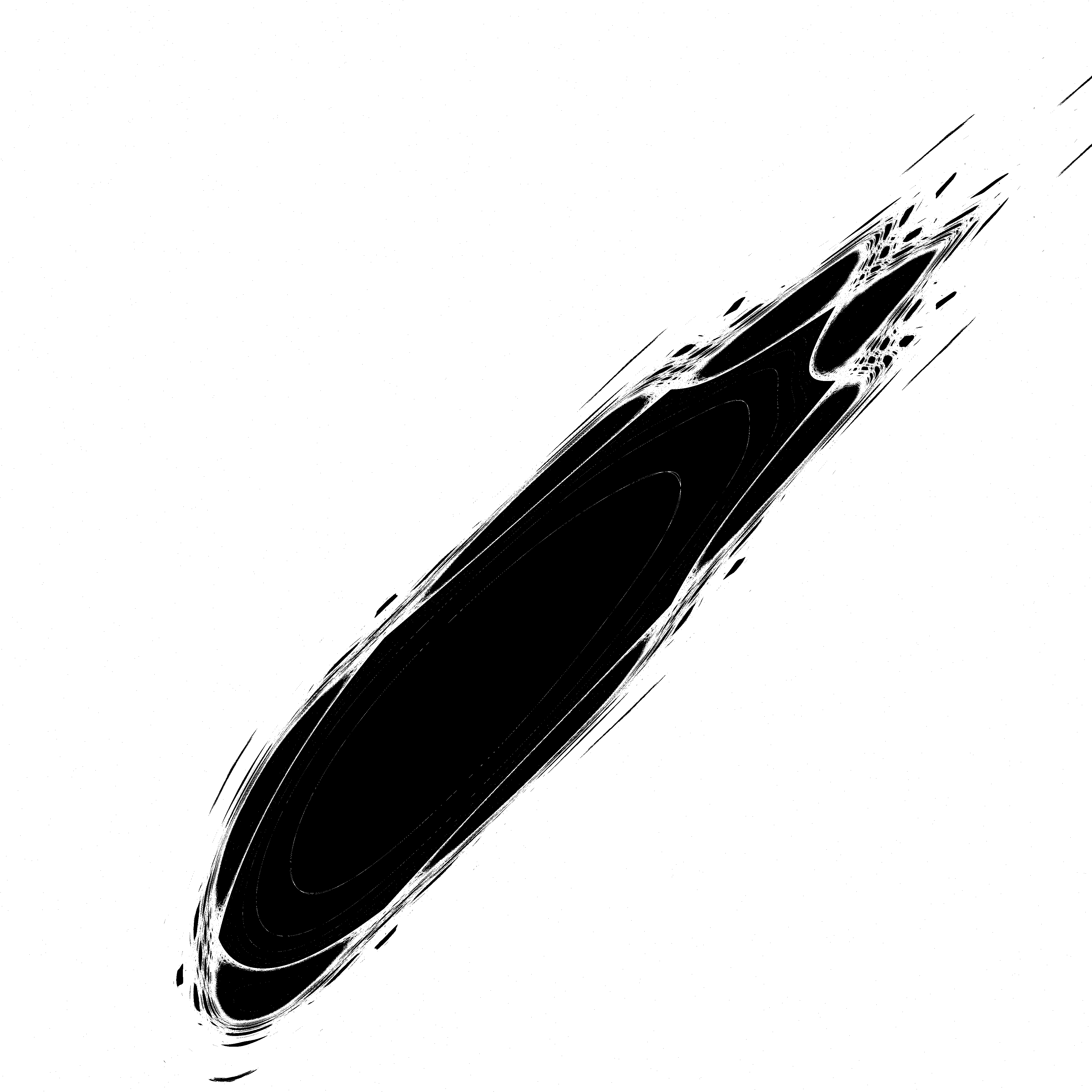}
        \subcaption{}
        \label{subfig:accelerator_mode_phase_space_c}
    \end{subfigure}
    \begin{subfigure}[b]{0.33\textwidth}
        \includegraphics[width = \textwidth,frame]{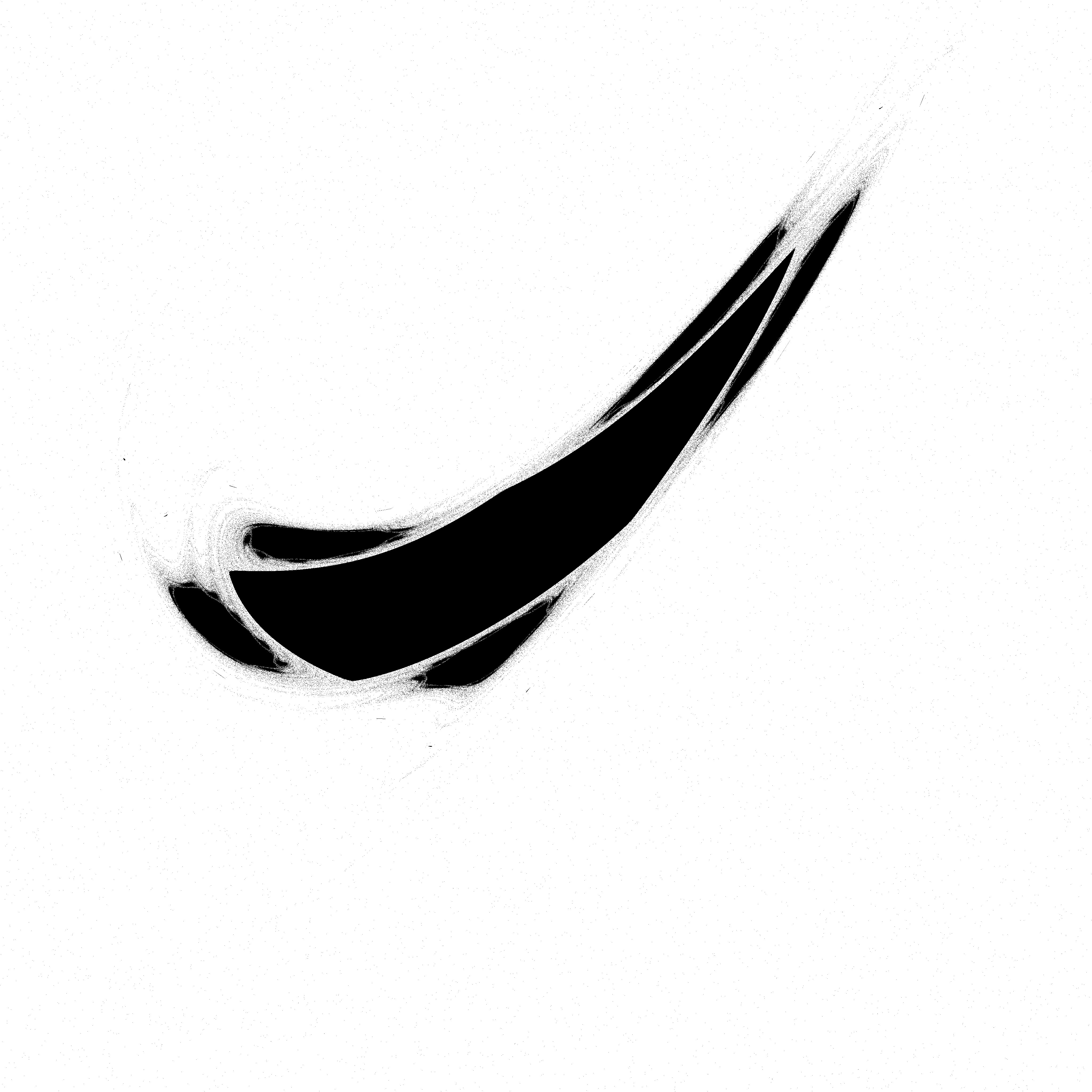}
        \subcaption{}
        \label{subfig:accelerator_mode_phase_space_d}
    \end{subfigure}
    \caption{Images depicting the standard map phase space at $k=6.9110$. \ref{subfig:accelerator_mode_phase_space_a} is the global phase space, \ref{subfig:accelerator_mode_phase_space_b} shows a magnified accelerator mode island in the range $(x,p)\in [0.13,0.24]\cross[-0.1,0.11]$, \ref{subfig:accelerator_mode_phase_space_c} is the lower left island in \ref{subfig:accelerator_mode_phase_space_b} and shows the range $[0.135,0.153]\cross[-0.085,-0.045]$, and \ref{subfig:accelerator_mode_phase_space_d} shows the small island visible at the bottom of c with range $[0.13825,0.1395]\cross[-0.085,-0.084]$.}
    \label{fig:accelerator_mode_phase_space}
\end{figure}

When $k$ is increased to $6.383$ we obtain the trajectories presented as subfigure \ref{subfig:accelerator_mode_trajectories_b}. We observe that both $x_0=-0.25$ and $x_0=-0.26$ appear to remain on accelerating trajectories for all time, while the points further away start on accelerating trajectories but again eventually deviate. This indicates that at this higher perturbation strength the accelerator mode is now a finite size phase space island instead of a single point. Note that once the island is a finite size the accelerating trajectories do not gain a uniform impulse on each iteration and instead wobble left and right in $x_n$ while never decelerating. 

As a brief aside we will compute the evolution of the accelerator mode shape as a function of $k$. To do this we compute the orbits of a grid of points from the $p=0$ axis for $T$ iterations. We then compute, for each trajectory, the first $n$ such that $p_{n+1}<p_n$, that is, the first time the trajectory decelerates, and is therefore no longer an accelerating solution. If a trajectory does not decelerate within $T$ iterations then we define its time of first deceleration to be $T$, for convenience. 

An example of the result of this calculation is presented a subfigure \ref{subfig:accelerator_mode_growth_a}. The set of points with the maximum time of first deceleration, that is which form a horizontal line on our scatter plot, are defined to be points in the accelerator mode island. Plotting all of the points we detected as inside the accelerator mode for various $k$ we obtain the fractal structure presented as subfigure \ref{subfig:accelerator_mode_growth_b}. From which we observe that as expected, the island is born at a single point at $k_1$ but quickly grows into a finite width island of accelerating solutions. The centre of the island shifts as $k$ is increased and after $k\approx 7.5$ the island no longer intersects with the $p=0$ axis. 

\begin{figure}
    \centering
    \begin{subfigure}[b]{0.49\textwidth}
        \includegraphics[width = \textwidth]{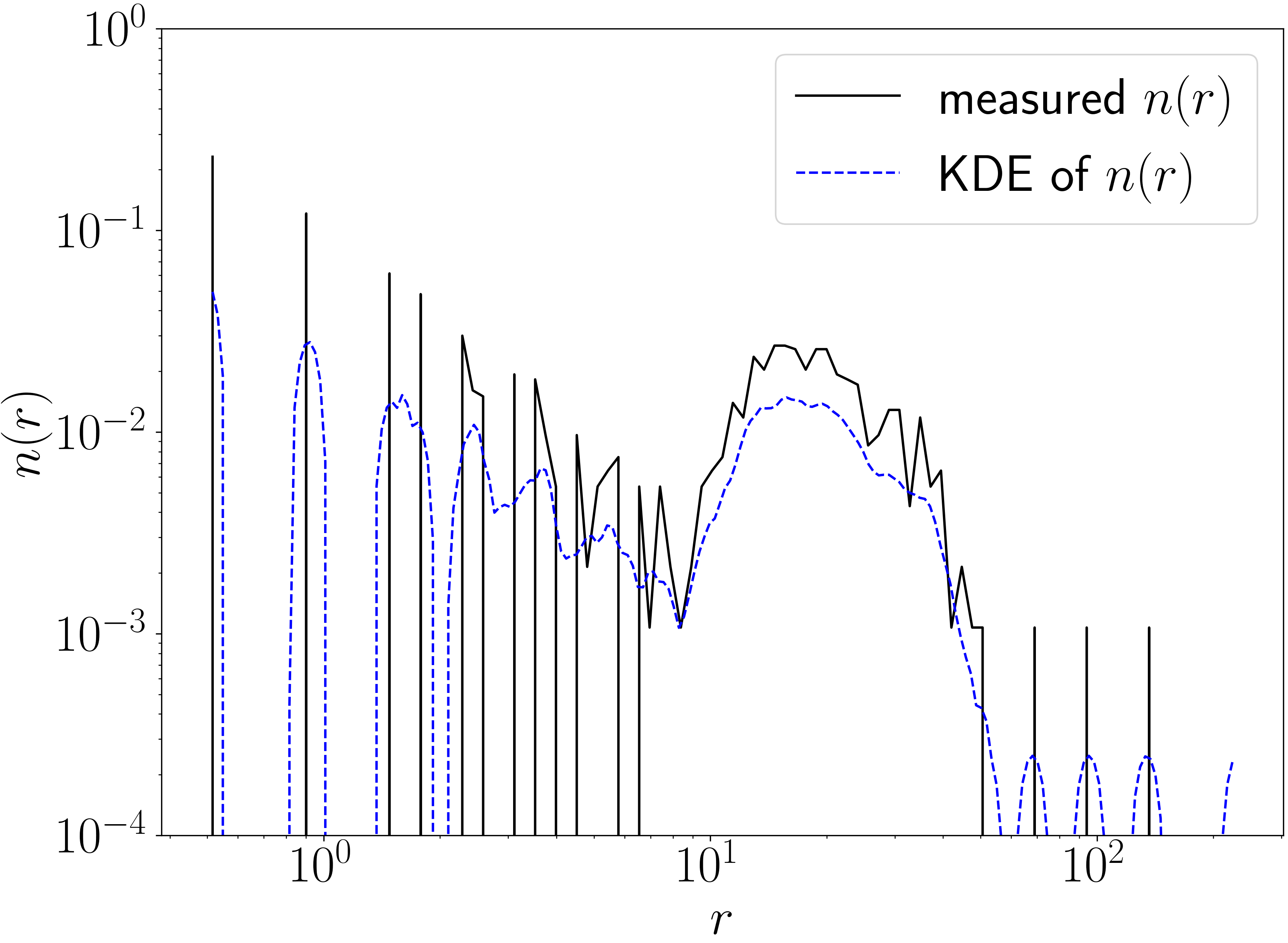}
        \subcaption{}
        \label{subfig:accelerator_mode_island_distributions_a}
    \end{subfigure}
    \begin{subfigure}[b]{0.49\textwidth}
        \includegraphics[width = \textwidth]{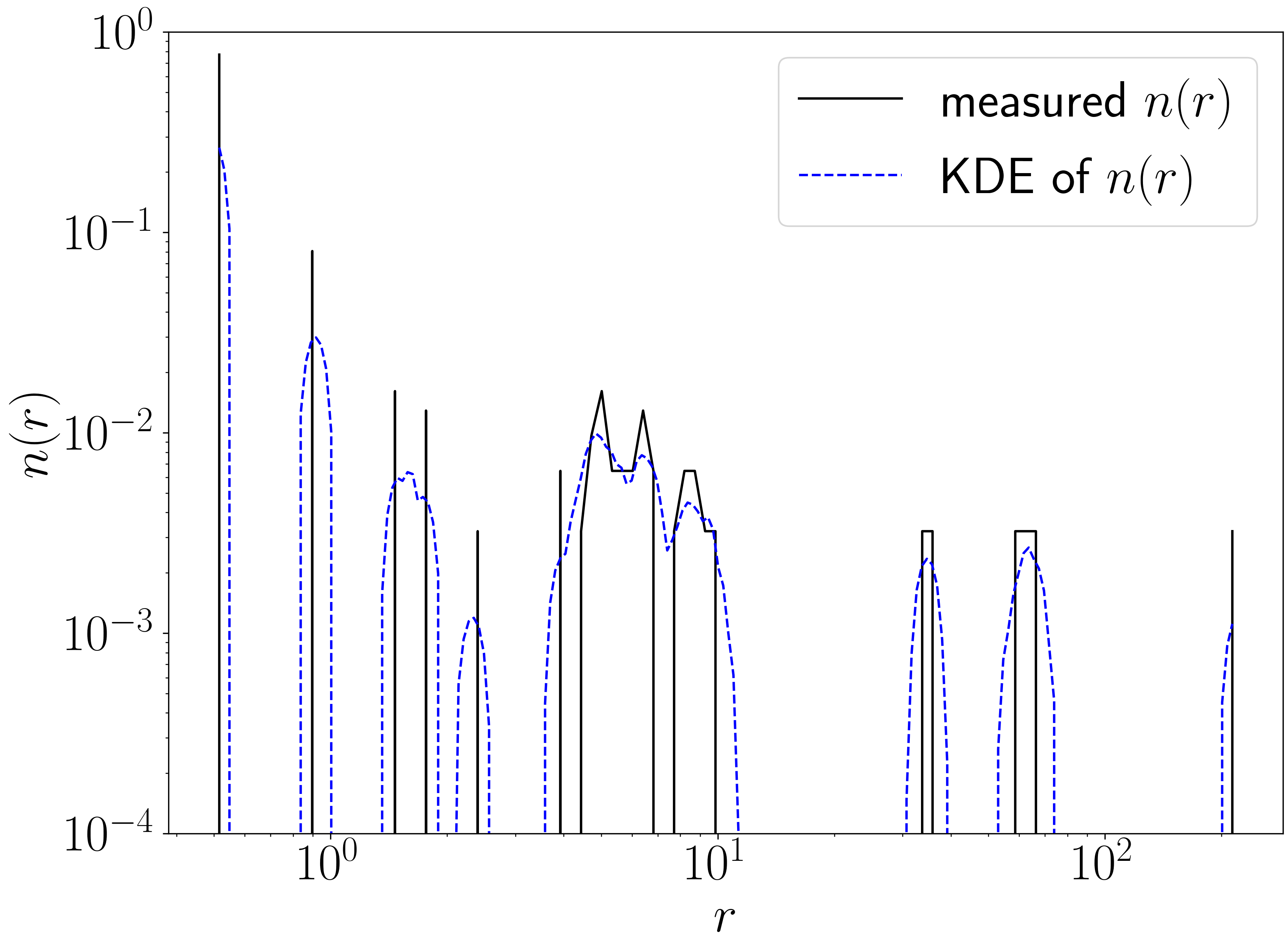}
        \subcaption{}
        \label{subfig:accelerator_mode_island_distributions_b}
    \end{subfigure}
    \caption{Island size distributions for the accelerator mode islands shown as subfigures \ref{subfig:accelerator_mode_phase_space_c} and \ref{subfig:accelerator_mode_phase_space_d}.}
    \label{fig:accelerator_mode_island_distributions}
\end{figure}

Having confirmed that these accelerator modes exist and constitute islands of nontrivial size we now proceed to image them and compute the size distribution of the islands. Figure \ref{fig:accelerator_mode_phase_space} presents binary images of the accelerator mode islands rendered using the WBA approach. We observe from subfigure \ref{subfig:accelerator_mode_phase_space_b} that each accelerator mode has a ``triangular'' structure with $1$ island in what we refer to as the zeroth generation, $3$ islands in the first generation, and more in the further generations \cite{Zaslavsky}. The distribution of islands in subfigures \ref{subfig:accelerator_mode_phase_space_c} and \ref{subfig:accelerator_mode_phase_space_d} is presented as subfigures \ref{subfig:accelerator_mode_island_distributions_a} and \ref{subfig:accelerator_mode_island_distributions_b} respectively. We observe that both the distributions have a similar contour in which at large $r$ they appear to be primarily composed of a continuous distribution of islands with a single mode but at smaller $r$ the distribution breaks apart into many disconnected peaks each which increase in height monotonically as $r$ decreases. This set of several small peaks is indicative of the very complicated islands around islands structure associated to an accelerator mode. It is this structure to which the anomalous diffusion of particles near accelerator modes is attributed \cite{Zaslavsky}. 


\section{The correlation with anomalous transport}

We observed above that the self-similar island hierarchy around an accelerator mode generates a distinct sequence of peaks in the birth time distribution. We also claimed that this is associated to the phenomenon of anomalous transport. We will now provide some evidence for this claim.

\begin{figure}[t]
    \centering
    \begin{subfigure}[b]{0.49\textwidth}
        \includegraphics[width = \textwidth]{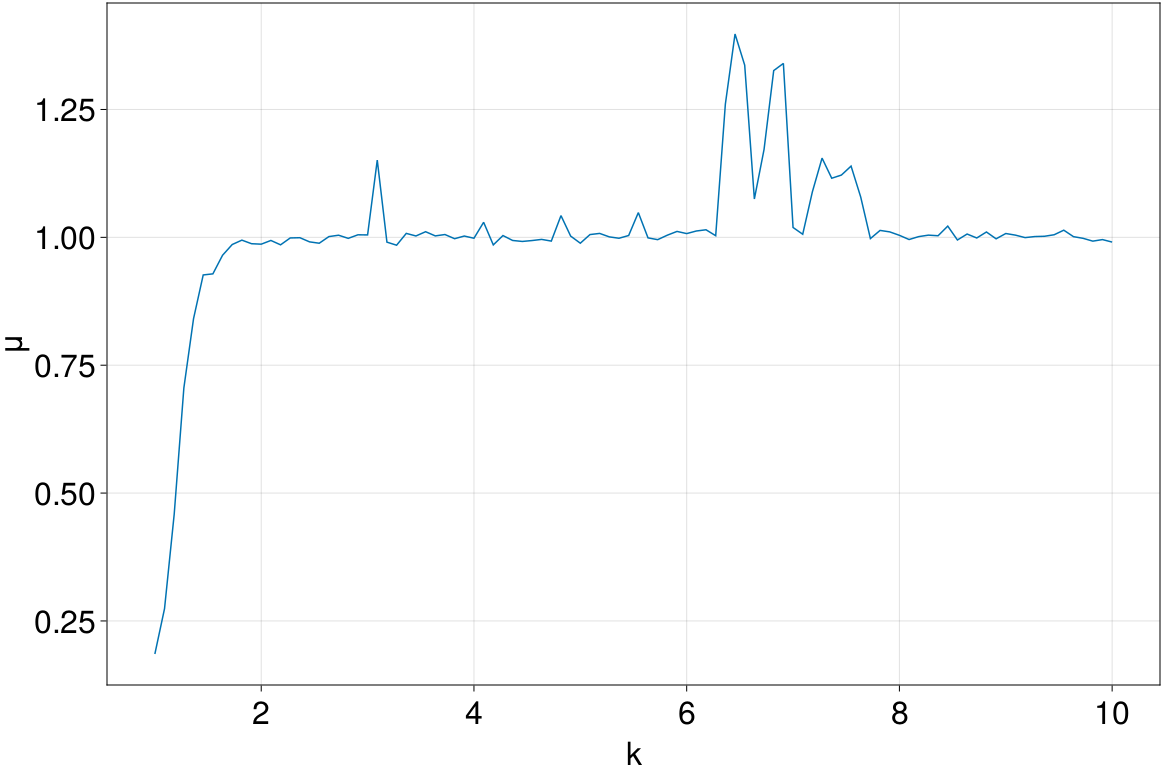}
        \subcaption{}
        \label{subfig:standard_map_transport_exponent_a}
    \end{subfigure}
    \begin{subfigure}[b]{0.49\textwidth}
        \includegraphics[width = \textwidth]{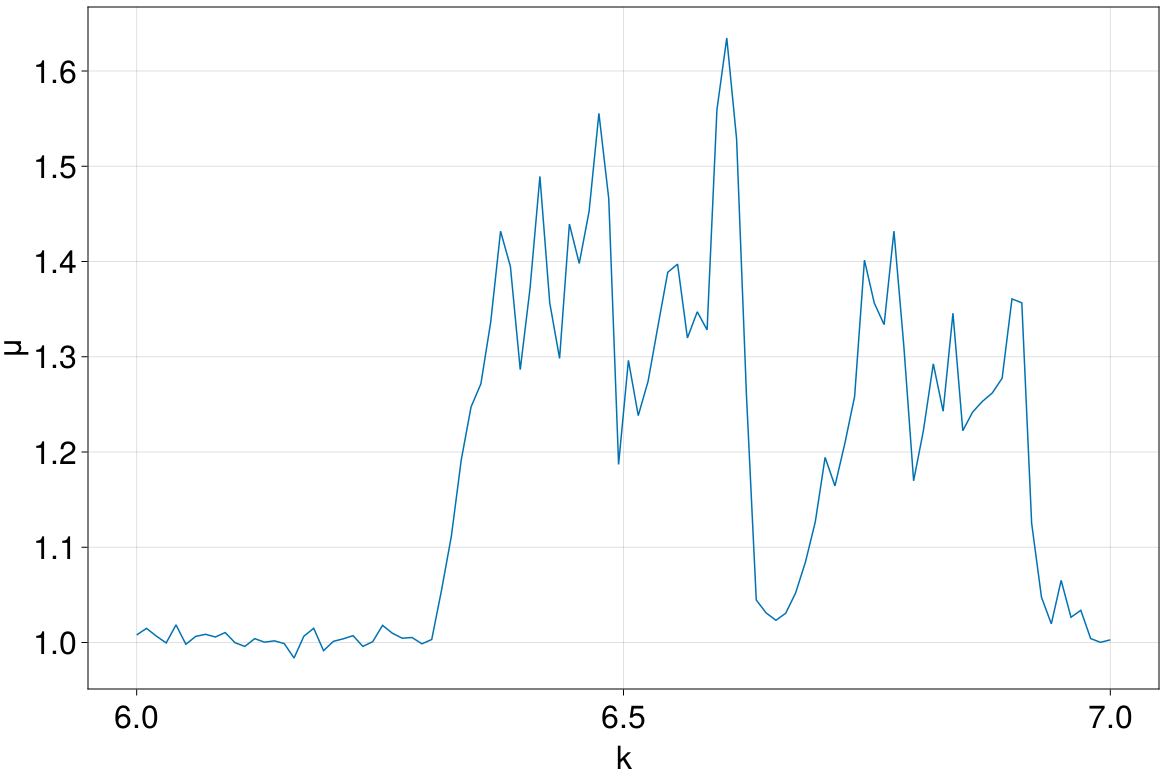}
        \subcaption{}
        \label{subfig:standard_map_transport_exponent_b}
    \end{subfigure}
    \caption{Anomalous transport exponents for the standard map computed over an average of $10^5$ stochastic trajectories. \ref{subfig:standard_map_transport_exponent_a} presents the scan over the full domain and \ref{subfig:standard_map_transport_exponent_b} is a higher resolution scan localised around the $k$ values containing an accelerator mode.}
    \label{fig:standard_map_transport_exponent}
\end{figure}

Anomalous transport describes any motion which is non-diffusive. That is, which has different transport properties to that of brownian motion. Suppose we have a time series $r(t)$, then the square displacement of this time series is denoted by
\begin{equation*}
    r^2(t) = (r(t)-r(0))^2\,.
\end{equation*}
We can define the mean squared displacement as an ensemble average $\langle r^2(t)\rangle$ where the average is taken over some family of similar trajectories. In our case we are concerned with the average across all initial points in our phase space which lie on stochastic orbits. For brownian motion it is straightforward to show that the means squared position displacement behaves as $\langle r^2(t)\rangle \propto t$ in the limit that $t\rightarrow \infty$ \cite{KassibrakisEtAl,huang2009introduction}. As such we describe any system which scales at large time differently to that of diffusion as displaying anomalous transport. We quantify this concept with the notion of an anomalous transport exponent $\mu$ defined by $\langle r^2(t)\rangle \propto t^\mu$ for $t\rightarrow \infty$. We recognise that $\mu=1$ describes classical diffusion, and call $\mu<1$ and $\mu>1$ sub-diffusive and super-diffusive transport respectively. We observed earlier that the projection of a standard map orbit onto the momentum degree of freedom looks like a brownian walk and so we will quantify the anomalous transport of the standard map by computing $\langle p_n^2\rangle \propto n^\mu$ and extracting $\mu$. We specifically do not want to compute $\langle x_n^2\rangle$ because this mean square deviation cannot grow arbitrarily large since the $x$ is restricted to a circle $S^1$ rather than an infinite line\footnote{Even lifting from $S^1\rightarrow \mathbb{R}$ does not actually help here because the step size always grows as the particle momentum increases creating strongly non-brownian motion.}. 

Extracting the anomalous transport exponent for the standard map over a range of $k$ values we obtain the plots presented as Figure \ref{fig:standard_map_transport_exponent}. We observe from subfigure \ref{subfig:standard_map_transport_exponent_a} that the most common behaviour when $k>2$ is $\mu\approx 1$ indicating that the transport is usually approximately diffusive. However, we observe that for $k<2$, $\mu < 1$ and therefore the transport is strongly sub-diffusive in this regime. This is because, as we observed earlier, the remnants of broken KAM toruses form nested layers of islands which appear in the topological data as the broad and unstructured birth time distributions shown in Figure \ref{fig:standard_map_island_distribution}. These islands oppose the transport of orbits in momentum space and so create sub-diffusive transport. In contrast near $k>6.5$ the transport is super-diffusive, this is due to the presence of the accelerator mode which appeared in the topological data as a hierarchical structure in the birth time distribution as was shown in Figure \ref{fig:accelerator_mode_island_distributions}. It is reasonable that these accelerator modes should create super-diffusive behaviour since when an orbit approaches an accelerator mode it will stick to it and follow a long flight in the momentum space\footnote{Recall that this was visible as medium length straight line sub-trajectories in Figure \ref{fig:accelerator_mode_trajectories}.}. We expect then that the accelerator modes should increase the rate at which our orbits explore the momentum space and hence create super-diffusive transport. The self-similar structure of the accelerator mode was responsible for creating the hierarchy of peaks in the birth time distribution, and this same structure is responsible for the island stickiness which causes the anomaly in the transport. Therefore it is our hypothesis that for any phase space in which a hierarchy of peaks is present in the birth time distribution the transport will be anomalous. However, substantial further research  is required to confirm this hypothesis as here we have only demonstrated a correlation between the birth time hierarchy and the anomalous transport, not proven a causative link between the two. 
 
\chapter{Detecting hidden renormalisation group transformations} 

\label{Chapter5} 

In recent years substantial attention has been paid to the use of the methods of computational topology for the detection of approximate symmetries \cite{DBLP:journals/tvcg/ThomasN11,bermingham2023planar}. In this chapter we propose the use of TDA to detect renormalisation group symmetries in Hamiltonian systems by analysing the toy model of a periodically perturbed pendulum near to a hyperbolic fixed point, that is, a saddle point of a Hamiltonian.

Suppose we have a $1\frac{1}{2}$D Hamiltonian of the form
\begin{equation}
    H(p,q;t) = H_0(p,q)+\varepsilon V(q,t)\,,
\end{equation}
where $H_0$ represents the unperturbed system, $\varepsilon >0$ describes the perturbation strength, and $V$ is a perturbation profile satisfying
\begin{equation}
    V(q,t+2\pi/\nu) = V(q,t)\,,
\end{equation}
for a particular $\nu \in \mathbb{R}$.


\section{Separatrix dynamics}

Near to the separatrix of $H_0$ we can write an approximation to the dynamics of the perturbed model called the separatrix map. A short derivation of this approximation is included in Chapter 2 of \cite{Zaslavsky}. The separatrix map itself is written as
\begin{equation}\label{separatrix_map}
    h_{n+1} = h_n+\epsilon M(n,\phi_n)\,,\,
    \phi_{n+1} = \phi_n+\frac{\pi \nu}{\omega_s A}\log\frac{B}{|h_{n+1}|}\,,
\end{equation}
where: $A$ and $B$ are constants which depend upon $H_0$, $h = (E-E_s)/E_s$ is the energy of the unperturbed orbit of the current point, $\phi$ is the associated \textit{angle}, and $M$ is a Melnikov integral. The value of this approximation is that it describes the dynamics of our Hamiltonian near to a hyperbolic fixed point when the system is perturbed. Therefore it encodes information about the structure of the phase space around the fixed point. We are concerned with the observation that the separatrix map, and hence the phase space structure, of a $1\frac{1}{2}$D Hamiltonian system is invariant under a particular renormalisation group transformation \cite{Zaslavsky}.

\subsection{Theoretical renormalisation group transform}

Renormalisation in dynamical systems is concerned with finding rescalings of the parameters of the system which leave the dynamics invariant \cite{guckenheimer2013nonlinear}. Here we specifically seek a rescaling of $\epsilon$ and $h$. Consider the transforms
\begin{equation}
    R_\epsilon:\epsilon\rightarrow \lambda \epsilon\,,\,  \ h\rightarrow \lambda h\,.
\end{equation}
The question we now ask is ``does there exist a choice of $\lambda$ such that the separatrix map is form invariant under this transform?'' the answer to which is ``yes''. To find the appropriate value of $\lambda$ we can substitute the transformed $h$ and $\epsilon$ into \eqref{separatrix_map} directly. The $h_{n+1}$ equation is automatically invariant but the $\phi_{n+1}$ equation becomes
\begin{equation}
    \phi_{n+1} = \phi_n+\frac{\pi \nu}{\omega_s A}\log\frac{B}{|h_{n+1}|}-\frac{\pi \nu}{\omega_s A}\log\lambda\,.
\end{equation}
Since $\phi$ is an angle it is taken modulo $2\pi$ so we will have preserved dynamics if
\begin{equation}
    -\frac{\pi \nu}{\omega_s A} = 2k\pi\,,
\end{equation}
for $k\in \mathbb{Z}$. It is sufficient to take $k=1$, since we will be able to reach all other valid renormalisations by repeated application of this case or its inverse. So we have that
\begin{equation}
    \lambda  = e^{\frac{2\omega_s A}{\nu}}\,,
\end{equation}
is the theoretical renormalisation rescaling.
        
The claim we will attempt to verify with TDA is that this renormalisation preserves the phase space topology near the saddle point. Specifically, if we move back out of action-angle coordinates our renormalisation on the separatrix map induces a renormalisation on the stroboscopic map itself. The full calculation here is somewhat technical and can be found in \cite{Zaslavsky}. We will instead simply state the main theorem below. 
\begin{thrm}\label{Renormalisation}
    The phase space topology near the saddle point, obtained as a stroboscopic map of the Hamiltonian $H_0+\epsilon V$ is preserved under the renormalisation transform
    \begin{equation*}
        R_\epsilon:\epsilon\rightarrow \lambda \epsilon,H_0\rightarrow \lambda H_0, t\rightarrow t+\frac{\pi}{\nu}\,,
    \end{equation*}
    \begin{equation*}
        q\rightarrow \lambda^{1/2}q,p\rightarrow \lambda^{1/2}p\,,
    \end{equation*}
    \begin{equation*}
        \lambda = e^{\frac{2\omega_s A}{\nu}}\,.
    \end{equation*}
\end{thrm}
Note that in Theorem \ref{Renormalisation} the term ``phase space topology'' refers to the number, size, and shape of islands in the stochastic region around the hyperbolic fixed point. Not all of this information is purely topological and so it would be more accurate to say the ``phase space geometry'' is approximately preserved. However, it is canonical in the literature to adopt the term ``phase space topology''.

The remainder of this chapter is concerned with numerically verifying Theorem \ref{Renormalisation} using TDA. We do this as an experimental mathematics exercise which will allow us to determine if it is feasible to utilise TDA to detect other renormalisation symmetries of similar type to Theorem \ref{Renormalisation} which are not as easy to establish theoretically. 


\section{Proposals for two detection procedures}

\begin{algorithm}
\caption{HomDistSEDT(N,T,H,H')}
\begin{algorithmic}[1]
\Require $N,T\in\mathbb{N}^+$ and $H,H'$ Hamiltonian dynamical systems equipped with Poincare maps or Stroboscopic maps
\State Render using WBA the binary image of the phase space of $H$ as in the \textit{BirthsDist} algorithm. Taking the Poincare (Stroboscopic) map of $H$ as the discrete dynamical system.
\State Compute the SEDT of this image and call it $SEDT(H)$.
\State Compute persistent homology $PH(SEDT(H))$ by sub-level set filtration.
\State Repeat $1-3$ with $H'$ to obtain $PH(SEDT(H'))$. 
\For{$n \in \mathbb{N}^+$ (a dimension)}
    \If{$PH_n(SEDT(H))$ and $PH_n(SEDT(H'))$ are both nontrivial}
        \State $d_n \gets d(PH_n(SEDT(H),PH_n(SEDT(H'))$ for a chosen distance $d$.
    \EndIf
\EndFor
\State \Return $(d_0,d_1,\ldots)$.
\end{algorithmic}
\end{algorithm}

So to numerically test Theorem \ref{Renormalisation} we need a procedure to compute and compare the phase space topologies of different Hamiltonian systems, since each choice of $\lambda$ corresponds to a different Hamiltonian with its own phase space. We propose two different approaches to this problem one using the SEDT methods from Chapter \ref{Chapter4} and another using the Vietoris-Rips methods from Chapter \ref{Chapter3}. The algorithm \textit{HomDistSEDT} below presents a procedure which is well defined for general Hamiltonian systems. In contrast \textit{HomDistVR} is only well defined for different transformations of the same system, this is because \textit{HomDistVR} relies upon the existence of a hyperbolic fixed point shared by both Hamiltonians. It is only the phase topology in the vicinity of said fixed point which \textit{HomDistVR} compares.

\begin{algorithm}
\caption{HomDistVR(T,H,H')}
\begin{algorithmic}[1]
\Require $T\in\mathbb{N}^+$ and $H,H'$ Hamiltonian dynamical systems equipped with Poincare maps or Stroboscopic maps $P$ and $P'$ respectively and which share a hyperbolic fixed point $x_h$.
\State Choose a point $x$ in a neighborhood of $x_h$.
\State Check if the orbit of $x$ induced by the Hamiltonian evolution of $H$ is chaotic by the WBA method.
\If{Orbit of $x$ under $P$ is chaotic}
\State  Generate the point cloud $X_T(x)$ and call it $X$
\Else
\State Return to 1 and choose a new point
\EndIf
\State Repeat 1-5 for $H'$, generating another point cloud $X'$ describing a chaotic trajectory in the vicinity of $x_h$.
\State Compute the Vietoris-Rips persistent homologies $PH(X)$ and $PH(X')$
\For{$n \in \mathbb{N}^+$ (a dimension)}
    \If{$PH_n(X)$ and $PH_n(X')$ are both nontrivial}
        \State $d_n \gets d(PH_n(X),PH_n(X'))$ for a chosen distance $d$.
    \EndIf
\EndFor
\State \Return $(d_0,d_1,\ldots)$.
\end{algorithmic}
\end{algorithm}

Both of these approaches are hypothetically applicable to our problem here. Our approach to which is as follows. We fix an initial unrenormalised Hamiltonian $H$. Then, for a range of renormalisation scales $\lambda$, we construct the renormalised Hamiltonian $H_\lambda$ defined as $H$ subject to the $R_\epsilon$ transform for renormalisation scale $\lambda$. We then compute $\textit{HomDist}(H,H_\lambda)$ for each $\lambda$ by either $\textit{HomDistSEDT}$ or $\textit{HomDistVR}$. We expect that the phase space topologies of $H$ and $H_\lambda$ will be ``closest'', that is the $\textit{HomDist}$ will be smallest, when the renormalisation scale $\lambda$ equals its theoretical value. So after computing $\textit{HomDist}(H,H_\lambda)$ for each $\lambda$ we search for local minima in the associated one parameter function.

As a technical note, to calculate the VR filtrations needed in \textit{HomDistVR} we require that the space in which our points are embedded is equipped with a metric. Formally our Hamiltonian phase spaces are not equipped with one. For our purposes here we will assume it is sufficient to equip them with the Euclidean metric with respect to the chosen generalised coordinates and momenta.

\subsection{Distances between persistence diagrams}

The \textit{HomDistSEDT} and \textit{HomDistVR} algorithms above are constructed for a generic distance $d$ between persistent homologies. In order to perform these calculations we need to specify how we can calculate such a distance. We will focus on two common options here, the Bottleneck and Wasserstein distances. Each of these distances have advantages and disadvantages and we will explore their effects when we benchmark our procedure on a test case below. 

\subsubsection{Bottleneck distance}

For two persistence diagrams $PD,PD'$ define the bottleneck distance $W_\infty(PD,PD')$ as 
\begin{equation}
    W_\infty(PD,PD') = \underset{\eta:PD\rightarrow PD'}{\inf}\underset{x\in PD'}{\sup}||x-\eta(x)||_\infty\,.
\end{equation}
where $\eta:PD\rightarrow PD'$ is a matching function which assigns birth-death pairs in $PD$ either to birth-death pairs in $PD'$ or to the diagonal. The Bottleneck distance is a metric on the space of persistence diagrams and has been adopted when demonstrating the stability of persistence diagrams to noise in the underlying datasets  \cite{EDELSBRUNNER}.

\subsubsection{Wasserstein distance}

The other common distance function of interest to use is the Wasserstein$-q$ distance which is defined as
\begin{equation}
    W_q(PD,PD') = \left[ \underset{\eta:PD\rightarrow PD'}{\inf}\sum_{x\in PD}||x-\eta(x)||^q_q \right]^{1/q}\,,
\end{equation}
where $q$ is referred to as the order or degree \cite{EDELSBRUNNER}. Also note $||\cdot||_q$ refers to the $L_q$ metric on the $(b,d)$ coordinates of the classes on a persistence diagram and that we are performing a $q$-th power summation. We will specifically utilise the case of $q=2$ below because this was observed to perform well empirically on our renormalisation scan datasets. 


\section{Benchmarking with the perturbed pendulum}

We will use a periodically perturbed nonlinear pendulum as a benchmark for example for the detection procedures discussed above. The Hamiltonian for this system is
\begin{equation}\label{PerturbedPendulumRenorm}
    H(p,q,t) = \frac{1}{2}p^2-\omega^2\cos(q)+\varepsilon \cos(kq - \nu t)\,.
\end{equation}
For comparison to Theorem \ref{Renormalisation} we have that $\omega_s = \omega$ and $A=\pi$ for this perturbed pendulum \cite{Zaslavsky}.

\begin{figure}
    \centering
    \includegraphics[width=0.4\textwidth]{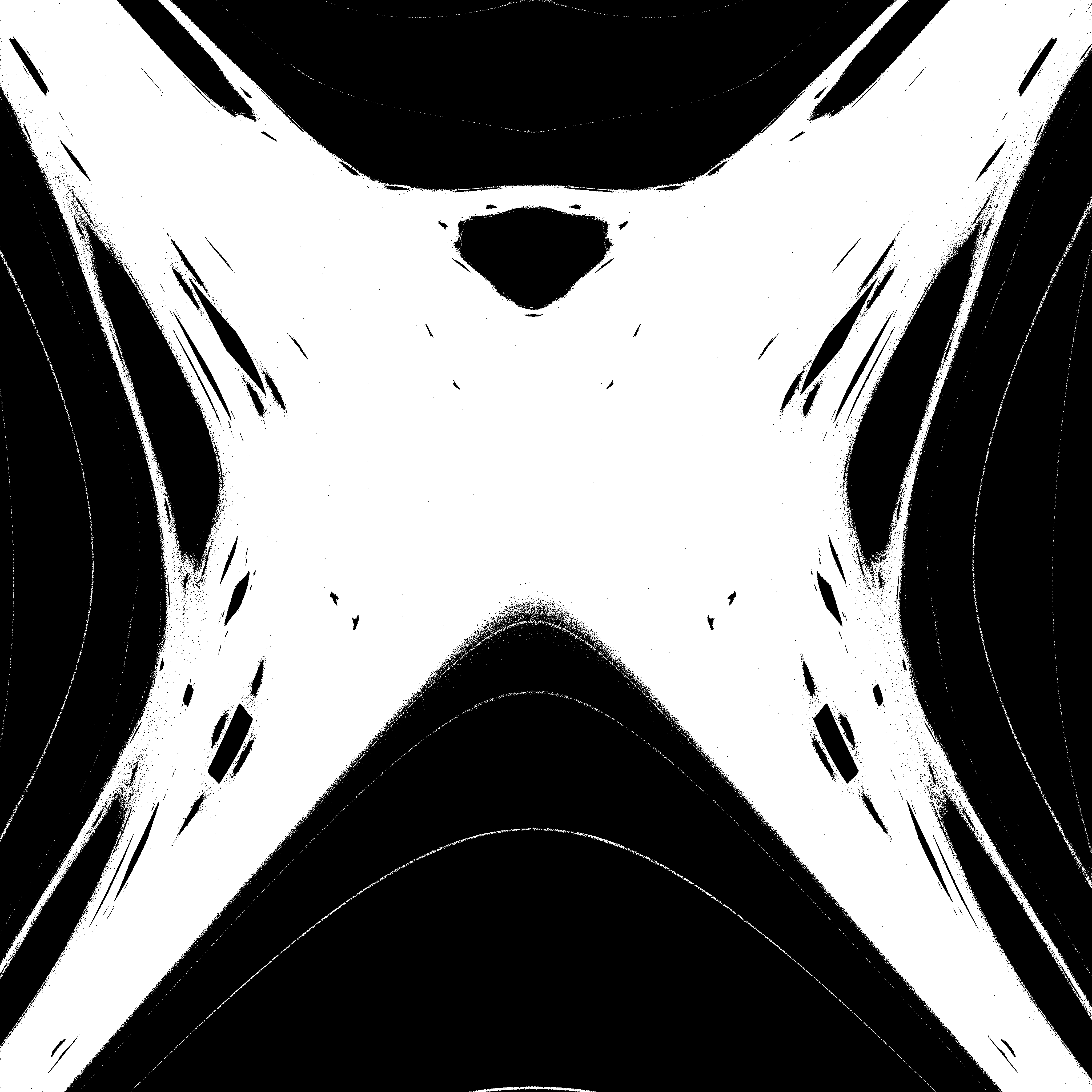}
    \caption{Neighborhood of the perturbed pendulum hyperbolic fixed point.}
    \label{fig:unrenormalisedPhaseSpace}
\end{figure}

\subsection{The SEDT method}

Firstly we will test the \textit{HomDistSEDT} approach. Note that our base Hamiltonian $H$ is \eqref{PerturbedPendulumRenorm} with parameters $\omega = 1$, $k=1$, $\varepsilon=0.01$, and $\nu=4.8$. The strobing frequency of the stroboscopic map is also taken to be $\nu$. For reference, Figure \ref{fig:unrenormalisedPhaseSpace} presents a binary image of the phase space in a neighborhood of a hyperbolic fixed point. We will refer to the phase space shown in this image as the \textit{base} phase space.

\begin{figure}
    \centering
    \begin{subfigure}[b]{0.24\textwidth}
        \includegraphics[width = \textwidth,frame]{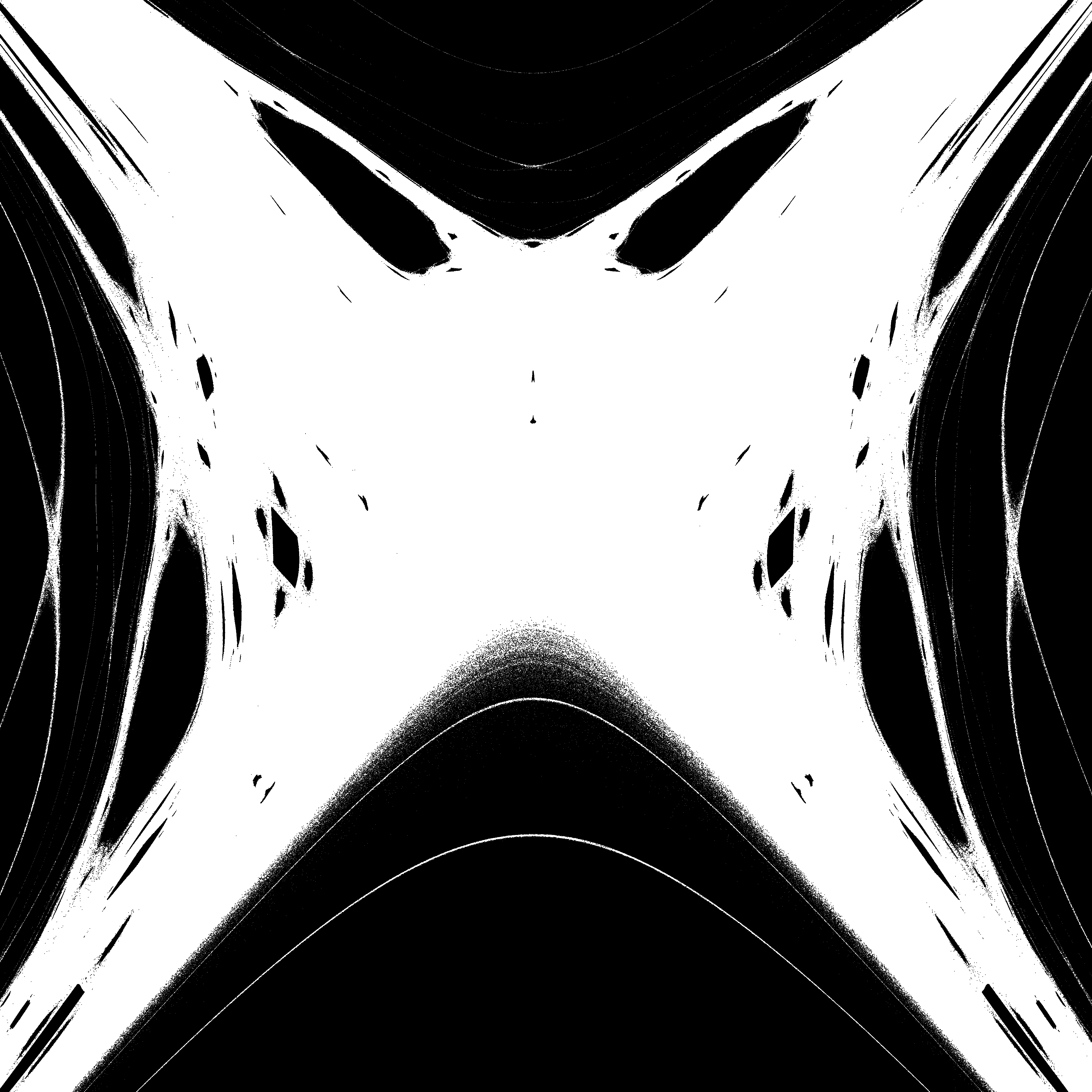}
        \subcaption{$\lambda = 1$}
    \end{subfigure}
    \begin{subfigure}[b]{0.24\textwidth}
        \includegraphics[width = \textwidth,frame]{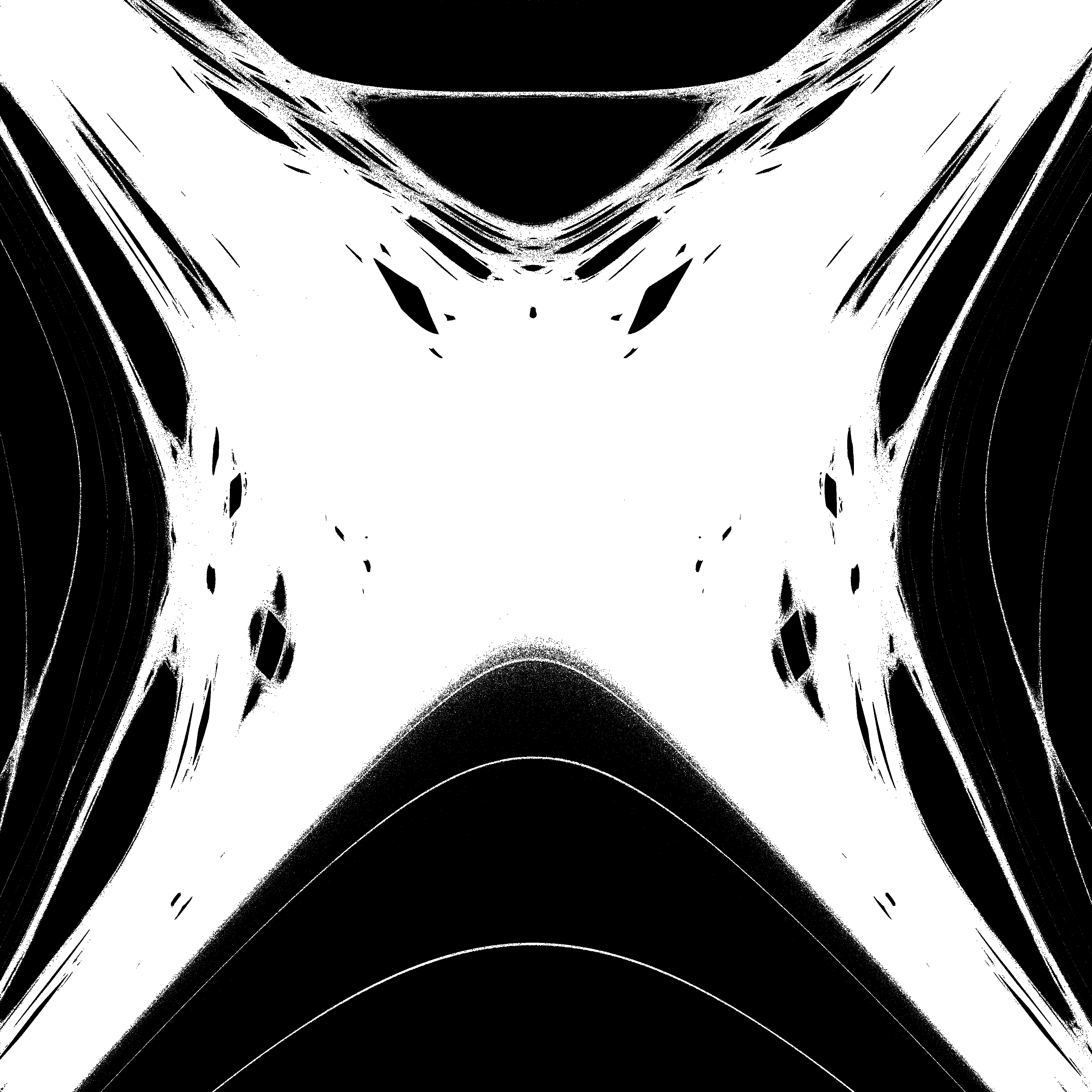}
        \subcaption{$\lambda=1.924$}
    \end{subfigure}
    \begin{subfigure}[b]{0.24\textwidth}
        \includegraphics[width = \textwidth,frame]{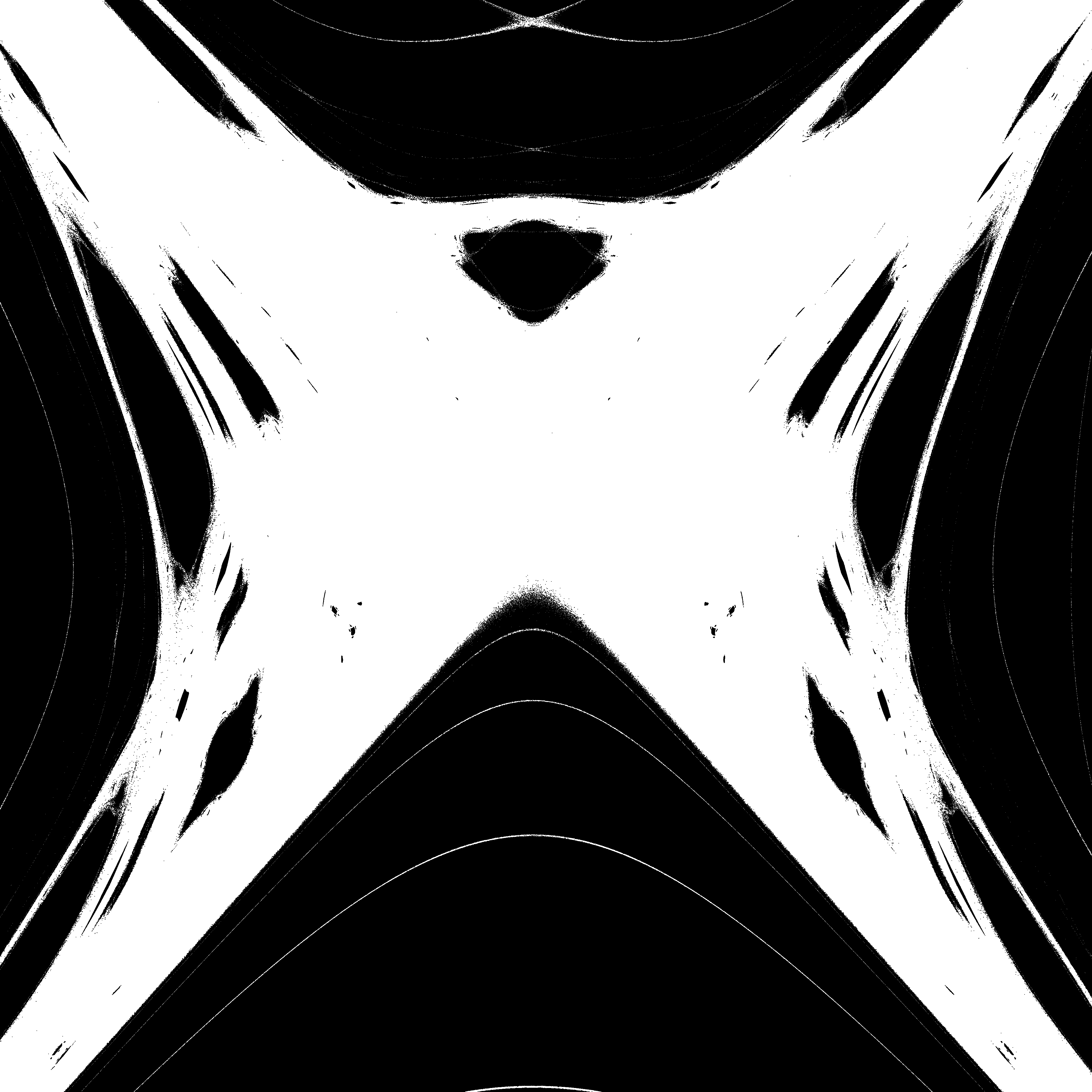}
        \subcaption{$\lambda=3.7024$}
        \label{subfig:lambda3.7024}
    \end{subfigure}
    \begin{subfigure}[b]{0.24\textwidth}
        \includegraphics[width = \textwidth,frame]{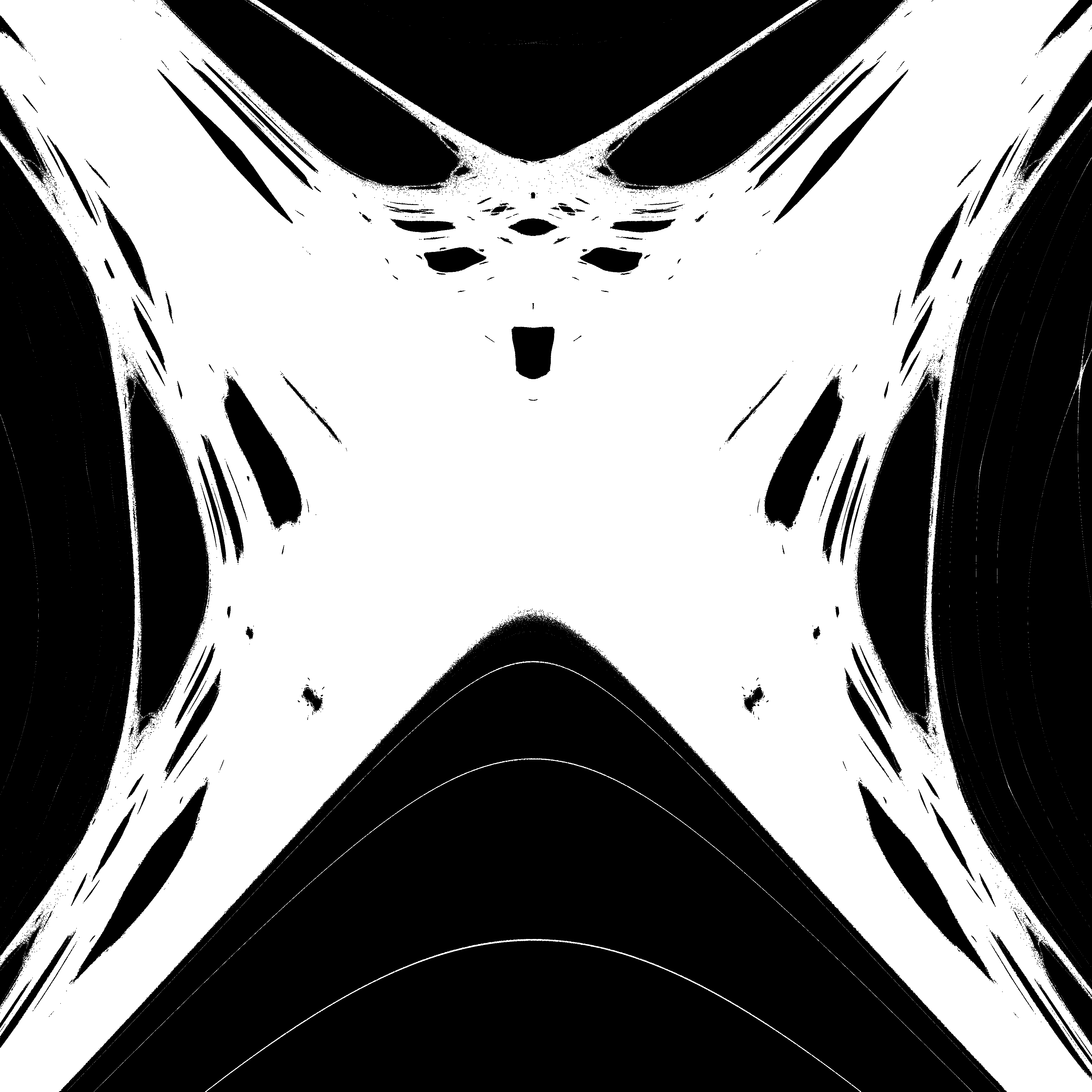}
        \subcaption{$\lambda=7.124$}
    \end{subfigure}
    \caption{Perturbed pendulum phase spaces after a renormalisation by scale parameter $\lambda$.}
    \label{fig:renormalisedPhaseSpace}
\end{figure}

As a first visual check on the renormalisation theorem consider Figure \ref{fig:renormalisedPhaseSpace} which presents the binary images of the phase space of the renormalised system $H_\lambda$ for four values of $\lambda$. Observe that the phase space with $\lambda=1$ is different to the base phase space Figure \ref{fig:unrenormalisedPhaseSpace}. This is the effect of the phase shift by $\pi/\nu$ applied as part of the $R_\epsilon$ renormalisation. On inspection, of the four phase spaces shown the case of subfigure \ref{subfig:lambda3.7024} appears most similar to the base case. This is what we would expect as $\lambda=3.7024$ is very close to the theoretical value of $\lambda$ predicted in Theorem \ref{Renormalisation}.

\subsubsection{Scanning over the bottleneck distance}

\begin{figure}
    \centering
    \begin{subfigure}[b]{0.49\textwidth}
        \includegraphics[width = \textwidth]{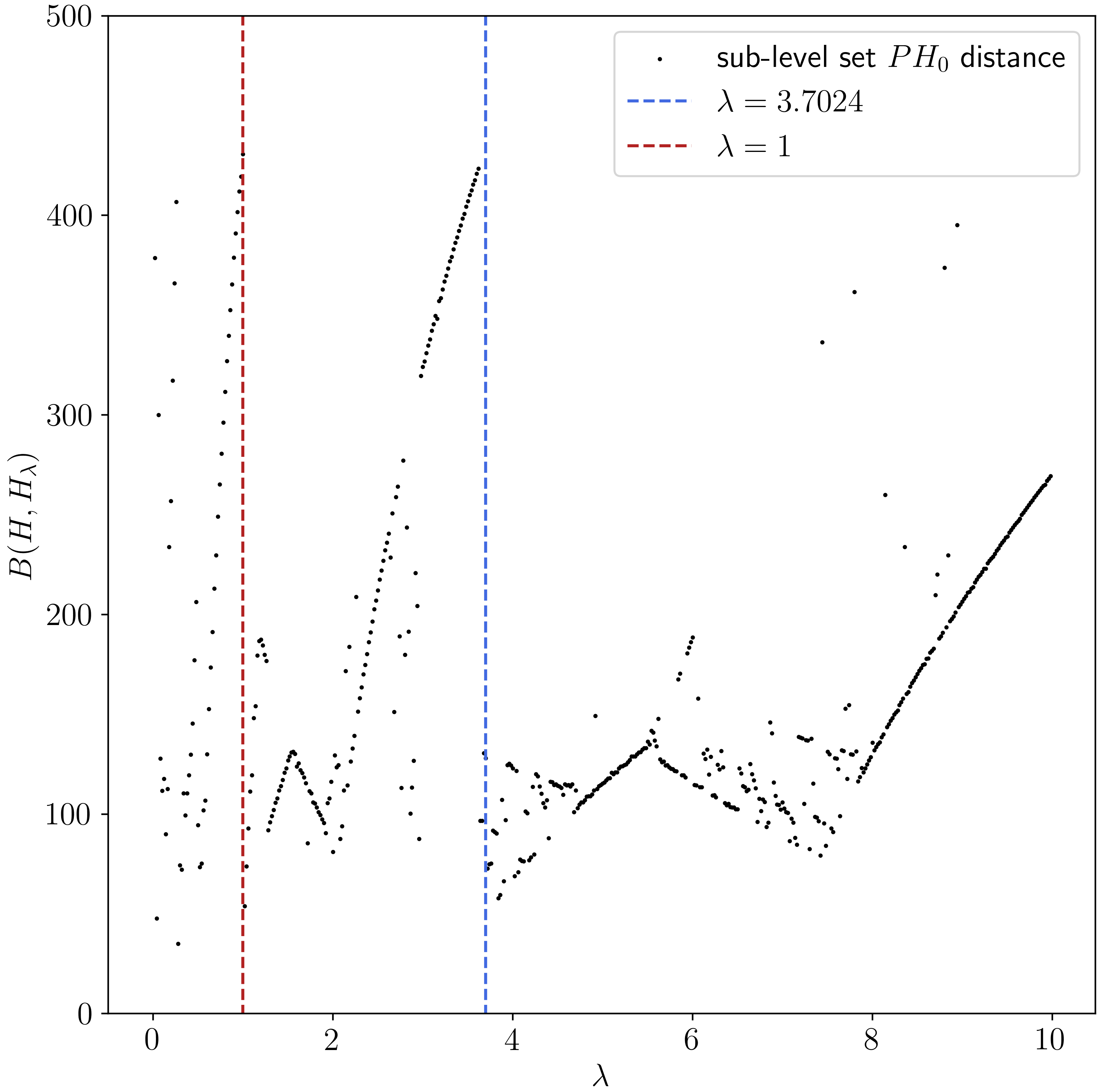}
        \subcaption{$PH_0(SEDT)$ case}
        \label{subfig:SEDTLinearScanPH0}
    \end{subfigure}
    \begin{subfigure}[b]{0.49\textwidth}
        \includegraphics[width = \textwidth]{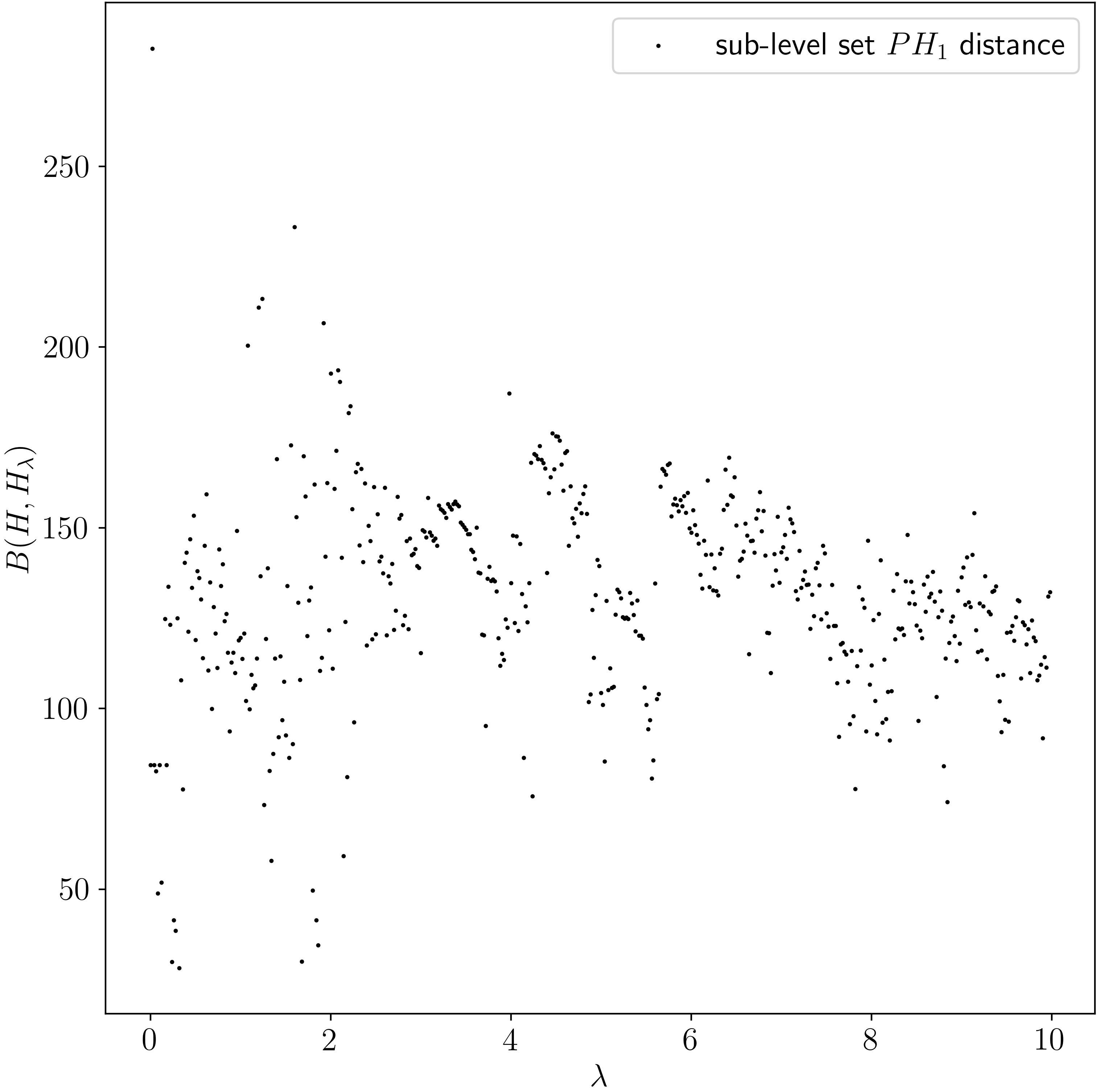}
        \subcaption{$PH_1(SEDT)$ case}
        \label{subfig:SEDTLinearScanPH1}
    \end{subfigure}
    \caption{Results of a scan over the bottleneck distance between the sub-level set persistent homology of the unrenormalised and renormalised Hamiltonians.}
    \label{fig:SEDTLinearScan}
\end{figure}

Figure \ref{fig:SEDTLinearScan} presents the results of scanning $\textit{HomDistSEDT}(H,H_\lambda)$ for the case in which $d$ is chosen to be the bottleneck distance. Recall that in the sub-level set homology of the SEDT, the $H_0$ persistence describes the size and separation of connected black and white regions, which in our case are the islands. These are the dominant feature of the phase space topology which we are interested in comparing and so we will focus on subfigure \ref{subfig:SEDTLinearScanPH0}. The key feature we highlight is the cluster of points at low $B(H,H_\lambda)$ following a sharp edge. This is highlighted by a vertical line at the theoretical value of $\lambda$. The existence of a cluster of low bottleneck distances around this value suggests that the phase space topology around $\lambda=3.7024$ is closer to the base case than the other $\lambda$ values in the vicinity. This agrees with our expectation that the phase space topology should be preserved under renormalisation with $\lambda=3.7024$. 

However, it would be unreasonable to claim this as sufficient evidence that we are accurately detecting the preservation of the phase space topology with TDA. This is because the blue line is not the actual global minimum of the scan. The actual global minimum occurs when $\lambda<1$ and is in rough agreement with the value $\lambda = 1/3.7024$, which is a valid renormalisation transform, since it is an odd integer power of $3.7024$. This suggests that we could improve our confidence in our detected renormalisation transform by searching not for the $\lambda$ which is the global minimum but instead searching for a series of local minima related by a power law. 

Finally, observe that the quasi-periodic rising and falling arcs visible in sub-figure \ref{subfig:SEDTLinearScanPH0} suggest that we are ``picking up on'' other self-similar structure in the phase space and so we need to isolate the influence of the topology in the immediate vicinity of the hyperbolic fixed point from other structures. If we look at the locations of the critical points which generate the most persistent $H_0$ classes in our sub-level set filtration, which have the largest effect on the bottleneck distance, we observe that the arcs in the distance plot are associated to the linear features visible in the top and bottom of the binary images of the phase space. These features are not local to the hyperbolic fixed point, and are effectively an imaging artifact, but tend to dominate the bottleneck distance due to their large geometric size. To minimise their impact and possibly improve our ability to detect the phase space symmetric we would prefer to adopt a different homological distance which does not weight these linear features as highly. This is the motivation behind testing the Wasserstein distance since it does not weight highly persistent features as heavily as the bottleneck distance and so we suspect it will be less susceptible to these linear image artifacts. 

\subsubsection{Scanning over a Wasserstein distance}

\begin{figure}
    \centering
    \begin{subfigure}[b]{0.49\textwidth}
        \includegraphics[width = \textwidth]{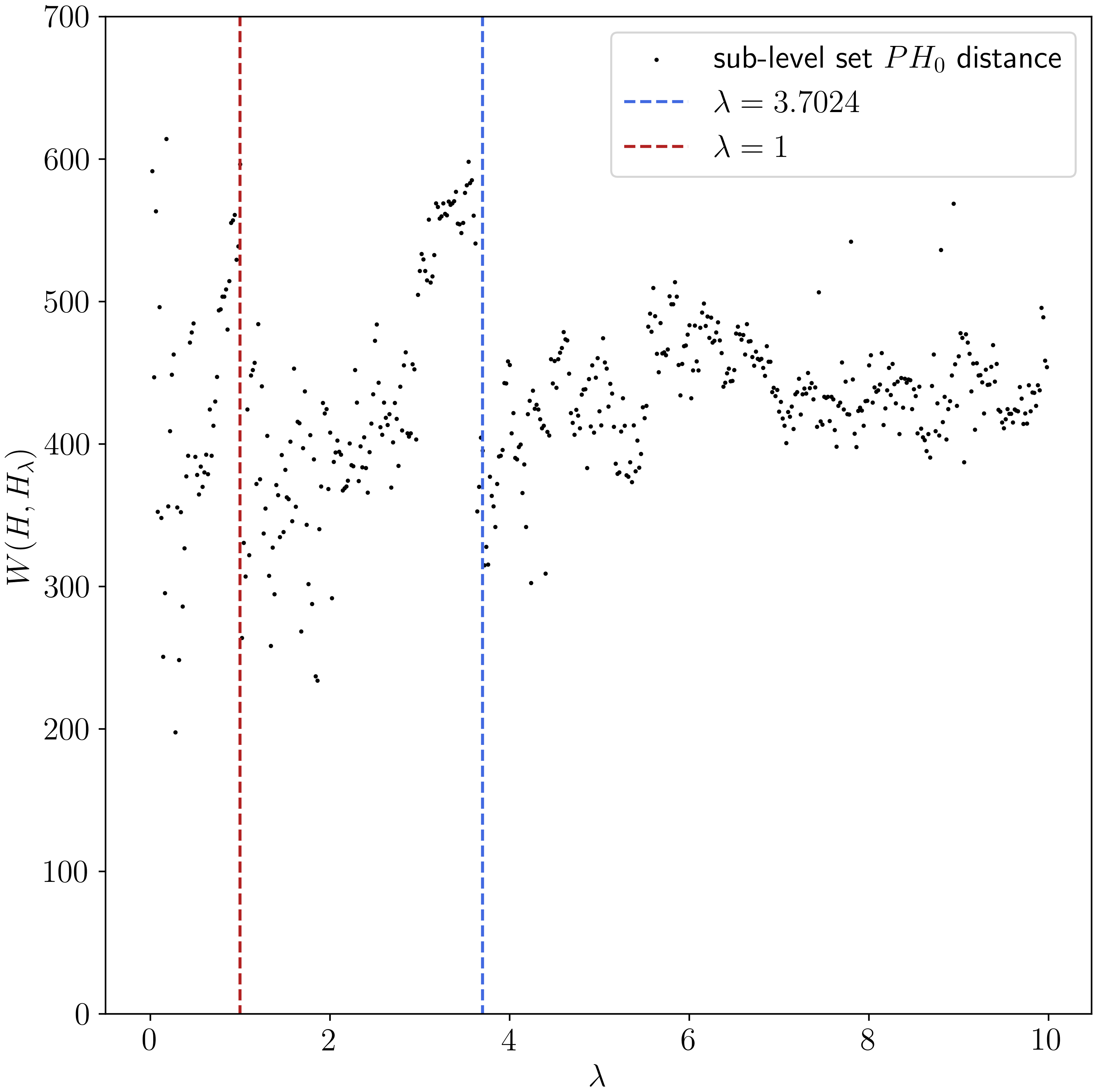}
        \subcaption{$PH_0(SEDT)$ case.}
        \label{subfig:SEDTLinearScanPH0Wasserstein}
    \end{subfigure}
    \begin{subfigure}[b]{0.49\textwidth}
        \includegraphics[width = \textwidth]{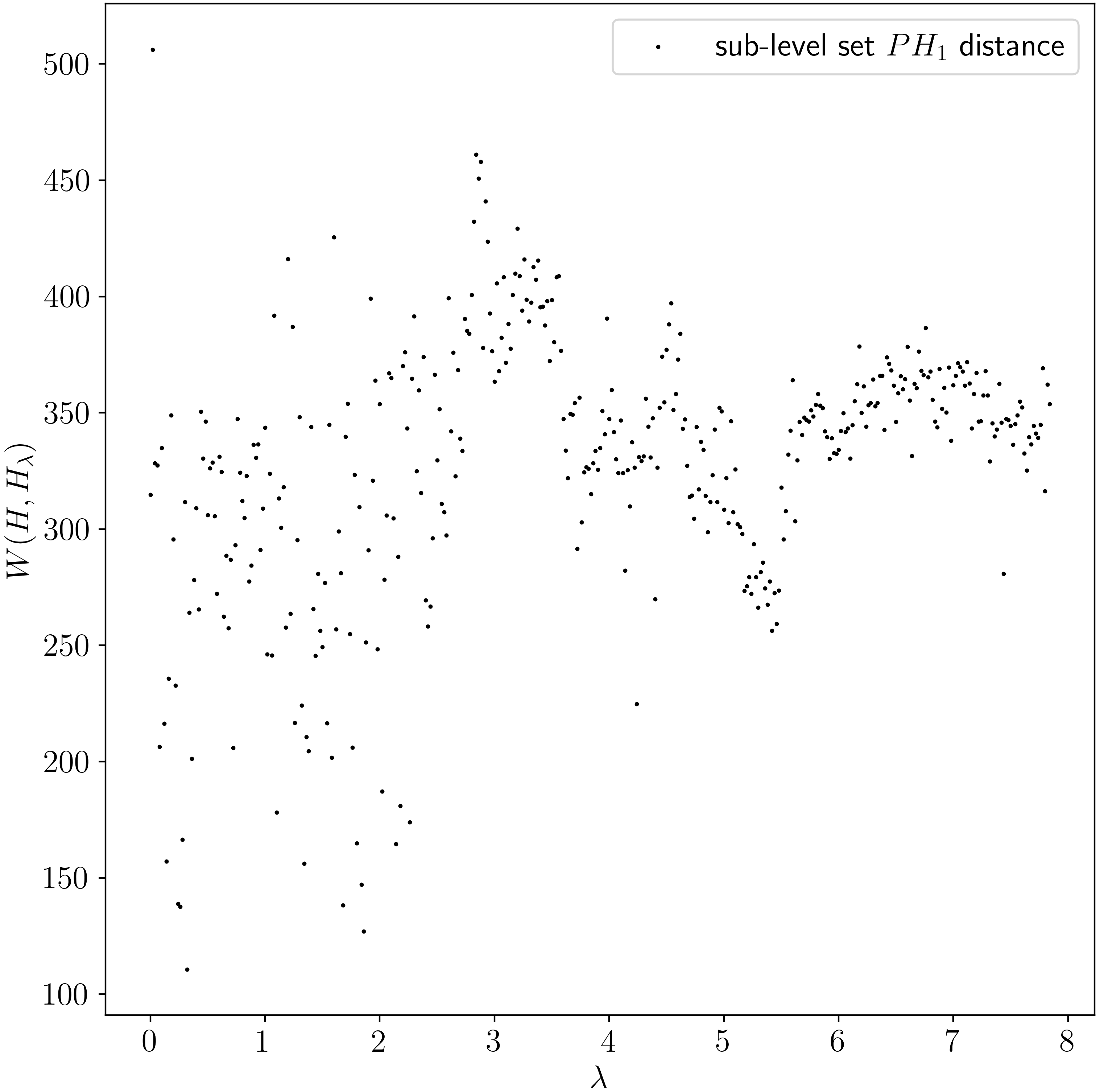}
        \subcaption{$PH_1(SEDT)$ case.}
        \label{subfig:SEDTLinearScanPH1Wasserstein}
    \end{subfigure}
    \caption{Results of a scan over the Wasserstein-$2$ distance between the sub-level set persistent homology of the unrenormalised and renormalised Hamiltonians.}
    \label{fig:SEDTLinearScanWasserstein}
\end{figure}

As an alternative we tested the effect of using Wasserstein-$2$ instead of the bottleneck distance for the same scan as above. The results of this scan are presented as Figure \ref{fig:SEDTLinearScanWasserstein}. The values of the $PH_0$ distance for this calculation are generally more tightly clustered because the Wasserstein distance weights less persistent features in the persistent homology more heavily than the Bottleneck distance\footnote{This is because the supremum in the definition of the bottleneck distance tends to favour more persistent points which will usually have the largest change between two diagrams. Persistent points have ``further to move'' without being annihilated by the diagonal.}. We again observe a small local minimum clustered above the theoretical value but also a larger cluster above $\lambda=1$. This suggests that the Wasserstein distance is sensitive to the renormalisation symmetry we are trying to confirm but more detailed experimentation and analysis is required to quantify this sensitivity and to confirm it with sufficient confidence. 

\subsection{The Rips approach}

We now test the \textit{HomDistVR} procedure using the same two distances as above. However, before we do so we will first highlight a problem with the procedure as defined above. We will then discuss and test two potential solutions.

\subsubsection{The problem of coverage}

A major problem for which TDA, and persistent homology specifically, was developed for was the identification of the topology of manifolds from finite approximations, really point samples. One of the challenges which was encountered can be described as the issue of ``incomplete coverage'' and this same problem poses us a challenge here. 

\begin{figure}
    \centering
    \begin{subfigure}[b]{0.24\textwidth}
        \includegraphics[width = \textwidth]{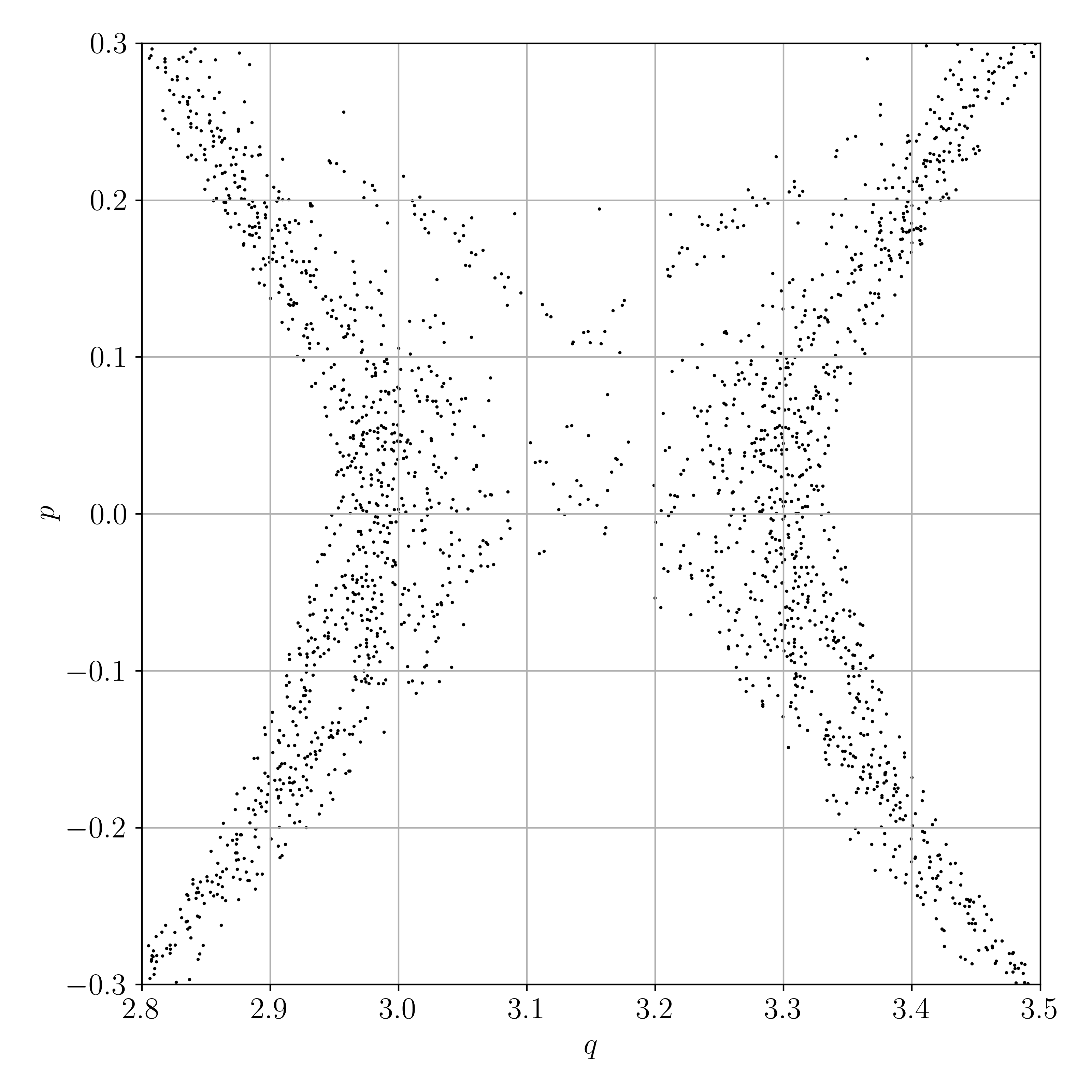}
        \subcaption{}
        \label{subfig:evidenceForTransportBarriersA}
    \end{subfigure}
    \begin{subfigure}[b]{0.24\textwidth}
        \includegraphics[width = \textwidth]{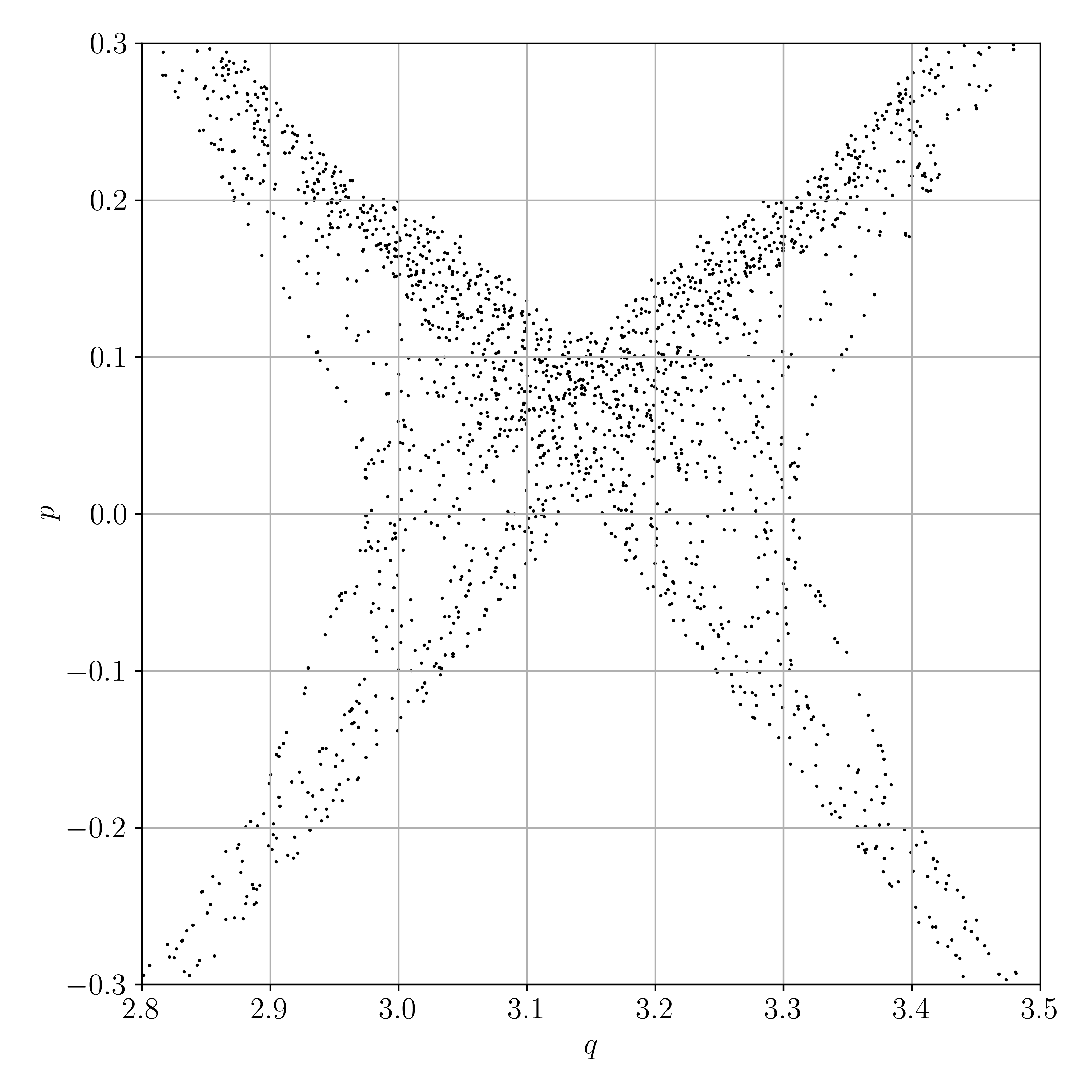}
        \subcaption{}
        \label{subfig:evidenceForTransportBarriersB}
    \end{subfigure}
    \begin{subfigure}[b]{0.24\textwidth}
        \includegraphics[width = \textwidth]{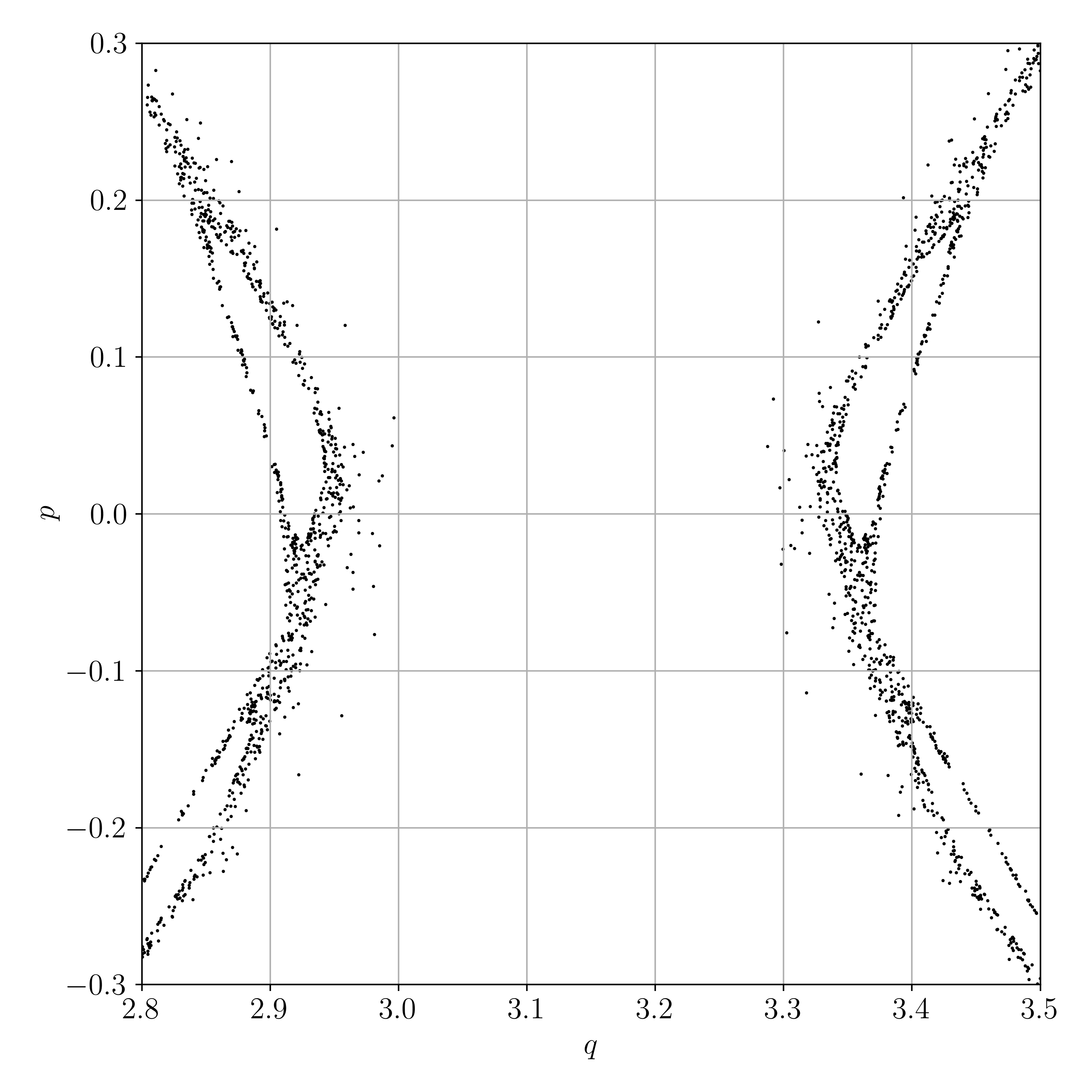}
        \subcaption{}
        \label{subfig:evidenceForTransportBarriersC}
    \end{subfigure}
    \begin{subfigure}[b]{0.24\textwidth}
        \includegraphics[width = \textwidth]{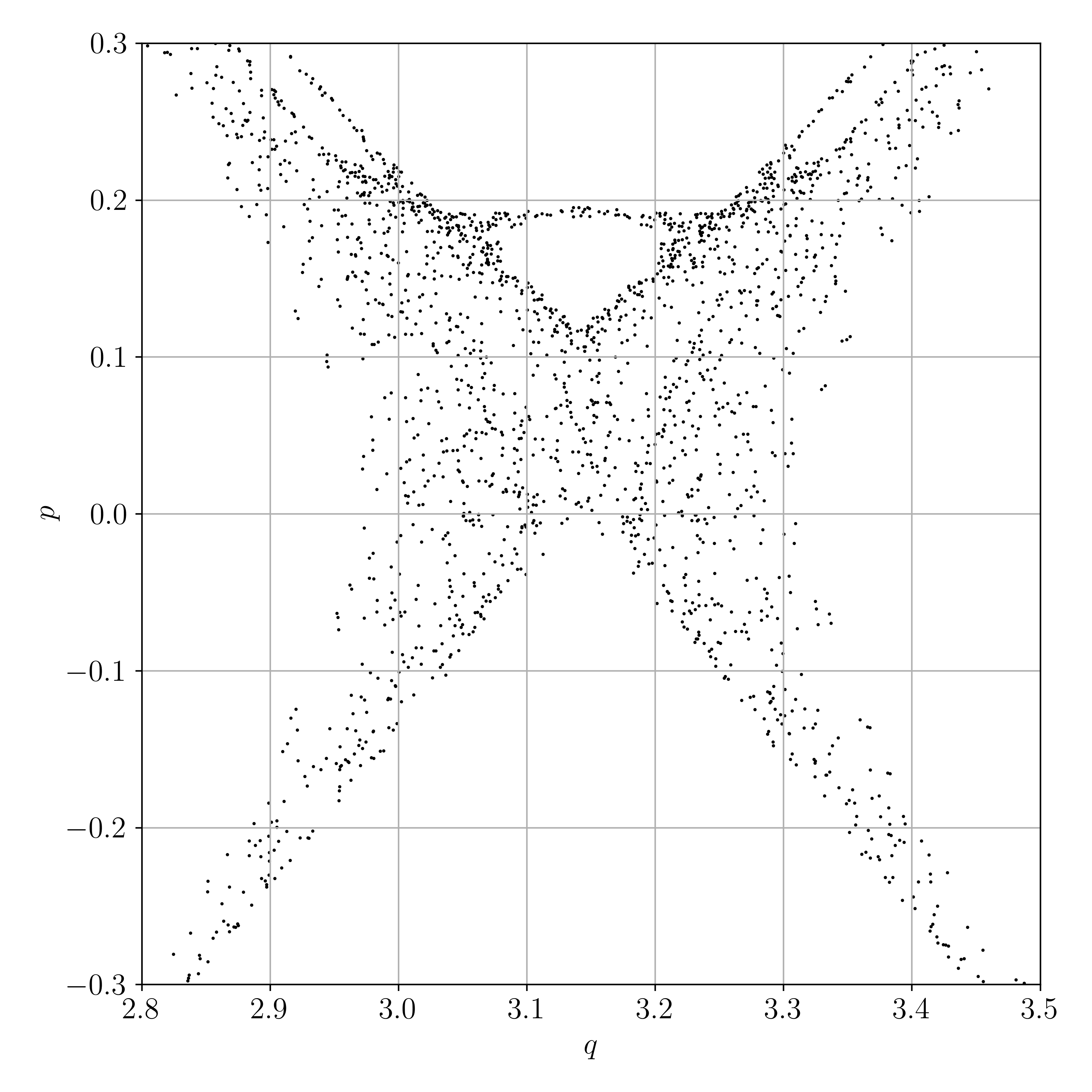}
        \subcaption{}
        \label{subfig:evidenceForTransportBarriersD}
    \end{subfigure}
    \caption{Orbits of four initial conditions in the stochastic region near to a hyperbolic fixed point for the perturbed pendulum. }
    \label{fig:evidenceForTransportBarriers}
\end{figure}

Consider Figure \ref{fig:evidenceForTransportBarriers} which presents four sub-sequences of $~20000$ points each taken from the same orbit in the stochastic layer of the perturbed pendulum. We observe that the fractal region of interest is not uniformly covered by each point cloud. Instead, the trajectory spends long periods of time in separated regions of the phase space. For example in subfigure \ref{subfig:evidenceForTransportBarriersC} the trajectory is struck in two bands on the left and right edges and does not reenter the central region for a long time. Similarly, observe that there  is an island in the top center of the image visible in subfigure \ref{subfig:evidenceForTransportBarriersD} but the other subsequences of the orbit do not explore this area.

This effect has two complementary causes. It is partly associated to the stickiness of the islands, as the trajectory can get stuck in the boundary layer of an island for a long time. However we can still observe incomplete coverage in a stochastic layer with few if any islands. This is a consequence of a famous observation about almost integrable Hamiltonian systems. The phase space contains remnants of broken KAM toruses which form Cantor sets called Cantori. The Cantor sets do not form total transport barriers like the full KAM toruses but instead form partial barriers through which the transport of phase space points is exponentially damped \cite{mackay1984transport}. There exists methods for detecting these Cantori and describing the statistical properties of transport through them but these methods rely on stronger dynamical information than we are concerned with here \cite{wiggins2013chaotic}. So, for the purposes of the \textit{HomDistVR} procedure we take that it is impossible to know if our region of interest contains Cantori and therefore if our trajectory will fail to explore the full stochastic layer. This reduces our confidence in the accuracy of the VR persistent homology as a measure of the topology and geometry of the stochastic layer around a hyperbolic fixed point. 

Despite the existence of partial barriers it is believed that any trajectory in a connected stochastic layer will eventually explore the entire region available to it\footnote{This statement is just the ergodic hypothesis for area-preserving maps.}. So it is reasonable to assume that our coverage gets better, that is we miss fewer and fewer phase space regions, for longer trajectories since in the limit of an infinite length orbit we must hit all phase space regions. So the solution to our problem seems simple, just take $T$ to be as large as required. However, this is not a particularly practical solution because the calculation of VR persistent homology on large, that is with $\order{10^5}$ points, is computationally challenging. Using \verb!Ripser! it is actually almost impossible to handle point clouds this large because the memory requirements of \verb!Ripser! grow too quickly with point cloud size. Using a method recently developed by Koyama \textit{et at} it is possible to compute the $H_1$ persistent homology for clouds with more that $10^5$ points in Euclidean spaces \cite{koyama2023reduced}. This method is applicable in our case because we equipped our phase space with an ad hoc Euclidean metric. So the first method we will test is using clouds with $10^5$ points\footnote{Which was the largest number of points the Author could reliably compute the homology of using Koyama's software.}. As an alternative we will try taking $\order{10}$ different point clouds each $2000$ points long for each $\lambda$. We can then compute the homology of each sub-cloud and construct an averaged distance between the renormalised and base phase spaces. 

\subsubsection{Testing the effect of large point clouds}

\begin{figure}
    \centering
    \begin{subfigure}[b]{0.49\textwidth}
        \includegraphics[width = \textwidth]{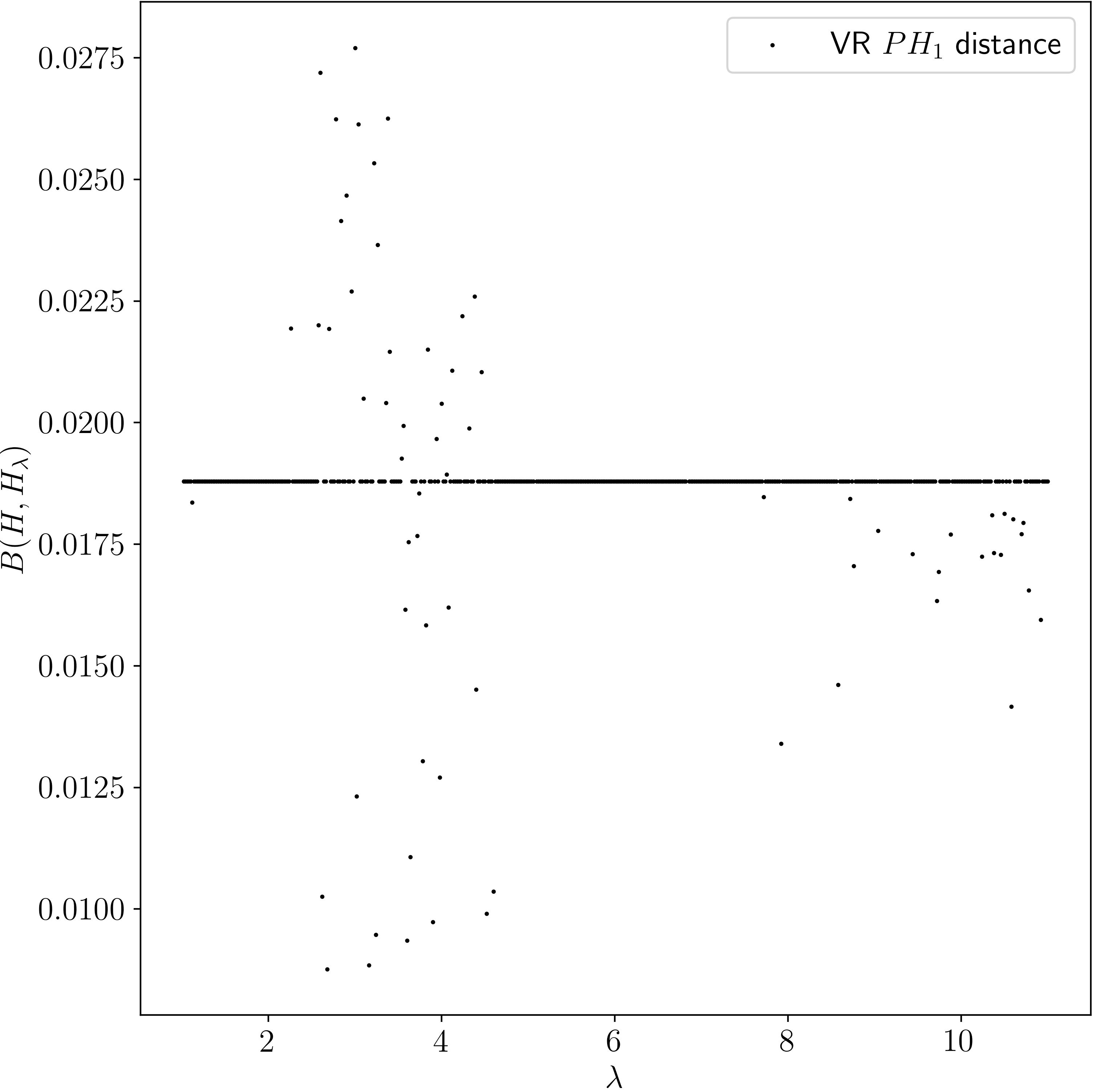}
        \subcaption{Bottleneck case.}
        \label{subfig:RipsLinearScanPH1BottlneckBigPointClouds}
    \end{subfigure}
    \begin{subfigure}[b]{0.49\textwidth}
        \includegraphics[width = \textwidth]{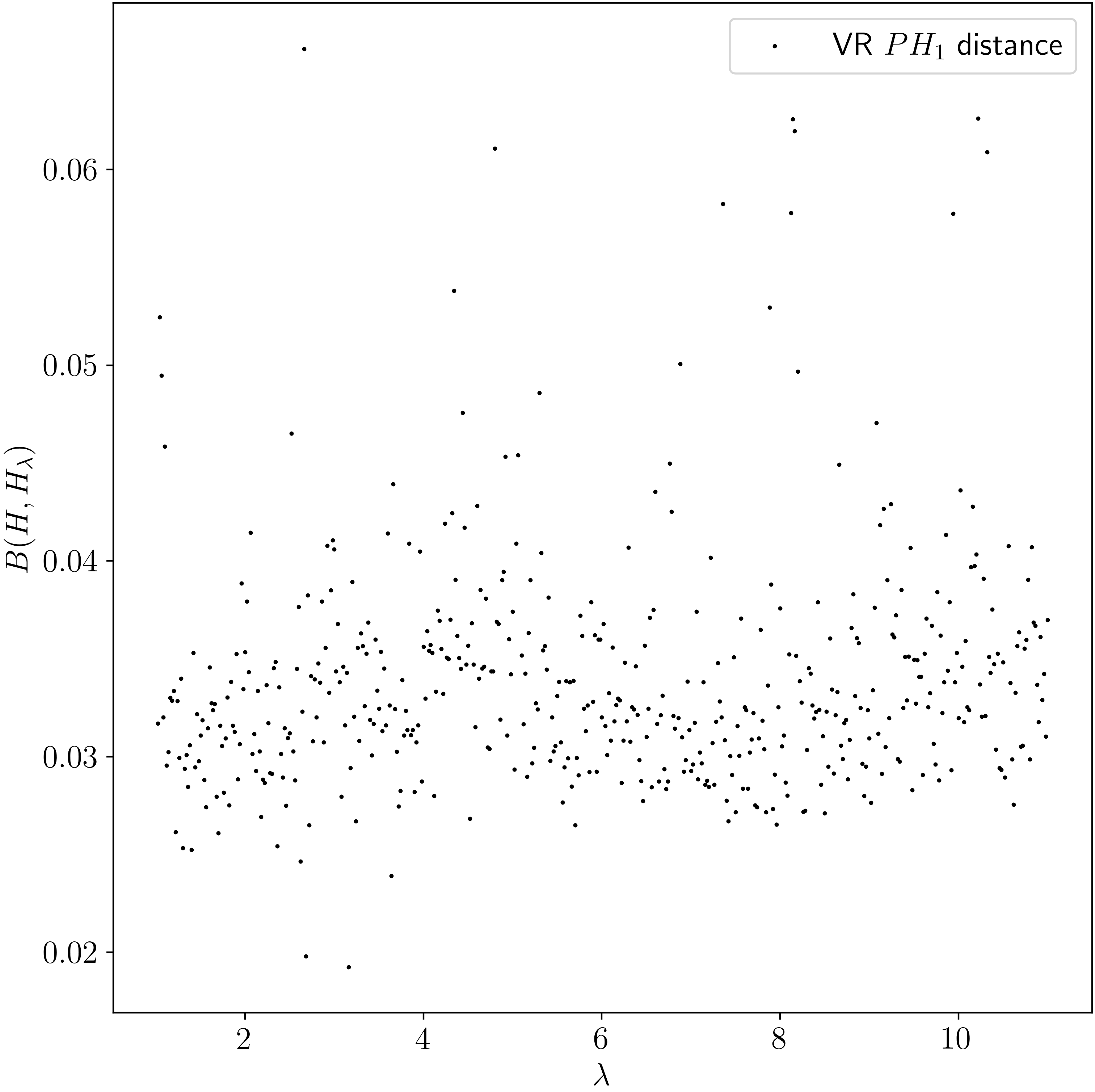}
        \subcaption{Wasserstein-$2$ case.}
        \label{subfig:RipsLinearScanPH1WassersteinBigPointClouds}
    \end{subfigure}
    \caption{Results of a scan over the Wasserstein-$2$ and Bottleneck distances between the VR persistent homology of the base and renormalised Hamiltonians for $10^5$ point orbits.}
    \label{fig:RipsLinearScanBigPointClouds}
\end{figure}

We computed $10^5$ point orbits of a stochastic trajectory in the neighborhood of the hyperbolic fixed point of the perturbed pendulum and then computed the VR persistent homology using Koyama's method. The Bottleneck and Wasserstein distances between the persistence diagrams for the point cloud of each renormalised Hamiltonian and the base Hamiltonian were then calculated. The results of this calculation are presented as Figure \ref{fig:RipsLinearScanBigPointClouds}. There is no clear minimum in either the Bottleneck or the Wasserstein-$2$ distance plots. This suggests that the \textit{HomDistVR} has failed to detect the relatively strong similarity in the phase space of the renormalised Hamiltonian when $\lambda \approx 3.7$ when compared to other values of $\lambda$. To understand this behaviour the Author inspected the point clouds in question and noted that with $10^5$ points it was still very common for some regions of the accessible phase space to be unexplored. This indicates that many more points are needed in each point cloud for the \textit{HomDistVR} procedure to be a practical method for the detection of the renormalisation symmetries of Hamiltonian phase spaces. 

\subsubsection{Testing the averaging procedure}

We now test the alternative procedure in which we compute the averaged distance between many smaller point clouds rather than the distance between single large point clouds. Specifically, suppose that we have $N$ base point clouds $\{X_1,\ldots, X_N\}$ and $M$ comparison point clouds $\{X'_0,\ldots X'_M\}$. Then we define a \textit{mean-minimum} distance\footnote{What we are defining here is not a formal metric mathematically and is intended only as a summary statistic.} between the point cloud families by
\begin{equation}
    d(\{X_1,\ldots, X_N\},\{X'_0,\ldots X'_M\}) = \frac{1}{M}\sum_{j=1}^{M}\min_{i\in \{1,\ldots,N\}}\big(d(X_i,X'_j)\big)\,.
\end{equation}
This is not the only reasonable proposal for such a measure of distance. The Author experimented with several alternatives and found empirically that this mean-minimum distance produced the strongest ``signal'' in the homological distance scan. Determining which distance measure is best would require further analysis. For the proof of concept we are conducting here we just take that our mean-minimum distance is acceptable.

\begin{figure}
    \centering
    \begin{subfigure}[b]{0.49\textwidth}
        \includegraphics[width = \textwidth]{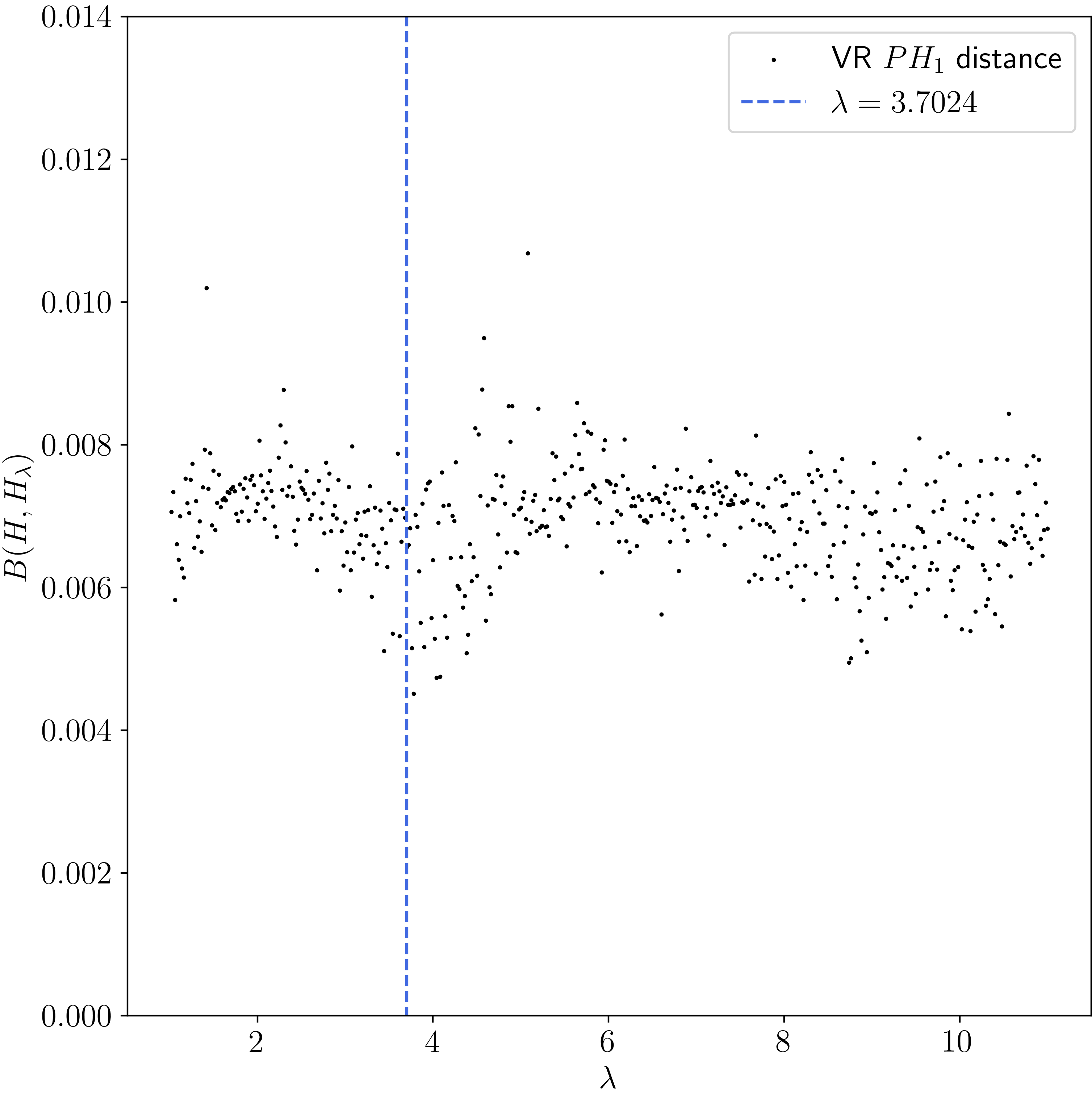}
        \subcaption{Bottleneck case.}
        \label{subfig:RipsLinearScanPH1Bottlneck}
    \end{subfigure}
    \begin{subfigure}[b]{0.49\textwidth}
        \includegraphics[width = \textwidth]{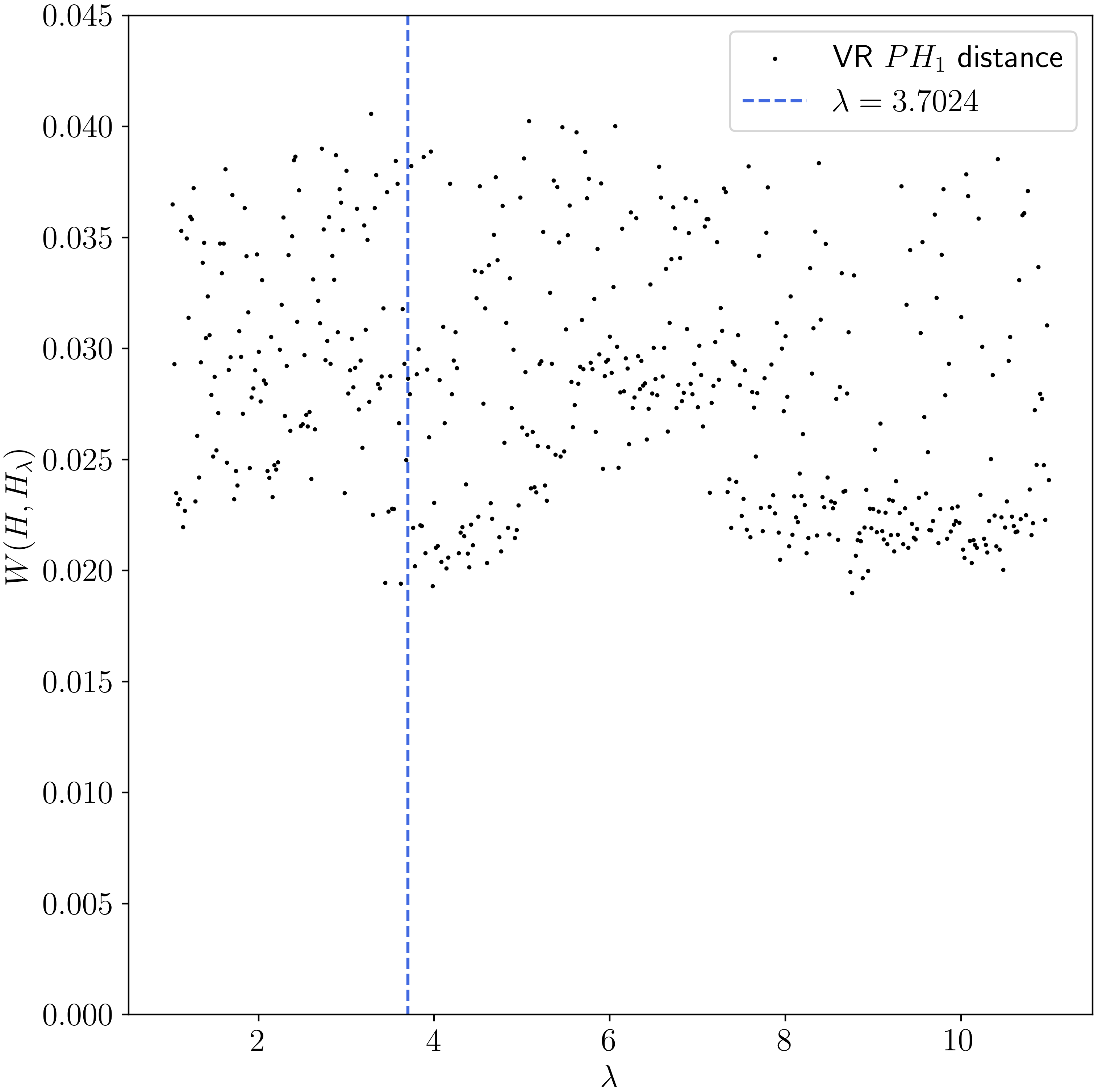}
        \subcaption{Wasserstein-$2$ case.}
        \label{subfig:RipsLinearScanPH1Wasserstein}
    \end{subfigure}
    \caption{Results of a scan over the Wasserstein-$2$ and Bottleneck mean-minimum distances between the VR persistent homology of the unrenormalised and renormalised Hamiltonians.}
    \label{fig:RipsLinearScan}
\end{figure}

So, the VR persistent homology of a set of $\order{10}$ point clouds of $2000$ points each, was computed for each $\lambda$ and the mean-minimum distance between each set with the base set was calculated, where the base set refers to a sample of $N$ trajectories generated using the same base Hamiltonian as above. The results for both the Bottleneck and Wasserstein-$2$ distance case are presented as Figure \ref{fig:RipsLinearScan}. Here we observe a stronger signal than in Figure \ref{fig:RipsLinearScanBigPointClouds} with minor clusters of lower mean-minimum distance around the theoretical value. This suggests that the mean-minimum approach is more sensitive to the renormalisation symmetry we are attempting to demonstrate than that of the large point cloud case. However said sensitivity is very weak and our evidence for the sensitivity is similarly very weak. 

\subsection{Some concluding remarks and a ``middle-ground'' proposal}

We have seen above that both the \textit{HomDistVR} and \textit{HomDistSEDT} approaches to renormalisation symmetry detection present major and somewhat complementary difficulties. For \textit{HomDistVR} there is no a priori method to ensure that the stochastic region is completely covered by our point clouds and so we are likely to miss important topological structure especially when a realistic number of points. This increases the spread of the points in our distance scan and therefore reduces our effective signal to noise ratio leaving only a very weak drop in the homological distance near the theoretical value for $\lambda$. In contrast, the \textit{HomDistSEDT} is guaranteed to cover the entire stochastic region, since it involved scanning over an entire subset of the phase space and identifying all stochastic points in a regular lattice over this subset. However, this procedure will also capture stochastic regions in the phase space which are disconnected from the hyperbolic fixed point and therefore are not associated to the renormalisation of the separatrix map. These disconnected regions dominate in the bottleneck distance scan and again decrease the effective sensitivity of the Wasserstein distance scan reducing our ability to observe a clear minimum and detect a symmetry. Colloquially we can describe our problems as ``the rips approach covers too little'' and ''the SEDT approach covers too much''. So it is reasonable to suggest that it is possible to construct a ``middle-ground'' approach using tools from both methods which will ``cover just enough''. 

Specifically, note that the orbit point clouds used in the Rips approach never leave the stochastic region in which we are interested. So, we can use one of these point clouds to filter a binary image of the phase space, as used in the SEDT approach. Suppose we compute the orbit of a point near to the hyperbolic fixed point for a very long time, $\order{10^6}$ points or more, so that the stochastic region will be as close to fully explored by the orbit as possible. Name said orbit point cloud $X$. We cannot calculate $PH(X)$ using the Vietoris-Rips complex in a reasonable length of time so we propose not attempting to calculate this. Instead, take a binary image of the phase space calculated with our WBA method and call the said image $B$. Each white pixel in $B$ indicates that a particular square region of phase space contains chaotic points. We just need to eliminate chaotic points which are not connected to the hyperbolic fixed point and suggest doing this by comparison with $X$. Specifically, for each white pixel in $p \in B$, pull its associated phase space region $U_p\subset \mathcal{M}$ and check if any of the points in the orbit lie in this region. If $U_p\cap X = \phi$ then we can be confident that the stochastic region indicated by $p$ is not in the component connected to the identity and so can be removed, by setting this pixel to black. This will create a new image $B'$ whose pixels are only white in the stochastic neighborhood of the fixed point. We can then apply the rest of the \textit{HomDistSEDT} procedure to the $B'$ image as before. However, at the time of writing this is only a concept and we have no evidence regarding its efficacy. 

\chapter{Conclusion} 

\label{Chapter6} 

Computational topology, and the related field TDA, were partially developed out of a need to produce new computational and mathematical tools which could provide insight into the geometry of chaotic dynamical systems. It is in this role that they are applicable to research in fusion science and general plasma dynamics. Specifically, as a tool for the characterisation of the geometric structures which affect chaotic motions in nonlinear plasma physics systems. In this thesis we have provided some evidence that these geometric structures can be captured by existing computational topology tools.

We now briefly review the major novel results presented in this thesis and suggest some possible directions for future research.

\section{Overview of results}

In Chapter \ref{Chapter3} we demonstrated that by calculating Vietoris-Rips persistent homology of the orbits of magnetic field lines under a Poincare map it is possible to automatically classify the field lines into different orbit classes. We demonstrated the use of this classification to both map the classes of field line orbits in a toy model of a tokamak and also to determine how the class of a specific field line changes as the parameters of the tokamak are varied.

In Chapter \ref{Chapter4} we developed a method for estimating the distribution of the size of islands in a Hamiltonian phase space using the sub-level set persistent homology to compute the size of connected regions in a binary image of the phase space. This size distribution affects the statistics of particle transport around the phase space and we demonstrated this in detail by considering the specific case of an accelerator mode island in the standard map. Computing the size distribution for a real model may provide insight into the transport statistics of said model.

Finally, in Chapter \ref{Chapter5} we proposed the use of TDA to detect renormalisation group transformations which leave the phase space topology invariant. We experimented with two different methods, one using VR persistent homology and another the sub-level set persistent homology, to detect such a symmetry both of which were only partially successful. Further research into this possible application of TDA to Hamiltonian chaos is needed to determine whether our inability to detect a clear symmetry is due to a fundamental inability for TDA to detect these renormalisation symmetries or if it is our current methods which are insufficient. 


\section{Some suggestions for further work}

There exists possible avenues for further research into the applications of computational topology to plasma physics discussed in this thesis. 

Our classification procedures presented in Chapter \ref{Chapter3} largely rely only on the $PH_1$ information of the orbits. In Section \ref{StochLayersWithIslands} we noted that we could incorporate some $PH_0$ information to automatically separate KAM toruses from stochastic layers. It may be possible to incorporate more information from $PH_0$ which is of value because the nearest neighbor distances between the points are contained in $PH_0$. However, more detailed analysis is needed to determine if this is feasible. Other possible extensions of our orbit classification research include: applying our classification procedure to the field configuration of a perturbed tokamak experiment rather only toy data, and also investigating adopting alternatives to the VR complex such as alpha complexes for the same calculation \cite{EDELSBRUNNER}.

We specifically investigated applying the island size distribution procedure to the standard map and therefore only calculated it in two dimensions. However, the SEDT and sub-level set persistent homology approach to void size extraction can be performed in higher dimensions. Its early historical applications were actually for the three dimensional case \cite{robins2016percolating}. Also, the WBA method we used to image the phase space is applicable in higher dimensions as well \cite{MEISS2021133048}. It would interesting to compute the island size distribution for a volume-preserving map instead of the area-preserving case investigated here.


\appendix 



\chapter{Exact expressions for the fields of simple current distributions} 

\label{AppendixA} 

In Chapter \ref{Chapter3} of this thesis we compute the orbits of field lines for a toy model of a tokamak magnetic field which is constructed from a circular current loop and an infinite line current. This model is chosen because there exists analytic expressions for the fields from these current configurations. In this appendix we list these expressions. 


\section{Circular current loop}

The magnetic field due to a current loop $C$ carrying current $I$ can be computed directly from the Biot-Savart law
\begin{equation}\label{Biot-Savart}
    \textbf{B}(\textbf{x}) = \frac{\mu_0}{4\pi}\int_C\frac{I d\textbf{l}'\cross(\textbf{x}-\textbf{x}')}{||\textbf{x}-\textbf{x}'||^3}\,,
\end{equation}
where $\mu_0$ refers to the magnetic permeability of free space. For the case of a circular loop centered on the $z$-axis in the $x-y$ plane with radius $r$ this integral can be taken exactly. Note that the field from a general circular loop can be determined from this simplified geometry by changing the frame of the observer \cite{Griffiths}. 

In cylindrical coordinates $(s,\theta,z)$ the $\hat{s}$ and $\hat{z}$ components of the field are given by
\begin{align}
    &B_s(s,z) = \frac{\mu_0 zI}{(2s)^{3/2}\pi \sqrt{r(A+1)}}\Bigg(\frac{A}{A-1}E\left(\frac{2}{A+1}\right)-K\left(\frac{2}{A+1}\right)\Bigg) \,,\\
    &B_z(s,z) = \frac{\mu_0 I}{(2s)^{3/2}\pi \sqrt{r(A+1)}}\Bigg(\frac{r-sA}{A-1}E\left(\frac{2}{A+1}\right)+sK\left(\frac{2}{A+1}\right)\Bigg) \,,
\end{align}
where $A$ is given by
\begin{equation}
    A = \frac{s^2+r^2+z^2}{2rs}\,,
\end{equation}
and $K$ and $E$ refer to the complete elliptic integrals of the first and second kinds respectively. Note that the exact definition of these functions varies between authors but we adopt the following definitions
\begin{align*}
    K(m) = \int_0^{2\pi}\frac{d\theta}{\sqrt{1-m\sin^2\theta}}\,,\\
    E(m) = \int_0^{2\pi}d\theta\sqrt{1-m\sin^2\theta}\,.
\end{align*}
Note that the $B_\theta$ component is zero everywhere for this field configuration.

\section{Straight line current}

We also need the field of a straight line current which can be computed either directly from the Biot-Savart or Ampere laws but can also be found in any text on electromagnetism \cite{Griffiths,Jackson}. For the case of a current $I$ along the $z$-axis the field can be written in cylindrical coordinates as 
\begin{equation}
    \textbf{B}(\textbf{x}) = \frac{\mu_0 I}{2\pi s}\hat{\theta}\,.
\end{equation}
Again, by shifting the frame of the observer the field due to any infinite line current can be determined directly.


\printbibliography[heading=bibintoc]

@book{Nakahara,
    author = {Nakahara, M.},
    title = {Geometry, Topology, and Phyics},
    publisher = {Taylor and Francis},
    year ={2003},
}

@book{Hatcher,
    author = {Allen Hatcher},
    title = {Algebraic Topology},
    year = {2000},
    publisher = {Cambridge University Press},
}

@book{Jackson,
    author={Jackson,John D.},
    year={1962},
    title={Classical Electrodynamics},
    publisher={J. Wiley},
    address={New York},
    edition = {Second},
}

@book{Griffiths,
    author={Griffiths,David J.},
    year={1999},
    title={Introduction to electrodynamics},
    publisher={Cambridge University Press},
    edition={3rd},
}

@article{Morrison2000,
    author = {Morrison, Philip},
    year = {2000},
    month = {06},
    pages = {2279-2289},
    title = {Magnetic field lines, Hamiltonian dynamics, and nontwist systems},
    volume = {7},
    journal = {Physics of Plasmas},
    doi = {10.1063/1.874062}
}

@article{KRAMAR201682,
    title = {Analysis of Kolmogorov flow and Rayleigh–Bénard convection using persistent homology},
    journal = {Physica D: Nonlinear Phenomena},
    volume = {334},
    pages = {82-98},
    year = {2016},
    note = {Topology in Dynamics, Differential Equations, and Data},
    issn = {0167-2789},
    doi = {https://doi.org/10.1016/j.physd.2016.02.003},
    url = {https://www.sciencedirect.com/science/article/pii/S0167278916000270},
    author = {Miroslav Kramár and Rachel Levanger and Jeffrey Tithof and Balachandra Suri and Mu Xu and     Mark Paul and Michael F. Schatz and Konstantin Mischaikow},
}

@article{MEISS2021133048,
    title = {Birkhoff averages and the breakdown of invariant tori in volume-preserving maps},
    journal = {Physica D: Nonlinear Phenomena},
    volume = {428},
    pages = {133048},
    year = {2021},
    issn = {0167-2789},
    doi = {https://doi.org/10.1016/j.physd.2021.133048},
    url = {https://www.sciencedirect.com/science/article/pii/S0167278921002050},
    author = {J.D. Meiss and E. Sander},
}

@article{SandersAndMeiss,
    title = {Birkhoff averages and rotational invariant circles for area-preserving maps},
    journal = {Physica D: Nonlinear Phenomena},
    volume = {411},
    pages = {132569},
    year = {2020},
    issn = {0167-2789},
    doi = {https://doi.org/10.1016/j.physd.2020.132569},
    url = {https://www.sciencedirect.com/science/article/pii/S016727891930836X},
    author = {E. Sander and J.D. Meiss},
}

@book{EDELSBRUNNER,
    author = {Edelsbrunner, Herbert and Harer, John},
    year = {2010},
    month = {01},
    pages = {},
    title = {Computational Topology: An Introduction},
    isbn = {978-0-8218-4925-5},
    doi = {10.1007/978-3-540-33259-6_7}
}

@book{GHRIST,
    author = {Ghrist, Robert},
    year = {2014},
    month = {07},
    pages = {},
    title = {Elementary Applied Topology},
    edition = {1.0},
    isbn = {978-1502880857},
    publisher = {Createspace},
}

@ARTICLE{DIAMORSE1,
    author={Delgado-Friedrichs, Olaf and Robins, Vanessa and Sheppard, Adrian},
    journal={IEEE Transactions on Pattern Analysis and Machine Intelligence}, 
    title={Skeletonization and Partitioning of Digital Images Using Discrete Morse Theory}, 
    year={2015},
    volume={37},
    number={3},
    pages={654-666},
    doi={10.1109/TPAMI.2014.2346172}
}

@ARTICLE{DIAMORSE,
    author={Robins, Vanessa and Wood, Peter John and Sheppard, Adrian P.},
    journal={IEEE Transactions on Pattern Analysis and Machine Intelligence}, 
    title={Theory and Algorithms for Constructing Discrete Morse Complexes from Grayscale Digital   Images}, 
    year={2011},
    volume={33},
    number={8},
    pages={1646-1658},
    doi={10.1109/TPAMI.2011.95}
}

@article{ChaoticFields,
  title = {Ubiquity of chaotic magnetic-field lines generated by three-dimensionally crossed wires in modern electric circuits},
  author = {Hosoda, M. and Miyaguchi, T. and Imagawa, K. and Nakamura, K.},
  journal = {Phys. Rev. E},
  volume = {80},
  issue = {6},
  pages = {067202},
  numpages = {4},
  year = {2009},
  month = {12},
  publisher = {American Physical Society},
  doi = {10.1103/PhysRevE.80.067202},
  url = {https://link.aps.org/doi/10.1103/PhysRevE.80.067202}
}

@book{Zaslavsky,
author = {Zaslavsky, George},
year = {2007},
month = {05},
pages = {1-315},
title = {The physics of Chaos in Hamiltonian systems, second edition},
isbn = {978-1-86094-795-7},
doi = {10.1142/P507}
}

@inbook{WhiteRaxWu,
  author    = "White, R. and Rax, J. and Wu, Y.",
  title     = "Transport, Chaos, and Plasma Physics",
  chapter   = "Transport near stochastic threshold",
  publisher = "World Scientific Publishing",
  year      = "1993",
  month     = "07",
  pages     = "153-164",
}

@InProceedings{KassibrakisEtAl,
    author="Kassibrakis, S.
    and Benkadda, S.
    and White, R. B.
    and Zaslavsky, G. M.",
    editor="Benkadda, Sadruddin
    and Zaslavsky, George M.",
    title="Nonuniversality of transport for the standard map",
    booktitle="Chaos, Kinetics and Nonlinear Dynamics in Fluids and Plasmas",
    year="1998",
    publisher="Springer Berlin Heidelberg",
    address="Berlin, Heidelberg",
    pages="403--438",
    isbn="978-3-540-69180-8"
}

@book{huang2009introduction,
  title={Introduction to statistical physics},
  author={Huang, Kerson},
  year={2009},
  publisher={CRC press}
}

@article{bobrowski2023universal,
  title={A universal null-distribution for topological data analysis},
  author={Bobrowski, Omer and Skraba, Primoz},
  journal={Scientific Reports},
  volume={13},
  number={1},
  pages={12274},
  year={2023},
  publisher={Nature Publishing Group UK London}
}

@inproceedings{ye1988signed,
  title={The signed Euclidean distance transform and its applications},
  author={Ye, Q-Z},
  booktitle={9th International conference on pattern recognition},
  pages={495--496},
  year={1988},
  organization={IEEE Computer Society}
}

@book{guckenheimer2013nonlinear,
  title={Nonlinear oscillations, dynamical systems, and bifurcations of vector fields},
  author={Guckenheimer, John and Holmes, Philip},
  volume={42},
  year={2013},
  publisher={Springer Science \& Business Media}
}

@book{arnol2013mathematical,
  title={Mathematical methods of classical mechanics},
  author={Arnol'd, Vladimir Igorevich},
  volume={60},
  year={2013},
  publisher={Springer Science \& Business Media}
}

@book{natiello2007user,
  title={The user's approach to topological methods in 3d dynamical systems},
  author={Natiello, Mario A and others},
  year={2007},
  publisher={World Scientific}
}

@book{hazeltine2003plasma,
  title={Plasma confinement},
  author={Hazeltine, Richard D and Meiss, James D},
  year={2003},
  publisher={Courier Corporation}
}

@incollection{mackay2020survey,
  title={Survey of Hamiltonian dynamics},
  author={MacKay, RS and Meiss, JD},
  booktitle={Hamiltonian Dynamical Systems},
  pages={3--19},
  year={2020},
  publisher={CRC Press}
}

@article{greene1979method,
  title={A method for determining a stochastic transition},
  author={Greene, John M},
  journal={Journal of Mathematical Physics},
  volume={20},
  number={6},
  pages={1183--1201},
  year={1979},
  publisher={American Institute of Physics}
}

@article{arnold2009small,
  title={Small denominators and problems of stability of motion in classical and celestial mechanics},
  author={Arnold, Vladimir I},
  journal={Collected Works: Representations of Functions, Celestial Mechanics and KAM Theory, 1957--1965},
  pages={306--412},
  year={2009},
  publisher={Springer}
}

@article{nunez2022topological,
  title={Topological data analysis of Lagrangian orbits in natural convection flows confined in a cylinder},
  author={N{\'u}{\~n}ez, Jos{\'e} and Gonz{\'a}lez, Ahtziri and Ramos, Eduardo},
  journal={Physical Review Fluids},
  volume={7},
  number={12},
  pages={123501},
  year={2022},
  publisher={APS}
}

@inproceedings{banesh2020topological,
  title={Topological analysis of magnetic reconnection in kinetic plasma simulations},
  author={Banesh, Divya and Lo, Li-Ta and Kilian, Patrick and Guo, Fan and Hamann, Bernd},
  booktitle={2020 IEEE Visualization Conference (VIS)},
  pages={6--10},
  year={2020},
  organization={IEEE}
}

@article{tempelman2020look,
  title={A look into chaos detection through topological data analysis},
  author={Tempelman, Joshua R and Khasawneh, Firas A},
  journal={Physica D: Nonlinear Phenomena},
  volume={406},
  pages={132446},
  year={2020},
  publisher={Elsevier}
}

@article{tymochko2020using,
  title={Using zigzag persistent homology to detect Hopf bifurcations in dynamical systems},
  author={Tymochko, Sarah and Munch, Elizabeth and Khasawneh, Firas A},
  journal={Algorithms},
  volume={13},
  number={11},
  pages={278},
  year={2020},
  publisher={MDPI}
}

@book{szeliski2022computer,
  title={Computer vision: algorithms and applications},
  author={Szeliski, Richard},
  year={2022},
  publisher={Springer Nature}
}

@article{zaslavsky1994renormalization,
  title={Renormalization group theory of anomalous transport in systems with Hamiltonian chaos},
  author={Zaslavsky, GM},
  journal={Chaos: An Interdisciplinary Journal of Nonlinear Science},
  volume={4},
  number={1},
  pages={25--33},
  year={1994},
  publisher={American Institute of Physics}
}

@article{mackay1984transport,
  title={Transport in Hamiltonian systems},
  author={MacKay, RS and Meiss, JD and Percival, IC},
  journal={Physica D: Nonlinear Phenomena},
  volume={13},
  number={1-2},
  pages={55--81},
  year={1984},
  publisher={Elsevier}
}

@book{wiggins2013chaotic,
  title={Chaotic transport in dynamical systems},
  author={Wiggins, Stephen},
  volume={2},
  year={2013},
  publisher={Springer Science \& Business Media}
}

@article{koyama2023reduced,
  title={Reduced Vietoris-Rips Complexes: New methods to compute Vietoris-Rips Persistent Homology},
  author={Koyama, Musashi Ayrton and Robins, Vanessa and Turner, Katharine and Memoli, Facundo},
  journal={arXiv preprint arXiv:2307.16333},
  year={2023}
}

@article{bermingham2023planar,
  title={Planar Symmetry Detection and Quantification using the Extended Persistent Homology Transform},
  author={Bermingham, Nicholas and Robins, Vanessa and Turner, Katharine},
  journal={arXiv:2307.08281},
  year={2023}
}

@book{miyamoto1997fundamentals,
  title={Fundamentals of plasma physics and controlled fusion},
  author={Miyamoto, Kenro},
  year={1997},
  publisher={Iwanami Book Service Center Tokyo}
}

@book{kaczynski2004computational,
  title={Computational homology},
  author={Kaczynski, Tomasz and Mischaikow, Konstantin Michael and Mrozek, Marian},
  volume={3},
  number={7},
  year={2004},
  publisher={Springer}
}

@book{siegel2012lectures,
  title={Lectures on celestial mechanics},
  author={Siegel, Carl L and Moser, J{\"u}rgen K},
  year={2012},
  chapter = {1},
  pages = {151-155},
  publisher={Springer Science \& Business Media}
}

@book{ott2002chaos,
  title={Chaos in dynamical systems},
  author={Ott, Edward},
  year={2002},
  chapter ={7},
  pages ={216-219},
  publisher={Cambridge university press}
}

@article{CHIRIKOV1979263,
title = {A universal instability of many-dimensional oscillator systems},
journal = {Physics Reports},
volume = {52},
number = {5},
pages = {263-379},
year = {1979},
issn = {0370-1573},
doi = {https://doi.org/10.1016/0370-1573(79)90023-1},
url = {https://www.sciencedirect.com/science/article/pii/0370157379900231},
author = {Boris V Chirikov}
}

@article{del1996area,
  title={Area preserving nontwist maps: periodic orbits and transition to chaos},
  author={del-Castillo-Negrete, D and Greene, JM and Morrison, PJ},
  journal={Physica D: Nonlinear Phenomena},
  volume={91},
  number={1-2},
  pages={1--23},
  year={1996},
  publisher={Elsevier}
}

@article{Das_2018,
doi = {10.1088/1361-6544/aa99a0},
url = {https://dx.doi.org/10.1088/1361-6544/aa99a0},
year = {2018},
month = {01},
publisher = {IOP Publishing},
volume = {31},
number = {2},
pages = {491},
author = {Suddhasattwa Das and James A Yorke},
title = {Super convergence of ergodic averages for quasiperiodic orbits},
journal = {Nonlinearity}
}

@misc{EUROfusion_2023, url={https://euro-fusion.org/}, journal={EUROfusion}, year={2023}, month={04}}

@article{deshmukh2021toward,
  title={Toward automated extraction and characterization of scaling regions in dynamical systems},
  author={Deshmukh, Varad and Bradley, Elizabeth and Garland, Joshua and Meiss, James D},
  journal={Chaos: An Interdisciplinary Journal of Nonlinear Science},
  volume={31},
  number={12},
  year={2021},
  publisher={AIP Publishing}
}

@article{garland2016exploring,
  title={Exploring the topology of dynamical reconstructions},
  author={Garland, Joshua and Bradley, Elizabeth and Meiss, James D},
  journal={Physica D: Nonlinear Phenomena},
  volume={334},
  pages={49--59},
  year={2016},
  publisher={Elsevier}
}

@phdthesis{robins2000computational,
  title={Computational topology at multiple resolutions: foundations and applications to fractals and dynamics},
  author={Robins, Vanessa},
  year={2000},
  school={University of Colorado at Boulder}
}

@article{mackay1983renormalization,
  title={A renormalization approach to invariant circles in area-preserving maps},
  author={MacKay, Robert S},
  journal={Physica D: Nonlinear Phenomena},
  volume={7},
  number={1-3},
  pages={283--300},
  year={1983},
  publisher={Elsevier}
}

@article{PerellaExistence,
title = "Existence of global symmetries of divergence-free fields with first integrals",
author = "David Perrella and Nathan Duignan and David Pfefferl{\'e}",
year = "2023",
month = may,
day = "24",
doi = "10.1063/5.0152213",
language = "English",
volume = "64",
journal = "Journal of Mathematical Physics",
issn = "0022-2488",
publisher = "American Institute of Physics",
number = "5",

}

@article{muldoon1993topology,
  title={Topology from time series},
  author={Muldoon, MR and MacKay, RS and Huke, JP and Broomhead, DS},
  journal={Physica D: Nonlinear Phenomena},
  volume={65},
  number={1-2},
  pages={1--16},
  year={1993},
  publisher={Elsevier}
}

@article{joffrin2003internal,
  title={Internal transport barrier triggering by rational magnetic flux surfaces in tokamaks},
  author={Joffrin, E and Challis, CD and Conway, GD and Garbet, X and Gude, A and G{\"u}nter, S and Hawkes, NC and Hender, TC and Howell, DF and Huysmans, GTA and others},
  journal={Nuclear fusion},
  volume={43},
  number={10},
  pages={1167},
  year={2003},
  publisher={IOP Publishing}
}

@article{robins2016percolating,
  title={Percolating length scales from topological persistence analysis of micro-CT images of porous materials},
  author={Robins, Vanessa and Saadatfar, Mohammad and Delgado-Friedrichs, Olaf and Sheppard, Adrian P},
  journal={Water Resources Research},
  volume={52},
  number={1},
  pages={315--329},
  year={2016},
  publisher={Wiley Online Library}
}

@article{DBLP:journals/tvcg/ThomasN11,
  author       = {Dilip Mathew Thomas and
                  Vijay Natarajan},
  title        = {Symmetry in Scalar Field Topology},
  journal      = {{IEEE} Trans. Vis. Comput. Graph.},
  volume       = {17},
  number       = {12},
  pages        = {2035--2044},
  year         = {2011},
  url          = {https://doi.org/10.1109/TVCG.2011.236},
  doi          = {10.1109/TVCG.2011.236},
  bibsource    = {dblp computer science bibliography, https://dblp.org}
}


\end{document}